\documentclass[prd,aps,showpacs,epsf,floats,onecolumn]{revtex4}%
\usepackage{amssymb}
\usepackage{amsfonts}
\usepackage{amsmath}
\usepackage{graphicx}%
\providecommand{\U}[1]{\protect\rule{.1in}{.1in}}

\begin{document}
\title{\textbf{The Information Geometry of Chaos}}
\author{Carlo Cafaro}
\affiliation{Department of Physics, State University of New York at Albany-SUNY, 1400
Washington Avenue, Albany, NY 12222, USA}

\begin{abstract}
In this Thesis, we propose a new theoretical information-geometric framework
(IGAC, Information Geometrodynamical Approach to Chaos) suitable to
characterize chaotic dynamical behavior of arbitrary complex systems. First,
the problem being investigated is defined; its motivation and relevance are
discussed. The basic tools of information physics and the relevant
mathematical tools employed in this work are introduced. The basic aspects of
Entropic Dynamics (ED) are reviewed. ED is an information-constrained dynamics
developed by Ariel Caticha to investigate the possibility that laws of physics
- either classical or quantum - may emerge as macroscopic manifestations of
underlying microscopic statistical structures. ED is of primary importance in
our IGAC. The notion of chaos in classical and quantum physics is introduced.
Special focus is devoted to the conventional Riemannian geometrodynamical
approach to chaos (Jacobi geometrodynamics) and to the Zurek-Paz quantum chaos
criterion of linear entropy growth. After presenting this background material,
we show that the ED formalism is not purely an abstract mathematical
framework, but is indeed a general theoretical scheme from which conventional
Newtonian dynamics is obtained as a special limiting case. The major elements
of our IGAC and the novel notion of information geometrodynamical entropy
(IGE) are introduced by studying two toy models. To illustrate the potential
power of our IGAC, one application is presented. An information-geometric
analogue of the Zurek-Paz quantum chaos criterion of linear entropy growth is
suggested. Finally, concluding remarks emphasizing strengths and weak points
of our approach are presented and possible further research directions are
addressed. At this stage of its development, IGAC remains an ambitious
unifying information-geometric theoretical construct for the study of chaotic
dynamics with several unsolved problems. However, based on our recent
findings, we believe it already provides an interesting, innovative and
potentially powerful way to study and understand the very important and
challenging problems of classical and quantum chaos.

\end{abstract}

\maketitle

\begin{center}
\bigskip\pagebreak

{\LARGE PREFACE}
\end{center}

My research at the Joseph Henry Department of Physics, SUNY at Albany, has
focused mainly on applications of geometric Clifford algebra to
electrodynamics and 3+1 gravity and on applications of information physics to
classical and quantum chaos. Thanks to my research, I became invited reviewer
of several international scientific journals, among the others: AIP Conf.
Proceedings in Mathematical and Statistical Physics, The Open Astronomy
Journal, Journal of Electromagnetic Waves and Applications, Progress in
Electromagnetics Research, Foundations of Physics and Mathematical Reviews (AMS).

During these years I worked under the precious supervision of Prof. Ariel
Caticha. Moreover, I collaborated on several projects with my friends and
colleagues Dr. Saleem Ali (SUNY at Albany), Prof. Salvatore Capozziello
(University of Napoli, Italy), Dr. Christian Corda (University of Pisa,
Italy), and Mr. Adom Giffin (SUNY at Albany). This thesis THE INFORMATION
GEOMETRY OF CHAOS reports exclusively results of some of my research on
applications of information physics to the study of chaos carried out between
January 2004 and May 2008 at the Physics Department in Albany. Sections of my
doctoral dissertation have been presented at International Conferences
(Paris-France, Saratoga-USA, Erice-Italy) and they have appeared in a series
of collaborative and/or single-authored conference proceedings and/or journal articles.

\pagebreak

\begin{center}
{\LARGE OVERVIEW}
\end{center}

In this doctoral dissertation, I consider two important questions: First, are
laws of physics practical rules to process information about the world using
geometrical methods? Are laws of physics rules of inference? Second, since a
unifying framework to describe chaotic dynamics in classical and quantum
domains is missing, is it possible to construct a new information-geometric
model, to develop new tools so that a unifying framework is provided or, at
least, new insights and new understandings are given? After setting the scene
of my thesis and after stating the problem and its motivations, I review the
basic elements of the maximum relative entropy formalism (ME method) and
recall the basics of Riemannian geometry with special focus to its application
to probability theory (this is known as Information Geometry, IG). IG and ME
are the fundamental tools that Prof. Ariel Caticha has used to build a form of
information-constrained dynamics on statistical manifolds to investigate the
possibility that Einstein's general theory of gravity (or any classical or
quantum theory of physics) may emerge as a macroscopic manifestation of an
underlying microscopic statistical structure. This dynamics is known in the
literature as Entropic Dynamics (ED). Therefore, since ED is an important
element of this thesis, I review the key-points of such dynamics, emphasizing
the most relevant points that I will use in my own information
geometrodynamical approach to chaos (IGAC). Of course, before introducing my
IGAC, I briefly review the basics of the conventional Riemannian
geometrodynamical approach to chaos and discuss the notion of chaos in
classical and quantum physics in general. After this background information, I
start with my original contributions.

\textbf{First}, two chaotic entropic dynamical models are considered. The
geometric structure of the statistical manifolds underlying these models is
studied. It is found that in both cases, the resulting metric manifolds are
negatively curved. Moreover, the geodesics on each manifold are described by
hyperbolic trajectories. A detailed analysis based on the Jacobi-Levi-Civita
equation for geodesic spread (JLC equation) is used to show that the
hyperbolicity of the manifolds leads to chaotic exponential instability. A
comparison between the two models leads to a relation among scalar curvature
of the manifold ($\mathcal{R}$), Jacobi field intensity ($J$) and information
geometrodynamical entropy (IGE, $\mathcal{S}$). I introduce the IGE entropy as
a brand new measure of chaoticity. These three quantities, $\mathcal{R}$, $J$,
and $\mathcal{S}$ are suggested as useful indicators of chaoticity. Indeed, in
analogy to the Zurek-Paz quantum chaos criterion of linear entropy growth, a
novel classical information-geometric criterion of linear IGE growth for
chaotic dynamics on curved statistical manifolds is presented.

\textbf{Second}, in collaboration with Prof. Ariel Caticha, I show that the ED
formalism is not purely an abstract mathematical framework; it is indeed a
general theoretical scheme where conventional Newtonian dynamics is obtained
as a special limiting case. Newtonian dynamics is derived from prior
information codified into an appropriate statistical model. The basic
assumption is that there is an irreducible uncertainty in the location of
particles so that the state of a particle is defined by a probability
distribution. The corresponding configuration space is a statistical manifold
the geometry of which is defined by the information metric. The trajectory
follows from a principle of inference, the method of Maximum Entropy. No
additional physical postulates such as an equation of motion, or an action
principle, nor the concepts of momentum and of phase space, not even the
notion of time, need to be postulated. The resulting entropic dynamics
reproduces the Newtonian dynamics of any number of particles interacting among
themselves and with external fields. Both the mass of the particles and their
interactions are explained as a consequence of the underlying statistical manifold.

\textbf{Third}, I extend my study of chaotic systems (information
geometrodynamical approach to chaos, IGAC) to an ED Gaussian model describing
an arbitrary system of $3N$ degrees of freedom. It is shown that the
hyperbolicity of a non-maximally symmetric $6N$-dimensional statistical
manifold $\mathcal{M}$ underlying the ED Gaussian model leads to linear
information-geometrodynamical entropy growth and to exponential divergence of
the Jacobi vector field intensity, quantum and classical features of chaos
respectively. As a special physical application, the information
geometrodynamical scheme is applied to investigate the chaotic properties of a
set of $n$-uncoupled three-dimensional anisotropic inverted harmonic
oscillators (IHOs) characterized by an Ohmic distributed frequency spectrum
and I show that the asymptotic behavior of the information-geometrodynamical
entropy is characterized by linear growth. Finally, the anisotropy of the
statistical manifold underlying such physical system and its relationship with
the spectrum of frequencies of the oscillators are studied.

Finally, I present concluding remarks emphasizing strengths and weak points of
my approach and I address possible further research directions.

\pagebreak

\begin{center}
{\LARGE Chapter 1: Background and Motivation}
\end{center}

I discuss the general setting of my thesis. I define the problem I am going to
study, its motivation and its relevance. Moreover, I describe briefly the
structure of the dissertation and, finally I state my original contributions

\section{Introduction}

In the orthodox scientific community, it is commonly accepted that laws of
physics reflect laws of nature. In this Thesis we uphold a different line of
thinking: laws of physics consist of rules designed for processing information
about the world for the purpose of describing and, to a certain extent,
understanding natural phenomena. The form of the laws of physics should
reflect the form of rules for manipulating information.

This unorthodox point of view has important consequences for the laws of
physics themselves: the identification of the relevant information-constraints
of a particular phenomenon implies that the rules for processing information
can take over, and thus, the laws of physics that govern the specific
phenomenon should be determined. In principle, the field of applicability of
the ideas explored in this Thesis is not limited to physics. This novel line
of thinking may be useful to various scientific disciplines since all subject
matters are investigated by applying the same universal methods and tools of reasoning.

One of our major line of research concerns the possibility that laws of
physics can be derived from rules of inference. In the case of a positive
answer, we expect that physics should use the same tools, the same kind of
language that has been found useful for inference. We are well aware that most
of our basic physics theories make use of the concepts of probability, of
entropy, and of geometry.

We know that the evidence supporting the fact that laws of physics can be
derived from rules of inference is already considerable. Indeed, most of the
formal structure of statistical mechanics can already be derived from
principles of inference \cite{jaynes}. Moreover, the derivation of quantum
mechanics as information physics is already quite developed \cite{catichaQM}.
Finally, it seems plausible that even Einstein's theory of gravity (GR,
general relativity) might be derived from a more fundamental statistical
geometrodynamics in analogy to the way in which thermodynamics can be derived
from a microscopic statistical mechanics \cite{catichaGR}.

It is known that any attempt to build a unified theory of all forces is
problematic: these problems arise from the difficulties of incorporating the
classical theory of gravity with quantum theories of electromagnetic, weak and
strong forces. Few situations, where gravitational and quantum phenomena
coexist in a non-trivial way, can be studied in some detail. It just happens
that in these situations (for instance, Hawking's black hole evaporation
\cite{waldino}) thermodynamics plays an essential role. These considerations
lead to believe that concepts such as \emph{entropy should play a key-role in
any successful theoretical construct attempting to unify all fundamental
interactions in nature}.

The relation between physics and nature is more complicated than has been
usually assumed. An explicit admission of such a statement is represented by
the recognition that statistical physics and quantum physics are theories of
inference. This scientific awareness is leading to a gradual acceptance of the
fact that the task of theoretical physics is that of putting some order in our
understanding of natural phenomena, not to write the ultimate equations of the universe.

In this Thesis, we want to carry on this line of thought and explore the
possibility that laws of physics become rules for the consistent manipulation
of information: laws of physics work not because they are laws of nature but
rather because they are laws of inference and they have been designed to work.
Therefore my first line of investigation will be the following:

\begin{description}
\item[First line of research] \textit{I would like to push the possibility to
extend the notion that laws of physics can be derived from rules of inference
into new unexplored territories. I would like to construct new entropic
dynamical models that reproduce recognizable laws of physics using merely the
tools of inference (probability, information geometry and entropy).}
\end{description}

However, I have to select the laws of physics that I want to reproduce.

It is known there is no unified characterization of chaos in classical and
quantum dynamics. In the Riemannian \cite{casetti} and Finslerian
\cite{cipriani} (a Finsler metric is obtained from a Riemannian metric by
relaxing the requirement that the metric be quadratic on each tangent space)
geometrodynamical approach to chaos in classical Hamiltonian systems, an
active field of research concerns the possibility of finding a rigorous
relation among the sectional curvature, the Lyapunov exponents, and the
Kolmogorov-Sinai dynamical entropy (i. e. the sum of positive Lyapunov
exponents) \cite{kawabe}. The largest Lyapunov exponent characterizes the
degree of chaoticity of a dynamical system and, if positive, it measures the
mean instability rate of nearby trajectories averaged along a sufficiently
long reference trajectory. Moreover, it is known that classical chaotic
systems are distinguished by their exponential sensitivity to initial
conditions and that the absence of this property in quantum systems has lead
to a number of different criteria being proposed for quantum chaos.
Exponential decay of fidelity, hypersensitivity to perturbation, and the
Zurek-Paz quantum chaos criterion of linear von Neumann's entropy growth
\cite{zurek} are some examples \cite{caves}. These criteria accurately predict
chaos in the classical limit, but it is not clear that they behave the same
far from the classical realm.

I chose the second line of research below for the following three reasons: i)
lack of a unifying understanding of chaotic phenomena in classical and quantum
physics; ii) test the potential mathematical power of these
information-geometric tools at my disposal; test the potential predicting
power of entropic dynamical models.

\begin{description}
\item[Second line of research] \textit{I would like to push the possibility to
derive, explain and understand classical and quantum criteria of chaos from
rules of inference. I would like to construct new entropic chaotic dynamical
models that reproduce recognizable laws of mathematical-physics using merely
the tools of inference (probability, information geometry and entropy).}
\end{description}

\section{Problems under investigation, their relevance and original
contributions}

First I review the basic elements of the maximum relative entropy formalism
(ME method) and recall the basics of Riemannian geometry with special focus to
its application to probability theory (this is known as Information Geometry,
IG). IG and ME are the fundamental tools that Prof. Ariel Caticha has used to
build a form of information-constrained dynamics on statistical manifolds to
investigate the possibility that Einstein's general theory of gravity (or any
classical or quantum theory of physics) may emerge as a macroscopic
manifestation of an underlying microscopic statistical structure. This
dynamics is known in the literature as Entropic Dynamics\ (ED). Therefore,
since ED is an important element of this thesis, I review the key-points of
such dynamics, emphasizing the most relevant points that I use in my own
information geometrodynamical approach to chaos (IGAC). Of course, before
introducing my IGAC, I briefly review the basics of the conventional
Riemannian geometrodynamics approach to chaos and discussed the notion of
chaos in physics in general. After this long background information that is
needed because of the originality and novelty of these topics, I begin with my
original contributions.

First, two chaotic entropic dynamical models are considered. The geometric
structure of the statistical manifolds underlying these models is studied. It
is found that in both cases, the resulting metric manifolds are negatively
curved. Moreover, the geodesics on each manifold are described by hyperbolic
trajectories. A detailed analysis based on the Jacobi-Levi-Civita equation for
geodesic spread (JLC equation) is used to show that the hyperbolicity of the
manifolds leads to chaotic exponential instability. A comparison between the
two models leads to a relation among scalar curvature of the manifold
($\mathcal{R}$), Jacobi field intensity ($J$) and information
geometrodynamical entropy (IGE, $\mathcal{S}_{\mathcal{M}}$). The IGE entropy
is proposed as a brand new measure of chaoticity.

\bigskip

\begin{description}
\item[First Contribution] \cite{cafaro1, cafaro2, cafaro3}: \textit{I suggest
that these three quantities, }$\mathcal{R}$\textit{, }$J$\textit{, and
}$\mathcal{S}_{\mathcal{M}}$\textit{\ are useful indicators of chaoticity for
chaotic dynamical systems on curved statistical manifolds. Furthermore, I
suggest a classical \ information-geometric criterion of linear information
geometrodynamical entropy growth in analogy with the Zurek-Paz quantum chaos
criterion.}
\end{description}

\bigskip

Second, in collaboration with Prof. Ariel Caticha, I show that the ED
formalism is not purely an abstract mathematical framework; it is indeed a
general theoretical scheme where conventional Newtonian dynamics is obtained
as a special limiting case.

\bigskip

\begin{description}
\item[Second Contribution] \cite{caticha-cafaro}: \textit{The reproduction of
the Newtonian dynamics from first principles of probable inference and
information geometric methods is another original contribution of my work}.
\end{description}

\bigskip

Third, I extend my study of chaotic systems (information geometrodynamical
approach to chaos, IGAC) to an ED Gaussian model describing an arbitrary
system of $3N$ degrees of freedom. It is shown that the hyperbolicity of a
non-maximally symmetric $6N$-dimensional statistical manifold $\mathcal{M}%
_{\mathcal{S}}$ underlying the ED Gaussian model leads to linear
information-geometrodynamical entropy growth and to exponential divergence of
the Jacobi vector field intensity, quantum and classical features of chaos
respectively. As a special physical application, the information
geometrodynamical scheme is applied to investigate the chaotic properties of a
set of $n$-uncoupled three-dimensional anisotropic inverted harmonic
oscillators (IHOs) characterized by an Ohmic distributed frequency spectrum
and I show that the asymptotic behavior of the information-geometrodynamical
entropy is characterized by linear growth. Finally the anisotropy of the
statistical manifold underlying such physical system and its relationship with
the spectrum of frequencies of the oscillators are studied.

\bigskip

\begin{description}
\item[Third Contribution] \cite{cafaro4} \textit{I compute the asymptotic
temporal behavior of the information geometrodynamical entropy of a set of
}$n$\textit{-uncoupled three-dimensional anisotropic inverted harmonic
oscillators (IHOs) characterized by an Ohmic distributed frequency spectrum
and I suggest the classical information-geometric analogue of the Zurek-Paz
quantum chaos criterion in its classical reversible limit}.
\end{description}

\bigskip

I am aware that several points in my IGAC need deeper understanding and
analysis, however I hope that my work convincingly shows that:

\begin{description}
\item[Point 1] Laws of physics are deeply geometrical because they are
practical rules to process information about the world and geometry is the
most natural tool to carry out that task. The notion that laws of physics are
not laws of nature but rules of inference seems outrageous but cannot be
simply dismissed. Indeed, it deserves serious attention and further research.

\item[Point 2] This is a novel and unorthodox research area and there are many
risks and criticisms \cite{cafaro5}. I believe the information
geometrodynamical approach to chaos may be useful in providing a unifying
criterion of chaos of both classical and quantum varieties, thus deserving
further research and developments.
\end{description}

\bigskip\pagebreak

\begin{center}
{\LARGE Chapter 2: Maximum entropy methods and information geometry}
\end{center}

I review the basic information physics and mathematical tools employed in my
work. First, I review the major aspects of the Maximum Entropy method (ME
method), a unique general theory of inductive inference. Second, I recall some
basic elements of conventional Riemannian differential geometry useful for the
understanding of standard approaches to the geometrical study of chaos.
Finally, I introduce the basics of Riemannian geometry applied to probability
theory, namely Information Geometry (IG). IG and ME are the basic tools that I
need to introduce in order to allow the reader to follow the description of
the constrained information dynamics on curved statistical manifolds (entropic
dynamics, ED).

\section{Introduction}

Inference is the process of drawing conclusions from available information.
Information is whatever constraints rational beliefs \cite{caticha(aip07)}.

When the information available is sufficient to make unequivocal, unique
assessments of truth we speak of making deductions: on the basis of this or
that information we deduce that a certain proposition is true. The method of
reasoning leading to deductive inferences is called logic. Situations where
the available information is insufficient to reach such certainty lie outside
the realm of logic. In these cases we speak of making a probable inference,
and the method of reasoning is probability theory. An alternative name is
"\textit{inductive inference}". The word "induction" refers to the process of
using limited information about a few special cases to draw conclusions about
more general situations.

The main goal of inductive inference is to update from a prior probability
distribution to a posterior distribution when new relevant information becomes
available. Updating methods should be systematic and objective. The most
important updating methods are the \textit{Bayesian updating method}
\cite{sivia} and the ME method \cite{caticha(aip07), shore, skilling,
caticha(aip04), caticha(book), caticha-giffin(aip06)}.

The \textit{ME method} is a generalization of Jaynes' method of maximum
entropy, \textit{MaxEnt method }\cite{jaynes}. Jaynes' method of maximum
entropy is a method to assign probabilities on the basis of partial testable
information. Testable information is sufficient to make a prediction and
predictions can be tested. MaxEnt arises as a rule to assign a probability
distribution, however it can be extended to a full-fledged method for
inductive inference. The extended method will henceforth be abbreviated as ME.
In Jaynes's MaxEnt method, it is shown that statistical mechanics and thus
thermodynamics are theories of inference. MaxEnt can be interpreted as a
special case of ME when one updates from a uniform prior using the
Gibbs-Shannon entropy.

The nature of the information being processed dictates the choice between the
Bayesian updating method and the ME method. Bayes' theorem \cite{jaynes}
should be used to update our beliefs about the values of certain quantities
$\theta$ on the basis of information about the observed values of other
quantities $x$- the data- and of the known relation between them- the
conditional distribution $p\left(  x|\theta\right)  $. If $p\left(
\theta\right)  $ are the prior beliefs, the updated or posterior distribution
is given by $p\left(  \theta|x\right)  \propto p\left(  \theta\right)
p\left(  x|\theta\right)  $. The Bayesian method of updating is a consequence
of the product rule for probabilities and therefore it is limited to
situations where it makes sense to define the joint probability of $x$ and
$\theta$, $p\left(  x\text{, }\theta\right)  $. On the other hand, the ME
method is designed for updating from a prior to a posterior probability
distribution when the information to be processed takes the form of
constraints on the family of acceptable posterior distributions. Although the
terms "prior" and "posterior" are normally used only in the context of Bayes'
theorem, we will adopt the same terminology when using the ME method since we
are concerned with the same goal of processing information to update from a
prior to a posterior probability distribution. As a final remark, we point out
that in general it is meaningless to use Bayes' theorem to process information
in the form of constraints, and conversely, it is meaningless to process data
using ME. However, there are special cases where the same piece of information
can be both interpreted as data and as constraint, In such cases, both methods
can be used and it can be shown that they agree.

\section{What is the Maximum Relative Entropy Formalism?}

Consider a multidimensional discrete or continuous variable $x$. Assume the
prior probability distribution $q\left(  x\right)  $ describes the uncertainty
about $x$. When new relevant information becomes available, our goal is to
update from $q\left(  x\right)  $ to a posterior probability distribution
$P(x) $. Information appears in the form of constraints and usually could be
given in terms of expected values. The problem consists in selecting the
proper $p(x)$ among all those posterior probability distributions within the
family defined by the available relevant constraints.

Skilling \cite{skilling} suggested that in order to select the posterior
$p(x)$ it seems reasonable to rank the candidate distributions in order of
increasing preference. The ranking must be transitive: if distribution $p_{1}
$ is preferred over distribution $p_{2}$, and $p_{2}$ is preferred over
$p_{3}$, then $p_{1}$ is preferred over $p_{3}$. To each $p(x)$ is assigned a
real number $S[p]$, which we will henceforth call entropy, in such a way that
if $p_{1}$ is preferred over $p_{2}$, then $S[p_{1}]>S[p_{2}] $. The
probability distribution that maximizes the entropy $S[p]$ will be the
selected posterior distribution $P\left(  x\right)  $. Therefore, it becomes
evident to conclude that the Maximum Entropy method (ME) is a variational
method involving entropies which are real numbers.

Moreover, to define the ranking scheme, a functional form of $S[p]$ must be
chosen. Recall that the purpose of the ME method is to update from priors to
posteriors and that the ranking scheme must depend on the particular prior $q
$ and therefore the entropy $S$ must be a functional of both $p$ and $q$. Thus
the entropy $S[p$, $q]$ produces a ranking of the distributions $p$ relative
to the given prior $q$: $S[p$, $q]$ is the entropy of $p$ relative to $q$.
Accordingly $S[p$, $q]$ is commonly called relative entropy. The modifier
"relative" is redundant and will be dropped since all entropies are relative,
even when relative to a uniform distribution. Moreover, since we deal with
incomplete information the ME method cannot be deductive: the method must be
inductive. Therefore, we may find useful to use those special cases where we
know what the preferred distribution should be to eliminate those entropy
functionals $S[p$, $q]$ that fail to provide the right update. In general, the
known special cases are called the axioms of the theory. Since they define
what makes one distribution preferable over another, they play a very crucial
role in the ME updating method.

In what follows, we will briefly present the three axioms of the ME method.
The axioms do not refer to merely three cases; any induction from such a weak
foundation would hardly be reliable. The reason the axioms are convincing and
so constraining is that they refer to three infinitely large classes of known
special cases. Additional details and proofs are given in
\cite{caticha(aip04), caticha-giffin(aip06)}.

\textbf{Axiom 1: Locality.} \textit{Local information has local effects}.

Assume the information to be processed does not refer to a special subdomain
$\mathcal{D}$ of the space $\mathcal{X}$ of $x$'s. The PMU (\textbf{Principle
of Minimal Updating} (PMU): \textit{Beliefs should be updated only to the
extent required by the new information}) requires we do not change our minds
about $\mathcal{D}$ in the absence of any new available information about
$\mathcal{D}$. Therefore, we design the inference method so that the prior
probability of $x$ conditional on $x\in$ $\mathcal{D}$, $q(x|\mathcal{D})$, is
not updated. The selected conditional posterior is $P(x|\mathcal{D}%
)=q(x|\mathcal{D})$. The consequence of axiom $1$ is that non-overlapping
domains of $x$ contribute additively to the entropy. Dropping multiplicative
factors and additive terms that do not affect the overall ranking, the entropy
functional becomes%
\begin{equation}
S[p\text{, }q]=\int dx\mathcal{F}(p(x)\text{, }q(x)\text{, }x)\text{,}%
\end{equation}
where $\mathcal{F}$ is some unknown function.

\textbf{Axiom 2: Coordinate invariance.} \textit{The system of coordinates
carries no information}.

Any of a variety of coordinate systems can be used to label the points $x$.
The freedom to change coordinates should not affect the ranking of the
distributions. The consequence of axiom $2$ is that $S[p$, $q]$ can be written
in terms of coordinate invariants such as $dxm(x)$ and $p(x)/m(x)$, and
$q(x)/m(x)$:%
\begin{equation}
S[p\text{, }q]=\int dxm\left(  x\right)  \Phi\left(  \frac{p\left(  x\right)
}{m\left(  x\right)  }\text{,}\frac{q\left(  x\right)  }{m\left(  x\right)
}\right)  \text{.}%
\end{equation}
(Multiplicative factors and additive terms that do not affect the overall
ranking have been dropped.) Thus the unknown function $\mathcal{F}(p(x)$,
$q(x)$, $x)$ has been replaced by $m\left(  x\right)  \Phi\left(
\frac{p\left(  x\right)  }{m\left(  x\right)  }\text{,}\frac{q\left(
x\right)  }{m\left(  x\right)  }\right)  $. The unknown functions become now
the density $m(x)$ and the function $\Phi$ with two arguments. Next we
determine the density $m(x)$ by invoking the locality axiom $1$ once again.

\textbf{Axiom 1 (special case):} \textit{When there is no new information
there is no reason to change one's mind}.

The domain $\mathcal{D}$ in axiom $1$ coincides with the whole space
$\mathcal{X}$ when no new information is available. The conditional
probabilities $q(x|\mathcal{D})=q(x|\mathcal{X})=q(x)$ should not be updated
and the selected posterior distribution coincides with

the prior, $P(x)=q(x)$. The consequence is that $S[p$, $q]$ becomes,%
\begin{equation}
S[p\text{, }q]=\int dxq(x)\Phi\left(  \frac{p(x)}{q(x)}\right)  \text{.}%
\end{equation}
\textbf{Axiom 3: Consistency for independent subsystems.} \textit{When a
system is composed of subsystems that are known to be independent it should
not matter whether the inference procedure treats them separately or jointly.}

Assume the information on two independent subsystems $1$ and $2$ that are
treated separately is such that the prior distributions $q_{1}(x_{1})$ and
$q_{2}(x_{2})$ are respectively updated to $P_{1}(x_{1})$ and $P_{2}(x_{2})$.
When treated as a single system the joint prior is $q_{1}(x_{1})q_{2}(x_{2})$
and the family of potential posteriors is $p(x_{1}$, $x_{2})=p_{1}(x_{1}%
)p_{2}(x_{2})$. The entropy functional must be such that the selected
posterior is $P_{1}(x_{1})P_{2}(x_{2})$. The consequence of axiom $3$ for this
particular case of just two subsystems is that entropies are restricted to the
one-parameter family given by%
\begin{equation}
S_{\eta}[p\text{, }q]=\frac{1}{\eta\left(  \eta+1\right)  }\left[  1-\int
dxp\left(  x\right)  \left(  \frac{p(x)}{q(x)}\right)  ^{\eta}\right]
\text{.} \label{glre}%
\end{equation}
Multiplicative factors and additive terms that do not affect the overall
ranking scheme can be freely chosen. The $\eta=0$ case reproduces the usual
logarithmic relative entropy,%
\begin{equation}
S[p\text{, }q]=-\int dxp(x)log\left(  \frac{p(x)}{q(x)}\right)  \label{lre}%
\end{equation}
[Use $y^{\eta}=\exp\left(  \eta\log y\right)  \approx1+\eta\log y$ in
(\ref{glre}) and let $\eta$ $\rightarrow0$ to get (\ref{lre}).]

In \cite{caticha-giffin(aip06)}\ it was argued that the index $\eta$ has to be
the same for all systems. Consistency requires that $\eta$ must be a universal
constant. From the success of statistical mechanics as a theory of inference
it was inferred that the value of this constant must be $\eta$ $=0 $ leading
to the logarithmic entropy, eq.(\ref{lre}).

In conclusion, the ME updating method can be summarized as follows
\cite{caticha(aip07)}:

\textbf{The ME method}: \textit{The objective is to update from a prior
distribution }$q$\textit{\ to a posterior distribution }$P(x)$\textit{\ given
the information that the posterior lies within a certain family of
distributions }$p$\textit{. The selected posterior }$P(x)$\textit{\ is that
which maximizes the entropy }$S[p$\textit{, }$q]$\textit{. Since prior
information is valuable, the functional }$S[p$\textit{, }$q]$\textit{\ has
been chosen so that beliefs are updated only to the extent required by the new
information. No interpretation for }$S[p$\textit{, }$q]$\textit{\ is given and
none is needed.}

\section{Elements of Riemannian Geometry}

In this section I briefly recall some essential concepts and notations of
Riemannian differential geometry which are used in this dissertation. The
present section is only meant to facilitate the reader to follow the work
presented in the next Chapters, so that my discussion will not be a rigorous
treatment of the subject. For a more elaborate discussion, I refer the reader
to references in \cite{casetti}, to textbooks of general relativity
\cite{landau, wald, de felice} or to a more mathematically oriented
introductions to the subject given in \cite{do carmo}. Finally, a
comprehensive and rigorous treatment, which goes far beyond what is needed to
follow the exposition in this Thesis, can be found in Kobayashi and Nomizu
\cite{kobayashi}. As a side remark, I would like to emphasize that each work
appearing in reference \cite{casetti} has deeply shaped my own personal point
of view concerning the relevance of geometry in the study of chaos. However,
the author's main concern in \cite{casetti} was the investigation of
\emph{classical chaos in the Riemannian and Finslerian geometric frameworks}.
My personal objective is to investigate both \emph{classical and quantum
aspects of chaoticity in a hybrid information-geometric framework} being aware
that, thus far, no application of Finslerian geometry to probability theory is
available in the literature \cite{private}.

\subsection{Notes on Riemannian manifolds}

A differentiable manifold $\mathcal{M}$ is a set that can be covered with a
collection, either finite or denumerable, of \textit{charts}, such that each
point of $\mathcal{M}$ is represented at least on one chart, and the different
charts are differentiably connected to each other. A chart is a set of
coordinates on the manifold, i.e., it is a set of $n$ real numbers $(x_{1}$, .
. . , $x_{n})$ which denote the "position"\ of a point on the manifold. The
number $n$ of coordinates of a chart is the same for each connected part of
the manifold (and for the whole manifold if the latter is connected, i.e., it
cannot be split in two disjoint parts which are still manifolds); such a
number is called the \textit{dimension} of the manifold $\mathcal{M}$. The
union of the charts on $\mathcal{M}$ is called an \textit{atlas} of
$\mathcal{M}$.

\subsubsection{Tangent vectors and tensors}

A possible way to define a vector is using curves on the manifold
$\mathcal{M}$. Given a curve $\gamma$ in $\mathcal{M}$, represented in local
coordinates by the parametric equations $\theta=\phi(t)$, we define a tangent
vector at $P\in\mathcal{M}$ as the velocity vector of the curve in $P$, i.e.,%
\begin{equation}
v=\dot{\gamma}=\underset{t\rightarrow0}{\lim}\frac{\phi\left(  t\right)
-\phi\left(  0\right)  }{t}\text{, }\phi\left(  0\right)  =P\text{,}%
\end{equation}
so that the $n$ components of the tangent vector $v$ are given by%
\begin{equation}
v^{i}=\frac{d\phi^{i}}{dt}\text{.}%
\end{equation}
The set of all the tangent vectors of $\mathcal{M}$ in $P$ is a linear space,
referred to as the \textit{tangent space} of $\mathcal{M}$ in $P$, and denoted
by $T_{P}\mathcal{M}$. Each tangent space is isomorphic to an $n$-dimensional
Euclidean space. Given a chart $(x_{1}$, . . . , $x_{n})$ in a neighborhood of
$P$, a basis $(X_{1}$, . . . , $X_{n})$ of $T_{P}\mathcal{M}$ can be defined,
so that a generic vector $v$ is expressed as a sum of the $X_{i}$'s weighted
by its components,%
\begin{equation}
v=v^{i}X_{i}\text{.}%
\end{equation}
The basis $\{X_{i}\}$ is called a coordinate basis of $T_{P}\mathcal{M}$, and
its vectors $X_{i}$ are often denoted \ by $\partial/\partial x_{i}$ (the
origin of this notation is in the fact that vectors can be defined as
directional derivatives on $\mathcal{M}$). The basis depends on the chart:
choosing another chart, $(x_{1}^{\prime}$, . , $x_{n}^{\prime})$, we get
another basis $\{X_{i}^{\prime}\}$. The components of $v$ in the two different
bases are connected by the following rule,
\begin{equation}
v^{\prime i}=v^{j}\frac{\partial x^{\prime i}}{\partial x^{j}}\text{,}
\label{vtr}%
\end{equation}
referred to as the \textit{vector transformation rule}. Indeed, one can define
a vector as a quantity whose components transform according to (\ref{vtr}).
The union of all the tangent spaces $T_{P}\mathcal{M}$ of the manifold
$\mathcal{M}$,%
\begin{equation}
T\mathcal{M=}\underset{P\in\mathcal{M}}{\mathcal{\cup}}T_{P}\mathcal{M}%
\text{,}%
\end{equation}
is a $2n$-dimensional manifold and is referred to as the \textit{tangent
bundle} of $\mathcal{M}$.

A vector field $V$ on $\mathcal{M}$ is an assignment of a vector $v_{P}$ at
each point $P\in\mathcal{M}$. If $f$ is a smooth function,%
\begin{equation}
V(f)|P=v_{P}(f)
\end{equation}
is a real number for each $P\in\mathcal{M}$, i.e., $v(f)$ is a function on
$\mathcal{M}$. If such a function is smooth, $V$ is called a \textit{smooth
vector field} on $\mathcal{M}$. The curves $\phi(t)$ which satisfy the
differential equations%
\begin{equation}
\dot{\phi}=V(\phi(t))
\end{equation}
are called the \textit{trajectories} of the field $V$ , and the mapping
$\phi_{t}:\mathcal{M}\rightarrow\mathcal{M}$ which maps any point $P$ of
$\mathcal{M}$ along the trajectory of $V$ emanating from $P$ is called the
\textit{flow} of $V$ . Given two vector fields $V$, $W$, one can define the
\textit{commutator} as the vector field $[V$, $W]$ such that%
\begin{equation}
\lbrack V\text{, }W](f)=V(W(f))-W(V(f))\text{,}%
\end{equation}
i.e., in terms of the local components,%
\begin{equation}
\lbrack V\text{, }W]^{j}=V^{i}\frac{\partial W^{j}}{\partial x^{i}}-W^{i}%
\frac{\partial V^{j}}{\partial x^{i}}\text{.} \label{commutator}%
\end{equation}
We note that, if $\{X_{i}\}$ is a coordinate basis,%
\begin{equation}
\lbrack X_{i}\text{, }X_{j}]=0\text{ }\forall i\text{, }j\text{,}%
\end{equation}
and that, conversely, given $n$ nonvanishing and commuting vector fields that
are linearly independent, there always exists a chart for which these vector
fields are a coordinate basis.

Tangent vectors are not the only vector-like quantities that can be defined on
a manifold $\mathcal{M}$: there are also\textit{\ cotangent }vectors, which
can be defined as follows. Let us recall that the \textit{dual space}
$V^{\ast}$ of a vector space $V$ is the space of \textit{linear} maps from $V
$ to the real numbers. Given a basis of $V$ , $\{u_{i}\}$, a basis of
$V^{\ast}$, \{$u^{i\ast}$\}, called the \textit{dual basis}, is defined by%
\begin{equation}
u^{i\ast}(u_{j})=\delta_{j}^{i}\text{.}%
\end{equation}
The dual space of $T\mathcal{M}$, $T^{\ast}\mathcal{M}$, is called the
\textit{cotangent bundle} of $\mathcal{M}$. Its elements are called
\textit{cotangent vectors}, or sometimes \textit{covariant vectors }(while the
tangent vectors are sometimes denoted as \textit{contravariant vectors}). The
dual basis elements are usually denoted as $dx^{1}$,...,$dx^{n}$, i.e.,
$dx^{i}$ is such that $dx^{i}(\partial/\partial x^{j})=\delta_{j}^{i}$. The
components $\omega_{i}$ of cotangent vectors transform according to the rule%
\[
\omega_{i}^{\prime}=\omega_{j}\frac{\partial x^{j}}{\partial x^{\prime i}%
}\text{,}%
\]
to be compared with (\ref{vtr}). The common rule is to use subscripts to
denote the components of dual vectors and superscripts for those of vectors.

A $(k$, $l)$-\textit{tensor} $T$ over a vector space $V$ is a multilinear map%
\begin{equation}
T:\underset{k-\text{times}}{\left(  V^{\ast}\times....\times V^{\ast}\right)
}\times\underset{l-\text{times}}{\left(  V\times.........\times V\right)
}\mapsto%
\mathbb{R}%
\end{equation}
i.e., acting on $k$ dual vectors and $l$ vectors, $T$ yields a number, and it
does so in such a manner that if we fix all but one of the vectors or dual
vectors, it is a linear map in the remaining variable. A $(0$, $0)$ tensor is
a scalar, a $(0$, $1)$ tensor is a vector, and a $(1$, $0)$ tensor is a dual
vector. The space $\mathcal{T}(k$, $l)$ of the tensors of type $(k$, $l)$ is a
linear space; a $(k$, $l)$-tensor is defined once its action on $k$ vectors of
the dual basis and on $l$ vectors of the basis is known, and since there are
$n^{k}n^{l}$ independent ways of choosing these basis vectors, $\mathcal{T}%
(k$, $l)$ is a $n^{k+l}$-dimensional linear space. Two natural operations can
be defined on tensors. The first one is called \textit{contraction} with
respect to the $i$-th (dual vector) and the $j$-th (vector) arguments and is a
map%
\begin{equation}
C:\mathcal{T}(k\text{, }l)\ni T\mapsto CT\in\mathcal{T}(k-1\text{, }l-1)
\end{equation}
defined by%
\begin{equation}
CT=\underset{\sigma=1}{\overset{n}{\sum}}T\left(  \text{....,}\underset{i}%
{v}^{\sigma\ast}\text{,.. ; ..,}\underset{j}{v_{\sigma}}\text{ ,...}\right)
\text{.}%
\end{equation}
The contracted tensor $CT$ is independent of the choice of the basis, so that
the contraction is a well-defined, invariant, operation. The second operation
is the \textit{tensor product}, which maps an element $\mathcal{T}(k$, $l)$
$\times$\ $\mathcal{T}(k^{\prime}$, $l^{\prime})$ into an element of
$\mathcal{T}(k+k^{\prime}$, $l+l^{\prime})$, i.e., two tensors $T$ and
$T^{\prime}$ into a new tensor, denoted by $T\otimes T^{\prime}$, defined as
follows: given $k+k^{\prime}$ dual vectors $v^{1\ast}$, . . . ,
v$^{k+k^{\prime}\ast}$ and $l+l^{\prime}$ vectors $w_{1}$, . . .
,$w_{l+l^{\prime}}$ , then%
\begin{equation}
T\otimes T^{\prime}(v^{1\ast}\text{,..., }v^{k+k^{\prime}\ast}\text{; }%
w_{1}\text{,...,}w_{l+l^{\prime}})=T(v^{1\ast}\text{,..., }v^{k\ast}\text{;
}w_{1}\text{,...,}w_{l})T^{\prime}(v^{k+1\ast}\text{,..., }v^{k+k^{\prime}%
\ast}\text{; }w_{l+1}\text{,...,}w_{l+l^{\prime}})\text{.}%
\end{equation}
The tensor product allows one to construct a basis for $\mathcal{T}(k$, $l)$
starting from a basis $\{v_{%
\mu
}\}$ of $V$ and its dual basis $\{v^{\nu\ast}\}$: such a basis is given by the
$n^{k+l}$ tensors $\{v_{%
\mu
_{1}}\otimes$\textperiodcentered$\ $\textperiodcentered$\ $\textperiodcentered
$\otimes v_{%
\mu
_{k}}\otimes v^{v_{1}\ast}\otimes$\textperiodcentered$\ $\textperiodcentered
$\ $\textperiodcentered$\otimes v^{v_{l}\ast}\}$. Thus, every tensor $T\in$
$\mathcal{T}(k$, $l)$ allows a decomposition%
\begin{equation}
T=\overset{n}{\underset{\mu_{1}\text{,.., }\nu_{l}=1}{\sum}}%
T_{v_{1............}\nu_{l}}^{\mu_{1}...\mu_{k}}v_{%
\mu
_{1}}\otimes\text{\textperiodcentered}\ \text{\textperiodcentered
}\ \text{\textperiodcentered}\otimes v_{%
\mu
_{k}}\otimes v^{v_{1}\ast}\otimes\text{\textperiodcentered}%
\ \text{\textperiodcentered}\ \text{\textperiodcentered}\otimes v^{v_{l}\ast
}\text{;}%
\end{equation}
the numbers $T_{v_{1....}\nu_{l}}^{\mu_{1}...\mu_{k}}$ are called the
\textit{components} of $T$ in the basis $\{v_{%
\mu
}\}$. The components of the contracted tensor $CT$ are%
\begin{equation}
\left(  CT\right)  _{v_{1....}\nu_{l-1}}^{\mu_{1}...\mu_{k-1}}%
=T_{v_{1.....\sigma.........}\nu_{l}}^{\mu_{1}..\sigma...\mu_{k}}%
\end{equation}
and, the components of the tensor product $T\otimes T^{\prime}$ are%
\begin{equation}
\left(  T\otimes T^{\prime}\right)  _{v_{1.........}\nu_{l+l^{\prime}}}%
^{\mu_{1}...\mu_{k+k^{\prime}}}=T_{v_{1.........}\nu_{l}}^{\mu_{1}....\mu_{k}%
}T_{v_{l+1..............}\nu_{l+l^{\prime}}}^{\prime\mu_{k+1}....\mu
_{k+k^{\prime}}}%
\end{equation}
All these results are valid for a generic vector space, so that they hold in
particular for the vector spaces of the tangent bundle $T\mathcal{M}$ of
$\mathcal{M}$, over which tensors (and, analogously to vector fields, tensor
fields) can be defined exactly as above.

\subsubsection{Metrics on Riemannian manifolds}

The length element $ds^{2}$ (the infinitesimal square distance, the metric) on
$\mathcal{M}$ can be defined at each point $P\in\mathcal{M}$ by means of a
$(0$, $2)$-tensor $g$, provided it is \textit{symmetric}, i.e., $g(v$,
$w)=g(w$, $v)$, and \textit{nondegenerate}, i.e., $g(v$, $w)=0$ $\forall v\in
T_{P}\mathcal{M}$ if and only if $w=0$. In fact, a $g$ with these properties
induces on the tangent bundle $T\mathcal{M}$ a nondegenerate quadratic form
(called the \textit{scalar product}),%
\begin{equation}
g:\left(  T\mathcal{M}\times\ T\mathcal{M}\right)  \ni\left(  v\text{,
}w\right)  \mapsto g\left(  v\text{, }w\right)  =\left\langle v\text{,
}w\right\rangle \in R\text{.} \label{quadratic form}%
\end{equation}
Then it is possible to measure lengths on the manifold. A manifold
$\mathcal{M}$, equipped with a scalar product, is called a (pseudo)Riemannian
manifold, and the scalar product is referred to as a (pseudo)Riemannian
structure on $\mathcal{M}$. If the quadratic form (\ref{quadratic form}) is
positive-definite, then one speaks of a (proper) Riemannian metric. In the
latter case the squared length element is always positive. For instance, one
can define the length of a curve as%
\begin{equation}
l\left(  \gamma\right)  =\underset{\gamma}{\int}\sqrt{\left\langle \dot
{\gamma}\text{, }\dot{\gamma}\right\rangle }dt\text{.}%
\end{equation}
The curves $\gamma$ which are extremals of the length functional are called
the \textit{geodesics} of $\mathcal{M}$.

In a coordinate basis, we can expand the metric $g$ as%
\begin{equation}
g=g_{ij}dx^{i}\otimes dx^{j}\text{,}%
\end{equation}
so that one defines the invariant (squared) length element on the manifold, in
local coordinates, as%
\begin{equation}
ds^{2}=g_{ij}dx^{i}dx^{j}\text{.}%
\end{equation}
The scalar product of two vectors $v$ and $w$ is given, in terms of $g$, by%
\begin{equation}
\left\langle v\text{, }w\right\rangle =g_{ij}v^{i}w^{j}=v_{j}w^{j}=v^{i}%
w_{i}\text{.}%
\end{equation}
In the above equation we have made use of the fact that $g$ establishes a
one-to-one correspondence between vectors and dual vectors, i.e., in
components,%
\begin{equation}
g_{ij}v^{j}=v_{i}\text{.}%
\end{equation}
For this reason, the components of the inverse metric $g^{-1}$ are simply
denoted by $g^{ij}$ , instead of $(g^{-1})^{ij}$ , and allow to pass from dual
vector (covariant) components to vector (contravariant) components:%
\begin{equation}
g^{ij}v_{j}=v^{i}\text{.}%
\end{equation}
This operation of raising and lowering the indices can be applied not only to
vector, but also to tensor components. This allows us to pass from $(k$, $l)$
tensor components to the corresponding $(k+1$, $l-1)$ tensor components and
vice versa. What does not change in the operation is the sum $k+l$ which is
called the \textit{rank} (or the \textit{order}) of the tensor.

\subsection{Covariant differentiation on Riemannian manifolds}

Differential calculus on Riemannian manifolds is complicated by the fact that
ordinary derivatives do not map vectors into vectors, i.e., the ordinary
derivatives of the components of a vector $w$, $dw^{i}/dt$, taken for instance
at a point $P$ along a given curve $\gamma(t)$, are not the components of a
vector in $T_{P}\mathcal{M}$, because they do not transform according to the
rule (\ref{vtr}). The geometric origin of this fact is that the
\textit{parallel transport} of a vector from a point $P$ to a point $Q$ on a
non-Euclidean manifold depends on the path chosen to join $P$ and $Q$. Since
in order to define the derivative of a vector at $P$, we have to move that
vector from $P$ to a neighboring point along a curve and then
parallel-transport it back to the original point in order to measure the
difference, we need a definition of parallel transport to define a derivative;
conversely, given a (consistent) derivative, i.e., a derivative which maps
vectors into vectors, one could define the parallel transport by imposing that
a vector is parallel transported along a curve if its derivative along the
curve is zero. The two ways are conceptually equivalent: we follow the first
way, by introducing the notion of a \textit{connection }and then using it to
define the derivative operator. Such a derivative will be referred to as the
\textit{covariant derivative}.

A (linear) \textit{connection }on the manifold $\mathcal{M}$ is a map $\nabla$
such that, given two vector fields (one could also consider tensor fields, but
for the sake of simplicity we define connections using vectors) $A$ and $B$,
it yields a third field $\nabla_{A}B$ with the following properties:

1. $\nabla_{A}B$ is bilinear in $A$ and $B$, i.e., $\nabla_{A}\left(  \alpha
B+\beta C\right)  $ $=\alpha\nabla_{A}B+\beta\nabla_{A}C$ and $\nabla_{\alpha
A+\beta B}C=\alpha\nabla_{A}C+\beta\nabla_{B}C$;

2. $\nabla_{f(A)}B=f(\nabla_{A}B)$;

3. (Leibnitz rule) $\nabla_{A}f(B)=(\partial_{A}f)B+f(\nabla_{A}B)$, where
$\partial_{A}$ is the ordinary directional derivative in the direction of $A$.

The \textit{parallel transport} of a vector $V$ along a curve $\gamma$, whose
tangent vector field is $\dot{\gamma}$, is then defined as the (unique) vector
field $W(t)=W(\gamma(t))$ along $\gamma(t)$ such that

1. $W(0)=V$ ;

2. $\nabla_{\dot{\gamma}}W=0$ along $\gamma$.

The notion of covariant derivative now immediately follows: the
\textit{covariant derivative} $DV/dt$ of $V$ along $\gamma$ is given by the
vector field%
\begin{equation}
\frac{DV}{dt}=\nabla_{\dot{\gamma}}V\text{.} \label{covariant derivative}%
\end{equation}
On the basis of equation (\ref{covariant derivative}), with a certain abuse of
language, one often refers to $\nabla_{X}Y$ as the covariant derivative of $Y$
along $X$, where $X$ and $Y$ are generic vector fields. Among all the possible
linear connections, and given a metric $g$, there is one and only one which
(i) is \textit{symmetric}, i.e.,%
\begin{equation}
\nabla_{X}Y-\nabla_{Y}X=[X\text{, }Y]\text{ }\forall X\text{,}Y\text{,}
\label{sym}%
\end{equation}
and (ii) conserves the scalar product, i.e., the scalar product of two
\textit{parallel} vector fields $P$ and $P^{\prime}$ is constant along
$\gamma$,%
\begin{equation}
\frac{d}{dt}\left\langle P\text{, }P^{\prime}\right\rangle \equiv0\text{.}%
\end{equation}
Such a linear connection is obviously the natural one on a Riemannian
manifold, and is referred to as the \textit{Levi-Civita} (or
\textit{Riemannian}) connection. Whenever we refer to a \textit{covariant
derivative} without any specification, we mean the covariant derivative
induced by the Riemannian connection. The components of the Riemannian
connection $\nabla$ with respect to a coordinate basis $\{X_{i}\}$ are the
\textit{Christoffel symbols}, given by%
\begin{equation}
\Gamma_{jk}^{i}=\left\langle dx^{i}\text{, }\nabla_{X_{j}}X_{k}\right\rangle
\end{equation}
and are given, in terms of the derivatives of the components of the metric, by
the following formula%
\begin{equation}
\Gamma_{jk}^{i}=\frac{1}{2}g^{im}\left(  \partial_{j}g_{km}+\partial_{k}%
g_{mj}-\partial_{m}g_{jk}\right)  \text{,}%
\end{equation}
where $\partial_{i}=\partial/\partial x^{_{i}}$. The expression in local
coordinates of the covariant derivative (\ref{covariant derivative}) of a
vector field $V$ is then%
\begin{equation}
\frac{DV^{i}}{dt}=\frac{dV^{i}}{dt}+\Gamma_{jk}^{i}\frac{dx^{j}}{dt}%
V^{k}\text{.} \label{cv}%
\end{equation}

\subsubsection{Geodesic Equation}

The \textit{geodesics} are defined as the curves of extremal length on the
manifold and can also be defined as \textit{self-parallel curves}, i.e.,
curves such that the tangent vector $\dot{\gamma}$ is always parallel
transported. Thus geodesics are the curves $\gamma\left(  t\right)  $ which
satisfy the equation (referred to as the \textit{geodesic equation})%
\begin{equation}
\frac{D\dot{\gamma}}{dt}=0
\end{equation}
whose expression in local coordinates follows from (\ref{cv}), and is%
\begin{equation}
\frac{d^{2}x^{i}}{dt^{2}}+\Gamma_{jk}^{i}\frac{dx^{j}}{dt}\frac{dx^{k}}%
{dt}=0\text{,} \label{geodesic equation}%
\end{equation}
provided $t$ is an affine parameter. Since the norm of the tangent vector
$\dot{\gamma}$ of a geodesic is constant, $|d\gamma/dt|=c$, the arc length of
a geodesic is proportional to the parameter:%
\begin{equation}
s\left(  t\right)  =\overset{t_{2}}{\underset{t_{1}}{\int}}dt\left\vert
\frac{d\gamma}{dt}\right\vert =c\left(  t_{2}-t_{1}\right)  \text{.}%
\end{equation}
When the parameter is actually the arc length, i.e., $c=1$, we say that the
geodesic is \textit{normalized}. Whenever we consider a geodesic, we assume it
is normalized, if not explicitly stated otherwise. This means that equations
(\ref{geodesic equation}) are nothing but the Euler-Lagrange equations for the
length functional along a curve $\gamma\left(  s\right)  $ parametrized by the
arc length,%
\begin{equation}
l\left(  \gamma\right)  =\underset{\gamma}{\int}ds\text{.}%
\end{equation}
Given a congruence of geodesics $\gamma\left(  s\right)  $ on $\mathcal{M}$,
there exists a unique vector field $G$ on $T\mathcal{M}$ such that its
trajectories are $(\gamma\left(  s\right)  $, $\dot{\gamma}\left(  s\right)
)$. Such a vector field $G$ is called the \textit{geodesic field }and its flow
$(\gamma\left(  s\right)  $, $\dot{\gamma}\left(  s\right)  )$ the
\textit{geodesic flow} on $\mathcal{M}$.

\subsection{The curvature tensor}

A way of measuring how much a Riemannian manifold $(\mathcal{M}$, $g)$
deviates from being Euclidean is by use of the \textit{curvature tensor}. This
quantity, also known as the \textit{Riemann-Christoffel tensor}, is a tensor
of order $4$ defined as%
\begin{equation}
R\left(  X\text{, }Y\right)  =\nabla_{X}\nabla_{Y}-\nabla_{Y}\nabla_{X}%
-\nabla_{\left[  X\text{, }Y\right]  }\text{,} \label{ct}%
\end{equation}
where $\nabla$ is the Levi-Civita connection of $\mathcal{M}$. Observe that if
$\mathcal{M}=%
\mathbb{R}
^{N}$, then $R(X$, $Y)=0$ for all the pairs of tangent vectors $X$, $Y$,
because of the commutativity of the ordinary derivatives. In addition, $R$
measures the noncommutativity of the covariant derivative: in fact, if we
choose a coordinate system $\{x_{1}$,...,$x_{n}\}$, we have, since $\left[
\frac{\partial}{\partial x_{i}}\text{, }\frac{\partial}{\partial x_{j}%
}\right]  =0$,
\begin{equation}
R\left(  \frac{\partial}{\partial x_{i}}\text{, }\frac{\partial}{\partial
x_{j}}\right)  =\nabla_{\partial/\partial x_{i}}\nabla_{\partial/\partial
x_{j}}-\nabla_{\partial/\partial x_{j}}\nabla_{\partial/\partial x_{i}%
}\text{.}%
\end{equation}
In local coordinates, the components of the Riemann curvature tensor
(considered here as a $(1$, $3)$-tensor) are given by%
\begin{equation}
R_{jkl}^{i}=\frac{\partial\Gamma_{jl}^{i}}{\partial x^{k}}-\frac
{\partial\Gamma_{kl}^{i}}{\partial x^{j}}+\Gamma_{jl}^{r}\Gamma_{kr}%
^{i}-\Gamma_{kl}^{r}\Gamma_{jr}^{i}\text{.}%
\end{equation}
Thus, given a metric $g$, the curvature $R$ is uniquely defined. A manifold
$(\mathcal{M}$, $g)$ is called flat when the curvature tensor vanishes.

Given a positive function $f^{2}$, a \textit{conformal transformation} is the
transformation%
\begin{equation}
(\mathcal{M}\text{, }g)\rightarrow(\mathcal{M}\text{, }\tilde{g})\text{;
}\tilde{g}=f^{2}g\text{,}%
\end{equation}
where $g$ is the metric tensor. Two Riemannian manifolds are said
\textit{conformally related} if they are linked by a conformal transformation.
In particular, a manifold is $(\mathcal{M}$, $g)$ \textit{conformally flat} if
it is possible to find a conformal transformation that sends $g$ into a flat
metric. Conformally flat manifolds exhibit some remarkable simplifications for
the calculation of the curvature tensor components (see \cite{goldberg}).

Closely related to the curvature tensor is the sectional --- or Riemannian ---
curvature, which we define now. Let us consider two vectors $u$, $v$ $\in
T_{P}\mathcal{M}$, and let us put%
\begin{equation}
\left\vert u\wedge v\right\vert =\left(  \left\vert u\right\vert
^{2}\left\vert v\right\vert ^{2}-\left\langle u\text{, }v\right\rangle
\right)  ^{\frac{1}{2}}\text{,}%
\end{equation}
which is the area of the two-dimensional parallelogram determined by $u$ and
$v$. If $|u\wedge v|\neq0$ the vectors $u$, $v$ span a two-dimensional
subspace $\pi\subset T_{P}\mathcal{M}$. We define the \textit{sectional
curvature} at the point $P$ relative to $\pi$, as the quantity:%
\begin{equation}
K\left(  P\text{; }u\text{, }v\right)  =K\left(  P\text{, }\pi\right)
=\frac{\left\langle R\left(  u\text{, }v\right)  u\text{, }v\right\rangle
}{\left\vert u\wedge v\right\vert ^{2}} \label{sectional curvature}%
\end{equation}
which can be shown to be independent of the choice of the two vectors $u$,
$v\in\pi$. In local coordinates, (\ref{sectional curvature}) becomes%
\begin{equation}
K\left(  P\text{; }u\text{, }v\right)  =R_{ijkl}\frac{u^{i}v^{j}u^{k}v^{l}%
}{\left\vert u\wedge v\right\vert ^{2}}\text{.}%
\end{equation}
The knowledge of $K$ for the $N(N-1)$ planes $\pi$ spanned by a maximal set of
linearly independent vectors completely determines $R$ at $P$.

If dim$\left(  \mathcal{M}\right)  =2$ then $K$ coincides with the Gaussian
curvature of the surface, i.e., with the product of the reciprocals of two
curvature radii.

A manifold is called \textit{isotropic} if $K(P$, $\pi)$ does not depend on
the choice of the plane $\pi$. A remarkable result --- Schur's theorem
\cite{do carmo} --- is that in this case $K$ is also constant, i.e. it does
not depend on the point $P$ either.

Some \textquotedblleft averages\textquotedblright\ of the sectional curvatures
are very important. The \textit{Ricci curvature} $K_{R}$ at $P$ in the
direction $v$ is defined as the sum of the sectional curvatures at $P$
relative to the planes determined by $v$ and the $N-1$ directions orthogonal
to $v$, i.e., if $\{e_{1}$,...,$e_{N-1}$, $v=e_{N}\}$ is an orthonormal basis
of $T_{P}\mathcal{M}$ and $\pi_{i}$ is the plane spanned by $v$ and $e_{i}$,%
\begin{equation}
K_{R}\left(  P\text{, }v\right)  =\underset{i=1}{\overset{N-1}{\sum}}K\left(
P\text{, }\pi_{i}\right)  \text{.}%
\end{equation}
The\textit{\ scalar curvature} $\mathcal{R}$ at P is the sum of the $N$ Ricci
curvatures at $P$,%
\begin{equation}
\mathcal{R}\left(  P\right)  =\underset{i=1}{\overset{N}{\sum}}K_{R}\left(
P\text{, }e_{i}\right)  \text{.}%
\end{equation}
In terms of the components of the curvature tensor, such curvatures can be
defined as follows (in the following formulae, we drop the dependence on $P$,
because it is understood that the components are local quantities). We first
define the \textit{Ricci tensor} as the two-tensor whose components, $R_{ij}$,
are obtained by contracting the first and the third indices of the Riemann
tensor,%
\begin{equation}
R_{ij}=R_{ikj}^{k}\text{;}%
\end{equation}
then,%
\begin{equation}
K_{R}\left(  v\right)  =R_{ij}v^{i}v^{j}\text{.} \label{saturation}%
\end{equation}
The right hand side of (\ref{saturation}) is called "saturation"\ of $R_{ij}$
with $v$. The scalar curvature can be obtained as the trace of the Ricci
tensor,%
\begin{equation}
\mathcal{R}=R_{i}^{i}\text{.}%
\end{equation}
In the case of a \textit{constant curvature} --- or isotropic --- manifold,
the components of the Riemann curvature tensor have the remarkably simple form%
\begin{equation}
R_{ijkl}=K\left(  g_{ik}g_{jl}-g_{il}g_{jk}\right)  \text{,}%
\end{equation}
where $K$ is the constant sectional curvature, so that the components of the
Ricci tensor are%
\begin{equation}
R_{ij}=Kg_{ij}\text{,}%
\end{equation}
and all the above defined curvatures are constants, and are related by%
\begin{equation}
K=\frac{1}{N-1}K_{R}=\frac{1}{N\left(  N-1\right)  }\mathcal{R}\text{.}%
\end{equation}

\subsection{The Jacobi-Levi-Civita equation}

A brief derivation of the JLC (Jacobi-Levi-Civita) equation is presented in
this subsection. We will proceed as follows: first, we will define the
geodesic separation vector field $J$, then we will show that the field $J$ is
actually a Jacobi field, i.e., obeys the Jacobi equation.

Let us define a \textit{geodesic congruence} as a family of geodesics
$\{\gamma(s)=\gamma(s$, $\tau)|\tau\in%
\mathbb{R}
\}$ issuing from a neighborhood $\mathcal{I}$ of a manifold point, smoothly
dependent on the parameter $\tau$, and let us fix a reference geodesic
$\bar{\gamma}\left(  s\text{, }\tau_{0}\right)  $. Denote then by $\dot
{\gamma}(s)$ the vector field tangent to $\bar{\gamma}$ in $s$, i.e., the
velocity vector field whose components are%
\begin{equation}
\dot{\gamma}^{i}=\frac{dx^{i}}{ds}\text{,} \label{tv}%
\end{equation}
and by $J(s)$ the vector field tangent in $\tau_{0}$ to the curves $\gamma
_{s}\left(  \tau\right)  $ for a fixed $s$, i.e., the vector field of
components%
\begin{equation}
J^{i}=\frac{dx^{i}}{d\tau}\text{.} \label{gsf}%
\end{equation}
The field $J$ will be referred to as the \textit{geodesic separation field},
and measures the distance between nearby geodesics. Let us now show that $J$
is a Jacobi field. First of all, we notice that the field $J$ commutes with
$\dot{\gamma}$, i.e., $[\dot{\gamma}$, $J]=0$. In fact, from the definition of
the commutator (\ref{commutator}) and from the definitions of $J$,
(\ref{gsf}), and of $\dot{\gamma}$, (\ref{tv}), we have%
\begin{equation}
\left[  \dot{\gamma}\text{, }J\right]  ^{i}=\dot{\gamma}^{j}\frac{\partial
J^{i}}{\partial x^{j}}-J^{j}\frac{\partial\dot{\gamma}^{i}}{\partial x^{j}%
}=\frac{\partial x^{j}}{\partial s}\frac{\partial J^{i}}{\partial x^{j}}%
-\frac{\partial x^{j}}{\partial\tau}\frac{\partial\dot{\gamma}^{i}}{\partial
x^{j}}=\frac{\partial J^{i}}{\partial s}-\frac{\partial\dot{\gamma}^{i}%
}{\partial\tau}\text{,}%
\end{equation}
and using again (\ref{gsf}) and (\ref{tv}), we find that%
\begin{equation}
\frac{\partial J^{i}}{\partial s}=\frac{\partial}{\partial s}\frac{\partial
x^{i}}{\partial\tau}=\frac{\partial}{\partial\tau}\frac{\partial x^{i}%
}{\partial s}=\frac{\partial\dot{\gamma}^{i}}{\partial\tau}\text{,}%
\end{equation}
so that $[\dot{\gamma}$, $J]=0$. Now, let us compute the second covariant
derivative of the field $J$, $\nabla_{\dot{\gamma}}^{2}J$. First of all, let
us recall that our covariant derivative comes from a Levi-Civita connection,
which is symmetric (see (\ref{sym})), so that%
\begin{equation}
\nabla_{\dot{\gamma}}J-\nabla_{J}\dot{\gamma}=\left[  \dot{\gamma}\text{,
}J\right]  \text{,}%
\end{equation}
and having just shown that $[\dot{\gamma}$, $J]=0$, we can write%
\begin{equation}
\nabla_{\dot{\gamma}}J=\nabla_{J}\dot{\gamma}\text{.}%
\end{equation}
Now, using this result, and the fact that $\nabla_{\dot{\gamma}}\dot{\gamma
}=0$ because $\bar{\gamma}$ is a geodesic, we can write%
\begin{equation}
\nabla_{\dot{\gamma}}^{2}J=\nabla_{\dot{\gamma}}\nabla_{\dot{\gamma}}%
J=\nabla_{\dot{\gamma}}\nabla_{J}\dot{\gamma}=\left[  \nabla_{_{\dot{\gamma}}%
}\text{, }\nabla_{J}\right]  \dot{\gamma}\text{,}%
\end{equation}
from which, using the definition of the curvature tensor (\ref{ct}) and,
again, the vanishing of the commutator $[\dot{\gamma}$, $J]$, we get%
\begin{equation}
\nabla_{\dot{\gamma}}^{2}J=R\left(  \dot{\gamma}\text{, }J\right)  \dot
{\gamma}\text{,} \label{JLC equation}%
\end{equation}
which is nothing but the Jacobi equation written in compact notation. In an
explicit way the (\ref{JLC equation}) can be written as,%
\begin{equation}
\frac{D^{2}J^{i}}{ds^{2}}+R_{jkl}^{i}\frac{dq^{i}}{ds}J^{k}\frac{dq^{l}}{ds}=0
\end{equation}
It is worth noticing that the normal component $J_{\perp}$ of $J$, i.e., the
component of $J$ orthogonal to $\dot{\gamma}$ along the geodesic $\gamma$, is
again a Jacobi field, since we can always write $J=J_{\perp}+\lambda
\dot{\gamma}$: one immediately finds then that the velocity $\dot{\gamma}$
satisfies the Jacobi equation, so that $J_{\perp}$ must obey the same
equation. This can allow us to restrict ourselves to the study of the normal
Jacobi fields.

\section{What is Information Geometry?}

In the present section, I describe some of the basics of information geometry
(IG). IG is the result of applying conventional Riemannian geometry to
probability theory. Although interest in this subject can be traced back to
the late 1960's, it reached maturity only through the work of Amari in the
1980's \cite{amari}. The development of the field of information geometry can
only be said to have just begun.

\subsection{Notes on Information Geometry}

IG began as an investigation of the natural differential geometric structure
possessed by families of probability distributions. As a rather simple
example, consider the set of normal distributions with mean $\mu$ and variance
$\sigma^{2}$:%
\begin{equation}
p\left(  x\text{; }\mu\text{, }\sigma\right)  =\frac{1}{\sqrt{2\pi\sigma^{2}}%
}\exp\left(  -\frac{\left(  x-\mu\right)  ^{2}}{2\sigma^{2}}\right)  \text{.}%
\end{equation}
By specifying $\left(  \mu\text{, }\sigma\right)  $ we determine a particular
normal distribution, and therefore the set may be viewed as a $2$-dimensional
space (manifold) which has $\left(  \mu\text{, }\sigma\right)  $ as a
coordinate system. However, this is not an Euclidean space, but rather a
Riemannian space with a metric which naturally follows from the underlying
properties of the probability distributions. Probability distributions are the
fundamental element over which fields such as statistics, stochastic
processes, and information theory are developed. Therefore, the geometric
structure of the space of probability distributions must play a fundamental
role in these information sciences. In fact, considering statistical
estimation from a differential geometric viewpoint has provided statistics
with a new analytic tool which has allowed several previously open problems to
be solved; information geometry has already established itself within the
field of statistics. In the fields of information theory, stochastic
processes, and systems, information geometry is being currently applied to
allow the investigation of hitherto unexplored possibilities.

The utility of information geometry, however, is not limited to these fields.
It has, for example, been applied productively to areas such as statistical
physics and mathematical theory underlying neural networks.

From a mathematical point of view, quantum mechanics may be constructed as an
extension of probability theory, and it is possible to generalize many
concepts in probability theory to their quantum equivalents. The framework of
information geometry for statistical models may also be extended to the
quantum mechanical setting. A variety of important works related to
differential geometrical aspects of quantum mechanics have so far been made by
many researchers. However, the study of quantum information geometry has just
started and we are far from getting its whole perspective at present.

One of the fundamental questions information geometry helps to answer is:
"Given two probability distributions, is it possible to define a notion of
"distance"\ between them?". This chapter, which introduces some of the basic
concepts of information geometry, does not presuppose any knowledge of the
theory of probability and distributions. Unfortunately however, it does
require some knowledge of Riemannian geometry.

\subsection{Families of probability distributions as statistical manifolds}

For our purposes, a probability distribution over some field (or set)
$\mathcal{X}$ is a distribution
\begin{equation}
p:\mathcal{X\ni}x\mapsto p\left(  x\right)  \in%
\mathbb{R}
^{+}%
\end{equation}
such that%
\begin{equation}
\underset{\mathcal{X}}{\int}dxp\left(  x\right)  =1\text{ and }\underset
{\mathcal{S}}{\int}dxp\left(  x\right)  >0\text{ }\forall\text{ subset
}\mathcal{S\subset X}\text{.}%
\end{equation}
In what follows, we will consider families of such distributions. In most
cases these families will be parametrized by a set of continuous parameters
$\theta=\left\{  \theta^{\mu}\right\}  _{\mu=1\text{,..,}n}$ that take values
in some open interval $\mathcal{I}_{\theta}\subseteq%
\mathbb{R}
^{n}$ and we write $p\left(  x|\theta\right)  $ to denote members of the
family. For any fixed $\theta$, $p_{\theta}$ is the mapping from $\mathcal{X}$
to $%
\mathbb{R}
$ with%
\begin{equation}
p_{\theta}:\mathcal{X\ni}x\mapsto p_{\theta}\left(  x\right)  \equiv p\left(
x|\theta\right)  \in%
\mathbb{R}
\text{.}%
\end{equation}
In information geometry, one extends a family of distributions $\mathcal{F}$,%
\begin{equation}
\mathcal{F}=\left\{  p_{\theta}|\text{ }\theta\in\mathcal{I}_{\theta}\subseteq%
\mathbb{R}
^{n}\right\}  \text{,}%
\end{equation}
to a manifold $\mathcal{M}$ such that the points in $\mathcal{M}$ are in a
\textit{one to one} relation with the distributions in $\mathcal{F}$. In doing
so, one hopes to gain some insight into the structure of such a family. For
example, one might hope to discover a reasonable measure of "nearness" of two
distributions in the family. Having made the link between families of
distributions and manifolds, one can try to identify which objects in the
language of distributions naturally correspond to objects in the language of
manifolds and vice versa. The most important objects in the language of
manifolds are tangent vectors. The tangent space $\mathcal{T}_{\theta}$ at the
point in $\mathcal{M}$ with coordinates $\left\{  \theta^{\mu}\right\}
_{\mu=1\text{,..,}n}$ is seen to be isomorphic to the vector space spanned by
the model parameters (function from $\mathcal{X}$ to $%
\mathbb{R}
$) $\frac{\partial\log p\left(  x|\theta\right)  }{\partial\theta^{\mu}}$.
This space is called $\mathcal{T}_{\theta}^{\left(  1\right)  }$. Therefore, a
vector field $v\left(  \theta\right)  \in\mathcal{T}(\mathcal{M})$,%
\begin{equation}
v:\mathcal{M\ni}\theta\mapsto v\left(  \theta\right)  =v^{\mu}\left(
\theta\right)  \hat{e}_{\mu}\in\mathcal{T}(\mathcal{M})\text{,}%
\end{equation}
is equivalent to a model parameter $v_{\theta}\left(  \cdot\right)
\in\mathcal{T}^{\left(  1\right)  }(\mathcal{M})$,%
\begin{equation}
v_{\theta}\left(  x\right)  =v^{\mu}\left(  \theta\right)  \frac{\partial\log
p\left(  x|\theta\right)  }{\partial\theta^{\mu}}\text{.}
\label{1-representation}%
\end{equation}
Just as $\mathcal{T}(\mathcal{M})$ is the space of continuously differentiable
mappings that assigns some vector $v\left(  \theta\right)  \in\mathcal{T}%
_{\theta}$ to each point $\theta\in\mathcal{M}$, $\mathcal{T}^{\left(
1\right)  }(\mathcal{M})$ assigns a model parameter $v_{\theta}\in
\mathcal{T}_{\theta}^{\left(  1\right)  }$.

In view of the above equivalence, we will not find necessary to distinguish
between the vector field $v$ and the corresponding random variable $v_{\theta
}\left(  \cdot\right)  $. Equation (\ref{1-representation}) is called the
$1$-\textit{representation} of the vector field $v$. It is clearly possible to
use some other basis of functionals of $p\left(  x|\theta\right)  $ instead of
$\frac{\partial\log p\left(  x|\theta\right)  }{\partial\theta^{\mu}}$. Our
present choice has the advantage that the $1$-\textit{representation }of a
vector has zero expectation value,%
\begin{equation}
E\left[  \frac{\partial\log p\left(  x|\theta\right)  }{\partial\theta^{\mu}%
}\right]  \equiv\underset{\mathcal{X}}{\int}dxp\left(  x|\theta\right)
\frac{\partial\log p\left(  x|\theta\right)  }{\partial\theta^{\mu}}%
=\frac{\partial}{\partial\theta^{\mu}}\underset{\mathcal{X}}{\int}dxp\left(
x|\theta\right)  =0\text{.} \label{ev1}%
\end{equation}
Using other functionals can be useful, and in fact the $1$%
-\textit{representation} turns out to be just one member of the family of
$\alpha$-\textit{representations }$\left(  \alpha\neq1\right)  $.

\subsection{Distances between distributions: the Fisher-Rao information
metric}

In several applications we are interested in distances between distributions.
For example, given a distribution $p\in\mathcal{M}$ and a submanifold
$\mathcal{S\subseteq M}$ we may wish to find the distribution $p^{\prime}%
\in\mathcal{S}$ that is "nearest" to $p$ in some sense. In order to fulfill
such a task we need a notion of \textit{distance on manifolds of
distributions}. In other words, we need a metric.

Consider a family of probability distributions $p\left(  x|\theta\right)  $
defined in terms of some parameters $\theta^{\mu}$ with $\mu=1$,..., $n$. The
space of these distributions constitutes a manifold, the points of which are
the distributions and, the parameters $\theta^{\mu}$ are convenient set of
coordinates. The structure of such manifolds is studied by introducing
conventional differential geometrical notions. Ultimately, the problem is to
quantify the extent to which we can distinguish between two neighboring
probability distributions $p\left(  x|\theta\right)  $ and $p\left(
x|\theta+d\theta\right)  $. If $d\theta$ is small enough the distributions
overlap considerably, it is easy to confuse them and we are inclined to say
that the distributions are near. More specifically we seek a real positive
number that provides a quantitative measure of the extent to which the two
distributions can be distinguished. When we interpret this measure of
distinguishability as a distance- the information metric- the manifold of
distributions immediately acquires a geometric structure and we can proceed to
study it using the mathematical techniques of differential geometry. It
appears that the introduction of geometrical methods is the natural way to
study spaces of probability distributions, to study how one changes one's mind
and effectively moves in such a space as a result of acquiring information.
Perhaps, this is the reason why the models we develop to describe the world
are so heavily geometrical.

We are looking for a quantitative measure of the extent that two probability
distributions $p\left(  x|\theta\right)  $ and $p\left(  x|\theta
+d\theta\right)  $ can be distinguished. An appealing and intuitive way to
approach this problem is the following. Consider the relative difference,%
\begin{equation}
\frac{p\left(  x|\theta+d\theta\right)  -p\left(  x|\theta\right)  }{p\left(
x|\theta\right)  }=\frac{\partial\log p\left(  x|\theta\right)  }%
{\partial\theta^{\mu}}d\theta^{\mu}\text{.}%
\end{equation}
It might seem that the expected value of the relative difference is a good
candidate. However, it is not because it vanishes identically (see
(\ref{ev1})),%
\begin{equation}
\int dxp\left(  x|\theta\right)  \frac{\partial\log p\left(  x|\theta\right)
}{\partial\theta^{\mu}}d\theta^{\mu}=d\theta^{\mu}\frac{\partial}%
{\partial\theta^{\mu}}\int dxp\left(  x|\theta\right)  =0\text{.}%
\end{equation}
Instead, the variance does not vanish and therefore it is a good choice,%
\begin{equation}
dl^{2}=\int dxp\left(  x|\theta\right)  \frac{\partial\log p\left(
x|\theta\right)  }{\partial\theta^{\mu}}\frac{\partial\log p\left(
x|\theta\right)  }{\partial\theta^{\nu}}d\theta^{\mu}d\theta^{\nu}%
\overset{\text{def}}{=}g_{\mu\nu}d\theta^{\mu}d\theta^{\nu}\text{.}
\label{line}%
\end{equation}
This is the measure of distinguishability for which we are searching; a small
value of $dl^{2}$ means the points $\theta$ and $\theta+d\theta$ are difficult
to distinguish. The matrix $g_{\mu\nu}$ is called the Fisher information
matrix \cite{fisher}. Thus far, no notion of distance has been introduced on
the space of states. In general, it is said that the reason it is difficult to
distinguish between two points is that they happen to be too close together.
However, it is very tempting to invert the logic and assert that the two
points $\theta$ and $\theta+d\theta$ must be very close together because they
are difficult to distinguish. Moreover, notice that $dl^{2}$ is positive since
it is a variance and it vanishes only when $d\theta$ vanishes. Therefore it is
natural to interpret $g_{\mu\nu}$ as the metric tensor of a Riemannian space
\cite{rao}. It is known as the Fisher-Rao information metric. This metric is a
suitable metric for manifolds of distributions and it is given by,%
\begin{equation}
g_{\mu\nu}\left(  \theta\right)  =E\left[  \partial_{\mu}l\left(
\theta\right)  \partial_{\nu}l\left(  \theta\right)  \right]  \equiv\int
dxp\left(  x|\theta\right)  \frac{\partial\log p\left(  x|\theta\right)
}{\partial\theta^{\mu}}\frac{\partial\log p\left(  x|\theta\right)  }%
{\partial\theta^{\nu}} \label{info-metric}%
\end{equation}
with $\partial_{\mu}\equiv\frac{\partial}{\partial\theta^{\mu}}$ and $l\left(
\theta\right)  \equiv\log p\left(  x|\theta\right)  $. Notice that $E\left[
\partial_{\mu}l\left(  \theta\right)  \partial_{\nu}l\left(  \theta\right)
\right]  =-E\left[  \partial_{\mu}\partial_{\nu}l\left(  \theta\right)
\right]  $. Obviously, the information metric also defines an inner product:
for two vector fields $v$ and $w$, we have%
\begin{equation}
\left\langle v\text{, }w\right\rangle \overset{\text{def}}{=}g_{\mu\nu}\left(
\theta\right)  v_{\theta}^{\mu}w_{\theta}^{\nu}=E\left[  v_{\theta}^{\mu
}\partial_{\mu}l\left(  \theta\right)  w_{\theta}^{\nu}\partial_{\nu}l\left(
\theta\right)  \right]  =E\left[  v_{\theta}w_{\theta}\right]  \text{.}%
\end{equation}
Rao recognized that $g_{\mu\nu}$ is a metric in the space of probability
distributions and this recognition gave rise to the subject of information
geometry \cite{amari}. This heuristic argument presents a disadvantage, namely
it does not make explicit a crucial property of the Fisher-Rao metric: except
for an overall multiplicative constant this Riemannian metric is unique
\cite{cencov, campbell}. The coordinates $\theta$ are quite arbitrary and one
can freely switch from one set to another. It is then easy to check that
$g_{\mu\nu}$ are the components of a tensor, that is, the distance $dl^{2}$ is
an invariant, a scalar. Incidentally, $dl^{2}$ is also dimensionless. At first
sight, the definition (\ref{info-metric}) may seem rather ad hoc. However, it
has been proven \cite{amari, rao, corcuera} to be unique in having the
following very appealing properties:

\begin{enumerate}
\item $g_{\mu\nu}\left(  \theta\right)  $ is invariant under
reparametrizations of the sample space $\mathcal{X}$;

\item $g_{\mu\nu}\left(  \theta\right)  $ is covariant under
reparametrizations of the manifold $\mathcal{M}$ (the parameter space).
\end{enumerate}

The uniqueness of the Fisher-Rao information metric is the most remarkable
aspect about this metric: except for a constant scale factor, it is the only
Riemannian metric that adequately takes into account that points of the
manifold are not structureless; that is, that they are probability
distributions. As said before, a proof of such a result is given in
\cite{cencov}. Once the information metric is given, then connection
coefficients, curvatures and other aspects of the geometry can be computed in
the conventional way.

\subsection{Volume elements in curved statistical manifolds}

Once the distances among probability distributions have been assigned, a
natural next step is to obtain measures for extended regions in the space of distributions.

Consider an $n$-dimensional volume of the statistical manifold $\mathcal{M}%
_{s}$ of distributions $p\left(  x|\theta\right)  $ labelled by parameters
$\theta^{\mu}$ with $\mu=1$,..., $n$. The parameters $\theta^{\mu}$ are
coordinates for the point $P$ and in these coordinates it may not be obvious
how to write down an expression for a volume element $dV_{\mathcal{M}_{s}}$.
However, within a sufficiently small region (volume element) any curved space
looks flat. Curved spaces are "locally flat". The idea then is rather simple:
within that very small region, we should use Cartesian coordinates and the
metric takes a very simple form, it is the identity matrix $\delta_{\mu\nu}$.
In locally Cartesian coordinates $\chi^{\alpha}$ the volume element is simply
given by the product%
\begin{equation}
dV_{\mathcal{M}_{s}}=d\chi^{1}d\chi^{2}\text{.....}d\chi^{n}\text{,}%
\end{equation}
which, in terms of the old coordinates, is%
\begin{equation}
dV_{\mathcal{M}_{s}}=\left\vert \frac{\partial\chi}{\partial\theta}\right\vert
d\theta^{1}d\theta^{2}\text{.....}d\theta^{n}=\left\vert \frac{\partial\chi
}{\partial\theta}\right\vert d^{n}\theta\text{.}%
\end{equation}
The problem consists in calculating the Jacobian $\left\vert \frac
{\partial\chi}{\partial\theta}\right\vert $ of the transformation that takes
the metric $g_{\mu\nu}$ into its Euclidean form $\delta_{\mu\nu}$.

Let the new coordinates be defined by $\chi^{\mu}=\Xi^{\mu}\left(  \theta
^{1}\text{,...., }\theta^{n}\right)  $. A small change in $d\theta$
corresponds to a small change in $d\chi$,%
\begin{equation}
d\chi^{\mu}=X_{m}^{\mu}d\theta^{m}\text{ where }X_{m}^{\mu}\overset
{\text{def}}{=}\frac{\partial\chi^{\mu}}{\partial\theta^{m}}\text{,}%
\end{equation}
and the Jacobian is given by the determinant of the matrix $X_{m}^{\mu}$,%
\begin{equation}
\left\vert \frac{\partial\chi}{\partial\theta}\right\vert =\left\vert
\det\left(  X_{m}^{\mu}\right)  \right\vert \text{.}%
\end{equation}
The distance between two neighboring points is the same whether we compute it
in terms of the old or the new coordinates,%
\begin{equation}
dl^{2}=g_{\mu\nu}d\theta^{\mu}d\theta^{\nu}=\delta_{\alpha\beta}d\chi^{\alpha
}d\chi^{\beta}\text{.}%
\end{equation}
Therefore the relation between the old and the new metric is,%
\begin{equation}
g_{\mu\nu}=\delta_{\alpha\beta}X_{\mu}^{\alpha}X_{\nu}^{\beta}\text{.}
\label{metric relation}%
\end{equation}
Taking the determinant of (\ref{metric relation}), we obtain%
\begin{equation}
g\overset{\text{def}}{=}\det\left(  g_{\mu\nu}\right)  =\left[  \det\left(
X_{\mu}^{\alpha}\right)  \right]  ^{2}%
\end{equation}
and, therefore%
\begin{equation}
\left\vert \det\left(  X_{\mu}^{\alpha}\right)  \right\vert =\sqrt{g}\text{.}%
\end{equation}
Finally, we have succeeded in expressing the volume element totally in terms
of the coordinates $\theta$ and the known metric $g_{\mu\nu}\left(
\theta\right)  $,%
\begin{equation}
dV_{\mathcal{M}_{s}}=\sqrt{g}d^{n}\theta\text{.}%
\end{equation}
The volume of any extended region on the manifold is given by,%
\begin{equation}
V_{\mathcal{M}_{s}}=\int dV_{\mathcal{M}_{s}}=\int\sqrt{g}d^{n}\theta\text{.}%
\end{equation}
As a final remark, note that $\sqrt{g}d^{n}\theta$ is a scalar quantity and
therefore is invariant under general coordinate transformations,
$\theta\rightarrow\theta^{\prime}$, preserving orientation. The square root of
the metric tensor transforms as \cite{xavier},%
\begin{equation}
\sqrt{g\left(  \theta\right)  }\overset{\theta\rightarrow\theta^{\prime}%
}{\rightarrow}\left\vert \frac{\partial\theta^{\prime}}{\partial\theta
}\right\vert \sqrt{g\left(  \theta^{\prime}\right)  } \label{pre1}%
\end{equation}
and the flat infinitesimal volume element $d^{n}\theta$ transforms as
\begin{equation}
d^{n}\theta\overset{\theta\rightarrow\theta^{\prime}}{\rightarrow}\left\vert
\frac{\partial\theta}{\partial\theta^{\prime}}\right\vert d^{n}\theta^{\prime
}\text{.} \label{pre2}%
\end{equation}
Thus, from (\ref{pre1}) and (\ref{pre2}), we obtain%
\begin{equation}
\sqrt{g\left(  \theta\right)  }d^{n}\theta\overset{\theta\rightarrow
\theta^{\prime}}{\rightarrow}\sqrt{g\left(  \theta^{\prime}\right)  }%
d^{n}\theta^{\prime}\text{.} \label{pre3}%
\end{equation}

Equation (\ref{pre3}) implies that the infinitesimal statistical volume
element is invariant under general coordinate transformations that preserve
orientation, that is with positive Jacobian.

\section{ME and IG at work: a simple example}

In what follows, I will briefly illustrate a couple of examples where the ME
method is employed. Further examples can be found in \cite{sivia}.

Let the microstates of a physical system be labelled by $x$, and let $m\left(
x\right)  dx$ be the number of microstates in the range $dx$. We assume that a
state of the system, a macrostate, is defined by the known expected values
$F^{\mu}$ of some $n_{F}$ variables $f^{\mu}\left(  x\right)  $ with $\mu
=1$,...., $n_{F}$,%
\begin{equation}
\left\langle f^{\mu}\left(  x\right)  \right\rangle =\int dxp\left(  x\right)
f^{\mu}\left(  x\right)  =F^{\mu}\text{.} \label{expected values}%
\end{equation}
This limited information will certainly not be sufficient to answer all
questions that one could conceivably ask about the system. Choosing the right
set of variables $\left\{  f^{\mu}\right\}  $ is perhaps the most difficult
problem in statistical mechanics. A crucial assumption is that
(\ref{expected values}) is not just any random information, instead it is the
right information for our purposes.

It is convenient to think of each state as a point in an $n_{F}$-dimensional
statistical manifold; the numerical values $F^{\mu}$ associated to each point
form a convenient set of coordinates. The ME method allows us to associate a
probability distribution to each point in the space of states. The probability
distribution $p\left(  x|F\right)  $ that best reflects the prior information
contained in $m\left(  x\right)  $, updated by the information $F^{\mu}$, is
obtained by maximizing the relative logarithmic entropy%
\begin{equation}
S\left[  p|m\right]  =-\int dxp\left(  x\right)  \log\left(  \frac{p\left(
x\right)  }{m\left(  x\right)  }\right)
\end{equation}
subject to the constraints (\ref{expected values}) and to the normalization
constraint $\int dxp\left(  x\right)  =1$. Upon setting $\delta S=0$ where $S$
is given by%
\begin{equation}
S\equiv S\left[  p|m\right]  -\lambda\left(  \int dxp\left(  x\right)
-1\right)  -\lambda_{\mu}\left(  \int dxp\left(  x\right)  f^{\mu}\left(
x\right)  -F^{\mu}\right)
\end{equation}
and using the normalization constraint, we obtain%
\begin{equation}
p\left(  x|F\right)  =\frac{m\left(  x\right)  }{Z\left(  \lambda\right)
}e^{-\lambda_{\mu}f^{\mu}(x)}\text{.}%
\end{equation}
The partition function $Z\left(  \lambda\right)  $ and the Lagrange
multipliers $\lambda_{\mu}$ are defined as,%
\begin{equation}
Z\left(  \lambda\right)  =\int dxm\left(  x\right)  e^{-\lambda_{\mu}f^{\mu
}(x)}\text{ and }-\frac{\partial\log Z\left(  \lambda\right)  }{\partial
\lambda_{\mu}}=F^{\mu}\text{.}%
\end{equation}
The maximized value of entropy is given by%
\begin{equation}
S\left(  F\right)  =S\left[  p|m\right]  |_{p=p\left(  x|F\right)  }=\log
Z\left(  \lambda\right)  +\lambda_{\mu}F^{\mu}\text{.}%
\end{equation}
For the sake of clarity, let us consider the following special case: assume
the normalization and information constraints are given by
\begin{equation}
\int dxp\left(  x\right)  =1\text{, }\int dxp\left(  x\right)  \left(
x-\mu\right)  ^{2}=\sigma^{2}%
\end{equation}
where $\mu=\int dxxp\left(  x\right)  $. Upon maximizing $S\left[  p\right]
=-\int dxp\left(  x\right)  \log p\left(  x\right)  $ (where we have assumed a
uniform prior, $m\left(  x\right)  =1$) subject to the above constraints, we
obtain \
\begin{equation}
p(x|\mu\text{, }\sigma)=\frac{1}{\sqrt{2\pi\sigma^{2}}}\exp\left(
-\frac{(x-\mu)^{2}}{2\sigma^{2}}\right)  \text{.} \label{Gaussian}%
\end{equation}
The probability distribution $p(x|\mu$, $\sigma)$ in (\ref{Gaussian}) is the
well-known Gaussian probability distribution. As a last step, let us calculate
the Fisher-Rao information metric associated with a statistical manifold of
Gaussians. It is known that \cite{amari},%
\begin{equation}
g_{ij}=\left(  \frac{\partial^{2}S\left(  \theta^{\prime}|\theta\right)
}{\partial\theta_{i}^{\prime}\partial\theta_{j}^{\prime}}\right)
_{\theta^{\prime}=\theta} \label{info metric}%
\end{equation}
where,%
\begin{equation}
S\left(  \theta^{\prime}|\theta\right)  =\int_{-\infty}^{+\infty}dxp\left(
x|\theta^{\prime}\right)  \log\left[  \frac{p\left(  x|\theta^{\prime}\right)
}{p\left(  x|\theta\right)  }\right]  \text{ with }\theta^{\prime}%
\equiv\left(  \mu\text{, }\sigma\right)  \text{.} \label{rel entropy}%
\end{equation}
Substituting (\ref{Gaussian}) in (\ref{rel entropy}), after some algebra, we
obtain%
\begin{equation}
S\left(  \theta^{\prime}|\theta\right)  =\log\left(  \frac{\sigma}%
{\sigma^{\prime}}\right)  +\frac{\sigma^{\prime2}}{2\sigma^{2}}+\frac{\left(
\mu^{\prime}-\mu\right)  ^{2}}{2\sigma^{2}}\text{.} \label{expl rel entropy}%
\end{equation}
Finally, substituting (\ref{expl rel entropy}) in (\ref{info metric}), we get%
\begin{equation}
g_{ij}=\left(
\begin{array}
[c]{cc}%
\frac{1}{\sigma_{2}^{2}} & 0\\
0 & \frac{2}{\sigma_{2}^{2}}%
\end{array}
\right)
\end{equation}
with a line element $ds^{2}$ given by%
\begin{equation}
ds^{2}=g_{ij}d\theta^{i}d\theta^{j}=\frac{1}{\sigma^{2}}d\mu^{2}+\frac
{2}{\sigma^{2}}d\sigma^{2}\text{.}%
\end{equation}
The Gaussian distribution is quite remarkable, it applies to a wide variety of
problems such as the distribution of errors affecting experimental data, the
distribution of velocities of molecules in gases and liquids, the distribution
of fluctuations of thermodynamical quantities, and so on. Basically, it arises
whenever we deal with macroscopic variable that is the result of adding a
large number of small independent contributions. Gaussian distributions can be
derived as the distributions that codify information about the mean (first
moment) and the variance (second moment) while remaining maximally ignorant
about everything else. Gaussian distributions are successful when third and
higher moments are irrelevant.

\pagebreak

\begin{center}
{\LARGE Chapter 3: Information-constrained dynamics, part I: Entropic
Dynamics}
\end{center}

In this Chapter, I describe (following reference \cite{catichaED}) the basic
aspects of "Entropic Dynamics", a theoretical construct developed to
investigate the possibility that Einstein's general theory of gravity (or more
generally, any classical or quantum theory of physics) may emerge as a
macroscopic manifestation of an underlying microscopic statistical structure.
ED is the starting point of the major contribution of this thesis (the IGAC,
information geometrodynamical approach to chaos) and therefore I must revisit
it. I emphasize the most relevant points that I will use in my own IGAC and
suggest further improvements that, indeed, will be the object of Chapters 6
and 7.

\section{Introduction}

In Caticha's Entropic Dynamics \cite{catichaED}, \ the author explores the
possibility that the laws of physics might be laws of inference rather than
laws of nature. He explores what sort of dynamics can one derive from
well-established rules of inference. Specifically, given relevant information
codified in the initial and the final states, the problem is to study the
trajectory that the system is expected to follow. It turns out the solution to
this problem follows from a principle of inference, the principle of maximum
entropy, and not from a principle of physics. The entropic dynamics derived
this way exhibits some remarkable formal similarities with other generally
covariant theories such as general relativity.

Dynamics is the study of changes occurring in nature. For instance,
thermodynamics deals with changes between states of equilibrium and addresses
the question of which final states can be reached from any given initial
state. Mechanics is the study of changes known as motion, chemistry considers
chemical reactions, quantum mechanics deals with transitions between quantum
states, and so on. In all of these examples, the objective is predicting or
explaining the observed changes on the basis of relevant available information
that is codified in the "states" of the system. In some cases the final state
can be predicted with certainty, in others the information available is
incomplete and we can only assign probabilities. Thermodynamics holds a very
special place among all these forms of dynamics. Thanks to the development of
statistical mechanics by Maxwell, Boltzmann, Gibbs and others, and thanks to
Jaynes' work \cite{jaynes}, thermodynamics became the first clear example of a
fundamental physical theory that could be derived from general principles of
probable inference. An appropriate choice of which states one is considering,
plus well-known principles of inference \cite{note 1}, namely, consistency,
objectivity, universality and honesty lead to the derivation of the entire
theory of thermodynamics. These principles lead to a unique set of rules for
processing information: these are the rules of probability theory \cite{cox}
and the method of maximum entropy \cite{jaynes, shore}. There are strong
indications that quantum mechanics can be deduced from principles of inference
\cite{caticha pla}. Many features of the theory follow from the correct
identification of the subject matter plus general principles of inference. If
the "fundamental" theory of quantum mechanics can be derived in this way, then
it is possible that other forms of dynamics might ultimately reflect laws of
inference rather than laws of nature. The fundamental equations of change, or
motion, or evolution would follow from probabilistic and entropic arguments in
the case that dynamics reflects laws of inference. The discovery of new
dynamical laws would be reduced to the discovery of what is the necessary
information for carrying out correct inferences. This is a very important
point. Unfortunately, this search for the right variables has always been and
remains to this day the major obstacle in the understanding of new phenomena.

In his work (ED), Caticha explores the possible connection between the
fundamental laws of physics and the theory of probable inference. He explores
the possibility that dynamics can be derived from inference and rather than
starting with a known dynamical theory and attempting to derive it, he
proceeds in the opposite direction. His ED arises simply from well-established
rules of inference! In next section, the notation is introduced, the space of
states is defined, and a brief review concerning the introduction of a natural
quantitative measure of the change involved in going from one state to another
turns the space of states into a metric space \cite{caticha IED} is presented.
(Such metric structures have been found useful in statistical inference, where
the subject is known as Information Geometry \cite{amari}, and in physics, to
study both equilibrium \cite{weinhold} and nonequilibrium thermodynamics
\cite{balian}.) Typically, once the kinematics appropriate to a certain motion
has been selected, one proceeds to define the dynamics by additional
postulates. This is precisely the option Caticha wanted to avoid: in the
dynamics developed here there are no such postulates. The equations of motion
follow from an assumption about what information is relevant and sufficient to
predict the motion.

In a previous work (irreversible entropic dynamics) \cite{caticha IED},
Caticha considered a similar problem. Assuming that the system evolves from a
given initial state to other states, he studied the trajectory that the system
is expected to follow. In this problem, the existence of a trajectory is
assumed and, in addition, it is assumed that information about the initial
state is sufficient to determine it. The application of a principle of
inference, the method of maximum entropy (ME), to the only information
available, the initial state and the recognition that motion occurred leads to
the dynamical law. The resulting "entropic" dynamics is very simple: the
system moves irreversibly and continuously along the entropy gradient.
However, the question of whether the actual trajectory is the expected one
remains unanswered and it depends on whether the information encoded in the
initial state happened to be sufficient for prediction. For many systems more
information is needed, even for those for which the dynamics is reversible. In
the reversible case, assuming that the system evolves from a given initial
state to a given final state, the objective is to study what trajectory is the
system expected to follow. Again, it is implicitly assumed that there is a
trajectory, that in moving from one state to another the system will pass
through a continuous set of intermediate states. Again, the equation of motion
follows from a principle of inference, the principle of maximum entropy, and
not from a principle of physics. (For a brief account of the ME method in a
form that is convenient for our current purpose see previous Chapter). In the
resulting "entropic" dynamics, the system moves along a geodesic in the space
of states. The geometry of the space of states is curved and possibly quite
complicated. Important features of this entropic dynamics are explored in
section 4.

\section{The Fisher-Rao information metric}

In this section, a quantitative description of the notion of change is briefly
reviewed (for more details see \cite{caticha IED}). First, change can be
measured by distinguishability since the larger the change involved in going
from one state to another, the easier it is to distinguish between them. Next,
using the ME method one assigns a probability distribution to each state. This
transforms the problem of distinguishing between two states into the problem
of distinguishing between the corresponding probability distributions. The
extent to which one distribution can be distinguished from another is given by
the distance between them as measured by the Fisher-Rao information metric
\cite{amari, fisher, rao}. Thus, change is measured by distinguishability
which is measured by distance. Let the microstates of a physical system be
labelled by $x$, and let $m(x)dx$ be the number of microstates in the range
$dx$. We assume that a state of the system (i.e., a macrostate) is defined by
the expected values $\theta^{\mu} $ of some $n_{\Theta}$ appropriately chosen
variables $\Theta^{\mu}\left(  x\right)  $ ($\mu=1$, $2$, . . . , $n_{\theta}%
$),
\begin{equation}
\left\langle \Theta^{\mu}\left(  x\right)  \right\rangle =\int dxp\left(
x\right)  \Theta^{\mu}\left(  x\right)  =\theta^{\mu}\text{ with }\mu=1\text{,
}2\text{,..., }n_{\theta}\text{.} \label{constraint equation}%
\end{equation}
A very important assumption is that the selected variables codify all the
information relevant to answering the particular questions under
investigation. Again, we emphasize there is no systematic procedure to choose
the right variables. The selection of relevant variables is made on the basis
of intuition guided by experiment. Essentially, it is a matter of trial and
error. The variables should include those that can be controlled or observed
experimentally, but there are cases where others must also be included. For
instance, the success of equilibrium thermodynamics originates from the fact
that a few variables are sufficient to describe a static situation, and being
few, these variables are easy to identify. On the other hand, in fluid
dynamics the selection is more difficult. One must include many more
variables, such as the local densities of particles, momentum, and energy,
that are neither controlled nor usually observed. The states of the system
form an $n_{\theta}$-dimensional manifold with coordinates given by the
numerical values $\theta^{\mu}$. A probability distribution $p(x|\theta)$ is
associated with each state. In order to obtain the distribution that best
reflects the prior information contained in $m(x)$ updated by the information
$\theta$, we maximize the logarithmic relative entropy%
\begin{equation}
S\left[  p|m\right]  =-\int dxp\left(  x\right)  \log\left(  \frac{p\left(
x\right)  }{m\left(  x\right)  }\right)
\end{equation}
subject to the constraints (\ref{constraint equation}). The distribution
obtained this way is%
\begin{equation}
p\left(  x|\theta\right)  =\frac{1}{Z}m\left(  x\right)  \exp\left[
\lambda_{\mu}\theta^{\mu}\left(  x\right)  \right]  \text{,}%
\end{equation}
where the partition function $Z$ and the Lagrange multipliers $\lambda_{\mu}$
are given by%
\begin{equation}
Z\left(  \lambda\right)  =\int dxm\left(  x\right)  \exp\left[  \lambda_{\mu
}\theta^{\mu}\left(  x\right)  \right]  \text{ and }-\frac{\partial\log
Z\left(  \lambda\right)  }{\partial\lambda_{\mu}}=\theta^{\mu}\text{.}%
\end{equation}
Furthermore, the change involved in going from state $\theta$ to the state
$\theta+d\theta$ can be measured by the extent to which the two distributions
can be distinguished. As discussed in \cite{amari}, except for an overall
multiplicative constant, the measure of distinguishability is given by the
"distance" $d\ell$ between $p(x|\theta)$ and $p(x|\theta+d\theta)$,%
\begin{equation}
dl^{2}=g_{\mu\nu}\left(  \theta\right)  d\theta^{\mu}d\theta^{\nu}\text{,}%
\end{equation}
where%
\begin{equation}
g_{\mu\nu}\left(  \theta\right)  =\int dxp(x|\theta)\frac{\partial\log
p(x|\theta)}{\partial\theta^{\mu}}\frac{\partial\log p(x|\theta)}%
{\partial\theta^{\nu}}%
\end{equation}
is the Fisher-Rao metric \cite{fisher, rao}. This metric is unique, it is the
\emph{only} Riemannian metric that adequately reflects the fact that the
states $\theta$ are probability distributions, not "structureless points".

\section{Entropic dynamics}

The main objective of ED\ is deriving the expected trajectory of the system,
assuming it evolves from a given initial state to a given final state. The
entropic dynamical framework implicitly assumes that there exists a trajectory
or, stated otherwise, that large changes are the result of a continuous
succession of very many small changes. Therefore, the problem of studying
large changes becomes the much simpler problem of studying small changes.
Focusing on small changes and assuming that the change in going from the
initial state $\theta_{i}$ to the final state $\theta_{f}=\theta_{i}%
+\Delta\theta$ is small enough, the distance $\Delta l$ between such states
becomes
\begin{equation}
\Delta l^{2}=g_{\mu\nu}\left(  \theta\right)  \Delta\theta^{\mu}\Delta
\theta^{\nu}\text{.}%
\end{equation}
In what follows, we explain how to find which states are expected to lie on
the trajectory between $\theta_{i}$ and $\theta_{f}$. First, in going from the
initial to the final state the system must pass through a halfway point, that
is, a state $\theta$ that is equidistant from $\theta_{i}$ and $\theta_{f}$.
Chosen the halfway state, the expected trajectory of the system is determined.
Indeed, there is nothing special about halfway states. Similarly, we could
have argued that in going from the initial to the final state the system must
first traverse a third of the way, that is, it must pass through a state that
is twice as distant from $\theta_{f}$ as it is from $\theta_{i}$. In general,
the system must pass through an intermediate states $\theta_{\xi}$ such that,
having already moved a distance $dl$ away from the initial $\theta_{i}$, there
remains a distance $\xi dl$ to be covered to reach the final $\theta_{f}$ .
Halfway states have $\xi$ $=1$, "third of the way" states have $\xi$ $=2$, and
so on. Each different value of $\xi$ provides a different criterion to select
the trajectory. If there are several ways to determine an (assumed) existing
trajectory, consistency requires that all these ways should agree. The
selected trajectory must be independent of $\xi$. Stated otherwise, the main
ED problem becomes the following: "Initially, the system is in state
$p(x|\theta_{i})$ and new information is given to us. The system has moved to
one of the neighboring states in the family $p(x|\theta_{\xi})$. The problem
becomes selecting the proper $p(x|\theta_{\xi})$". This new formulation of the
ED problem is precisely the kind of problem to be tackled using the ME method.
Following \cite{caticha fluctuations} and what was reported in Chapter 2, we
recall that the ME method is a method for processing information. It allows us
to go from an old set of beliefs, described by the prior probability
distribution, to a new set of beliefs, described by the posterior
distribution, when the available information is just a specification of the
family of distributions from which the posterior must be selected \cite{note
2}. Usually, this family of posteriors is defined by the known expected values
of some relevant variables. This is not necessary and the
information-constraints need not be linear functionals. In ED, constraints are
defined geometrically. Whenever one contemplates using the ME method, it is
important to specify which entropy should be maximized. The selection of a
distribution $p(x|\theta)$ requires that the entropies to be considered must
be of the form%
\begin{equation}
S\left[  p|q\right]  =-\int dxp\left(  x|\theta\right)  \log\left(
\frac{p\left(  x|\theta\right)  }{q\left(  x\right)  }\right)  \text{.}
\label{juve}%
\end{equation}
Equation (\ref{juve}) defines the entropy of $p\left(  x|\theta\right)  $
relative to the prior $q(x)$. The interpretation of $q(x)$ as the prior
follows from the logic behind the ME method itself. As a side remark,
following reference \cite{caticha fluctuations}, I would like to recall that
in the absence of new information there is no reason to change one's mind. The
selected posterior distribution should coincide with the prior distribution
when there are no constraints. Since the distribution that maximizes $S\left[
p|q\right]  $ subject to no constraints is $p\propto q$, we must set $q(x)$
equal to the prior. That said, let us return to our ED problem. Assuming we
know that the system is initially in state $p(x|\theta_{i})$ and we are not
given the information that the system moved. Therefore, we have no reason to
believe that any change has occurred. The prior $q(x)$ should be chosen so
that the maximization of $S\left[  p|q\right]  $ subject to no constraints
leads to the posterior $p=$ $p(x|\theta_{i})$. The correct choice is $q(x)=$
$p(x|\theta_{i})$. Instead, assuming we know that the system is initially in
state $p(x|\theta_{i})$ and we are given the information that the system moved
to one of the neighboring states in the family $p(x|\theta_{\xi})$, then the
correct selection of the posterior probability distribution is obtained by
maximizing the entropy%
\begin{equation}
S\left[  \theta|\theta_{i}\right]  =-\int dxp(x|\theta)\log\left(
\frac{p(x|\theta)}{p(x|\theta_{i})}\right)  \text{,}%
\end{equation}
subject to the constraint $\theta=\theta_{\xi}$. For the sake of simplicity,
let us write $\theta_{\xi}$ $=$ $\theta_{i}$ $+d\theta$ and $\theta_{f}%
=\theta_{i}+\Delta\theta$ so that $S[\theta_{\xi}|\theta_{i}]$ becomes%
\begin{equation}
S\left[  \theta_{i}+d\theta|\theta_{i}\right]  =-\frac{1}{2}g_{\mu\nu}\left(
\theta\right)  d\theta^{\mu}d\theta^{\nu}\text{,}%
\end{equation}
and the distances $dl_{i}$ and $dl_{f}$ from $\theta_{\xi}$ to $\theta_{i} $
and $\theta_{f}$ are defined as%
\begin{equation}
dl_{i}^{2}=g_{\mu\nu}\left(  \theta\right)  d\theta^{\mu}d\theta^{\nu
}\text{and }dl_{f}^{2}=g_{\mu\nu}\left(  \theta\right)  \left(  \Delta
\theta^{\mu}-d\theta^{\mu}\right)  \left(  \Delta\theta^{\nu}-d\theta^{\nu
}\right)  \text{.} \label{constraints squared}%
\end{equation}
In order to maximize $S[\theta_{i}+d\theta|\theta_{i}]$ under variations of
$d\theta$ subject to the constraint
\begin{equation}
\xi dl_{i}=dl_{f}\text{,} \label{constraint}%
\end{equation}
we introduce a Lagrange multiplier $\lambda$,%
\begin{equation}
\delta\left(  -\frac{1}{2}g_{\mu\nu}\left(  \theta\right)  d\theta^{\mu
}d\theta^{\nu}+\lambda\left(  \xi^{2}dl_{i}^{2}-dl_{f}^{2}\right)  \right)
=0\text{.} \label{minimi}%
\end{equation}
After some algebra, it can shown that%
\begin{equation}
d\theta^{\mu}=\eta\Delta\theta^{\mu}\text{ with }\eta\equiv\left(  1-\xi
^{2}-\frac{1}{2\lambda}\right)  ^{-1}\text{.}%
\end{equation}
The multiplier $\lambda$ and the quantity $\eta$ are determined substituting
back into the constraint (\ref{constraint}). From (\ref{constraints squared})
we obtain $dl_{i}=\eta\Delta l$ and $dl_{f}=(1-\eta)\Delta l$, and therefore
\cite{note 3}%
\begin{equation}
\eta=\frac{1}{1+\xi}\text{ and }\lambda=\frac{1}{2\xi\left(  1+\xi\right)
}\text{.}%
\end{equation}
Therefore, the intermediate state $\theta_{\xi}$ selected by the maximum
entropy method must satisfy the following relation%
\begin{equation}
dl_{i}+dl_{f}=\Delta l\text{.} \label{napoli}%
\end{equation}
The geometrical interpretation of (\ref{napoli}) is straightforward: the
triangle defined by the points $\theta_{i}$, $\theta_{\xi}$, and $\theta_{f}$
degenerates into a straight line. This is sufficient to determine a short
segment of the trajectory: all intermediate states lie on the straight line
between $\theta_{i}$ and $\theta_{f}$. The generalization beyond short
trajectories is immediate: if any three nearby points along a curve lie on a
straight line the curve is a \emph{geodesic}. This result is independent of
the arbitrarily chosen value $\xi$ so the potential consistency problem we
mentioned before does not arise. Summarizing, the answer to the ED problem is
the following \cite{catichaED}: "The expected trajectory is the geodesic that
passes through the given initial and final states". As a final remark, we
would like to point out that in ED the motion is predicted on the basis of a
"principle of inference", the principle of maximum entropy, and not from a
"principle of physics". ED is derived in an unusual way and one should expect
some unusual features. Indeed, unusual features arise as soon as one asks any
question concerning time. For example, ED determines the vector tangent to the
trajectory $\frac{d\theta^{\mu}}{dl}$, but not the actual "velocity"
$\frac{d\theta}{dl}$. This becomes clear since there is no reference to an
external time $t$ nowhere in the ED problem nor in any implicit background
information. In order to find a relation between the distance $l$ along the
trajectory and the external time $t$, additional information is required. In
conventional forms of dynamics (ED\ is not a conventional form of dynamics)
this information is implicitly encoded in a "principle of physics", in the
Hamiltonian which fixes the evolution of a system relative to external clocks.
However, the ED problem does not mention any external universe. The only clock
available is the system itself, and the problem becomes that of deciding how
this clock should be read. For instance, one of the variables $\theta^{\mu}$
could be chosen as a clock variable and it could be arbitrarily called
intrinsic time. Intrinsic time should be defined so that motion looks simple.
Intrinsic time $\tau$ may be considered as quantified change. A natural
definition for the intrinsic time $\tau$ consists in stipulating that the
system moves with unit velocity, then $\tau$ is given by the distance $l$
itself, $d\tau=dl$. A very special consequence of this definition of intrinsic
time is that intervals between events along the trajectory are not known a
priori. Intervals are determined only after the equations of motion are solved
and the actual trajectory is determined. ED\ shares this peculiar feature with
the theory of General Relativity (GR). In GR, as in ED, there is no reference
to an external time. For instance, it is known that in GR the proper time
interval along any curve between the initial and final three-dimensional
geometries of space representing the given initial and final states is only
determined after solving the Einstein equations of motion \cite{bsw}. A
serious impediment in understanding the classical theory of gravity is caused
by the absence of an external time \cite{york}, since there is no clear
understanding about which variables represent the true gravitational degrees
of freedom. This absence of an external time gives rise to problems also in
formulating a quantum theory of gravity \cite{kuchar}, because of difficulties
in defining equal-time commutators. In the following section, following
reference \cite{catichaED}, we point out some further formal similarities
between ED and GR by presenting a Lagrangian and Hamiltonian formulation of
Entropic Dynamics.

\section{Canonical formalism for entropic dynamics}

ED can be derived from an "action" principle. The trajectory of the system is
a geodesic and therefore the "action" is the length itself%
\begin{equation}
\mathcal{J}\left[  \theta\right]  =\overset{\chi_{f}}{\underset{\chi_{i}}%
{\int}}d\chi\mathcal{L}\left(  \theta\text{, }\dot{\theta}\right)  \text{,}
\label{ED action}%
\end{equation}
where $\chi$ is an arbitrary parameter along the trajectory. The Lagrangian
$\mathcal{L}\left(  \theta\text{, }\dot{\theta}\right)  $ is given by%
\begin{equation}
\mathcal{L}\left(  \theta\text{, }\dot{\theta}\right)  =\left(  g_{\mu\nu}%
\dot{\theta}^{\mu}\dot{\theta}^{\nu}\right)  ^{\frac{1}{2}}\text{ and }%
\dot{\theta}^{\mu}=\frac{d\theta^{\mu}}{d\chi}\text{.}%
\end{equation}
The action $\mathcal{J}\left[  \theta\right]  $ is invariant under
reparametrizations $\theta\left(  \chi\right)  \rightarrow\theta\left(
f\left(  \chi\right)  \right)  $ provided the end points are left unchanged,
$f\left(  \chi_{i}\right)  =\chi_{i}$ and $f\left(  \chi_{f}\right)  =\chi_{f}
$ . Indeed, when the transformation is infinitesimal, $f\left(  \chi\right)
=\chi+\varepsilon\left(  \chi\right)  $, the corresponding change in the
action $\delta\mathcal{J}$ becomes,%
\begin{equation}
\delta\mathcal{J}=\left(  g_{\mu\nu}\dot{\theta}^{\mu}\dot{\theta}^{\nu
}\right)  ^{\frac{1}{2}}\varepsilon\left(  \chi\right)  |_{\chi_{i}}^{\chi
_{f}}\text{.} \label{gigione}%
\end{equation}
This change in equation (\ref{gigione}) vanishes provided $\varepsilon\left(
\chi_{i}\right)  =\varepsilon\left(  \chi_{f}\right)  =0$. As a side remark,
we would like to point out that ED shares with GR the fact that both are
generally covariant theories. Furthermore, as emphasized in \cite{claudio},
there is an important distinction between the symmetries of a generally
covariant theory such as GR and the internal symmetries of a proper gauge
theory. The action of a generally covariant theory (such as ED) is invariant
under those reparametrizations that are restricted to map the boundary onto
itself. For proper internal gauge transformations there are no such restrictions.

Recall that the standard Hamilton's principle of least action for a
nonrelativistic particle demands extremizing the action%
\begin{equation}
\underset{t_{i}}{\overset{t_{f}}{\int}}dt\left(  \frac{m}{2}\delta_{ij}%
\frac{dx^{i}}{dt}\frac{dx^{j}}{dt}-U\left(  x\right)  \right)
\end{equation}
where $t$ is "physical" time, and the interval between initial and final
states $t_{f}-t_{i}$ is given. Furthermore, recall that Jacobi's principle of
least action for a particle with energy $E$ moving in a potential $U(x)$
determines the trajectory by extremizing the action%
\begin{equation}
\mathcal{J}\left[  x\right]  =\overset{\chi_{f}}{\underset{\chi_{i}}{\int}%
}d\chi\left(  2m\delta_{ij}\frac{dx^{i}}{d\chi}\frac{dx^{j}}{d\chi}\right)
^{\frac{1}{2}}\left(  E-U\left(  x\right)  \right)  ^{\frac{1}{2}}\text{.}%
\end{equation}
In Jacobi's principle there is no reference to any time $t$. The time interval
between initial and final states is not given, and the parameter $\chi$ is
unphysical and arbitrary. The determination of the temporal evolution along
the trajectory requires an additional constraint,%
\begin{equation}
\frac{m}{2}\delta_{ij}\frac{dx^{i}}{dt}\frac{dx^{j}}{dt}+U\left(  x\right)
=E\text{.}%
\end{equation}
Therefore, we emphasize that the ED action (\ref{ED action}) is an action of
the Jacobi type, not of the Hamiltonian type. The natural choice for a
supplementary constraint that defines $\tau$ and determines the evolution
along the trajectory is given by%
\begin{equation}
g_{\mu\nu}\frac{d\theta^{\mu}}{d\tau}\frac{d\theta^{\nu}}{d\tau}=1\text{.}
\label{conditio}%
\end{equation}
It is known that GR is also described by a Jacobi-type action \cite{brown} and
this leads to an additional formal similarity between GR and ED. In order to
explore this similarity further it is convenient to construct the canonical
Hamiltonian version of Jacobi's action. The canonical momenta $\pi_{\mu}$ are
defined as%
\begin{equation}
\pi_{\mu}=\frac{\partial\mathcal{L}}{\partial\dot{\theta}^{\mu}}=\frac
{g_{\mu\nu}\dot{\theta}^{\nu}}{\left(  g_{\alpha\beta}\dot{\theta}^{\alpha
}\dot{\theta}^{\beta}\right)  ^{\frac{1}{2}}}%
\end{equation}
and have unit magnitude,%
\begin{equation}
g^{\mu\nu}\pi_{\mu}\pi_{\nu}=1\text{.} \label{momenta magnitude}%
\end{equation}
The canonical Hamiltonian $\mathcal{H}_{\text{can}}\left(  \theta\text{, }%
\pi\right)  $ vanishes identically,%
\begin{equation}
\mathcal{H}_{\text{can}}\left(  \theta\text{, }\pi\right)  =\dot{\theta}^{\mu
}\pi_{\mu}-\mathcal{L}\left(  \theta\text{, }\dot{\theta}\right)
\equiv0\text{,}%
\end{equation}
because the Lagrangian is homogeneous of first degree in the $\dot{\theta}$'s.
From a physics point of view, this is evident since the generator of time
evolution (Hamiltonian) can be expected to vanish whenever there is no
external time with respect to which the system could possibly evolve. We are
led to consider the canonical action%
\begin{equation}
\overset{\chi_{f}}{\underset{\chi_{i}}{\int}}d\chi\left(  \dot{\theta}^{\mu
}\pi_{\mu}-\mathcal{H}_{\text{can}}\right)  =\overset{\chi_{f}}{\underset
{\chi_{i}}{\int}}d\chi\dot{\theta}^{\mu}\pi_{\mu}\text{,}%
\end{equation}
subject to the constrained variations of the momenta $\pi_{\mu}$
(\ref{momenta magnitude}). Therefore, the correct variational principle
requires to extremize the action $I\left[  \theta\text{, }\pi\text{,
}N\right]  $ given by%
\begin{equation}
I\left[  \theta\text{, }\pi\text{, }N\right]  =\overset{\chi_{f}}%
{\underset{\chi_{i}}{\int}}d\chi\left[  \dot{\theta}^{\mu}\pi_{\mu}-Nh\left(
\theta\text{, }\pi\right)  \right]
\end{equation}
where%
\begin{equation}
h\left(  \theta\text{, }\pi\right)  =\frac{1}{2}g^{\mu\nu}\pi_{\mu}\pi_{\nu
}-\frac{1}{2} \label{h function}%
\end{equation}
and $N\left(  \chi\right)  $ are Lagrange multipliers that enforce the
constraint
\begin{equation}
h\left(  \theta\text{, }\pi\right)  =0 \label{h constraint}%
\end{equation}
for each value of $\chi$. For later convenience, the overall factor of $1/2$
in (\ref{h function}) is introduced (it amounts to rescaling $N$). Variation
of $I\left[  \theta\text{, }\pi\text{, }N\right]  $ with respect to $\theta$,
$\pi$, and $N$ leads to the equations of motion,%
\begin{equation}
\dot{\pi}_{\mu}=-N\frac{\partial h}{\partial\theta^{\mu}}\text{, }\dot{\theta
}^{\mu}=N\frac{\partial h}{\partial\pi^{\mu}}\text{,}%
\end{equation}
and (\ref{h constraint}). Of course, there is no equation of motion for $N$
and it must be determined from the constraint. It follows that,%
\begin{equation}
N=\left(  g_{\mu\nu}\dot{\theta}^{\mu}\dot{\theta}^{\nu}\right)  ^{\frac{1}%
{2}}\text{,}%
\end{equation}
which, using the supplementary condition (\ref{conditio}), implies%
\begin{equation}
d\tau=Nd\chi\text{.}%
\end{equation}
The lapse function would be the analogue of $N$ in GR. Accordingly, the
analogue of (\ref{h constraint}) in GR is called the Hamiltonian constraint.
The quantity $N$ describes the increase of "intrinsic" time per unit increase
of the unphysical parameter $\chi$. In terms of $\tau$ the equations of motion
are given by%
\begin{equation}
\frac{d\pi_{\mu}}{d\tau}=-\frac{\partial h}{\partial\theta^{\mu}}\text{ and
}\frac{d\theta^{\mu}}{d\tau}=\frac{\partial h}{\partial\pi_{\mu}}\text{.}%
\end{equation}
In generally covariant theories (such as GR and ED) there is no canonical
Hamiltonian (it vanishes identically) but there are constraints.
\emph{Information-constraints play the role of generators of evolution and
change in the unorthodox entropic dynamics.}

\section{Conclusions}

\ In this Chapter, following reference \cite{catichaED}, we tried to describe
the philosophy underlying the ED theoretical construct. We emphasized that
entropic dynamics is formally similar to other generally covariant theories:
the dynamics is reversible, the trajectories are geodesics, the system
supplies its own notion of an intrinsic time, the motion can be derived from a
variational principle that turns out to be of the form of Jacobi's action
principle rather than the more familiar principle of Hamilton. Furthermore, we
pointed out that \emph{"the canonical Hamiltonian formulation of ED is an
example of a constrained information-dynamics where the
information-constraints play the role of generators of evolution"}. In
conclusion, as a general remark, it would be worthwhile emphasizing that a
reasonable physical theory must satisfy two key requirements: the first is
that it must provide us with a set of mathematical models, the second is that
the theory must identify real physical systems to which the models might
possibly apply. The ED proposed in \cite{catichaED} satisfies the first
requirement, but it fails with respect to the second. There are formal
similarities with GR and whether Einstein's theory of gravity will in the end
turn out to be an example of ED is at this point no more than a speculation. A
more definite answer may be achieved once the still unsettled problem of
identifying those variables that describe the true degrees of freedom of the
gravitational field is resolved \cite{york, kuchar}.

In what follows in this Thesis, I briefly explain the main idea that allowed
Caticha and I to make progress in such ED theoretical construct. It is known
that Caticha's ultimate goal is to develop statistical geometrodynamics. The
problem of GR is twofold: one is how geometry evolves, and the other is how
particles move in a given geometry. My work on chaos focuses on how particles
move in a given geometry and neglects the other problem (the evolution of the
geometry). The realization that there exist two separate and distinct problems
was a turning point in my research and lead to an unexpected result that I
present in the next Chapters. Especially, in Chapter 6, I will argue that ED
may lead to conventional physical theories such as Newtonian dynamics.

\pagebreak

\begin{center}
{\LARGE Chapter 4: The notion of chaos in physics}
\end{center}

The notion of chaos in classical and quantum physics is introduced. The
Zurek-Paz quantum chaos criterion of linear entropy growth (von Neumann
entropy) is described. Moreover, I briefly review the basics of the
conventional Riemannian geometric approach to chaos. Finally, the notion of
Kolmogorov-Sinai dynamical entropy and that of Lyapunov exponents is introduced.

\section{Classical Chaotic Dynamics}

Since the time of Laplace and until rather recently most physicists believed
that, given dynamic equations and initial conditions, the behavior of any
macroscopic system can be reliably predicted. This confidence in the
deterministic nature of classical physics amazingly coexisted with the
experience of a large number of phenomena indicating the opposite: fluid
turbulence, various kinds of plasma instabilities, games of chances (roulette,
for example), and so on. The lack of predictability in this cases was
attributed to "inessential" features, such as uncertainty of the initial
conditions, the influence of uncontrolled external disturbances, or the
participation of a very large number of degrees of freedom, which made
predictions practically impossible.

Although the discovery of quantum mechanics destroyed the belief in the
deterministic nature of physical systems at the microscopic scale, this was
thought to be of no consequence for macroscopic systems that are well
described by classical physics. Meanwhile, the understanding of the limited
predictive capabilities of classical physics gathered force, especially due to
the advent of electronic computers that permitted the systematic study of
nonlinear dynamical systems. The importance of the remarkable work of H.
Poincar\'{e} \cite{poincare1}, who recognized the non-integrability of even
simple dynamical systems and their chaotic properties, was not immediately
recognized. Whereas, M. Born \cite{born}, for example, believed that the lack
in long-term predictability of classical motions could be attributed to errors
in initial data, the famous study of Fermi, Pasta and Ulam \cite{fermi1}
surprisingly showed that not all nonlinear dynamical systems are characterized
by stochastic behavior.

Eventually, mainly through the efforts of mathematicians, a qualitatively new
concept of the nature of nonintegrable dynamical systems, namely local
instability of the trajectories of the majority of nonlinear systems, was
established. Numerous simple, apparently deterministic dynamical systems are
characterized by extremely irregular and ultimately unpredictable motion,
exclusively governed by the internal dynamics of the system. This chaotic
behavior, which is in no way associated with the influence of external noise
or uncertainty in the initial conditions, can be truly called "dynamical
stochasticity". Dynamical chaos is characteristic of many nonlinear dynamical
systems in different branches of physics and other sciences: astronomy,
chemistry, biology, meteorology, end even econometrics. In fact, chaotic
behavior in mesoscopic systems is often the rule rather than the exception.
This statement has a precise mathematical foundation in the famous Siegel
theorem \cite{siegel}: \textit{the nonitegrable Hamiltonians are dense among
all analytic Hamiltonians (Hamiltonians that are infinitely differentiable and
are locally described by a convergent power series }\cite{zampieri}\textit{),
but the integrable Hamiltonians are not}. This allows us to say that
nonintegrable systems are more abundant than the integrable ones.

Nonlinear dynamics was developed to its present form mainly by the efforts of
mathematicians, such as Poincar\'{e}, Birkhoff, Siegel, Kolmogorov, Arnold,
Moser, and Sinai.

As physicists we must be grateful for their far-sighted contributions, and we
still have to explore the full implications of their deep insight into the
physical world.

At the beginning of the last century, H. Poincar\'{e} \cite{poincare} observed
that a fully deterministic dynamics does not necessarily imply explicit
predictions on the evolution of a dynamical system. This can be considered a
milestone in the approach to the study of dynamical chaos. The content of the
work by H. Poincar\'{e} and J. Hadamard is much more conceptually deep and
subtle than I have resumed here. Anyway, chaos as an effect of instability of
orbits in dynamical systems has remained for a long time a sort of pure
mathematical subject.

Only in the fifties, the Kolmogorov-Arnold-Moser theorem (KAM) \cite{kam} and
the numerical experiment on a chain of nonlinearly coupled oscillators by
Fermi-Pasta-Ulam \cite{fermi} have stressed again the fundamental relevance of
dynamical chaos not only on a mathematical, but also on a physical ground.
Later on, the works by E. Lorenz \cite{lorenz}, M. Henon and C. Heiles
\cite{henon} and B. V. Chirikov \cite{chirikov} have provided new insights on
the origin of chaotic behaviors in dissipative as well as in conservative
systems. The main conceptual improvement is the observation that dynamical
chaos is not necessarily a consequence of the many degrees of freedom present
in a system; on the other hand such a system, at the same time, can display,
under certain conditions, ordered and very complex behaviors.

It is known that nonabelian gauge theories show a chaotic behavior in the
classical limit. This manifests itself mainly in the rapid divergence of gauge
fields configurations initially adjacent in configuration space which leads to
a kind of saturation; a resemblance of a thermal state. Chaotic phenomena play
an important role in the world of fundamental interactions, contributing to
properties such as quark confinement, chiral symmetry breaking, and particle
reactions at very high energy \cite{biro}

\subsection{Integrable Dynamical Systems}

The use of the $2n$-dimensional Hamiltonian phase space $\left\{  \left(
q_{i}\text{, }p_{i}\right)  \text{; }i=1\text{,..., }n\right\}  $, where
$q_{i}$, $p_{i}$ are the generalized coordinates and momenta, respectively,
provides the ideal framework for the discussion of the concepts of
integrability and local instability of trajectories. Assume that the
Hamiltonian $\mathcal{H}\left(  q_{i}\text{, }p_{i}\right)  $ is given in
terms of analytic functions of the $q_{i}$ and $p_{i}$. From Hamiltonian
equations%
\begin{equation}
\dot{q}_{i}=\frac{\partial\mathcal{H}}{\partial p_{i}}\text{, }\dot{p}%
_{i}=-\frac{\partial\mathcal{H}}{\partial q_{i}} \label{hequations}%
\end{equation}
it then follows Liouville's theorem:%
\begin{equation}
\frac{d\rho}{dt}=\frac{\partial\rho}{\partial t}+\underset{i}{\sum}\left(
\frac{\partial\rho}{\partial q_{i}}\dot{q}_{i}+\frac{\partial\rho}{\partial
p_{i}}\dot{p}_{i}\right)  =0\text{,}%
\end{equation}
where $\rho\left(  p\text{, }q\right)  $ is the constant phase space
distribution along any trajectory in phase space. It is worthwhile mentioning
that this particular property of Hamiltonian dynamics does not prevent the
occurrence of dynamical stochastic motion (chaotic motion).

The integration of (\ref{hequations}) depends on the existence, and
identification, of so-called \textit{integrals of motion}. In the Hamiltonian
formalism, the time-dependence of dynamical quantities is conveniently
expressed in terms of Poisson brackets. For some function $f\left(
q_{i}\text{, }p_{i}\text{, }t\text{ }\right)  $ the total time derivative upon
using (\ref{hequations}) is given by%
\begin{equation}
\frac{df}{dt}=\frac{\partial f}{\partial t}+\left\{  \mathcal{H}\text{,
}f\right\}  \equiv\frac{\partial f}{\partial t}+\underset{i=1}{\overset
{n}{\sum}}\left(  \frac{\partial\mathcal{H}}{\partial p_{i}}\frac{\partial
f}{\partial q_{i}}-\frac{\partial\mathcal{H}}{\partial q_{i}}\frac{\partial
f}{\partial p_{i}}\right)  \text{,}%
\end{equation}
where $\left\{  H\text{, }f\right\}  $ is called the Poisson bracket of
$\mathcal{H}$ and $f$. If the function $f$ is explicitly time-independent, it
is an \textit{integral} (or constant) \textit{of motion} if its Poisson
bracket with the Hamiltonian $\mathcal{H}$ vanishes. Evidently, for explicitly
time-independent Hamiltonians, $\mathcal{H}$ itself is a constant of motion,
i.e. the total energy of the system is conserved.

A Hamiltonian system is said to be \textit{integrable}, if there exist
precisely $n$ independent \textit{isolating} integrals of motion $I_{i}$:%
\begin{equation}
I_{i}\left(  p_{1}\text{,..., }p_{n}\text{; }q_{1}\text{,.., }q_{n}\right)
=C_{i}\text{, }i=1\text{,.,}n
\end{equation}
where $C_{i}$ denotes the constant value of $I_{i}$. The independence of the
quantities $I_{i}$ can be defined in terms of their mutual Poisson brackets:%
\begin{equation}
\left\{  I_{i}\text{, }I_{j}\right\}  =0\text{ }\forall i\text{, }j\text{.}%
\end{equation}
Usually, $\mathcal{H}$ is taken as one of the constants of motion; this then
implies the time-independence of the $I_{i}$: $\left\{  \mathcal{H}\text{,
}I_{i}\right\}  =0$.

Examples of integrable dynamical systems are all systems with a single degree
of freedom described by an analytic Hamiltonian $\mathcal{H}\left(  q\text{,
}p\right)  $ and all systems with $n$ degrees of freedom that are described by
linear equations of motion. Such systems can be reduced to $n$ decoupled
normal modes by linear transformation. There are examples of nonlinear
dynamical systems that are integrable. A well-known example of a nonlinear
integrable system is the so-called \textit{Toda chain }\cite{toda}%
\textit{\ }corresponding to a chain of particles coupled by exponential
two-body potentials. For a closed chain of three particles, the Toda
Hamiltonian is:%
\begin{equation}
\mathcal{H=}\frac{1}{2}\left(  p_{1}^{2}+p_{2}^{2}+p_{3}^{2}\right)
+e^{-\left(  q_{1}-q_{3}\right)  }+e^{-\left(  q_{2}-q_{1}\right)
}+e^{-\left(  q_{3}-q_{2}\right)  }-3\text{.}%
\end{equation}
Another familiar example of a completely integrable system is the two-body
Kepler problem. Besides the constants of motion determined by space-time
symmetry (energy and angular momentum), there exists a third integral of
motion, the Runge-Lenz vector (as a result of a hidden dynamical $O\left(
4\right)  $ symmetry). The existence of this symmetry is the reason why the
(elliptical) Kepler orbits are closed and the perihelion does not precess.

\subsection{Non-integrable (chaotic) dynamical system}

A dynamical system that is not integrable is called non-integrable. For
instance, turbulent dynamical systems are non-integrable. However, turbulence
and chaos are not synonyms! Turbulence is an example of a
\textit{spatio-temporal complexity}. Spatio-temporal complex dynamics is not
yet completely understood, both from an experimental and a theoretical point
of view. Turbulent motions are indeed chaotic, but chaotic motions need not be
turbulent. Chaos may involve only a small number of degrees of freedom, that
is it can be narrow band in space and/or time. There are numerous examples of
chaotic systems characterized by \textit{temporal complexity} but spatial
simplicity, like the Lorenz's system. Turbulence is different because it is
always complex both in space and time. Turbulent motions are not time
reversible even though it is governed by classical mechanics, i. e. Newton's
dynamics, which is time reversible.

\subsubsection{Chaos in the weather: the Lorenz model}

Sensitivity to initial conditions is what causes the seemingly unpredictable,
long-term evolution of chaotic motion, because even a tiny error in the
measurement of the initial conditions of a real dynamical system leads rapidly
to a lack of predictability of its long-term behavior. As we cannot measure
any real dynamical system with infinite precision, the long term prediction of
chaotic motion in such systems is impossible, even if we know their equations
of motion exactly.

The sensitive dependence on initial conditions of chaotic systems is more
popularly known as the \emph{butterfly effect }\cite{addison}. This phenomenon
was first discovered by Edward Lorenz during his investigation into a system
of coupled ordinary differential equations (ODEs) used as a simplified model
of 2D thermal convection. known as Rayleigh-Benard convection
\cite{landau(fluids)}. These equations are now called the Lorenz equations, or
Lorenz model.

In the Rayleigh-Benard convection there are two confining plates, a bottom
plate at temperature $T_{b}$ and a top plate at temperature $T_{t}<T_{b}$. For
small temperature differences between the two plates, heat is conducted
through the stationary fluid between the plates. However, when $T_{b}-T_{t}$
becomes large enough, buoyancy forces within the heated fluid overcome
internal fluid viscosity and a pattern of counter-rotating, steady
recirculating vortices is set up between the plates. Lorenz noticed that, in
his simplified mathematical model of Rayleigh-Benard convection, very small
differences in the initial conditions blew up and quickly led to enormous
differences in the final behavior. Lorenz reasoned that if this type of
behavior could occur in such a simple dynamical system, then it may also be
possible in a much more complex physical system involving convection: the
weather system. Thus, a very small perturbation, caused for instance by a
butterfly flapping its wings, would lead rapidly to a complete change in
future weather patterns. The Lorenz equations are%
\begin{equation}
\dot{x}=-\sigma\left(  x-y\right)  \text{, }\dot{y}=-xz+rx-y\text{, }\dot
{z}=xy-bz\text{.}%
\end{equation}
This system has two nonlinearities, the $xz$ term and the $xy$ term, and
exhibits both periodic and chaotic motion depending upon the values of the
control parameters $\sigma$, $r$ and $b$. The parameter $\sigma$ is the
Prandtl number which relates the energy losses within the fluid due to
viscosity to those due to thermal conduction; $r$ corresponds to the
dimensionless measure of the temperature difference between the plates known
as the Rayleigh number; $b$ is related to the ratio of the vertical height of
the fluid layer to the horizontal extent of the convective rolls within it.
Note also that the variables $x$, $y$, $z$ are not spatial coordinates but
rather represent the convective overturning \cite{greenT}, horizontal
temperature variation, and vertical temperature variation respectively.

\section{What is Quantum Chaos?}

The problem of quantum chaos arose from the attempts to understand the very
peculiar phenomenon of classical dynamical chaos in terms of quantum
mechanics. Preliminary investigations immediately unveiled a very deep
difficulty related to the fact that the two crucial properties of classical
mechanics necessary for dynamical chaos to occur (continuous spectrum of
motion and continuous phase space) are violated in quantum mechanics. Indeed,
the energy and frequency spectra of any quantum motion, bounded in phase
space, are always discrete. According to the existing theory of dynamical
systems such motion corresponds to the limiting case of regular motion. The
ultimate origin of this fundamental quantum property is discreteness of the
phase space itself or, in modern mathematical language, a non-commutative
geometry of the latter. This is the very basis of all quantum physics directly
related to the fundamental uncertainty principle which implies a finite size
of an elementary phase-space cell,%
\begin{equation}
\Delta p\cdot\Delta x\simeq\hslash\text{ (per freedom).}%
\end{equation}
The naive resolution of this difficulty would be the absence of any quantum
chaos. For this reason it was even proposed to use the term "quantum chaology"
\cite{berry87} which essentially means the study of the absence of chaos in
quantum mechanics. If the above conclusions were true, a sharp contradiction
would arise with the correspondence principle which requires the transition
from quantum to classical mechanics for all phenomena including the new one:
dynamical chaos. Does this really mean a failure of the correspondence
principle as some authors insist \cite{ford}? If it were so quantum chaos
would, indeed, be a great discovery since it would mean that classical
mechanics is not the limiting case of quantum mechanics but a distinct theory.
Unfortunately, there exists a less radical (but also interesting and
important) resolution of this difficulty.

A recent breakthrough in the understanding of quantum chaos has been achieved,
particularly, due to a new philosophy which, either explicitly or implicitly,
is generally accepted; namely the whole physical problem of quantum dynamics
is considered as divided into two qualitatively different parts:

\begin{description}
\item[1] proper quantum dynamics as described by specific dynamical variables,
the wavefunction $\psi\left(  t\right)  $; and

\item[2] quantum measurement including the recording of the result and hence
the collapse of the wavefunction $\psi\left(  t\right)  $.
\end{description}

The first part is described by some deterministic equation, for example, the
Schrodinger equation and naturally belongs to the general theory of dynamical
systems. The problem is well posed and this allows for extensive studies.

The second part still remains very vague to the extent that there is no common
agreement even on the question whether this is a real physical problem or an
ill-posed one so that the Copenhagen interpretation of quantum mechanics gives
satisfactory answers to all the admissible questions. In any event, there
exists as yet no dynamical description of quantum measurement including the
$\psi$-collapse.

The absence of a classical-like chaos is true for the above mentioned first
part of quantum dynamics only. Quantum measurement as far as the result is
concerned, is a random process: all quantum measurements to date are
thermodynamically irreversible. They are accompanied by an increase in entropy
of the measured system together with the measuring apparatus. This
irreversibility can be used to locate the boundary between the classical and
quantum domains. However, there are good reasons to believe that this
randomness can be interpreted as a particular manifestation of dynamical chaos
\cite{percival}.

The separation of the first part of quantum dynamics, which is very natural
from a mathematical point of view, was introduced by Schrodinger who, however,
certainly underestimated the importance of the second part in physics.

\subsection{The Zurek-Paz Quantum Chaos Criterion}

In what follows, I will briefly describe the Zurek-Paz quantum chaos criterion
of von Neumann's linear entropy growth.

\subsubsection{Introduction}

It is now well-known that nonlinearity in classical systems generically leads
to chaotic behavior. A necessary, but not sufficient, characteristic of this
is a sensitive dependence of the orbits on initial conditions. Within this
context the quantum analogue does not exist. Classical and quantum mechanics,
however, are not very different when the dynamics of classical systems is
rephrased in terms of the linear Liouville equation for phase space
distributions. The sensitive dependence on initial conditions is mirrored, in
this formulation, by the linear increase with time of the coarse-grained Gibbs
entropy which is also known as the Shannon entropy. The quantum analogue of
the Shannon entropy is the von Neumann entropy $\mathcal{S}$ $\equiv
\mathcal{S}_{\text{von Neumann}}$ defined by%
\begin{equation}
\mathcal{S=}-tr(\rho\log\rho)\text{,}%
\end{equation}
$\rho$ being the density matrix of the system. For a Hamiltonian system
unitary evolution implies%
\[
\frac{d\mathcal{S}}{dt}=0\text{.}%
\]
An ingredient which is crucial in quantum mechanics is measurement.
Classically measurement on a system can be made such that there is an
arbitrarily small disturbance on the system. In quantum mechanics this is not
the case as can be appreciated from the Heisenberg uncertainty relations. The
role of the environment in the quantum evolution of chaotic systems is
well-known \cite{sarkar}. This line of thought suggests that a more natural
way to restore the quantum classical correspondence is to consider physical
systems not as being isolated from the rest of the universe but as undergoing
constant and varied interactions with it. This is an inescapable fact of life
and must be accounted for, at least approximately, in any attempt to describe
the real quantum behavior of microscopic and macroscopic objects \cite{zurek}.
The destruction of phase coherence in a quantum system because of the
continuous monitoring of its state by internal \cite{kolovsky}\ and external
\cite{zurek} degrees of freedom is a process known as \emph{decoherence}. In
other words, decoherence is the loss of phase coherence between the set of
preferred quantum states in the Hilbert space of the system due to the
interaction with the environment.

The coupling of a quantum system to an environment turns on the decoherence
process which leads to the emergence of classicality. In conclusion,
classicality is an emergent property of an open quantum system (during a
measurement process, information gets transformed from quantum to classical).

\subsubsection{A toy model: the inverted harmonic oscillator}

In a rigorous examination of the entropy approach to the classical-quantum
correspondence problem Zurek and Paz \cite{zurek} have considered the
completely tractable model of an inverted harmonic oscillator coupled to a
high temperature (harmonic) bath. The Hamiltonian of the combined system is%
\begin{equation}
\mathcal{H}_{\text{tot}}=\mathcal{H}_{\text{system}}+\mathcal{H}%
_{\text{environment}}+\mathcal{H}_{\text{interaction}}%
\end{equation}
where $\mathcal{H}_{\text{system}}$ is the inverted oscillator Hamiltonian
\begin{equation}
\mathcal{H}_{\text{system}}\left(  p\text{, }q\right)  =\frac{p^{2}}{2}%
-\frac{\lambda^{2}q^{2}}{2}\text{,} \label{iho}%
\end{equation}
$\mathcal{H}_{\text{environment}}$ is the Hamiltonian of the chosen bath with
canonical commutation relations $\left[  q_{n}\text{, }p_{m}\right]
=i\hbar\delta_{nm}$,%
\begin{equation}
\mathcal{H}_{\text{environment}}\left(  p_{n}\text{, }q_{n}\right)
=\underset{n}{\sum}\left(  \frac{p_{n}^{2}}{2}+\frac{\omega_{n}^{2}q_{n}^{2}%
}{2}\right)  \text{,}%
\end{equation}
and $\mathcal{H}_{\text{interaction}}$ is the Hamiltonian of interaction
describing the (possibly time-dependent) coupling of the inverted harmonic
oscillator, through its position variable of each of the environmental
oscillators,%
\begin{equation}
\mathcal{H}_{\text{interaction}}\left(  q\text{, }q_{n}\right)  =-qc\left(
t\right)  \underset{n}{\sum}q_{n}\text{.}%
\end{equation}
The potential energy function in (\ref{iho}) is an inverted parabola with its
apex at the origin. It is a model of instability in classical mechanics and
the phase space dynamics governed by this Hamiltonian is an excellent model of
a hyperbolic fixed point \cite{giannoni}. It is certainly not a chaotic system
--- it lacks the folding requirement --- but the parameter $\lambda$ is
analogous to a Lyapunov exponent in a genuinely chaotic system. This is
because it induces the exponential rate of divergence (convergence) of nearby
points on the unstable (stable) manifold in its $2$-dimensional phase space.
These linear manifolds intersect at the origin, i.e. at the only fixed point
of the dynamics. Let us consider the time evolution of a particle moving in
the inverted oscillator potential $V(q)=-\frac{\lambda^{2}q^{2}}{2}$, but now
let us also consider it to be coupled through its position $q$ to the position
variables of each oscillator in the infinite set which we use as a model of a
thermal bath at a high temperature. Further choosing the distribution of
frequencies of this set to be of an Ohmic type \cite{zurek1} it is possible to
derive a master equation for the reduced density matrix, $\rho_{r}$, which
describes the state of the particle at any time. The quantum Liouville
equation reads \cite{zurek, caldeira}%
\begin{equation}
\frac{\partial\rho_{r}}{\partial t}=\frac{1}{i\hbar}\left[  H\text{, }\rho
_{r}\right]  -\gamma\left(  x-y\right)  \left(  \frac{\partial}{\partial
x}-\frac{\partial}{\partial y}\right)  \rho_{r}-\frac{D}{\hbar^{2}}\left(
x-y\right)  ^{2}\rho_{r}\text{,}%
\end{equation}
where $\rho_{r}\left(  x\text{, }y\right)  =\left\langle x\left\vert \rho
_{r}\right\vert y\right\rangle $ is the reduced density matrix in the position
representation, $D=2m\gamma k_{B}T$ and $\gamma$ describes the strength of
coupling to the environment and serves as a dissipation parameter. Upon making
the weak coupling assumption of $\gamma<<1$, Zurek and Paz have solved the
equation corresponding to the above for the Wigner function. This task is made
considerably easier by the fact that the form of the potential for the
inverted oscillator implies that all the quantum correction terms vanish
identically and the quantum Liouville equation in Wigner's representation
becomes%
\begin{equation}
\frac{\partial W}{\partial t}=-\lambda q\frac{\partial W}{\partial p}%
-p\frac{\partial W}{\partial q}+D\frac{\partial^{2}W}{\partial p^{2}}\text{.}
\label{Wigner}%
\end{equation}
For more general potentials, $3$rd or higher order derivatives of $V(q)$ would
appear in (\ref{Wigner}). On calculating the rate of change of von Neumann
entropy it can be shown that \cite{zurek}%
\begin{equation}
\frac{dS\left(  t\right)  }{dt}\approx\lambda\left(  1+ae^{-2\lambda
t}\right)  ^{-1}\overset{t\rightarrow\infty}{\rightarrow}\lambda
\end{equation}
that is,%
\begin{equation}
S\left(  t\right)  \overset{t\rightarrow\infty}{\approx}\lambda t
\label{zurek-paz criterion}%
\end{equation}
where $S\left(  t\right)  =-tr\left(  \rho_{r}\left(  t\right)  \log\rho
_{r}\left(  t\right)  \right)  $ is the von Neumann entropy of the system,
$\rho_{r}\left(  t\right)  $ is the reduced density matrix of the system at
time $t$ and $\lambda$ is the same as in (\ref{iho}). Here $a$ is a constant
dependent on the initial choice of density matrix. The quantum entropy
production rate is determined by the classical instability parameter $\lambda
$. Given that the classical Lyapunov exponent to which $\lambda$ is analogous
is equal to the Kolmogorov-Sinai (KS) entropy of the system, this is indeed a
remarkable characterization. It suggests that after a time, a quantum,
classically chaotic system loses information to the environment at a rate
determined entirely by the rate at which the classical system loses
information as a result of its dynamics, namely, the KS entropy. Notice that
the KS entropy is not really an entropy. It is an entropy per unit time, or an
"entropy rate". Further details on this last remark will appear at the end of
this Chapter.

The inverted harmonic oscillator is intended as a model of instability and, in
fact, the dynamical behavior in phase space is dominated by a hyperbolic point
at the origin. The unstable and stable directions and the rate at which
initial phase space distributions expand and contract in these directions,
respectively, are determined by $\lambda$. In this sense we call $\lambda$ an
instability parameter analogous to a Lyapunov exponent in a classical chaotic
system. Indeed, at any point on a trajectory the sum of the Lyapunov exponents
is zero. For a chaotic trajectory there must be at least two nonzero Lyapunov exponents.

The Zurek-Paz conjecture is not free from criticisms \cite{miller}. Equation
(\ref{zurek-paz criterion}) has been derived with the help of various
simplifying assumptions. Their conjecture is supported by the fact that
hyperbolic points, such as that exhibited at the origin in the inverted
oscillator model, are ubiquitous in the phase space of a chaotic system. The
inverted harmonic oscillator model is, therefore, a possible representation of
the local behavior in chaotic classical evolution.

However, there are a number of reasons why we should question any conclusions
drawn as to the implications for a real chaotic system based on so simple a
model. First, there are no quantum corrections to the Wigner function
evolution for this quadratic potential (derivatives of third and higher order
vanish, and with them the quantum corrections). The model does not allow for
these influences on the dynamics, which, though small in the presence of an
environment in comparison to the classical terms, nonetheless are generally
always present. Moreover, Zurek and Paz let the dissipation parameter
$\gamma\rightarrow0$ while keeping $D=2m\gamma k_{B}T$ constant. This means,
essentially, increasing the environmental temperature and/or the mass of the
particle. However, the dissipation term can never be dropped completely since
this would entail setting $\gamma=0$ identically, implying decoupling from the
environment and hence $D=0$ too. A third and related model-dependent
assumption concerns the choice of a thermal bath as environment. Assumed are
such features as a very large or even infinite number of degrees of freedom in
the bath, the special choice of the density of frequencies of the oscillators
which comprise it and the independent, non-interacting nature of these oscillators.

Moreover, phase-space considerations may be addressed as well. The stable and
unstable manifolds associated with all hyperbolic points in Hamiltonian
chaotic systems intersect one another and those intersection points are
associated with other hyperbolic points (hyperbolic points are sometimes
called saddle points; they are fixed points in phase space with the property
that the trajectories are hyperbolae around them \cite{tel}). In this way
homoclinic and heteroclinic points are formed (homoclinic points are
intersections of the stable and unstable manifolds of the same cycle point
while heteroclinic points are intersections of the stable and unstable
manifolds of different cycle points \cite{tel, arecchi}). The stable and
unstable manifolds of the inverted oscillator intersect only at the hyperbolic
origin in phase space. Clearly, therefore, the effect that the complicated
distribution of homoclinic points might have on the open dynamics is not taken
into account. Neither, of course, is the effect of heteroclinic points.

The inverted oscillator model is not chaotic. It has fixed stable and unstable
directions which intersect at the origin of phase space, the only fixed point
of the dynamics. Thus, it does not take into account the effect of elliptic
points, homoclinic points, heteroclinic points, stable islands or cantori
\cite{tel} (cantori are invariant Cantor sets in the irregular or stochastic
region of phase space remaining after destruction of KAM\ surfaces and create
partial barriers to transport in chaotic regions) on the open quantum dynamics
\cite{tabor}. In short, it is not a good model of the extremely complex mixed
phase spaces in which trajectories of generic Hamiltonian systems evolve.
Importantly, too, the inverted oscillator model does not take into account the
folding mechanism which, along with stretching, characterizes classical chaos.
Indeed, the fixed direction of the stable and unstable manifolds are quite
inadequate to represent the rapid change in the direction of the local stable
and unstable manifolds typically seen in genuinely chaotic systems. In
ignoring this essential ingredient of chaos one is, in effect, ignoring the
fact that the directions of squeezing and contraction change rapidly along a
typical trajectory. Farini et al. \cite{farini} have illustrated the dangers
of ignoring the folding effect by studying a driven particle in a quartic
double well potential in the absence of an environment (see \cite{habib} for a
study with an environment).

Notwithstanding these objections, however, the inverted oscillator remains a
tractable model of instability both for a closed system and for an open system
in the presence of an environment. As such, it deserves attention for the
insights it might give regarding the qualitative and maybe quantitative
behavior of genuine, open quantum analogs of classically chaotic systems.

The purpose of studying such an elementary system is to build up some degree
of intuition as to the behavior of quantum chaotic systems coupled to an
environment. A priori the claims of applicability of an inverted oscillator to
modeling a chaotic system should be treated with caution. To a certain degree,
the results for the oscillator can serve as a guide to actual quantum behavior
in chaotic systems. There is indeed some value in using the inverted harmonic
oscillator as a toy model of instability in open quantum systems. The
conjectures that follow from it should, however, be tested in more systems
that are classically chaotic.

\section{Geometry and Chaos}

The investigations on the occurrence of regular and chaotic behavior in
$N$-dimensional dynamical systems are performed with a variety of methods and
mathematical tools. Recently, this ensemble widened with the inclusion of the
Riemannian and Finslerian geometric approaches \cite{casetti, cipriani}.

As other new approaches, this tool has been suggested and applied to the study
of stability properties of general dynamical systems, in the hope to bring the
phenomenological analysis of their possibly chaotic behavior back to an
inquiry directed towards an explanation, at least qualitative, of the
mechanisms responsible for the onset of chaos. Within the framework of the
geometrical picture, this explanation was sought through a possible link
between a change in the curvature properties of the underlying manifold and a
modification of the qualitative dynamical behavior of the system. Within the
Hamiltonian approach, the ingredients needed to make chaos lie basically on
the presence of \textit{stretching} and \textit{folding} of dynamical
trajectories; i.e., in the existence of a s\textit{trong dependence on initial
conditions}, which, together with a bound on the extension of the phase space,
yield to a substantial unpredictability on the long time evolution of a
system. Usually, the strong dependence is detected looking at the occurrence
of an exponentially fast increase of the separation between initially
arbitrarily close trajectories. To have true chaos, this last property must be
however supplemented by the \textit{compactness }of the ambient space where
dynamical trajectories live, this simply in order to discard trivial
exponential growths due to the unboundedness of the volume at disposal of the
dynamical system. Stated otherwise, the folding is necessary in order to have
a dynamics actually able to mix the trajectories, making practically
impossible, after a finite interval of time, to discriminate between
trajectories which were very nearby each other at the initial time. When the
space isn't compact, even in presence of strong dependence on initial
conditions, it could be possible, in some instances (though not always), to
distinguish among different trajectories originating within a small distance
and then evolved subject to exponential instability. In the geometric
description of dynamics, the recipe to find chaos is essentially the same,
with some minor differences, which nevertheless prove sometimes to be very
relevant for the understanding of the qualitative behavior of the system. When
the geometrization procedure is accomplished, the study of dynamical
trajectories is brought back to the analysis of a geodesic flow on a suitable
manifold $\mathcal{M}$. In order to define chaos, $\mathcal{M}$ should be
compact, and geodesics on it have to deviate exponentially fast.

\section{Riemannian geometrization of Hamiltonian dynamics}

Consider a classical Hamiltonian dynamical systems with $N$ degrees of
freedom, confined in a finite volume (usually systems defined on a lattice are
considered), whose Hamiltonian is of the form%
\begin{equation}
H=\frac{1}{2}\overset{N}{\underset{i=1}{\sum}}p_{i}^{2}+U\left(
q_{1}\text{,..., }q_{N}\right)  \text{,} \label{hamiltonian}%
\end{equation}
where the $q$'s and the $p$'s are, respectively, the coordinates and the
conjugate momenta of the system. Our emphasis is on systems with a large
number of degrees of freedom. The dynamics of the system (\ref{hamiltonian})
is defined in the $2N$-dimensional phase space spanned by the $q$'s and the
$p$'s. It is possible to relate the dynamical and the statistical properties
of the system (\ref{hamiltonian}) with the geometrical and topological
properties of the phase space where the dynamical trajectories of the system
live \cite{casetti}. It turns out that as long as we consider Hamiltonians of
the form (\ref{hamiltonian}) we can restrict ourselves to the study of the
geometry and the topology of the $N$-dimensional configuration space
(actually, an enlarged configuration space with two extra dimensions may also
be considered) without losing information. In fact, the dynamical trajectories
can be seen as geodesics of the configuration space, provided the latter has
been endowed with a suitable metric.

A Hamiltonian system whose kinetic energy is a quadratic form in the
velocities is referred to as a natural Hamiltonian system. Every Newtonian
system, that is a system of particles interacting through forces derived from
a potential, i.e. of the form (\ref{hamiltonian}), belongs to this class. The
trajectories of a natural system can be seen as geodesics of a suitable
Riemannian manifold. This classical result is based on a variational
formulation of dynamics. In fact Hamilton's principle states that the motions
of a Hamiltonian system are the extrema of the functional (Hamiltonian action
$\mathcal{I}$)%
\begin{equation}
\mathcal{I}=\int\mathcal{L}dt
\end{equation}
where $\mathcal{L}$ is the Lagrangian function of the system, and the
geodesics of a Riemannian manifold are the extrema of the length functional%
\begin{equation}
l=\int ds
\end{equation}
where $s$ is the arc-length parameter. Once a connection between length and
action is established, by means of a suitable choice of the metric, it will be
possible to identify the geodesics with the physical trajectories.

\subsection{Geometry and dynamics}

Even if we restrict ourselves to the case of natural systems, the Riemannian
formulation of classical dynamics is far from unique. There are many possible
choices for the ambient space and its metric. The most commonly known choice
--- dating back to the nineteenth century --- is the so-called Jacobi metric
on the configuration space of the system. Actually this was the geometric
framework of Krylov's work \cite{krilov}. There are other possibilities, for
instance a metric originally introduced by Eisenhart on an enlarged
configuration space-time \cite{eisenhart}, but this will not be discussed
here. The choice of the metric to be used will be dictated mainly by
convenience. These choices certainly do not contain all the possibilities of
geometrizing conservative dynamics. For instance, with regard to systems whose
kinetic energy is not quadratic in the velocities --- the classical example is
a particle subject to conservative as well as velocity-dependent forces, such
as the Lorentz force --- it is impossible to give a Riemannian geometrization,
but becomes possible in the more general framework of a Finsler geometry
\cite{cipriani}. However, we will not consider this here, and restrict
ourselves to standard Hamiltonian systems.

\subsubsection{The Jacobi metric tensor}

Consider an autonomous dynamical system, i.e., a system with interactions
which do not explicitly depend on time, whose Lagrangian can be written as%
\begin{equation}
\mathcal{L}=T-U=\frac{1}{2}m_{ij}\dot{q}_{i}\dot{q}_{j}-U\left(
q_{1}\text{,....,}q_{N}\right)  \text{,}%
\end{equation}
where the dot stands for a derivative with respect to the parameter on which
the $q$'s depend (such a parameter is the time $t$ here, but could also be the
arc-length $s$).

The Hamiltonian $\mathcal{H}=T+U$ is an integral of motion, whose value, the
energy $E$, is a conserved quantity. Hence Hamilton's principle can be cast in
Maupertuis' form \cite{arnold}: the natural motions of the system are the
stationary paths in the configuration space $\Gamma$ for the functional%
\begin{equation}
\mathcal{F}=\underset{\gamma\left(  t\right)  }{\int}p_{i}dq^{i}%
=\underset{\gamma\left(  t\right)  }{\int}\frac{\partial\mathcal{L}}%
{\partial\dot{q}^{i}}\dot{q}^{i}dt
\end{equation}
among all the isoenergetic curves, i.e. the curves $\gamma\left(  t\right)  $
connecting the initial and final points parametrized so that the Hamiltonian
$\mathcal{H}(p$, $q)$ is a constant equal to the energy $E$. The fact that the
curves must be isoenergetic with energy $E$ implies that the accessible part
of the configuration space is not the whole $\Gamma$, but only the subspace
$\Gamma_{E}\subset\Gamma$ defined by%
\begin{equation}
\Gamma_{E}=\left\{  q\in\Gamma:U\left(  q\right)  \leq E\right\}  \text{.}%
\end{equation}
In fact a curve $\gamma^{\prime}$ that lies outside $\Gamma_{E}$ will never be
parametrizable in such a way that the energy is $E$, because $\gamma^{\prime}$
will then pass through points where $U>E$ and the kinetic energy is positive.

The kinetic energy $T$ is a homogeneous function of degree two in the
velocities, hence Euler's theorem implies that%
\begin{equation}
2T=\dot{q}^{i}\frac{\partial\mathcal{L}}{\partial\dot{q}^{i}}\text{,}%
\end{equation}
and Maupertuis' principle reads as%
\begin{equation}
\delta\mathcal{F}=\delta\int2Tdt=0\text{.} \label{mp}%
\end{equation}
The configuration space $\Gamma$ of a dynamical system with $N$ degrees of
freedom has a differentiable manifold structure, and the Lagrangian
coordinates ($q_{1}$, . . . , $q_{N}$) can be regarded as local coordinates on
$\Gamma$. The latter becomes a Riemannian manifold once a proper metric is
defined. Consider systems of the form (\ref{hamiltonian}) where the kinetic
energy matrix is given by $m_{ij}$. If we write%
\begin{equation}
g_{ij}=2\left[  E-U\left(  q\right)  \right]  m_{ij}\text{,} \label{mt}%
\end{equation}
then, recalling that $T=\frac{1}{2}m_{ij}\frac{dq^{i}}{dt}\frac{dq^{j}}{dt}$,
equation (\ref{mp}) becomes%
\begin{equation}
0=\delta\int2Tdt=\delta\int\left(  g_{ij}\dot{q}^{i}\dot{q}^{j}\right)
^{\frac{1}{2}}dt=\delta\int ds\text{,}%
\end{equation}
so that the motions are the geodesics of $\Gamma$ provided $ds$ is the
arc-length element, i.e., the metric on $\Gamma$ is given by the tensor whose
components are just the $g_{ij}$ defined in (\ref{mt}). This metric is
referred to as the \textit{Jacobi metric}, and its arc-length element is%
\begin{equation}
ds^{2}\equiv g_{ij}dq^{i}dq^{j}=2\left[  E-U\left(  q\right)  \right]
\frac{dq^{i}}{dt}\frac{dq_{i}}{dt}dt^{2}=4\left[  E-U\left(  q\right)
\right]  ^{2}dt^{2}\text{.} \label{le}%
\end{equation}
The geodesic equations written in the local coordinate frame ($q^{1}$, . . . ,
$q^{N}$) are
\begin{equation}
\frac{D\dot{\gamma}}{ds}\equiv\frac{d^{2}q^{i}}{ds^{2}}+\Gamma_{jk}^{i}%
\frac{dq^{j}}{ds}\frac{dq^{k}}{ds}=0\text{,} \label{geodesic equation(4)}%
\end{equation}
where $D/ds$ is the covariant derivative along the curve $\gamma\left(
s\right)  $, $\dot{\gamma}=dq/ds$ is the velocity vector of the geodesic and
the $\Gamma_{jk}^{i}$ are the Christoffel symbols. Using the definition of the
Christoffel symbols, it is straightforward to show that
(\ref{geodesic equation(4)}) becomes%
\begin{equation}
\frac{d^{2}q^{i}}{ds^{2}}+\frac{1}{2\left(  E-U\right)  }\left[
2\frac{\partial\left(  E-U\right)  }{\partial q_{j}}\frac{dq^{j}}{ds}%
\frac{dq^{i}}{ds}-g^{ij}\frac{\partial\left(  E-U\right)  }{\partial q_{j}%
}g_{km}\frac{dq^{k}}{ds}\frac{dq^{m}}{ds}\right]  =0\text{,}%
\end{equation}
whence, using (\ref{le}), Newton's equations are recovered,%
\begin{equation}
\frac{d^{2}q^{i}}{dt^{2}}=-\frac{\partial U}{\partial q_{i}}\text{.}%
\end{equation}
Note that the Jacobi metric is obtained by a conformal change of the kinetic
energy metric $m_{ij}$. In fact the general result for the Riemannian
geometrization of natural Hamiltonian dynamics \cite{ong} states that given a
dynamical system on a Riemannian manifold $(\Gamma$, $m)$, i.e., a dynamical
system whose Lagrangian is%
\begin{equation}
\mathcal{L}=\frac{1}{2}m_{ij}\dot{q}_{i}\dot{q}_{j}-U\left(  q\right)
\text{,}%
\end{equation}
then it is always possible to find a conformal transformation of the metric,%
\begin{equation}
g_{ij}=\Phi\left(  q\right)  m_{ij}%
\end{equation}
such that the geodesics of $(\Gamma$, $g)$ are the trajectories of the
original dynamical system; this transformation is defined by%
\begin{equation}
\Phi\left(  q\right)  =E-U\left(  q\right)  \text{.}%
\end{equation}

\subsection{Stability and curvature}

The study of the stability of the trajectories of a dynamical system finds a
natural framework in the geometrization of the dynamics since it links the
latter with the stability of the geodesics; this stability is completely
determined by the curvature of the manifold, as shown below.

Studying the stability of the dynamics means determining the evolution of
perturbations of a given trajectory. This implies that one should follow the
evolution of the linearized (tangent) flow along the reference trajectory. For
a Newtonian system, writing the perturbed trajectory as%
\begin{equation}
\tilde{q}^{k}\left(  t\right)  =q^{k}\left(  t\right)  +\eta^{k}\left(
t\right)  \text{,}%
\end{equation}
substituting this expression in the equations of motion%
\begin{equation}
\ddot{q}^{k}+\frac{\partial U\left(  q\right)  }{\partial q^{k}}=0\text{,}%
\end{equation}
and retaining terms up to first order in the $\eta$'s, one finds that the
perturbation obeys the so-called \textit{tangent dynamics equation} which
reads as%
\begin{equation}
\ddot{\eta}^{k}+\left(  \frac{\partial^{2}U\left(  q\right)  }{\partial
q^{k}\partial q^{l}}\right)  _{q^{k}=q^{k}\left(  t\right)  }\eta
^{l}=0\text{.}%
\end{equation}
This equation should be solved together with the dynamics in order to
determine the stability or instability of the trajectory: when the norm of the
perturbations grows exponentially, the trajectory is unstable, otherwise it is
stable. Let us now translate the stability problem into geometric language. By
writing, in close analogy to what has been done above in the case of dynamical
systems, a perturbed geodesic as%
\begin{equation}
\tilde{q}^{k}\left(  s\right)  =q^{k}\left(  s\right)  +J^{k}\left(  s\right)
\text{,}%
\end{equation}
and then inserting this expression in the equation for the geodesics
(\ref{geodesic equation(4)}), one finds that the evolution of the perturbation
vector $J$ is given by the following equation:%
\begin{equation}
\frac{D^{2}J^{k}}{ds^{2}}+R_{lmn}^{k}\frac{dq^{l}}{ds}J^{m}\frac{dq^{n}}%
{ds}=0\text{,} \label{jlc}%
\end{equation}
where $R_{lmn}^{k}$ are the components of the Riemann curvature tensor.
Equation (\ref{jlc}) is referred to as the Jacobi equation, and the tangent
vector field $J$ as the Jacobi field. This equation was first studied by
Levi-Civita and is also often referred to as the equation of Jacobi and Levi-Civita.

The remarkable fact is that the evolution of $J$ --- and then the stability or
instability of the geodesic --- is completely determined by the curvature of
the manifold. Therefore, if the metric is induced by a physical system, as in
the case of Jacobi or Eisenhart metrics, such an equation links the stability
or instability of the trajectories to the curvature of the ambient manifold.

\subsection{Mechanical manifolds and curvature}

In this subsection, we have to give explicit expressions for the curvature of
the mechanical manifolds, i.e., of those manifolds whose Riemannian structure
is induced by the dynamics via the Jacobi or the Eisenhart metric.

We already observed that the Jacobi metric is a conformal deformation of the
kinetic-energy metric, whose components are given by the kinetic energy matrix
$m_{lm}$ . In the case of systems whose kinetic energy matrix is diagonal,
this means that the Jacobi metric is conformally flat. This greatly simplifies
the computation of curvatures. It is convenient to define then a symmetric
tensor $A_{lm}$ whose components are \cite{ong}%
\begin{equation}
A_{lm}=\frac{N-2}{4\left(  E-U\right)  ^{2}}\left[  2\left(  E-U\right)
\partial_{l}\partial_{m}U+3\partial_{l}U\partial_{m}U-\frac{\delta_{lm}}%
{2}\left\vert \nabla U\right\vert ^{2}\right]  \text{,}%
\end{equation}
where $U$ is the potential, $E$ is the energy, and $\nabla$ and $\left\vert
{}\right\vert $ stand for the Euclidean gradient and norm, respectively. The
curvature of $(\Gamma_{E}$, $g_{J})$ can be expressed through $A_{lm}$. In
fact, the components of the Riemann tensor are%
\begin{equation}
R_{ijkm}=\frac{1}{N-2}\left[  A_{jk}\delta_{im}-A_{jm}\delta_{ik}+A_{im}%
\delta_{jk}-A_{ik}\delta_{jm}\right]  \text{.}%
\end{equation}
By contraction of the first and third indices, we obtain the Ricci tensor,
whose components are%
\begin{equation}
R_{lm}=\frac{N-2}{4\left(  E-U\right)  ^{2}}\left[  2\left(  E-U\right)
\partial_{l}\partial_{m}U+3\partial_{l}U\partial_{m}U\right]  +\frac
{\delta_{lm}}{4\left(  E-U\right)  ^{2}}\left[  2\left(  E-U\right)  \Delta
U-\left(  N-4\right)  \left\vert \nabla U\right\vert ^{2}\right]  \text{,}%
\end{equation}
and by a further contraction we obtain the scalar curvature
\begin{equation}
\mathcal{R}=\frac{N-1}{4\left(  E-U\right)  ^{2}}\left[  2\left(  E-U\right)
\Delta U-\left(  N-6\right)  \left\vert \nabla U\right\vert ^{2}\right]
\text{,}%
\end{equation}
where $\Delta=\nabla^{2}$. To summarize, the dynamical trajectories of a
Hamiltonian system of the form (\ref{hamiltonian}) can be seen as geodesics of
the configuration space once a suitable metric is defined. The general
relationship which holds between dynamical and geometrical quantities
regardless of the precise choice of the metric can be sketched as follows:%
\begin{equation}%
\begin{tabular}
[c]{ll}%
\textbf{Dynamics} & \textbf{Geometry}\\
$t$-time & $s$-arclength\\
$U$-potential energy & $g$-metric\\
$\partial U$-force & $\Gamma$-Christoffel symbols\\
$\partial^{2}U$, $\left(  \partial U\right)  ^{2}$-"curvature" of the
potential & $\mathcal{R}$-curvature of the manifold
\end{tabular}
\end{equation}
Furthermore, the stability of the dynamical trajectories can be mapped onto
the stability of the geodesics, which is completely determined by the
curvature of the manifold.

\subsection{Integrability and Killing Vectors}

In the Riemannian geometrodynamical approach to chaos (Jacobi
geometrodynamics), the strategy consists in making use of the Hamiltonian
formulation of the dynamical system and then in reducing the dynamics to a
geodesic flow. This reduction is performed at the level of the least action
principle. The most fascinating feature of this approach is that the problem
(often very complicated) of dynamics is reduced to geometrical properties of a
single object --- the manifold on which geodesic flow is induced. In the
Jacobi reformulation, all of the dynamical information is collected into a
single geometric object in which all the available manifest symmetries are retained.

For example, the sensitive dependence of trajectories on initial conditions,
which is a key ingredient of chaos, can be investigated starting from the
equation of geodesic equation. The integrability of the system, instead, is
connected with the existence of Killing vectors and tensors on this manifold
\cite{biesiada, uggla}. Consider an $n$-dimensional manifold $\mathcal{M}$
with metric tensor $g_{\mu\nu}\left(  x\right)  $. Any $4$-vector $\xi
_{\alpha}\left(  x\right)  $ that satisfies the Killing equation%
\begin{equation}
\mathcal{L}_{\xi}g\equiv\xi_{\alpha\text{; }\beta\text{ }}+\xi_{\beta\text{;
}\alpha\text{ }}=0 \label{Killing equation}%
\end{equation}
is said to form a Killing vector of the metric $g_{\mu\nu}\left(  x\right)  $.
The quantity $\mathcal{L}_{\xi}g$ is the Lie derivative of the metric tensor
$g$ along $\xi$ while the semi-colon in (\ref{Killing equation}) represents
the standard covariant derivative on curved manifolds. For the Lie derivative
$\mathcal{L}_{\xi}g$ to vanish, requires that the geometry to be unchanged as
one moves in the $\xi$-direction, that is, $\xi$ represents a direction of
symmetry of $\mathcal{M}$. \emph{Killing vectors provide the first integrals
of the geodesic equation}. Killing equations in general are enormously
difficult to solve. Construction of Killing vectors is a relatively simple
task in the case of conformally flat spaces (which is the case in the majority
of mechanical problems). For instance, a compact manifold $\mathcal{M}$ with
negative Ricci curvatures has no nontrivial Killing vector field (Theorem of
Bochner, \cite{frankel}). As a simple example, consider the Poincar\'{e} upper
half plane $\mathcal{M}_{\text{Poincar\'{e}}}=\left\{  \left(  x\text{,
}y\right)  :y>0\right\}  $ with the Poincar\'{e} line element $ds^{2}=\frac
{1}{y^{2}}\left(  dx^{2}+dy^{2}\right)  $. Since the metric coefficients are
independent of $x$, $\xi\equiv\frac{\partial}{\partial x}$ is a Killing vector
field. The vector $\xi$ has a length $\left\Vert \xi\right\Vert $ that tends
to infinity as we approach the $x$-axis $\left(  y\rightarrow0\right)  $.

An $n$-dimensional manifold $\mathcal{M}$ is said to be maximally symmetric if
it has $\frac{n\left(  n+1\right)  }{2}$ Killing vectors. The most familiar
examples of maximally symmetric spaces of Euclidean signatures are the
$n$-dimensional Euclidean space $%
\mathbb{R}
^{n}$ and the $n$-dimensional spheres $S^{n}$. \ For Euclidean signatures, the
flat maximally symmetric spaces are planes or appropriate higher-dimensional
generalizations, while the positively curved ones are spheres. Maximally
symmetric Euclidean spaces of negative curvature are hyperboloids. There are,
of course, maximally symmetric manifolds with Lorentzian signatures. The
maximally symmetric spacetime with $\mathcal{R}=0$ is simply Minkowski space.
The positively curved maximally symmetric spacetime is called de Sitter space,
while that with negative curvature is labeled anti-de Sitter space. In
particular, any space with vanishing curvature tensor is maximally symmetric;
the converse, however, is not true. If a manifold is maximally symmetric, the
curvature is the same everywhere and the same in every direction. Hence, if we
know the curvature of a maximally symmetric space at one point, we know it
everywhere. Indeed, there are only a small number of possible maximally
symmetric spaces; they are classified by scalar curvature $\mathcal{R}$ (which
will be constant everywhere), the dimensionality $n$, the metric signature,
and perhaps some discrete pieces of information relating to the global topology.

In any maximally symmetric space $\mathcal{M}$ , at any point, in any
coordinate system,%
\begin{equation}
R_{\mu\nu\rho\sigma}=\frac{\mathcal{R}}{n\left(  n-1\right)  }\left(
g_{\mu\rho}g_{\nu\sigma}-g_{\mu\sigma}g_{\nu\rho}\right)  \text{.}%
\end{equation}
It may be useful to introduce the Weyl projective curvature tensor $W_{\mu
\nu\rho\sigma}$ defined as,%
\begin{equation}
W_{\mu\nu\rho\sigma}\overset{\text{def}}{=}R_{\mu\nu\rho\sigma}-\frac
{\mathcal{R}}{n\left(  n-1\right)  }\left(  g_{\mu\rho}g_{\nu\sigma}%
-g_{\mu\sigma}g_{\nu\rho}\right)  \text{.}%
\end{equation}
The Weyl projective curvature tensor should not to be confused with Weyl's
conformal curvature tensor \cite{de felice}. Weyl's projective tensor
$W_{\mu\nu\rho\sigma}$ measures the deviation from isotropy of a given
manifold. For an isotropic manifold $W_{\mu\nu\rho\sigma}=0$. As a final
remark, we emphasize that for a maximally symmetric manifold (isotropic
manifold), the following relations among the scalar curvature $\mathcal{R}$,
Ricci curvature tensor $R_{\mu\nu}$ and Gaussian scalar curvature
$\mathcal{K}$ follow \cite{de felice},%
\begin{equation}
\mathcal{R}\overset{\text{def}}{=}R_{\mu\nu\rho\sigma}g^{\mu\rho}g^{\nu\sigma
}=\underset{\rho\neq\sigma}{%
{\displaystyle\sum}
}\mathcal{K}\left(  e_{\rho}\text{, }e_{\sigma}\right)  =n\left(  n-1\right)
\mathcal{K}\text{,} \label{chicobombo}%
\end{equation}
and,%
\begin{equation}
R_{\mu\nu}=\left(  n-1\right)  \mathcal{K}g_{\mu\nu}\text{.}%
\end{equation}
The quantities $\mathcal{K}\left(  e_{\rho}\text{, }e_{\sigma}\right)  $ in
(\ref{chicobombo}) are the sectional curvatures of planes spanned by pairs of
orthonormal basis elements. For two arbitrary vectors $a=a^{\mu}e_{\mu}$ and
$b=b^{\nu}e_{\nu}$, the sectional curvature $\mathcal{K}\left(  a\text{,
}b\right)  $ is defined as \cite{weinberg},%
\begin{equation}
\mathcal{K}\left(  a\text{, }b\right)  \overset{\text{def}}{=}\frac{R_{\mu
\nu\rho\sigma}a^{\mu}b^{\nu}a^{\rho}b^{\sigma}}{\left(  g_{\mu\sigma}%
g_{\nu\rho}-g_{\mu\rho}g_{\nu\sigma}\right)  a^{\mu}b^{\nu}a^{\rho}b^{\sigma}%
}\text{.}%
\end{equation}

\section{Lyapunov exponents and the Kolmogorov-Sinai entropy}

In strict mathematical terms, chaotic motion is defined in terms of the
long-term exponential divergence of neighboring trajectories in phase space.
Neighboring orbits of integrable systems, on the other hand, are either
performing stable oscillations around each other or diverge at most as a
finite power of time.

The rate of exponential divergence is quantitatively measured by the positive
Lyapunov exponents $\lambda_{i}>0$. These exponents can be introduced in the
context of general dynamical systems governed by the first-order differential
equations,%
\begin{equation}
\dot{x}_{i}=F_{i}\left(  x_{1}\text{,..., }x_{N}\right)  \text{,
}i=1\text{,...,}N\text{.} \label{fode}%
\end{equation}
Given a solution $\tilde{x}_{i}\left(  t\right)  $ of (\ref{fode}), we can
linearize the equations of motion around this reference orbit and obtain a set
of linear differential equations for the deviations $\delta x_{i}\left(
t\right)  =x_{i}\left(  t\right)  -\tilde{x}_{i}\left(  t\right)  $:%
\begin{equation}
\frac{d\left(  \delta x_{i}\left(  t\right)  \right)  }{dt}=\underset
{j=1}{\overset{N}{\sum}}\delta x_{j}\left(  t\right)  \left(  \frac{\partial
F_{i}}{\partial x_{j}}\right)  _{x_{j}=\tilde{x}_{j}\left(  t\right)
}\text{.}%
\end{equation}
The length of the vector $\delta\vec{x}\left(  t\right)  $,%
\begin{equation}
d\left(  t\right)  =\left(  \underset{i\text{, }j=1}{\overset{N}{\sum}}\left[
\delta_{ij}\delta x_{i}\left(  t\right)  \delta x_{j}\left(  t\right)
\right]  \right)  ^{\frac{1}{2}}=\left(  \underset{k=1}{\overset{N}{\sum}%
}\left[  \delta x_{j}\left(  t\right)  \right]  ^{2}\right)  ^{\frac{1}{2}%
}\text{,}%
\end{equation}
provides a measure of the divergence of the two neighboring trajectories
$\tilde{x}_{i}\left(  t\right)  $ and $x_{i}\left(  t\right)  $. The maximal
Lyapunov exponent $\lambda_{1}$ is defined as the long-time average of its
logarithmic growth rate:%
\begin{equation}
\lambda_{1}=\underset{t\rightarrow\infty}{\lim}\underset{d\left(  0\right)
\rightarrow0}{\lim}\frac{1}{t}\log\left(  \frac{d\left(  t\right)  }{d\left(
0\right)  }\right)  \text{.}%
\end{equation}
The double limit is required because the accessible phase space is usually
bounded, and hence $d\left(  t\right)  $ cannot continue to grow forever,
given a fixed initial distance $d\left(  0\right)  $. Regions of phase space
for which $\lambda_{1}>0$ exhibit sensitive dependence on the initial
conditions. An infinitesimal change in the initial data results in macroscopic
deviations after a sufficiently long time:%
\begin{equation}
d\left(  t\right)  \overset{t\rightarrow\infty}{\approx}d\left(  0\right)
\exp\left(  \lambda_{1}t\right)  \text{.}%
\end{equation}
The exponential instability of motion means positive maximal Lyapunov exponent
$\lambda_{1}>0$.

An \textit{attractor} is a subset of the manifold $\mathcal{M}$ toward which
almost all sufficiently close trajectories converge asymptotically, covering
it densely as the time goes on \cite{tel}. \textit{Strange attractors} are
called \textit{chaotic attractors}. Chaotic attractors have at least one
finite positive Lyapunov exponent. On the other hand, random (noisy)
attractors have an infinite positive Lyapunov exponent, as no correlation
exists between one point on the trajectory and the next (no matter how close
they are).

If we calculate the Lyapunov exponent for orthogonal directions of maximum
divergence in phase space, we obtain a set of Lyapunov exponents $(\lambda
_{1}$,...., $\lambda_{n})$\ where $n$\ is the dimension of the phase space.
This set of Lyapunov exponents is known as the Lyapunov spectrum and is
usually ordered from the largest positive Lyapunov exponent $\lambda_{1}$,
down to the largest negative exponent, $\lambda_{n}$, i. e. maximum divergence
to maximum convergence.

The reason why the exponentially unstable motion is called chaotic is that
almost all trajectories are unpredictable in the following sense: according to
the Alekseev-Brudno theorem \cite{alekseev} in the algorithmic theory of
dynamical systems, the information $I\left(  t\right)  $ associated with a
segment of a trajectory of length $\left\vert t\right\vert $ is equal
asymptotically to \cite{pesin}%
\begin{equation}
h_{KS}=\underset{\left\vert t\right\vert \rightarrow\infty}{\lim}%
\frac{I\left(  t\right)  }{\left\vert t\right\vert }=\underset{j}{\sum}%
\lambda_{j}^{\left(  +\right)  } \label{KS entropy}%
\end{equation}
where $\underset{j}{\sum}\lambda_{j}^{\left(  +\right)  }$ is the sum of all
positive Lyapunov exponents and $h_{KS}$ is the so-called Kolmogorov-Sinai
dynamical entropy (KS entropy; indeed, $h_{KS}$ is not really an entropy but
an entropy per unit time, or an "entropy rate"), a statistical indicator of
chaos\textit{.} The quantity $I\left(  t\right)  $ in (\ref{KS entropy}) is
formally called the Kolmogorov algorithmic complexity \cite{kolmogorov-c} and,
in other words, we may say that the Alekseev-Brudno theorem states that
"\emph{the KS entropy measures the algorithmic complexity of classical
trajectories}" \cite{benattaman}. In computer science, the Kolmogorov
complexity $I_{\mathcal{U}}\left(  x\right)  $ of a string $x$ with respect to
a universal computer $\mathcal{U}$ (Turing machine) is defined as
\cite{benattaman},%
\begin{equation}
I_{\mathcal{U}}\left(  x\right)  \overset{\text{def}}{=}\underset{p}{\min
}\left\{  l\left(  x\right)  :\mathcal{U}\left(  p\right)  =x\right\}
\text{,}%
\end{equation}
and it represents the minimum length over all binary programs $p$ that print
$x$ and halt. Thus, $I_{\mathcal{U}}\left(  x\right)  $ is the shortest
description length of $x$ over all descriptions interpreted by the computer
$\mathcal{U}$. Equation (\ref{KS entropy}) shows that in order to predict each
new segment of a chaotic trajectory, one needs an additional information
proportional to the length of this segment and independent of the full
previous length of trajectory (trajectories correspond to infinitely long
strings $x$). This means that this information cannot be extracted from
observation of the previous motion, even an infinitely long one! If the
instability is not exponential but, for example, only a power law, then the
required information per unit time is inversely proportional to the full
previous length of the trajectory and, asymptotically, the prediction becomes
possible. The important condition $h_{KS}>0$, which characterizes chaotic
motion, is not invariant with respect to the change of time variable. All
Lyapunov exponents of an integrable system are zero. Conversely, the existence
of a single positive Lyapunov exponent demonstrates the nonintegrability of a
dynamical system.

Chaos is characterized by the positivity of at least one Lyapunov exponent of
the system and therefore the significance of the concept of chaos depends
essentially on the invariance of the Lyapunov exponents. It is known that
under Lorentz transformations, $\lambda$ and $h_{KS}$ change, but their
positivity is preserved for chaotic systems \cite{zheng}. Under Rindler
transformations \cite{birrell}, $\lambda$ and $h_{KS}$ change in such a way
that systems, which are chaotic for an accelerated Rindler observer, can be
nonchaotic for an inertial Minkowski observer. Therefore, the concept of chaos
is observer-dependent \cite{zheng}. Furthermore, a Lyapunov exponent (which is
a "per time" measure of having exponential separation of nearby trajectories
in "time") is not invariant under transformations of the "time" coordinate!
The fact that the Lyapunov exponents are strongly gauge dependent quantities
leads to additional problems connected with the characterization of chaos,
especially in general relativity (where a gauge invariant measure of
chaoticity is still missing). Therefore, the risk of inventing chaotic
solutions in an artificial way is present and, because of that, extra care is
needed in order to characterize "true chaos" \cite{rugh-le}.

\pagebreak

\begin{center}
{\LARGE Chapter 5: Curvature, entropy and Jacobi fields}
\end{center}

In this Chapter, we study chaos in the context of two models the dynamics of
which is entropic dynamics on curved statistical manifolds. Two chaotic
entropic dynamical models are considered. The geometric structure of the
statistical manifolds underlying these models is studied. It is found that in
both cases, the resulting metric manifolds are negatively curved. Moreover,
the geodesics on each manifold are described by hyperbolic trajectories. A
detailed analysis based on the Jacobi-Levi-Civita equation for geodesic spread
(JLC equation) is used to show that the hyperbolicity of the manifolds leads
to chaotic exponential instability. A comparison between the two models leads
to a relation among scalar curvature of the manifold ($\mathcal{R}$), Jacobi
field intensity ($J$) and information geometrodynamical entropy (IGE,
$S_{\mathcal{M}}$). The IGE is a convenient new tool we introduce to study
chaotic entropic dynamics. We propose the IGE entropy as a new measure of
chaoticity. These three quantities, $\mathcal{R}$, $J$, and $\mathcal{S}%
_{\mathcal{M}}$ are suggested as useful indicators of chaoticity on curved
statistical manifolds. Indeed, in analogy to the Zurek-Paz quantum chaos
criterion (in its classical reversible limit), a classical
information-geometric chaos criterion of linear IGE growth is suggested.

\section{Introduction}

Entropic Dynamics (ED) \cite{caticha-ED} is a theoretical framework
constructed on statistical manifolds to explore the possibility that the laws
of physics, either classical or quantum, might be laws of inference rather
than laws of nature. It is known that thermodynamics can be obtained by means
of statistical mechanics which can be considered a form of statistical
inference \cite{jaynes} rather than a pure "physical" theory. Indeed, even
some features of quantum physics can be derived from principles of inference
\cite{caticha-pla}. Recent research considers the possibility that Einstein's
theory of gravity is derivable from general principles of inductive inference
\cite{caticha-piombino}. Unfortunately, the search for the correct variables
that encode relevant information about a system is a major obstacle in the
description and understanding of its evolution.\ The manner in which relevant
variables are selected is not straightforward. This selection is made, in most
cases, on the basis of intuition guided by experiment. The Maximum relative
Entropy (ME) method \cite{caticha-giffin} is used to construct ED models. The
ME method is designed to be a tool of inductive inference. It is used for
updating from a prior to a posterior probability distribution when new
information in the form of constraints becomes available. We use known
techniques \cite{caticha-ED} to show that this principle leads to equations
that are analogous to equations of motion. Information is processed using ME
methods in the framework of Information Geometry (IG) \cite{amari} that is,
Riemannian geometry applied to probability theory. In our approach,
probability theory is a form of generalized logic of plausible inference. It
should apply in principle, to any situation where we lack sufficient
information to permit deductive reasoning.

In this Chapter, we focus on two special entropic dynamical models. In the
first model $\left(  \text{ED1}\right)  $, we consider an hypothetical system
whose microstates span a $2D$ space labelled by the variables $x_{1}\in%
\mathbb{R}
^{+}$ and $x_{2}\in%
\mathbb{R}
$. We assume that the only testable information pertaining to the quantities
$x_{1}$ and $x_{2}$ consists of the expectation values $\left\langle
x_{1}\rangle\text{, }\langle x_{2}\right\rangle $ and the variance $\Delta
x_{2}$. In the second model $\left(  \text{ED2}\right)  $, we consider a $2D$
space of microstates labelled by the variables $x_{1}\in%
\mathbb{R}
$ and $x_{2}\in%
\mathbb{R}
$. In in this case, we assume that the only testable information pertaining to
the quantities $x_{1}$ and $x_{2}$ consist of the expectation values
$\left\langle x_{1}\rangle\text{ and }\langle x_{2}\right\rangle $ and of the
variances $\Delta x_{1}$ and $\Delta x_{2}$. Our models may be extended to
more elaborate systems (highly constrained dynamics) where higher dimensions
are considered. However, for the sake of clarity, we restrict our
considerations to the above relatively simple cases. Given two known boundary
macrostates, we investigate the possible trajectories of systems on the
manifolds. The geometric structure of the manifolds underlying the models is
studied. The metric tensor, Christoffel connections coefficients, Ricci and
Riemann curvature tensors are calculated in both cases and it is shown that in
both cases the dynamics takes place on negatively curved manifolds. The
geodesics of the dynamical models are hyperbolic trajectories on the
manifolds. A detailed study of the stability of such geodesics is presented
using the equation of geodesic deviation (Jacobi equation). The notion of
statistical volume elements is introduced to investigate the asymptotic
behavior of a one-parameter family of neighboring geodesics. It is shown that
the behavior of geodesics on such manifolds is characterized by exponential
instability that leads to chaotic scenarios on the manifolds. These
conclusions are supported by the asymptotic behavior of the Jacobi vector
field intensity. Finally, a relation among entropy-like quantities,
instability and curvature in the two models is presented.

\section{Curved Statistical Manifolds}

In the case of ED1, a measure of distinguishability among the states of the
system is achieved by assigning a probability distribution $p\left(  \vec
{x}|\vec{\theta}\right)  $ to each state defined by expected values
$\theta_{1}^{\left(  1\right)  }$, $\theta_{1}^{\left(  2\right)  }$,
$\theta_{2}^{\left(  2\right)  }$ of the variables $x_{1}$, $x_{2}$ and
$\left(  x_{2}-\left\langle x_{2}\right\rangle \right)  ^{2}$. In the case of
ED2, one assigns a probability distribution $p\left(  \vec{x}|\vec{\theta
}\right)  $ to each state defined by expected values $\theta_{1}^{\left(
1\right)  }$, $\theta_{2}^{\left(  1\right)  }$, $\theta_{1}^{\left(
2\right)  }$, $\theta_{2}^{\left(  2\right)  }$ \ of the variables $x_{1}$,
$\left(  x_{1}-\left\langle x_{1}\right\rangle \right)  ^{2}$, $x_{2}$ and
$\left(  x_{2}-\left\langle x_{2}\right\rangle \right)  ^{2}$. The process of
assigning a probability distribution to each state provides the statistical
manifolds of the ED models with a metric structure. Specifically, the
Fisher-Rao information metric \cite{fisher} defined in (\ref{metric tensor})
is used to quantify the distinguishability of probability distributions
$p\left(  \vec{x}|\vec{\theta}\right)  $ that live on the manifold (the family
of distributions $\left\{  p^{(tot)}\left(  \vec{x}|\vec{\theta}\right)
\right\}  $ is as a manifold, each distribution $p^{(tot)}\left(  \vec{x}%
|\vec{\theta}\right)  $ is a point with coordinates $\theta^{i}$ where $i$
labels the macrovariables). As such, the Fisher-Rao metric assigns an IG to
the space of states.

\subsection{The Statistical Manifold $\mathcal{M}_{S_{1}}$}

Consider a hypothetical physical system evolving over a two-dimensional
space.\ The variables $x_{1}\in%
\mathbb{R}
^{+}$ and $x_{2}\in%
\mathbb{R}
$ label the $2D$ space of microstates of the system. We assume that all
information relevant to the dynamical evolution of the system is contained in
the probability distributions. For this reason, no other information (such as
external fields) is required. \ We assume that the only testable information
pertaining to the quantities $x_{1}$ and $x_{2}$ consists of the expectation
values $\left\langle x_{1}\rangle\text{, }\langle x_{2}\right\rangle $ and the
variance $\Delta x_{2}$. Therefore, these \ three expected values define the
$3D$ space of macrostates $\mathcal{M}_{S_{1}}$ of the ED1 model. Each
macrostate may be thought as a point of a three-dimensional statistical
manifold with coordinates given by the numerical values of the expectations
$\theta_{1}^{\left(  1\right)  }$, $\theta_{1}^{\left(  2\right)  }$,
$\theta_{2}^{\left(  2\right)  }$. The available information can be written in
the form of the following constraint equations,%
\begin{equation}%
\begin{array}
[c]{c}%
\left\langle x_{1}\right\rangle =\int_{0}^{+\infty}dx_{1}x_{1}p_{1}\left(
x_{1}|\theta_{1}^{\left(  1\right)  }\right)  \text{, }\left\langle
x_{2}\right\rangle =\int_{-\infty}^{+\infty}dx_{2}x_{2}p_{2}\left(
x_{2}|\theta_{1}^{\left(  2\right)  }\text{, }\theta_{2}^{\left(  2\right)
}\right)  \text{,}\\
\\
\Delta x_{2}=\sqrt{\left\langle \left(  x_{2}-\left\langle x_{2}\right\rangle
\right)  ^{2}\right\rangle }=\left[  \int_{-\infty}^{+\infty}dx_{2}\left(
x_{2}-\left\langle x_{2}\right\rangle \right)  ^{2}p_{2}\left(  x_{2}%
|\theta_{1}^{\left(  2\right)  }\text{, }\theta_{2}^{\left(  2\right)
}\right)  \right]  ^{\frac{1}{2}}\text{,}%
\end{array}
\label{info constraints}%
\end{equation}
where $\theta_{1}^{\left(  1\right)  }=\left\langle x_{1}\right\rangle $,
$\theta_{1}^{\left(  2\right)  }=\left\langle x_{2}\right\rangle $ and
$\theta_{2}^{\left(  2\right)  }=\Delta x_{2}$. The probability distributions
$p_{1}$ and $p_{2}$ are constrained by the conditions of normalization,%
\begin{equation}
\int_{0}^{+\infty}dx_{1}p_{1}\left(  x_{1}|\theta_{1}^{\left(  1\right)
}\right)  =1\text{, }\int_{-\infty}^{+\infty}dx_{2}p_{2}\left(  x_{2}%
|\theta_{1}^{\left(  2\right)  }\text{, }\theta_{2}^{\left(  2\right)
}\right)  =1\text{.} \label{norm constraints}%
\end{equation}
Information theory identifies the exponential distribution as the maximum
entropy distribution if only the expectation value is known. The Gaussian
distribution is identified as the maximum entropy distribution if only the
expectation value and the variance are known (see the simple example presented
in Chapter 2). ME methods allow us to associate a probability distribution
$p^{(tot)}\left(  \vec{x}|\vec{\theta}\right)  $ to each point in the space of
states. The distribution that best reflects the information contained in the
prior distribution $m\left(  \vec{x}\right)  $ updated by the constraints
$(\left\langle x_{1}\right\rangle $, $\left\langle x_{2}\right\rangle $,
$\Delta x_{2})$ is obtained by maximizing the relative entropy%
\begin{equation}
\left[  S\left(  \vec{\theta}\right)  \right]  _{\text{ED1}}=-\int
_{0}^{+\infty}\int_{-\infty}^{+\infty}dx_{1}dx_{2}p^{(tot)}\left(
\overset{\rightarrow}{x}|\overset{\rightarrow}{\theta}\right)  \log\left[
\frac{p^{(tot)}\left(  \vec{x}|\vec{\theta}\right)  }{m\left(  \overset
{\rightarrow}{x}\right)  }\right]  \text{,} \label{relative entropy}%
\end{equation}
where $m(\vec{x})\equiv m$ is the uniform prior probability distribution. The
prior $m\left(  \overset{\rightarrow}{x}\right)  $ is set to be uniform since
we assume the lack of initial available information about the system
(postulate of equal \textit{a priori} probabilities). Upon maximizing
(\ref{relative entropy}), given the constraints (\ref{info constraints}) and
(\ref{norm constraints}), we obtain%
\begin{equation}
p^{(tot)}\left(  \vec{x}|\vec{\theta}\right)  =p_{1}\left(  x_{1}|\theta
_{1}^{\left(  1\right)  }\right)  p_{2}\left(  x_{2}|\theta_{1}^{\left(
2\right)  }\text{, }\theta_{2}^{\left(  2\right)  }\right)  =\frac{1}{\mu_{1}%
}e^{-\frac{x_{1}}{\mu_{1}}}\frac{1}{\sqrt{2\pi\sigma_{2}^{2}}}e^{-\frac
{(x_{2}-\mu_{2})^{2}}{2\sigma_{2}^{2}}}\text{,} \label{tot prob}%
\end{equation}
where $\theta_{1}^{\left(  1\right)  }=\mu_{1}$, $\theta_{1}^{\left(
2\right)  }=\mu_{2}$ and $\theta_{2}^{\left(  2\right)  }=\sigma_{2}$. The
probability distribution (\ref{tot prob}) encodes the available information
concerning the system and $\mathcal{M}_{s_{1}}$ becomes,%
\begin{equation}
\mathcal{M}_{s_{1}}=\left\{  p^{(tot)}\left(  \vec{x}|\vec{\theta}\right)
=\frac{1}{\mu_{1}}e^{-\frac{x_{1}}{\mu_{1}}}\frac{1}{\sqrt{2\pi\sigma_{2}^{2}%
}}e^{-\frac{(x_{2}-\mu_{2})^{2}}{2\sigma_{2}^{2}}}\right\}  \text{,}%
\end{equation}
where $\vec{x}\in%
\mathbb{R}
^{+}\times%
\mathbb{R}
$ and $\vec{\theta}\equiv\left(  \mu_{1}\text{, }\mu_{2}\text{, }\sigma
_{2}\right)  $. Note that we have assumed uncoupled constraints between the
microvariables $x_{1}$ and $x_{2}$. In other words, we assumed that
information about correlations between the microvariables did not need to be
tracked. This assumption leads to the simplified product rule in
(\ref{tot prob}). Coupled constraints however, would lead to a generalized
product rule in (\ref{tot prob}) and to a metric tensor (\ref{metric tensor})
with non-trivial off-diagonal elements (covariance terms). Correlation terms
may be fictitious. They may arise for instance from coordinate
transformations. On the other hand, correlations may arise from interaction of
the system with external fields. Such scenarios would require more delicate analysis.

\subsubsection{The Metric Tensor on $\mathcal{M}_{s_{1}}$}

We cannot determine the evolution of microstates of the system since the
available information is insufficient.\ Instead we can study the distance
between two total distributions with parameters $(\mu_{1}$, $\mu_{2}$,
$\sigma_{2})$ and $(\mu_{1}+d\mu_{1}$, $\mu_{2}+d\mu_{2}$, $\sigma_{2}%
+d\sigma_{2})$. Once the states of the system have been defined, the next step
concerns the problem of quantifying the notion of change in going from the
state $\vec{\theta}$ to the state $\vec{\theta}+d\vec{\theta}$. For our
purpose a convenient measure of change is distance. The measure we seek is
given by the dimensionless "distance" $ds$ between $p(\vec{x}|\vec{\theta})$
and $p(\vec{x}|\vec{\theta}+d\vec{\theta})$:%
\begin{equation}
ds^{2}=g_{ij}d\theta^{i}d\theta^{j}\text{,} \label{line element}%
\end{equation}
where%
\begin{equation}
g_{ij}=\int d\vec{x}\text{ }p(\vec{x}|\vec{\theta})\frac{\partial\log
p(\vec{x}|\vec{\theta})}{\partial\theta^{i}}\frac{\partial\log p(\vec{x}%
|\vec{\theta})}{\partial\theta^{j}} \label{metric tensor}%
\end{equation}
is the Fisher-Rao information metric. Substituting (\ref{tot prob}) into
(\ref{metric tensor}) , the metric $g_{ij}$ on $\mathcal{M}_{s_{1}}$ becomes,%
\begin{equation}
\left(  g_{ij}\right)  _{\mathcal{M}_{s_{1}}}=\left(
\begin{array}
[c]{ccc}%
\frac{1}{\mu_{1}^{2}} & 0 & 0\\
0 & \frac{1}{\sigma_{2}^{2}} & 0\\
0 & 0 & \frac{2}{\sigma_{2}^{2}}%
\end{array}
\right)  \text{.} \label{info matrix}%
\end{equation}
Substituting (\ref{info matrix}) into (\ref{line element}), the "length"
element reads,%
\begin{equation}
\left(  ds^{2}\right)  _{\mathcal{M}_{s_{1}}}=\frac{1}{\mu_{1}^{2}}d\mu
_{1}^{2}+\frac{1}{\sigma_{2}^{2}}d\mu_{2}^{2}+\frac{2}{\sigma_{2}^{2}}%
d\sigma_{2}^{2}\text{.}%
\end{equation}
Notice that the metric structure of $\mathcal{M}_{s_{1}}$ is an emergent
structure and is not itself fundamental. It arises only after assigning a
probability distribution $p\left(  \vec{x}|\vec{\theta}\right)  $ to each
state $\vec{\theta}$.

\subsubsection{The Curvature of $\mathcal{M}_{s_{1}}$}

In this paragraph we calculate the statistical curvature $\mathcal{R}%
_{\mathcal{M}_{s_{1}}}$. This is achieved via application of standard
differential geometric methods to the space of probability distributions
$\mathcal{M}_{s_{1}}$. Recall the definitions of the Ricci tensor $R_{ij}$ and
Riemann curvature tensor $R_{\alpha\mu\nu\rho}$,
\begin{equation}
R_{ij}=g^{ab}R_{aibj}=\partial_{k}\Gamma_{ij}^{k}-\partial_{j}\Gamma_{ik}%
^{k}+\Gamma_{ij}^{k}\Gamma_{kn}^{n}-\Gamma_{ik}^{m}\Gamma_{jm}^{k}\text{,}
\label{ricci}%
\end{equation}
and%
\begin{equation}
R_{\mu\nu\rho}^{\alpha}=\partial_{\nu}\Gamma_{\mu\rho}^{\alpha}-\partial
_{\rho}\Gamma_{\mu\nu}^{\alpha}+\Gamma_{\beta\nu}^{\alpha}\Gamma_{\mu\rho
}^{\beta}-\Gamma_{\beta\rho}^{\alpha}\Gamma_{\mu\nu}^{\beta}\text{.}
\label{riemann}%
\end{equation}
The Ricci scalar $\mathcal{R}_{\mathcal{M}_{s_{1}}}$ is obtained from
(\ref{ricci}) or (\ref{riemann}) via appropriate contraction with the metric
tensor $g_{ij\text{ }}$in (\ref{info matrix}), namely%
\begin{equation}
\mathcal{R}=R_{ij}g^{ij}=R_{\alpha\beta\gamma\delta}g^{\alpha\gamma}%
g^{\beta\delta}\text{,} \label{scalar curvature}%
\end{equation}
where $g^{ik}g_{kj}=\delta_{j}^{i}$ so that $g^{ij}=\left(  g_{ij}\right)
^{-1}=$diag$(\mu_{1}^{2}$, $\sigma_{2}^{2}$, $\frac{\sigma_{2}^{2}}{2})$. The
Christoffel symbols $\Gamma_{ij}^{k}$ appearing in (\ref{ricci}) and
(\ref{riemann}) are defined by,%
\begin{equation}
\Gamma_{ij}^{k}=\frac{1}{2}g^{km}\left(  \partial_{i}g_{mj}+\partial_{j}%
g_{im}-\partial_{m}g_{ij}\right)  \text{.} \label{connection}%
\end{equation}
Substituting (\ref{info matrix}) into (\ref{connection}), we calculate the
non-vanishing components of the connection coefficients,%
\begin{equation}
\Gamma_{11}^{1}=-\frac{1}{\mu_{1}}\text{, }\Gamma_{22}^{3}=\frac{1}%
{2\sigma_{2}}\text{, }\Gamma_{33}^{3}=-\frac{1}{\sigma_{2}}\text{, }%
\Gamma_{23}^{2}=\Gamma_{32}^{2}=-\frac{1}{\sigma_{2}}\text{.}
\label{explicit connection}%
\end{equation}
By substituting (\ref{explicit connection}) in (\ref{ricci}) we determine the
Ricci tensor components,%
\begin{equation}
R_{11}=0\text{, }R_{22}=-\frac{1}{2\sigma_{2}^{2}}\text{, }R_{33}=-\frac
{1}{\sigma_{2}^{2}}\text{.} \label{ricci explicit}%
\end{equation}
The non-vanishing Riemann tensor component is,%
\begin{equation}
R_{2323}=-\frac{1}{\sigma_{2}^{4}}\text{.} \label{riemann explicit}%
\end{equation}
Finally, by substituting (\ref{ricci explicit}) or (\ref{riemann explicit})
into (\ref{scalar curvature}) and using $\left(  g_{ij}\right)  ^{-1}$ we
obtain the Ricci scalar,%
\begin{equation}
\mathcal{R}_{\mathcal{M}_{s_{1}}}=-1<0\text{.} \label{ricci scalar}%
\end{equation}
From (\ref{ricci scalar}) we conclude that $\mathcal{M}_{s_{1}}$ is a manifold
of constant negative $(-1)$ curvature. We remark that the scalar curvature of
$\mathcal{M}_{s_{1}}$ arises from the presence of the Gaussian distribution.
Instead, the exponential distribution does not contribute to $\mathcal{R}%
_{\mathcal{M}_{s_{1}}}$.

\subsection{The Statistical Manifold $\mathcal{M}_{S_{2}}$}

In this case we assume that the $2D$ space of microstates of the system is
labelled by the variables $x_{1}\in%
\mathbb{R}
$ and $x_{2}\in%
\mathbb{R}
$. We assume, as in subsection $\left(  2.1\right)  $, that all information
relevant to the dynamical evolution of the system is contained in the
probability distributions. Moreover, we assume that the only testable
information pertaining to the quantities $x_{1}$ and $x_{2}$ consist of the
expectation values $\left\langle x_{1}\rangle\text{ and }\langle
x_{2}\right\rangle $ and of the variances $\Delta x_{1}$ and $\Delta x_{2}$.
Therefore, these four expected values define the $4D$ space of macrostates
$\mathcal{M}_{S_{2}}$ of the ED2 model. Each macrostate may be thought as a
point of a four-dimensional statistical manifold with coordinates given by the
numerical values of the expectations $\theta_{1}^{\left(  1\right)  }$,
$\theta_{2}^{\left(  1\right)  }$, $\theta_{1}^{\left(  2\right)  }$,
$\theta_{2}^{\left(  2\right)  }$. We emphasize the fact that entropic dynamic
is not defined on the space of microstates but on the space of macrostates.
The available information can be written in the form of the following
constraint equations,%
\begin{equation}%
\begin{array}
[c]{c}%
\left\langle x_{1}\right\rangle =\int_{-\infty}^{+\infty}dx_{1}x_{1}%
p_{1}\left(  x_{1}|\theta_{1}^{\left(  1\right)  }\text{, }\theta_{2}^{\left(
1\right)  }\right)  \text{, \ }\left\langle x_{2}\right\rangle =\int_{-\infty
}^{+\infty}dx_{2}x_{2}p_{2}\left(  x_{2}|\theta_{1}^{\left(  2\right)
}\text{, }\theta_{2}^{\left(  2\right)  }\right)  \text{,}\\
\\
\Delta x_{1}=\sqrt{\left\langle \left(  x_{1}-\left\langle x_{1}\right\rangle
\right)  ^{2}\right\rangle }=\left[  \int_{-\infty}^{+\infty}dx_{1}\left(
x_{1}-\left\langle x_{1}\right\rangle \right)  ^{2}p_{1}\left(  x_{1}%
|\theta_{1}^{\left(  1\right)  }\text{, }\theta_{2}^{\left(  1\right)
}\right)  \right]  ^{\frac{1}{2}}\text{,}\\
\\
\Delta x_{2}=\sqrt{\left\langle \left(  x_{2}-\left\langle x_{2}\right\rangle
\right)  ^{2}\right\rangle }=\left[  \int_{-\infty}^{+\infty}dx_{2}\left(
x_{2}-\left\langle x_{2}\right\rangle \right)  ^{2}p_{2}\left(  x_{2}%
|\theta_{1}^{\left(  2\right)  }\text{, }\theta_{2}^{\left(  2\right)
}\right)  \right]  ^{\frac{1}{2}}\text{,}%
\end{array}
\label{info constraints (2)}%
\end{equation}
where $\theta_{1}^{\left(  1\right)  }=\left\langle x_{1}\right\rangle $,
$\theta_{2}^{\left(  1\right)  }=\Delta x_{1}$, $\theta_{1}^{\left(  2\right)
}=\left\langle x_{2}\right\rangle $ and $\theta_{2}^{\left(  2\right)
}=\Delta x_{2}$. The probability distributions $p_{1}$ and $p_{2}$ are
constrained by the conditions of normalization,%
\begin{equation}
\int_{-\infty}^{+\infty}dx_{1}p_{1}\left(  x_{1}|\theta_{1}^{\left(  1\right)
}\text{, }\theta_{2}^{\left(  1\right)  }\right)  =1\text{, }\int_{-\infty
}^{+\infty}dx_{2}p_{2}\left(  x_{2}|\theta_{1}^{\left(  2\right)  }\text{,
}\theta_{2}^{\left(  2\right)  }\right)  =1\text{.}
\label{norm constraints (2)}%
\end{equation}
The distribution that best reflects the information contained in the uniform
prior distribution $m\left(  \vec{x}\right)  \equiv m$ updated by the
constraints $(\left\langle x_{1}\right\rangle $, $\Delta x_{1}$, $\left\langle
x_{2}\right\rangle $, $\Delta x_{2})$ is obtained by maximizing the relative
entropy%
\begin{equation}
\left[  S\left(  \vec{\theta}\right)  \right]  _{\text{ED2}}=-\int_{-\infty
}^{+\infty}\int_{-\infty}^{+\infty}dx_{1}dx_{2}p^{(tot)}\left(  \overset
{\rightarrow}{x}|\overset{\rightarrow}{\theta}\right)  \log\left[
\frac{p^{(tot)}\left(  \vec{x}|\vec{\theta}\right)  }{m\left(  \overset
{\rightarrow}{x}\right)  }\right]  \text{.} \label{rel entrop}%
\end{equation}
Upon maximizing (\ref{rel entrop}), given the constraints
(\ref{info constraints (2)}) and (\ref{norm constraints (2)}), we obtain%
\begin{equation}
p^{(tot)}\left(  \vec{x}|\vec{\theta}\right)  =\frac{1}{\sqrt{2\pi\sigma
_{1}^{2}}}e^{-\frac{(x_{1}-\mu_{1})^{2}}{2\sigma_{1}^{2}}}\frac{1}{\sqrt
{2\pi\sigma_{2}^{2}}}e^{-\frac{(x_{2}-\mu_{2})^{2}}{2\sigma_{2}^{2}}}\text{.}
\label{prob tot (2)}%
\end{equation}
The probability distribution (\ref{prob tot (2)}) encodes the available
information concerning the system and $\mathcal{M}_{s_{2}}$ becomes,%
\begin{equation}
\mathcal{M}_{s_{2}}=\left\{  p^{(tot)}\left(  \vec{x}|\vec{\theta}\right)
=\frac{1}{\sqrt{2\pi\sigma_{1}^{2}}}e^{-\frac{(x_{1}-\mu_{1})^{2}}{2\sigma
_{1}^{2}}}\frac{1}{\sqrt{2\pi\sigma_{2}^{2}}}e^{-\frac{(x_{2}-\mu_{2})^{2}%
}{2\sigma_{2}^{2}}}\right\}  \text{,}%
\end{equation}
where $\vec{x}\in%
\mathbb{R}
\times%
\mathbb{R}
$ and $\vec{\theta}\equiv\left(  \mu_{1}\text{, }\sigma_{1}\text{, }\mu
_{2}\text{, }\sigma_{2}\right)  $.

\subsubsection{The Metric Tensor on $\mathcal{M}_{s_{2}}$}

Proceeding as in $\left(  2.1.1\right)  $, we determine the metric on
$\mathcal{M}_{s_{2}}$. Substituting (\ref{prob tot (2)}) into
(\ref{metric tensor}), the metric $g_{ij}$ on $\mathcal{M}_{s_{2}}$ becomes,%
\begin{equation}
\left(  g_{ij}\right)  _{\mathcal{M}_{s_{2}}}=\left(
\begin{array}
[c]{cccc}%
\frac{1}{\sigma_{1}^{2}} & 0 & 0 & 0\\
0 & \frac{2}{\sigma_{1}^{2}} & 0 & 0\\
0 & 0 & \frac{1}{\sigma_{2}^{2}} & 0\\
0 & 0 & 0 & \frac{2}{\sigma_{2}^{2}}%
\end{array}
\right)  \text{.} \label{info matrix (2)}%
\end{equation}
Finally, substituting (\ref{info matrix (2)}) into (\ref{line element}), the
"length" element reads,%
\begin{equation}
\left(  ds^{2}\right)  _{\mathcal{M}_{s_{2}}}=\frac{1}{\sigma_{1}^{2}}d\mu
_{1}^{2}+\frac{2}{\sigma_{1}^{2}}d\sigma_{1}^{2}+\frac{1}{\sigma_{2}^{2}}%
d\mu_{2}^{2}+\frac{2}{\sigma_{2}^{2}}d\sigma_{2}^{2}\text{.}%
\end{equation}

\subsubsection{The Curvature of $\mathcal{M}_{s_{2}}$}

Proceeding as in $\left(  2.1.2\right)  $, we calculate the statistical
curvature $\mathcal{R}_{\mathcal{M}_{s_{2}}}$ of $\mathcal{M}_{s_{2}}$. Notice
that $g^{ij}=\left(  g_{ij}\right)  ^{-1}=$diag$(\sigma_{1}^{2}$,
$\frac{\sigma_{1}^{2}}{2}$, $\sigma_{2}^{2}$, $\frac{\sigma_{2}^{2}}{2})$.
Substituting (\ref{info matrix (2)}) into (\ref{connection}), the
non-vanishing components of the connection coefficients become,%
\begin{equation}
\Gamma_{12}^{1}=\Gamma_{21}^{1}=-\frac{1}{\sigma_{1}}\text{, }\Gamma_{22}%
^{2}=-\frac{1}{\sigma_{1}}\text{, }\Gamma_{11}^{2}=\frac{1}{2\sigma_{1}%
}\text{, }\Gamma_{34}^{3}=\Gamma_{43}^{3}=-\frac{1}{\sigma_{2}}\text{, }%
\Gamma_{33}^{4}=\frac{1}{2\sigma_{2}}\text{, }\Gamma_{44}^{4}=-\frac{1}%
{\sigma_{2}}\text{.} \label{connection exp}%
\end{equation}
By substituting (\ref{connection exp}) in (\ref{ricci}) we determine the Ricci
tensor components,%
\begin{equation}
R_{11}=-\frac{1}{2\sigma_{1}^{2}}\text{, }R_{22}=-\frac{1}{\sigma_{1}^{2}%
}\text{, }R_{33}=-\frac{1}{2\sigma_{2}^{2}}\text{, }R_{44}=-\frac{1}%
{\sigma_{2}^{2}}\text{.} \label{ricci (2)}%
\end{equation}
The non-vanishing Riemann tensor components are,%
\begin{equation}
R_{1212}=-\frac{1}{\sigma_{1}^{4}}\text{, }R_{3434}=-\frac{1}{\sigma_{2}^{4}%
}\text{.} \label{riemann (2)}%
\end{equation}
Finally, by substituting (\ref{ricci (2)}) or (\ref{riemann (2)}) into
(\ref{scalar curvature}) and using $\left(  g_{ij}\right)  ^{-1}$, we obtain
the Ricci scalar,%
\begin{equation}
\mathcal{R}_{\mathcal{M}_{s_{2}}}=-2<0\text{.} \label{scalar curvature (2)}%
\end{equation}
From (\ref{scalar curvature (2)}) we conclude that $\mathcal{M}_{s_{2}}$ is a
manifold of uncoupled Gaussian probability distributions of constant negative
$(-2)$ curvature.

\section{The ED\ Models}

The dynamics can be derived from a standard principle of least action
(Maupertuis- Euler-Lagrange-Jacobi-type) \cite{caticha-ED, arnold}. The main
differences are that the dynamics being considered here are defined on a space
of probability distributions $\mathcal{M}_{s}$, not on an ordinary linear
space $V$. Also, the standard coordinates $q_{j}$ of the system are replaced
by statistical macrovariables $\theta^{j}$.

Given the initial macrostate and that the system evolves to a fixed final
macrostate, we investigate the expected trajectories of the ED models on
$\mathcal{M}_{s_{1}}$ and $\mathcal{M}_{s_{2}}$. The classical dynamics of a
particle can be derived from the principle of least action in the
Maupertuis-Euler-Lagrange-Jacobi form \cite{arnold},%
\begin{equation}
\delta J_{\text{Jacobi}}\left[  q\right]  =\delta\int_{s_{i}}^{s_{f}%
}ds\mathcal{F}\left(  q_{j}\text{, }\frac{dq_{j}}{ds}\text{, }s\text{,
}E\right)  =0\text{,}%
\end{equation}
where $q_{j}$ are the coordinates of the system, $s$ is an affine parameter
along the trajectory and $\mathcal{F}$ \ is a functional defined as%
\begin{equation}
\mathcal{F}\left(  q_{j}\text{, }\frac{dq_{j}}{ds}\text{, }s\text{, }E\right)
\equiv\left[  2\left(  E-U\right)  \right]  ^{\frac{1}{2}}\left(
\underset{j,k}{\sum}a_{jk}\frac{dq_{j}}{ds}\frac{dq_{k}}{ds}\right)
^{\frac{1}{2}}\text{.}%
\end{equation}
For a non-relativistic system, the energy $E$ is,%
\begin{equation}
E=T+U\left(  q\right)  =\frac{1}{2}\underset{j,k}{\sum}a_{jk}\left(  q\right)
\dot{q}_{j}\text{ }\dot{q}_{k}+U\left(  q\right)
\end{equation}
where the coefficients $a_{jk}$ are the reduced mass matrix coefficients and
$\overset{\cdot}{q}=\frac{dq}{ds}$ is the time derivative of the canonical
coordinate $q$. We now seek the expected trajectory of the system assuming it
evolves from $\theta_{old}^{\mu}=\theta^{\mu}$\ $\equiv\left(  \mu_{1}\left(
s_{i}\right)  \text{, }\mu_{2}\left(  s_{i}\right)  \text{, }\sigma_{2}\left(
s_{i}\right)  \right)  $ to $\theta_{new}^{\mu}=\theta^{\mu}+d\theta^{\mu
}\equiv\left(  \mu_{1}\left(  s_{f}\right)  \text{, }\mu_{2}\left(
s_{f}\right)  \text{, }\sigma_{2}\left(  s_{f}\right)  \right)  $. Such a
system moves along a geodesic in the space of states, which is a curved
manifold with the appropriately chosen metric \cite{caticha-ED}.\ Since the
trajectory of the system is a geodesic, the ED-action is itself the length;
that is,%
\begin{equation}
J_{ED}\left[  \theta\right]  =\int\left(  ds^{2}\right)  ^{\frac{1}{2}}%
=\int\left(  g_{ij}d\theta^{i}d\theta^{j}\right)  ^{\frac{1}{2}}=\int_{s_{i}%
}^{s_{f}}ds\left(  g_{ij}\frac{d\theta^{i}\left(  s\right)  }{ds}\frac
{d\theta^{j}\left(  s\right)  }{ds}\right)  ^{\frac{1}{2}}\equiv\int_{s_{i}%
}^{s_{f}}ds\mathcal{L}(\theta\text{, }\dot{\theta})
\end{equation}
where $\dot{\theta}=\frac{d\theta}{ds}$ and $\mathcal{L}(\theta$, $\dot
{\theta})$ is the Lagrangian of the system,%
\begin{equation}
\mathcal{L}(\theta\text{, }\dot{\theta})=(g_{ij}\dot{\theta}^{i}\dot{\theta
}^{j})^{\frac{1}{2}}\text{.}%
\end{equation}
A useful choice for $s$ is one satisfying the condition $g_{ij}\frac
{d\theta^{i}}{ds}\frac{d\theta^{j}}{ds}=1$. Therefore, we formally identify
the affine parameter $s$ with the temporal evolution parameter $\tau$,
$s\equiv\tau$. Performing a standard calculus of variations with $s\equiv\tau
$, we obtain%
\begin{equation}
\delta J_{ED}\left[  \theta\right]  =\int d\tau\left(  \frac{1}{2}%
\frac{\partial g_{ij}}{\partial\theta^{k}}\dot{\theta}^{i}\dot{\theta}%
^{j}-\frac{d\dot{\theta}^{k}}{d\tau}\right)  \delta\theta^{k}=0\text{,
}\forall\delta\theta^{k}\text{.} \label{delta J}%
\end{equation}
Note that from (\ref{delta J}), $\frac{d\dot{\theta}^{k}}{d\tau}=$ $\frac
{1}{2}\frac{\partial g_{ij}}{\partial\theta^{k}}\dot{\theta}^{i}\dot{\theta
}^{j}$. This differential equation shows that if $\frac{\partial g_{ij}%
}{\partial\theta^{k}}=0$ for a particular $k$ then the corresponding
$\dot{\theta}^{k}$ is conserved. This suggests to interpret $\dot{\theta}^{k}$
as momenta. Equations (\ref{delta J}) and (\ref{connection}) lead to the
geodesic equation,%
\begin{equation}
\frac{d^{2}\theta^{k}(\tau)}{d\tau^{2}}+\Gamma_{ij}^{k}\frac{d\theta^{i}%
(\tau)}{d\tau}\frac{d\theta^{j}(\tau)}{d\tau}=0\text{.}
\label{geodesic equation(5)}%
\end{equation}
Observe that (\ref{geodesic equation(5)}) are \textit{nonlinear}, second order
coupled ordinary differential equations. These equations describe a dynamics
that is reversible and their solution is the trajectory between an initial and
a final macrostate. The trajectory can be equally well traversed in both directions.

\subsection{Model ED1: Geodesics on $\mathcal{M}_{s_{1}}$}

We seek the explicit form of (\ref{geodesic equation(5)}) for ED1.
Substituting (\ref{explicit connection}) in (\ref{geodesic equation(5)}), we
obtain,%
\[
\frac{d^{2}\mu_{1}}{d\tau^{2}}-\frac{1}{\mu_{1}}\left(  \frac{d\mu_{1}}{d\tau
}\right)  ^{2}=0\text{,}%
\]%
\[
\frac{d^{2}\mu_{2}}{d\tau^{2}}-\frac{2}{\sigma_{2}}\frac{d\mu_{2}}{d\tau}%
\frac{d\sigma_{2}}{d\tau}=0\text{,}%
\]%
\begin{equation}
\frac{d^{2}\sigma_{2}}{d\tau^{2}}-\frac{1}{\sigma_{2}}\left[  \left(
\frac{d\sigma_{2}}{d\tau}\right)  ^{2}-\frac{1}{2}\left(  \frac{d\mu_{2}%
}{d\tau}\right)  ^{2}\right]  =0\text{.} \label{geodesic equation (1)}%
\end{equation}
Integrating this set of differential equations, we obtain
\[
\mu_{1}\left(  \tau\right)  =A_{1}\left[  \cosh\left(  \alpha_{1}\tau\right)
-\sinh\left(  \alpha_{1}\tau\right)  \right]  \text{,}%
\]%
\[
\mu_{2}\left(  \tau\right)  =\frac{B_{1}^{2}}{2\beta_{1}}\frac{1}{\cosh\left(
2\beta_{1}\tau\right)  -\sinh\left(  2\beta_{1}\tau\right)  +\frac{B_{1}^{2}%
}{8\beta_{1}^{2}}}+C_{1}\text{,}%
\]
\
\begin{equation}
\sigma_{2}\left(  \tau\right)  =B_{1}\frac{\left[  \cosh\left(  \beta_{1}%
\tau\right)  -\sinh\left(  \beta_{1}\tau\right)  \right]  }{\cosh\left(
2\beta_{1}\tau\right)  -\sinh\left(  2\beta_{1}\tau\right)  +\frac{B_{1}^{2}%
}{8\beta_{1}^{2}}}+C_{2}\text{.} \label{macrovariables1}%
\end{equation}
The integration constants arising from the exponential contribution to the
geodesic equations are $A_{1}=A_{1}\left(  \mu_{1}\left(  0\right)  \text{,
}\dot{\mu}_{1}\left(  0\right)  \text{ }\right)  =\mu_{1}\left(  0\right)  $
and $\alpha_{1}=\left(  \mu_{1}\left(  0\right)  \text{, }\dot{\mu}_{1}\left(
0\right)  \text{ }\right)  =-\frac{\dot{\mu}_{1}\left(  0\right)  }{\mu
_{1}\left(  0\right)  }\in%
\mathbb{R}
^{+}$ with $\dot{\mu}_{1}\left(  0\right)  =\left(  \frac{d\mu_{1}\left(
\tau\right)  }{d\tau}\right)  _{\tau=0}$. The integration constants arising
from the Gaussian contribution to the geodesic equations are $B_{1}%
=B_{1}\left(  \mu_{2}\left(  0\right)  \text{, }\sigma_{2}\left(  0\right)
\text{, }\dot{\mu}_{2}\left(  0\right)  \text{, }\dot{\sigma}_{2}\left(
0\right)  \text{ }\right)  $, $C_{1}=C_{1}\left(  \mu_{2}\left(  0\right)
\text{, }\sigma_{2}\left(  0\right)  \text{, }\dot{\mu}_{2}\left(  0\right)
\text{, }\dot{\sigma}_{2}\left(  0\right)  \text{ }\right)  $, $C_{2}%
=C_{2}\left(  \mu_{2}\left(  0\right)  \text{, }\sigma_{2}\left(  0\right)
\text{, }\dot{\mu}_{2}\left(  0\right)  \text{, }\dot{\sigma}_{2}\left(
0\right)  \text{ }\right)  $ and $\beta_{1}=\left(  \mu_{2}\left(  0\right)
\text{, }\sigma_{2}\left(  0\right)  \text{, }\dot{\mu}_{2}\left(  0\right)
\text{, }\dot{\sigma}_{2}\left(  0\right)  \text{ }\right)  $. In total, there
are the six \textit{real} integration constants since there are six initial
conditions: three for the initial values of the macrovariables labelling
points on $\mathcal{M}_{s_{1}}$ and three for the initial values of the first
derivative of the macrovariables. However, the normalization constraint
$\left(  g_{ij}\left(  \theta\right)  \frac{d\theta^{i}}{ds}\frac{d\theta^{j}%
}{ds}\right)  ^{\frac{1}{2}}=\left(  \dot{\theta}_{j}\dot{\theta}^{j}\right)
^{\frac{1}{2}}=1$ leads to five independent initial conditions.

Note that the set of equations (\ref{macrovariables1}) parametrizes the
evolution surface of \ the statistical submanifold $\mathit{m}_{s_{1}}$ of
$\mathcal{M}_{s_{1}}$,%
\begin{equation}
\mathit{m}_{_{s_{1}}}=\left\{  p^{(tot)}\left(  \vec{x}|\vec{\theta}\right)
\in\mathcal{M}_{s_{1}}\text{: }\vec{\theta}\text{ satisfy
(\ref{geodesic equation (1)})}\right\}  \text{.}%
\end{equation}
The set of points in $\mathit{m}_{_{s_{1}}}$ are the expected intermediate
macrostates of the system in its evolution from a given initial macrostate
$\vec{\theta}_{i}\equiv\left(  \mu_{1}\left(  0\right)  \text{, }\mu
_{2}\left(  0\right)  \text{, }\sigma_{2}\left(  0\right)  \right)  $ to a
given final macrostate $\vec{\theta}_{f}\equiv\left(  \mu_{1}\left(
\tau\right)  \text{, }\mu_{2}\left(  \tau\right)  \text{, }\sigma_{2}\left(
\tau\right)  \right)  $. \ In other words, the measure of $\mathit{m}%
_{_{s_{1}}}$ describes the portion of statistical volume in configuration
space accessed by the system in its information-constrained evolution between
two given points of the manifold $\mathcal{M}_{s_{1}}\supset\mathit{m}%
_{_{s_{1}}}$. A different set of initial conditions would lead to consider a
different submanifold $\mathit{m}_{_{s_{1}}}^{\prime}$, $\mathcal{M}_{s_{1}%
}\supset\mathit{m}_{_{s_{1}}}^{\prime}\neq\mathit{m}_{_{s_{1}}}$.

\subsection{Model ED2: Geodesics on $\mathcal{M}_{s_{2}}$}

We seek the explicit form of (\ref{geodesic equation(5)}) for ED2.
Substituting (\ref{connection exp}) in (\ref{geodesic equation(5)}), we
obtain,%
\[
\frac{d^{2}\mu_{1}}{d\tau^{2}}-\frac{2}{\sigma_{1}}\frac{d\mu_{1}}{d\tau}%
\frac{d\sigma_{1}}{d\tau}=0\text{,}%
\]%
\[
\frac{d^{2}\sigma_{1}}{d\tau^{2}}-\frac{1}{\sigma_{1}}\left[  \left(
\frac{d\sigma_{1}}{d\tau}\right)  ^{2}-\frac{1}{2}\left(  \frac{d\mu_{1}%
}{d\tau}\right)  ^{2}\right]  =0\text{,}%
\]%
\[
\frac{d^{2}\mu_{2}}{d\tau^{2}}-\frac{2}{\sigma_{2}}\frac{d\mu_{2}}{d\tau}%
\frac{d\sigma_{2}}{d\tau}=0\text{,}%
\]%
\begin{equation}
\frac{d^{2}\sigma_{2}}{d\tau^{2}}-\frac{1}{\sigma_{2}}\left[  \left(
\frac{d\sigma_{2}}{d\tau}\right)  ^{2}-\frac{1}{2}\left(  \frac{d\mu_{2}%
}{d\tau}\right)  ^{2}\right]  =0\text{.} \label{geodesic equations (2)}%
\end{equation}
Integrating this set of differential equations, we obtain%
\[
\mu_{1}\left(  \tau\right)  =\frac{A_{2}^{2}}{2\alpha_{2}}\frac{1}%
{\cosh\left(  2\alpha_{2}\tau\right)  -\sinh\left(  2\alpha_{2}\tau\right)
+\frac{A_{2}^{2}}{8\alpha_{2}^{2}}}+C_{1}\text{,}%
\]%
\[
\sigma_{1}\left(  \tau\right)  =A_{2}\frac{\left[  \cosh\left(  \alpha_{2}%
\tau\right)  -\sinh\left(  \alpha_{2}\tau\right)  \right]  }{\cosh\left(
2\alpha_{2}\tau\right)  -\sinh\left(  2\alpha_{2}\tau\right)  +\frac{A_{2}%
^{2}}{8\alpha_{2}^{2}}}+C_{2}\text{,}%
\]%
\[
\mu_{2}\left(  \tau\right)  =\frac{B_{2}^{2}}{2\beta_{2}}\frac{1}{\cosh\left(
2\beta_{2}\tau\right)  -\sinh\left(  2\beta_{2}\tau\right)  +\frac{B_{2}^{2}%
}{8\beta_{2}^{2}}}+C_{3}\text{,}%
\]%
\begin{equation}
\sigma_{2}\left(  \tau\right)  =B_{2}\frac{\left[  \cosh\left(  \beta_{2}%
\tau\right)  -\sinh\left(  \beta_{2}\tau\right)  \right]  }{\cosh\left(
2\beta_{2}\tau\right)  -\sinh\left(  2\beta_{2}\tau\right)  +\frac{B_{2}^{2}%
}{8\beta_{2}^{2}}}+C_{4}\text{.} \label{macrovariables (2)}%
\end{equation}
The eight integration constants $A_{2}$, $B_{2}$, $C_{1}$, $C_{2}$, $C_{3}$,
$C_{4}$, $\alpha_{2}$ and $\beta_{2}$ assume \textit{real} values and they are
not independent from each other. They satisfy the normalization condition
$\left(  g_{ij}\left(  \theta\right)  \frac{d\theta^{i}}{ds}\frac{d\theta^{j}%
}{ds}\right)  ^{\frac{1}{2}}=\left(  \dot{\theta}_{j}\dot{\theta}^{j}\right)
^{\frac{1}{2}}=1$. Furthermore, they are functions of the initial values of
the macrovariable $\vec{\theta}\equiv\left(  \mu_{1}\text{, }\mu_{2}\text{,
}\sigma_{1}\text{, }\sigma_{2}\right)  $ and of its first derivative
$\frac{d\vec{\theta}}{d\tau}\equiv\left(  \dot{\mu}_{1}\text{, }\dot{\mu}%
_{2}\text{, }\dot{\sigma}_{1}\text{, }\dot{\sigma}_{2}\right)  $.

Again, note that the set of equations (\ref{macrovariables (2)}) parametrizes
the evolution surface of \ the statistical submanifold $\mathit{m}_{s_{2}}$ of
$\mathcal{M}_{s_{2}}$,%
\begin{equation}
\mathit{m}_{_{s_{2}}}=\left\{  p^{(tot)}\left(  \vec{x}|\vec{\theta}\right)
\in\mathcal{M}_{s_{1}}\text{: }\vec{\theta}\text{ satisfy
(\ref{geodesic equations (2)})}\right\}  \text{.}%
\end{equation}
The set of points in $\mathit{m}_{_{s_{1}}}$ are the expected intermediate
macrostates of the system in its evolution from a given initial macrostate
$\vec{\theta}_{i}\equiv\left(  \mu_{1}\left(  0\right)  \text{, }\mu
_{2}\left(  0\right)  \text{, }\sigma_{1}\left(  0\right)  \text{, }\sigma
_{2}\left(  0\right)  \right)  $ to a given final macrostate $\vec{\theta}%
_{f}\equiv\left(  \mu_{1}\left(  \tau\right)  \text{, }\mu_{2}\left(
\tau\right)  \text{, }\sigma_{1}\left(  \tau\right)  \text{, }\sigma
_{2}\left(  \tau\right)  \right)  $. \ In other words, the measure of
$\mathit{m}_{_{s_{2}}}$ describes the portion of statistical volume in
configuration space accessed by the system in its information-constrained
evolution between two given points of the manifold $\mathcal{M}_{s_{2}}%
\supset\mathit{m}_{_{s_{2}}}$. A different set of initial conditions would
lead to consider a different submanifold $\mathit{m}_{_{s_{2}}}^{\prime}$,
$\mathcal{M}_{s_{2}}\supset\mathit{m}_{_{s_{2}}}^{\prime}\neq\mathit{m}%
_{_{s_{2}}}$.

\section{Chaotic Instability in the ED Models}

The Riemannian curvature of a manifold is closely connected with the behavior
of the geodesics on it, i.e., with the motion of the corresponding dynamical
system \cite{arnold}. If the Riemannian curvature of a manifold is positive
(as on a sphere or ellipsoid), then the nearby geodesics oscillate about one
another in most cases; whereas if the curvature is negative (as on the surface
of a hyperboloid of one sheet), geodesics rapidly diverge from one another.

\subsection{Instability in ED1}

In this subsection, the stability of ED1 is considered. It is shown that
neighboring trajectories are exponentially unstable under small perturbations
of initial conditions. In the rest of the Chapter, we only assume that in ED1
we have $\beta_{1}=\alpha_{1}=-\frac{\dot{\mu}_{1}\left(  0\right)  }{\mu
_{1}\left(  0\right)  }\equiv\alpha\in%
\mathbb{R}
^{+}$; in ED2, we assume we have $\alpha_{2}=\beta_{2}\equiv\alpha\in%
\mathbb{R}
^{+}$. We make this choice because we are selecting the quantity "$\alpha$" as
the common parameter (the one-parameter characterizing families of geodesics
on configuration spaces) labelling different geodesics on manifolds
$\mathcal{M}_{s_{1}}$ and $\mathcal{M}_{s_{2}}$ and our objective is to
compare the degree of chaoticity of both dynamics (ED1 and ED2) on their
respective underlying curved statistical manifolds.

At this stage of our discussion, the parameter "$\alpha$" is a quantity
extracted from the continuous entropic dynamical evolution equations
(\ref{macrovariables1}) and (\ref{macrovariables (2)}). Of course, a different
set of initial conditions would lead to a different parameter $\alpha$. Once
the choice is made, we assume that $\alpha$ is roughly constant over
accessible region of configuration space. This assumption would not be new in
the literature, see reference \cite{cavicchio} for example. In the following
sections, we will discover that $\alpha$ plays the role of a standard Lyapunov exponent.

We could show that $\alpha$ does play the role of a Lyapunov exponent already
at this stage of our discussion. Lyapunov exponents are asymptotic quantities:
they are defined in the limit as time approaches infinity. The finite Lyapunov
exponent in the direction $e\in%
\mathbb{R}
^{n}$ of a trajectory $x\left(  \tau\text{, }x_{0}\right)  $ satisfying the
differential equation $\dot{x}=A\left(  \tau\right)  x$ with $x\in%
\mathbb{R}
^{n}$ and with initial condition $x\left(  0\text{, }x_{0}\right)  =x_{0}$ is
defined as \cite{lyapunov},%
\begin{equation}
\lambda\left(  e\right)  \overset{\text{def}}{=}\underset{\tau\rightarrow
\infty}{\lim}\log\left[  \frac{\sqrt{\left\langle Xe\text{, }Xe\right\rangle
}}{\sqrt{\left\langle e\text{, }e\right\rangle }}\right]  \text{.}
\label{LE_carlo}%
\end{equation}
The brackets $\left\langle \cdot\text{, }\cdot\right\rangle $ in
(\ref{LE_carlo}) denote the standard scalar product in $%
\mathbb{R}
^{n}$, $X=X\left(  \tau\text{; }x\left(  \tau\text{, }x_{0}\right)  \right)  $
is the asymptotically regular fundamental matrix of the differential equation
$\dot{x}=A\left(  \tau\right)  x$ \cite{verhulst}. In the ED1 model, the
differential equation to consider is%
\begin{equation}
\frac{d\vec{\theta}_{\text{ED1}}\left(  \tau\right)  }{d\tau}=A_{\text{ED1}%
}\left(  \tau\right)  \vec{\theta}_{\text{ED1}}\left(  \tau\right)  \text{,}%
\end{equation}
where $\vec{\theta}_{\text{ED1}}\left(  \tau\right)  \equiv\left(  \mu
_{1}\left(  \tau\right)  \text{, }\mu_{2}\left(  \tau\right)  \text{, }%
\sigma_{2}\left(  \tau\right)  \right)  $ are given in (\ref{macrovariables1}%
). In the asymptotic limit, the $3\times3$ matrix $A_{\text{ED1}}\left(
\tau\right)  $ can be approximated by a diagonal matrix with constant
coefficients, $A_{\text{ED1}}\left(  \tau\right)  \overset{\tau\rightarrow
\infty}{\approx}$diag$\left(  \alpha_{1}\text{, }0\text{, }\beta_{1}\right)
$. A straightforward calculation would lead to an asymptotically regular
$3\times3$ fundamental matrix
\begin{equation}
X_{\text{ED1}}\left(  \tau\right)  \overset{\tau\rightarrow\infty}{\approx
}\text{diag}\left(  c_{1}\exp\left(  \alpha_{1}\tau\right)  \text{, }c_{2}%
\tau\text{, }c_{3}\exp\left(  \beta_{1}\tau\right)  \right)
\end{equation}
with $c_{i}\in%
\mathbb{R}
$, $\forall i=1$, $2$, $3$. Therefore, equation (\ref{LE_carlo}) would lead to
the following interesting result%
\begin{equation}
\lambda_{\max}\left(  e\right)  =\underset{%
\mathbb{R}
^{+}}{\max}\left\{  \alpha_{1}\text{, }\beta_{1}\right\}  \text{, }\forall
e\in%
\mathbb{R}
^{3}\text{.}%
\end{equation}
We have shown that, under our assumptions, the leading Lyapunov exponent
$\lambda_{\max}$ is given by $\alpha$.

\subsubsection{The Geodesic Length $\Theta_{\mathcal{M}_{s_{1}}}$}

Consider the one-parameter family of geodesics $\mathcal{F}_{G_{\mathcal{M}%
_{s_{1}}}}\left(  \alpha\right)  \equiv\left\{  \theta_{\mathcal{M}_{s_{1}}%
}^{\mu}\left(  \tau;\alpha\right)  \right\}  _{\alpha\in%
\mathbb{R}
^{+}}^{\mu=1\text{, }2\text{, }3}$ where $\theta_{\mathcal{M}_{s_{1}}}^{\mu}%
$are solutions of (\ref{geodesic equation (1)}), the "selector parameter"
$\alpha$ tells which geodesic is being considered and the affine parameter
$\tau$ tells where is the point being considered on a given geodesic. The
length of geodesics in $\mathcal{F}_{G_{\mathcal{M}_{s_{1}}}}\left(
\alpha\right)  $ is defined as,%
\begin{equation}
\Theta_{\mathcal{M}_{s_{1}}}\left(  \tau\text{; }\alpha\right)  \overset
{\text{def}}{=}%
{\displaystyle\int}
\left(  g_{ij}d\theta^{i}d\theta^{j}\right)  ^{\frac{1}{2}}=%
{\displaystyle\int\limits_{0}^{\tau}}
\left[  \frac{1}{\mu_{1}^{2}}\left(  \frac{d\mu_{1}}{d\tau^{\prime}}\right)
^{2}+\frac{1}{\sigma_{2}^{2}}\left(  \frac{d\mu_{2}}{d\tau^{\prime}}\right)
^{2}+\frac{2}{\sigma_{2}^{2}}\left(  \frac{d\sigma_{2}}{d\tau^{\prime}%
}\right)  ^{2}\right]  ^{\frac{1}{2}}d\tau^{\prime}\text{,} \label{qui}%
\end{equation}
where $g_{ij}=\left(  g_{ij}\right)  _{\mathcal{M}_{s_{1}}}$ is given in
(\ref{info matrix}). Substituting (\ref{macrovariables1}) in (\ref{qui}) and
considering the asymptotic expression of $\Theta_{\mathcal{M}_{s_{1}}}\left(
\tau\text{; }\alpha\right)  $, we obtain%
\begin{equation}
\Theta_{\mathcal{M}_{s_{1}}}\left(  \tau\rightarrow\infty\text{; }%
\alpha\right)  \equiv\Theta_{1}\left(  \tau\text{; }\alpha\right)
\overset{\tau\rightarrow\infty}{\approx}\alpha\tau\text{.}%
\end{equation}
In evaluating (\ref{qui}), we have not imposed the conventional normalization
condition $\left(  \dot{\theta}_{j}\dot{\theta}^{j}\right)  ^{\frac{1}{2}}=1$.
Indeed, in our case $\left(  \dot{\theta}_{j}\dot{\theta}^{j}\right)
^{\frac{1}{2}}=$constant; what matters is that we will use this very same
normalization constant in evaluating the length of geodesics in $\mathcal{F}%
_{G_{\mathcal{M}_{s_{2}}}}\left(  \alpha\right)  $ so that we may compare both
lengths using the same "meter" (statistical affine parameter $\tau$).

In order to roughly investigate the asymptotic behavior of two neighboring
geodesics labelled by the parameters $\alpha$ and $\alpha+\delta\alpha$, we
consider the following difference,%
\begin{equation}
\Delta\Theta_{1}\equiv\left\vert \Theta_{1}\left(  \tau\text{; }\alpha
+\delta\alpha\right)  -\Theta_{1}\left(  \tau\text{; }\alpha\right)
\right\vert =\left\vert \left(  \frac{\partial\Theta_{1}}{\partial\alpha
}\right)  _{\tau}\delta\alpha\right\vert \overset{\tau\rightarrow\infty
}{\approx}\left\vert \delta\alpha\right\vert \tau\text{.}%
\end{equation}
It is clear that $\Delta\Theta_{1}$ diverges, that is, the lengths of two
neighboring geodesics\ with slightly different parameters $\alpha$ and
$\alpha+\delta\alpha$ differ in a remarkable way as the evolution parameter
$\tau$ $\rightarrow\infty$. This hints at the onset of instability of the
hyperbolic trajectories on $\mathcal{M}_{s_{1}}$.

\subsubsection{Evolution of Volumes $V_{\mathcal{M}_{s_{1}}}$ on the
Statistical Manifold $\mathcal{M}_{s_{1}}$}

The instability of ED1 can be further explored by studying the behavior of the
one-parameter family of statistical volume elements $\mathcal{F}%
_{V_{\mathcal{M}_{s_{1}}}}\left(  \alpha\right)  \equiv\left\{  V_{\mathcal{M}%
_{s_{1}}}\left(  \tau\text{; }\alpha\right)  \right\}  _{\alpha}$. Recall that
$\mathcal{M}_{s_{1}}$ is the space of probability distributions $p^{(tot)}%
\left(  \vec{x}|\vec{\theta}\right)  $ labeled by parameters $\theta
_{1}^{\left(  1\right)  }$, $\theta_{1}^{\left(  2\right)  }$, $\theta
_{2}^{\left(  2\right)  }$. These parameters are the coordinates of the point
$p^{(tot)}$, and in these coordinates a $3D$ volume element $dV_{\mathcal{M}%
_{s_{1}}}$ reads%
\begin{equation}
dV_{\mathcal{M}_{s_{1}}}=\sqrt{g}d\theta_{1}^{\left(  1\right)  }d\theta
_{1}^{\left(  2\right)  }d\theta_{2}^{\left(  2\right)  }\equiv\sqrt{g}%
d\mu_{1}d\mu_{2}d\sigma_{2}\text{,}%
\end{equation}
where in the ED1 model here presented, $g=|\det\left(  g_{ij}\right)
_{\mathcal{M}_{s_{1}}}|=\frac{2}{\mu_{1}^{2}\sigma_{2}^{4}}$. Hence, the
volume element $dV_{\mathcal{M}_{s_{1}}}$ is given by,%
\begin{equation}
dV_{\mathcal{M}_{s_{1}}}=\frac{\sqrt{2}}{\mu_{1}\sigma_{2}^{2}}d\mu_{1}%
d\mu_{2}d\sigma_{2}\text{.}%
\end{equation}
The volume increase of an extended region of $\mathcal{M}_{s_{1}}$ is given
by,%
\begin{equation}
V_{\mathcal{M}_{s_{1}}}\left(  \tau\text{; }\alpha\right)  \overset
{\text{def}}{=}%
{\displaystyle\int\limits_{\vec{\theta}_{\mathcal{M}_{s_{1}}}\left(  0\text{;
}\alpha\right)  }^{\vec{\theta}_{\mathcal{M}_{s_{1}}}\left(  \tau\text{;
}\alpha\right)  }}
dV_{\mathcal{M}_{s_{1}}}=%
{\displaystyle\int\limits_{\mu_{1}\left(  0\right)  }^{\mu_{1}\left(
\tau\right)  }}
{\displaystyle\int\limits_{\mu_{2}\left(  0\right)  }^{\mu_{2}\left(
\tau\right)  }}
{\displaystyle\int\limits_{\sigma_{2}\left(  0\right)  }^{\sigma_{2}\left(
\tau\right)  }}
\frac{\sqrt{2}}{\mu_{1}\sigma_{2}^{2}}d\mu_{1}d\mu_{2}d\sigma_{2}\text{.}
\label{qui 1}%
\end{equation}
Integrating (\ref{qui 1}) using (\ref{macrovariables1}), we obtain%
\begin{equation}
V_{\mathcal{M}_{s_{1}}}\left(  \tau\text{; }\alpha\right)  =\frac{\tau}%
{\sqrt{2}}e^{\alpha\tau}-\frac{\ln A}{\sqrt{2}\alpha}e^{\alpha\tau}+\frac{\ln
A}{\sqrt{2}\alpha}\text{.}%
\end{equation}
The quantity that actually encodes relevant information about the stability of
neighboring volume elements is the average volume $\left\langle V_{\mathcal{M}%
_{s_{1}}}\left(  \tau\text{; }\alpha\right)  \right\rangle _{\tau}$,
\begin{equation}
\left\langle V_{\mathcal{M}_{s_{1}}}\left(  \tau\text{; }\alpha\right)
\right\rangle _{\tau}\overset{\text{def}}{=}\frac{1}{\tau}%
{\displaystyle\int\limits_{0}^{\tau}}
V_{\mathcal{M}_{s_{1}}}\left(  \tau^{\prime}\text{; }\alpha\right)
d\tau^{\prime}=\frac{1}{\tau}\left\{  \frac{1}{\sqrt{2}\alpha^{2}}\left(
\alpha\tau-1\right)  e^{\alpha\tau}-\frac{\ln A}{\sqrt{2}\alpha^{2}}%
e^{\alpha\tau}+\frac{\ln A}{\sqrt{2}\alpha}\tau\right\}  \text{.}%
\end{equation}
For convenience, let us rename $\left\langle V_{\mathcal{M}_{s_{1}}}\left(
\tau\text{; }\alpha\right)  \right\rangle _{\tau}\equiv\mathcal{V}%
_{\mathcal{M}_{s_{1}}}\left(  \tau\text{; }\alpha\right)  $. Therefore, the
asymptotic expansion of $\mathcal{V}_{\mathcal{M}_{s_{1}}}\left(  \tau\text{;
}\alpha\right)  $ for $\tau$ $\rightarrow\infty$ reads,%
\begin{equation}
\mathcal{V}_{\mathcal{M}_{s_{1}}}\left(  \tau\text{; }\alpha\right)
\overset{\tau\rightarrow\infty}{\approx}\frac{1}{\sqrt{2}\alpha}e^{\alpha\tau
}\text{.} \label{exp statistical volume}%
\end{equation}
This asymptotic evolution in (\ref{exp statistical volume}) describes the
exponential increase of average volume elements on $\mathcal{M}_{s_{1}}$. The
exponential instability characteristic of chaos forces the system to rapidly
explore large areas (volumes) of the statistical manifolds. It is interesting
to note that this asymptotic behavior appears also in the conventional
description of quantum chaos \cite{zurek-paz} where the entropy increases
linearly at a rate determined by the Lyapunov exponents \cite{ruelle}. The
linear entropy increase as a quantum chaos criterion was introduced by Zurek
and Paz. In our information-geometric approach a relevant variable that will
be useful for comparison of the two different degrees of instability
characterizing the two ED models is the relative entropy-like quantity defined
as,%
\begin{equation}
\mathcal{S}_{\mathcal{M}_{s_{1}}}\left(  \tau\text{; }\alpha\right)
\overset{\text{def}}{=}\log\mathcal{V}_{\mathcal{M}_{s_{1}}}\left(
\tau\text{; }\alpha\right)  \text{.} \label{IG entropy}%
\end{equation}
Substituting (\ref{exp statistical volume}) in (\ref{IG entropy}) and
considering the asymptotic limit $\tau\rightarrow\infty$, we obtain%
\begin{equation}
\mathcal{S}_{\mathcal{M}_{s_{1}}}\left(  \tau\text{; }\alpha\right)
\overset{\tau\rightarrow\infty}{\approx}\alpha\tau\text{.}
\label{asymptotic IG entropy}%
\end{equation}
Notice that the late-time limit $\left(  \tau\rightarrow\infty\right)  $ is
necessary to describe chaos \cite{bhattacharya}. Furthermore, the study of the
asymptotic behavior of average quantities is very common in chaotic dynamics
\cite{cavicchio, bhattacharya}. The entropy-like quantity $\mathcal{S}%
_{\mathcal{M}_{s_{1}}}$ in (\ref{asymptotic IG entropy}) may be interpreted as
the asymptotic limit of the natural logarithm of a statistical weight
$\mathcal{V}_{\mathcal{M}_{s_{1}}}\left(  \tau\text{; }\alpha\right)
\overset{\text{def}}{=}\left\langle V_{\mathcal{M}_{s_{1}}}\left(  \tau\text{;
}\alpha\right)  \right\rangle _{\tau}$. defined on $\mathcal{M}_{s_{1}}$.
Equation (\ref{asymptotic IG entropy}) is the information-geometric analog of
the Zurek-Paz chaos criterion. As a final remark, we would like to emphasize
that the connection between $\alpha$ and $\mathcal{S}_{\mathcal{M}_{s_{1}}%
}\left(  \tau\text{; }\alpha\right)  $ may be compared to the connection
between the KS entropy and the "physical" entropy (the entropy of the second
law of thermodynamics) \cite{latora}; the steps we used to construct the IGE
resemble the ones that Toda and Ikeda used to propose a quantal Lyapunov
exponent \cite{ikeda}.

\subsubsection{The Jacobi Vector Field $\vec{J}_{\mathcal{M}_{s_{1}}}$}

We study the behavior of the one-parameter family of neighboring geodesics
$\mathcal{F}_{G_{\mathcal{M}_{s_{1}}}}\left(  \alpha\right)  \equiv\left\{
\theta_{\mathcal{M}_{s_{1}}}^{\mu}\left(  \tau\text{; }\alpha\right)
\right\}  _{\alpha\in%
\mathbb{R}
^{+}}^{\mu=1\text{, }2\text{, }3}$ where,%
\begin{align}
\theta^{1}\left(  \tau\text{; }\alpha\right)   &  =\mu_{1}\left(  \tau\text{;
}\alpha\right)  =Ae^{\alpha\tau}\text{, }\theta^{2}\left(  \tau\text{; }%
\alpha\right)  =\mu_{2}\left(  \tau\text{; }\alpha\right)  =\frac{A^{2}%
}{2\alpha}\frac{1}{e^{-2\alpha\tau}+\frac{A^{2}}{8\alpha^{2}}}\text{,}%
\label{macro (1)}\\
& \nonumber\\
\text{ }\theta^{3}\left(  \tau\text{; }\alpha\right)   &  =\sigma_{2}\left(
\tau\text{; }\alpha\right)  =A\frac{e^{-\alpha\tau}}{e^{-2\alpha\tau}%
+\frac{A^{2}}{8\alpha^{2}}}\text{.}\nonumber
\end{align}
The relative geodesic spread is characterized by the Jacobi equation \cite{de
felice, mtw},%
\begin{equation}
\frac{D^{2}\left(  \delta\theta^{i}\right)  }{D\tau^{2}}+R_{kml}^{i}%
\frac{\partial\theta^{k}}{\partial\tau}\frac{\partial\theta^{l}}{\partial\tau
}\delta\theta^{m}=0 \label{JLC equation(5)}%
\end{equation}
where $i=1$, $2$, $3$ and,%
\begin{equation}
\delta\theta^{i}\equiv\delta_{\alpha}\theta^{i}\overset{\text{def}}{=}\left(
\frac{\partial\theta^{i}\left(  \tau\text{; }\alpha\right)  }{\partial\alpha
}\right)  _{\tau}\delta\alpha\text{.} \label{Jacobi intensity}%
\end{equation}
Equation (\ref{JLC equation(5)}) forms a system of three coupled ordinary
differential equations \textit{linear} in the components of the deviation
vector field (\ref{Jacobi intensity}) but\textit{\ nonlinear} in derivatives
of the metric (\ref{metric tensor}). It describes the linearized geodesic
flow: the linearization ignores the relative velocity of the geodesics. When
the geodesics are neighboring but their relative velocity is arbitrary, the
corresponding geodesic deviation equation is the so-called generalized Jacobi
equation \cite{chicone}. The nonlinearity is due to the existence of
velocity-dependent terms in the system.

Neighboring geodesics accelerate relative to each other with a rate directly
measured by the curvature tensor $R_{\alpha\beta\gamma\delta}$. The Riemann
tensor is, within this information-geometric framework, an indicator of the
strength of "statistical tidal forces". Multiplying both sides of
(\ref{JLC equation(5)}) by $g_{ij}$ and using the standard symmetry properties
of the Riemann curvature tensor, the geodesic deviation equation becomes,%
\begin{equation}
g_{ji}\frac{D^{2}\left(  \delta\theta^{i}\right)  }{D\tau^{2}}+R_{lmkj}%
\frac{\partial\theta^{k}}{\partial\tau}\frac{\partial\theta^{l}}{\partial\tau
}\delta\theta^{m}=0\text{.}%
\end{equation}
Recall that the covariant derivative $\frac{D^{2}\left(  \delta\theta^{\mu
}\right)  }{D\tau^{2}}$ in (\ref{JLC equation(5)}) is defined as,%
\begin{align}
\frac{D^{2}\delta\theta^{\mu}}{D\tau^{2}}  &  =\frac{d^{2}\delta\theta^{\mu}%
}{d\tau^{2}}+2\Gamma_{\alpha\beta}^{\mu}\frac{d\delta\theta^{\alpha}}{d\tau
}\frac{d\theta^{\beta}}{d\tau}+\Gamma_{\alpha\beta}^{\mu}\delta\theta^{\alpha
}\frac{d^{2}\theta^{\beta}}{d\tau^{2}}+\Gamma_{\alpha\beta,\nu}^{\mu}%
\frac{d\theta^{\nu}}{d\tau}\frac{d\theta^{\beta}}{d\tau}\delta\theta^{\alpha
}+\nonumber\\
&  +\Gamma_{\alpha\beta}^{\mu}\Gamma_{\rho\sigma}^{\alpha}\frac{d\theta
^{\sigma}}{d\tau}\frac{d\theta^{\beta}}{d\tau}\delta\theta^{\rho}%
\end{align}
and that the only non-vanishing Riemann tensor component is $R_{2323}%
=-\frac{1}{\sigma_{1}^{4}}$. Therefore, the three differential equations for
the geodesic deviation are,%
\begin{equation}
\frac{d^{2}\left(  \delta\theta^{1}\right)  }{d\tau^{2}}+2\Gamma_{11}^{1}%
\frac{d\theta^{1}}{d\tau}\frac{d\left(  \delta\theta^{1}\right)  }{d\tau
}+\partial_{1}\Gamma_{11}^{1}\left(  \frac{d\theta^{1}}{d\tau}\right)
^{2}\delta\theta^{1}=0\text{,} \label{one}%
\end{equation}%
\begin{align}
&  \frac{d^{2}\left(  \delta\theta^{2}\right)  }{d\tau^{2}}+2\left[
\Gamma_{23}^{2}\frac{d\theta^{3}}{d\tau}\frac{d\left(  \delta\theta
^{2}\right)  }{d\tau}+\Gamma_{32}^{2}\frac{d\theta^{2}}{d\tau}\frac{d\left(
\delta\theta^{3}\right)  }{d\tau}\right]  +\partial_{3}\Gamma_{23}^{2}\left(
\frac{d\theta^{3}}{d\tau}\right)  ^{2}\delta\theta^{2}+\Gamma_{32}^{2}%
\Gamma_{33}^{3}\left(  \frac{d\theta^{3}}{d\tau}\right)  ^{2}\delta\theta
^{2}\nonumber\\
&  =\frac{1}{g_{22}}R_{2323}\frac{d\theta^{2}}{d\tau}\frac{d\theta^{3}}{d\tau
}\delta\theta^{3}-\frac{1}{g_{22}}R_{2323}\left(  \frac{d\theta^{3}}{d\tau
}\right)  ^{2}\delta\theta^{2}\text{,} \label{two}%
\end{align}%
\begin{align}
&  \frac{d^{2}\left(  \delta\theta^{3}\right)  }{d\tau^{2}}+2\left[
\Gamma_{22}^{3}\frac{d\theta^{2}}{d\tau}\frac{d\left(  \delta\theta
^{2}\right)  }{d\tau}+\Gamma_{33}^{3}\frac{d\theta^{3}}{d\tau}\frac{d\left(
\delta\theta^{3}\right)  }{d\tau}\right]  +\partial_{3}\Gamma_{33}^{3}\left(
\frac{d\theta^{3}}{d\tau}\right)  ^{2}\delta\theta^{3}+\Gamma_{22}^{3}%
\Gamma_{23}^{2}\frac{d\theta^{3}}{d\tau}\frac{d\theta^{2}}{d\tau}\delta
\theta^{2}\nonumber\\
&  =\frac{1}{g_{33}}R_{2323}\frac{d\theta^{2}}{d\tau}\frac{d\theta^{3}}{d\tau
}\delta\theta^{2}-\frac{1}{g_{33}}R_{2323}\left(  \frac{d\theta^{2}}{d\tau
}\right)  ^{2}\delta\theta^{3}\text{.} \label{tre}%
\end{align}
Substituting (\ref{explicit connection}), (\ref{riemann explicit}) and
(\ref{macro (1)}) in equations (\ref{one}), (\ref{two}) and (\ref{tre}) and
considering the asymptotic limit $\tau\rightarrow\infty$, the geodesic
deviation equations become,%
\begin{equation}
\frac{d^{2}\left(  \delta\theta^{1}\right)  }{d\tau^{2}}+2\alpha\frac{d\left(
\delta\theta^{1}\right)  }{d\tau}+\alpha^{2}\delta\theta^{1}=0\text{,}%
\end{equation}%
\begin{equation}
\frac{d^{2}\left(  \delta\theta^{2}\right)  }{d\tau^{2}}+2\alpha\frac{d\left(
\delta\theta^{2}\right)  }{d\tau}+\frac{16\alpha^{2}}{A}e^{-\alpha\tau}%
\frac{d\left(  \delta\theta^{3}\right)  }{d\tau}+\left(  \alpha^{2}%
-\frac{8\alpha^{3}}{A}e^{-\alpha\tau}\right)  \delta\theta^{3}=0\text{,}
\label{five}%
\end{equation}%
\begin{equation}
\frac{d^{2}\left(  \delta\theta^{3}\right)  }{d\tau^{2}}+2\alpha\frac{d\left(
\delta\theta^{3}\right)  }{d\tau}+\left(  \alpha^{2}-\frac{32\alpha^{4}}%
{A^{2}}e^{-2\alpha\tau}\right)  \delta\theta^{3}-\frac{8\alpha^{2}}%
{A}e^{-\alpha\tau}\frac{d\left(  \delta\theta^{2}\right)  }{d\tau}%
-\frac{8\alpha^{3}}{A}e^{-\alpha\tau}\delta\theta^{2}=0\text{.} \label{six}%
\end{equation}
Neglecting the exponentially decaying terms in $\delta\theta^{3}$ in
(\ref{five}) and (\ref{six}) and assuming that,%
\begin{equation}
\underset{\tau\rightarrow\infty}{\lim}\left(  \frac{16\alpha^{2}}{A}%
e^{-\alpha\tau}\frac{d\left(  \delta\theta^{3}\right)  }{d\tau}\right)
=0\text{, }\underset{\tau\rightarrow\infty}{\lim}\left(  \frac{8\alpha^{2}}%
{A}e^{-\alpha\tau}\frac{d\left(  \delta\theta^{2}\right)  }{d\tau}\right)
=0\text{,}\underset{\tau\rightarrow\infty}{\lim}\left(  \frac{8\alpha^{3}}%
{A}e^{-\alpha\tau}\delta\theta^{2}\right)  =0\text{, } \label{limits}%
\end{equation}
the geodesic deviation equations finally become,%
\[
\frac{d^{2}\left(  \delta\theta^{1}\right)  }{d\tau^{2}}+2\alpha\frac{d\left(
\delta\theta^{1}\right)  }{d\tau}+\alpha^{2}\delta\theta^{1}=0\text{,}%
\]%
\[
\frac{d^{2}\left(  \delta\theta^{2}\right)  }{d\tau^{2}}+2\alpha\frac{d\left(
\delta\theta^{2}\right)  }{d\tau}+\alpha^{2}\delta\theta^{3}=0\text{,}%
\]%
\begin{equation}
\frac{d^{2}\left(  \delta\theta^{3}\right)  }{d\tau^{2}}+2\alpha\frac{d\left(
\delta\theta^{3}\right)  }{d\tau}+\alpha^{2}\delta\theta^{3}=0\text{.}
\label{simple JLC equation}%
\end{equation}
Note that in order to prove that our assumptions in (\ref{limits}) are
correct, we will check \textit{a posteriori }its consistency. Integrating the
system of differential equations (\ref{simple JLC equation}), we obtain%
\[
\delta\mu_{1}\left(  \tau\right)  =\left(  a_{1}+a_{2}\tau\right)
e^{-\alpha\tau}\text{,}%
\]%
\[
\delta\mu_{2}\left(  \tau\right)  =\left(  a_{3}+a_{4}\tau\right)
e^{-\alpha\tau}-\frac{1}{2\alpha}a_{5}e^{-2\alpha\tau}+a_{6}\text{,}%
\]%
\begin{equation}
\delta\sigma_{2}\left(  \tau\right)  =\left(  a_{3}+a_{4}\tau\right)
e^{-\alpha\tau}\text{,} \label{simplemacro}%
\end{equation}
where $a_{i}$, $i=1$,..., $6$ are integration constants. Note that conditions
(\ref{limits}) are satisfied and therefore our assumption are compatible with
the solutions obtained. Finally, consider the vector field components
$J^{k}\equiv\delta\theta^{k}$ defined in (\ref{Jacobi intensity}) and its
magnitude $J$, \ \
\begin{equation}
J^{2}=J^{i}J_{i}=g_{ij}J^{i}J^{j}\text{.} \label{j squared}%
\end{equation}
The magnitude $J$ is called the Jacobi field intensity. In our case
(\ref{j squared}) becomes,%
\begin{equation}
J_{\mathcal{M}_{s_{1}}}^{2}=\frac{1}{\mu_{1}^{2}}\left(  \delta\mu_{1}\right)
^{2}+\frac{1}{\sigma_{2}^{2}}\left(  \delta\mu_{2}\right)  ^{2}+\frac
{2}{\sigma_{2}^{2}}\left(  \delta\sigma_{2}\right)  ^{2}\text{.}
\label{j squared (1)}%
\end{equation}
Substituting (\ref{macro (1)}) and (\ref{simplemacro}) in (\ref{j squared (1)}%
), and keeping the leading term in the asymptotic expansion in $J_{\mathcal{M}%
_{s_{1}}}^{2}$, we obtain%
\begin{equation}
J_{\mathcal{M}_{s_{1}}}\overset{\tau\rightarrow\infty}{\approx}C_{\mathcal{M}%
_{s_{1}}}e^{\alpha\tau}\text{,} \label{asymptotic j}%
\end{equation}
where the constant coefficient $C_{\mathcal{M}_{s_{1}}}=$ $\frac{Aa_{6}^{3}%
}{2\sqrt{2}\alpha}$ encodes information about initial conditions and depends
on the model parameter $\alpha$. We conclude that the geodesic spread on
$\mathcal{M}_{s_{1}}$ is described by means of an \textit{exponentially}
\textit{divergent} Jacobi vector field intensity $J_{\mathcal{M}_{s_{1}}}$. It
is known that classical chaotic systems exhibit exponential sensitivity to
initial conditions. This characterization, quantified in terms of Lyapunov
exponents, is an important ingredient in any conventional definition of
classical chaos. In our approach, the quantity $\lambda_{J}\overset
{\text{def}}{=}\underset{\tau\rightarrow\infty}{\lim}\frac{1}{\tau}\log\left(
\frac{\left\Vert J\left(  \tau\right)  \right\Vert }{\left\Vert J\left(
0\right)  \right\Vert }\right)  $ with $J$ given in (\ref{asymptotic j}) would
play the role of the conventional Lyapunov exponents.

\subsection{Instability in ED2}

In this subsection, the instability of the geodesics on $\mathcal{M}_{s_{2}}$
is studied. We proceed as in section $\left(  5\text{. }4\text{. }1\right)  $.

\subsubsection{The Geodesic Length $\Theta_{\mathcal{M}_{s_{2}}}$}

Consider the one-parameter family of geodesics $\mathcal{F}_{G_{\mathcal{M}%
_{s_{2}}}}\left(  \alpha\right)  \equiv\left\{  \theta_{\mathcal{M}_{s_{2}}%
}^{\mu}\left(  \tau\text{; }\alpha\right)  \right\}  _{\alpha\in%
\mathbb{R}
^{+}}^{\mu=1,2,3,4}$ where $\theta_{\mathcal{M}_{s_{2}}}^{\mu}$are solutions
of (\ref{geodesic equations (2)}). The length of geodesics in $\mathcal{F}%
_{G_{\mathcal{M}_{s_{2}}}}\left(  \alpha\right)  $ is defined as,%
\begin{equation}
\Theta_{\mathcal{M}_{s_{2}}}\left(  \tau\text{; }\alpha\right)  \overset
{\text{def}}{=}%
{\displaystyle\int}
\left(  g_{ij}d\theta^{i}d\theta^{j}\right)  ^{\frac{1}{2}}=%
{\displaystyle\int\limits_{0}^{\tau}}
\left[  \frac{1}{\sigma_{1}^{2}}\left(  \frac{d\mu_{1}}{d\tau^{\prime}%
}\right)  ^{2}+\frac{2}{\sigma_{1}^{2}}\left(  \frac{d\sigma_{1}}%
{d\tau^{\prime}}\right)  ^{2}+\frac{1}{\sigma_{2}^{2}}\left(  \frac{d\mu_{2}%
}{d\tau^{\prime}}\right)  ^{2}+\frac{2}{\sigma_{2}^{2}}\left(  \frac
{d\sigma_{2}}{d\tau^{\prime}}\right)  ^{2}\right]  ^{\frac{1}{2}}d\tau
^{\prime}\text{,} \label{gl (2)}%
\end{equation}
where $g_{ij}=\left(  g_{ij}\right)  _{\mathcal{M}_{s_{2}}}$ defined in
(\ref{info matrix (2)}). Substituting (\ref{macrovariables (2)}) in
(\ref{gl (2)}) and considering the asymptotic limit of $\Theta_{\mathcal{M}%
_{s_{2}}}\left(  \tau\text{; }\alpha\right)  $ when $\tau\rightarrow\infty$,
we obtain%
\begin{equation}
\Theta_{\mathcal{M}_{s_{2}}}\left(  \tau\rightarrow\infty\text{; }%
\alpha\right)  \equiv\Theta_{2}\left(  \tau\text{; }\alpha\right)
\overset{\tau\rightarrow\infty}{\approx}2\alpha\tau\text{.} \label{qui2}%
\end{equation}
In evaluating (\ref{qui2}), we have not imposed the conventional normalization
condition $\left(  \dot{\theta}_{j}\dot{\theta}^{j}\right)  ^{\frac{1}{2}}=1$.
Indeed, in our case $\left(  \dot{\theta}_{j}\dot{\theta}^{j}\right)
^{\frac{1}{2}}=$constant; what matters is that we have used the very same
normalization constant used in evaluating the length of geodesics in
$\mathcal{F}_{G_{\mathcal{M}_{s_{1}}}}\left(  \alpha\right)  $ so that we may
compare both lengths using the same "meter" (statistical affine parameter
$\tau$). In order to roughly investigate the asymptotic behavior of two
neighboring geodesics labelled by the parameters $\alpha$ and $\alpha
+\delta\alpha$, we consider the following difference,%
\begin{equation}
\Delta\Theta_{2}\equiv\left\vert \Theta_{2}\left(  \tau\text{; }\alpha
+\delta\alpha\right)  -\Theta_{2}\left(  \tau\text{; }\alpha\right)
\right\vert =\left\vert \left(  \frac{\partial\Theta_{2}}{\partial\alpha
}\right)  _{\tau}\delta\alpha\right\vert \overset{\tau\rightarrow\infty
}{\approx}2\left\vert \delta\alpha\right\vert \tau\text{.}%
\end{equation}
It is clear that $\Delta\Theta_{2}$ diverges, that is the lengths of two
neighboring geodesics\ with slightly different parameters $\alpha$ and
$\alpha+\delta\alpha$ differ in a significant way as the evolution parameter
$\tau\rightarrow\infty$. This hints at the onset of instability of the
hyperbolic trajectories on $\mathcal{M}_{s_{2}}$.

\subsubsection{Evolution of Volumes $V_{\mathcal{M}_{s_{2}}}$ on the
Statistical Manifold $\mathcal{M}_{s_{2}}$}

The instability of ED2 can be explored by studying the behavior of the
one-parameter family of statistical volume elements $\mathcal{F}%
_{V_{\mathcal{M}_{s_{2}}}}\left(  \alpha\right)  \equiv\left\{  V_{\mathcal{M}%
_{s_{2}}}\left(  \tau\text{; }\alpha\right)  \right\}  _{\alpha}$. Recall that
$\mathcal{M}_{s_{2}}$ is the space of probability distributions
$p^{(\text{tot})}\left(  \vec{x}|\vec{\theta}\right)  $ labeled by parameters
$\theta_{1}^{\left(  1\right)  }$, $\theta_{2}^{\left(  1\right)  }$,
$\theta_{1}^{\left(  2\right)  }$, $\theta_{2}^{\left(  2\right)  }$. These
parameters are the coordinates of the point $p^{(\text{tot})}$, and in these
coordinates a $4D$ infinitesimal volume element $dV_{\mathcal{M}_{s_{2}}}$
reads,%
\begin{equation}
dV_{\mathcal{M}_{s_{2}}}=\sqrt{g}d\theta_{1}^{\left(  1\right)  }d\theta
_{2}^{\left(  1\right)  }d\theta_{1}^{\left(  2\right)  }d\theta_{2}^{\left(
2\right)  }\equiv\sqrt{g}d\mu_{1}d\sigma_{1}d\mu_{2}d\sigma_{2}\text{,}%
\end{equation}
where in the ED2 model here presented, $g=|\det\left(  g_{ij}\right)
_{\mathcal{M}_{s_{2}}}|=\frac{4}{\sigma_{1}^{4}\sigma_{2}^{4}}$. Hence, the
infinitesimal volume element $dV_{\mathcal{M}_{s_{2}}}$ is given by,%
\begin{equation}
dV_{\mathcal{M}_{s_{2}}}=\frac{2}{\sigma_{1}^{2}\sigma_{2}^{2}}d\mu_{1}%
d\sigma_{1}d\mu_{2}d\sigma_{2}\text{.}%
\end{equation}
The volume of an extended region of $\mathcal{M}_{s_{2}}$ is defined by,%
\begin{equation}
V_{\mathcal{M}_{s_{2}}}\left(  \tau\text{; }\alpha\right)  \overset
{\text{def}}{=}%
{\displaystyle\int\limits_{\vec{\theta}_{\mathcal{M}_{s_{2}}}\left(  0\text{;
}\alpha\right)  }^{\vec{\theta}_{\mathcal{M}_{s_{2}}}\left(  \tau\text{;
}\alpha\right)  }}
dV_{\mathcal{M}_{s_{2}}}=%
{\displaystyle\int\limits_{\mu_{1}\left(  0\right)  }^{\mu_{1}\left(
\tau\right)  }}
{\displaystyle\int\limits_{\sigma_{1}\left(  0\right)  }^{\sigma_{1}\left(
\tau\right)  }}
\text{ }%
{\displaystyle\int\limits_{\mu_{2}\left(  0\right)  }^{\mu_{2}\left(
\tau\right)  }}
{\displaystyle\int\limits_{\sigma_{2}\left(  0\right)  }^{\sigma_{2}\left(
\tau\right)  }}
\frac{2}{\sigma_{1}^{2}\sigma_{2}^{2}}d\mu_{1}d\sigma_{1}d\mu_{2}d\sigma
_{2}\text{,} \label{statistical volume (2)}%
\end{equation}
Integrating (\ref{statistical volume (2)}) and using (\ref{macrovariables (2)}%
), we obtain%
\begin{equation}
V_{\mathcal{M}_{s_{2}}}\left(  \tau\text{; }\alpha\right)  =\frac{A^{2}%
}{2\alpha^{2}}e^{2\alpha\tau}-\frac{A^{2}}{2\alpha^{2}}\text{.}%
\end{equation}
The accessible volume, on average, on the configuration space $\mathcal{M}%
_{s_{2}}$ is $\left\langle V_{\mathcal{M}_{s_{2}}}\left(  \tau\text{; }%
\alpha\right)  \right\rangle _{\tau}$,
\begin{equation}
\left\langle V_{\mathcal{M}_{s_{2}}}\left(  \tau\text{; }\alpha\right)
\right\rangle _{\tau}\overset{\text{def}}{=}\frac{1}{\tau}V_{\mathcal{M}%
_{s_{2}}}\left(  \tau^{\prime}\text{; }\alpha\right)  d\tau^{\prime}%
=\frac{A^{2}}{4\alpha^{3}}\frac{e^{2\alpha\tau}}{\tau}-\frac{A^{2}}%
{2\alpha^{2}}%
\end{equation}
For convenience, let us rename $\left\langle V_{\mathcal{M}_{s_{2}}}\left(
\tau\text{; }\alpha\right)  \right\rangle _{\tau}\overset{\text{def}}%
{=}\mathcal{V}_{\mathcal{M}_{s_{2}}}\left(  \tau\text{; }\alpha\right)  $.
Therefore, the asymptotic expansion of $\mathcal{V}_{\mathcal{M}_{s_{2}}%
}\left(  \tau\text{; }\alpha\right)  $ for $\tau$ $\rightarrow\infty$ reads,%
\begin{equation}
\mathcal{V}_{\mathcal{M}_{s_{2}}}\left(  \tau\text{; }\alpha\right)
\overset{\tau\rightarrow\infty}{\approx}\frac{A^{2}}{4\alpha^{3}}%
\frac{e^{2\alpha\tau}}{\tau}\text{.} \label{asymptotic volume (2)}%
\end{equation}
In analogy to (\ref{IG entropy}) we introduce,%
\begin{equation}
\mathcal{S}_{\mathcal{M}_{s_{2}}}\left(  \tau\text{; }\alpha\right)
\overset{\text{def}}{=}\log\mathcal{V}_{\mathcal{M}_{s_{2}}}\left(
\tau\text{; }\alpha\right)  \text{.} \label{IG entropy (2)}%
\end{equation}
Substituting (\ref{asymptotic volume (2)}) in (\ref{IG entropy (2)}) and
considering its asymptotic limit, we obtain%
\begin{equation}
\mathcal{S}_{\mathcal{M}_{s_{2}}}\left(  \tau\text{; }\alpha\right)
\overset{\tau\rightarrow\infty}{\approx}2\alpha\tau\text{.} \label{IG (2)}%
\end{equation}

\subsubsection{The Jacobi Vector Field $\vec{J}_{\mathcal{M}_{s_{2}}}$}

We proceed as in $\left(  5\text{. }4\text{. }1\text{. }3\right)  $. Study the
behavior of the one-parameter $\left(  \alpha\right)  $ family of neighboring
geodesics on $\mathcal{M}_{s_{2}}$, $\left\{  \theta^{i}\left(  \tau\text{;
}\alpha\right)  \right\}  _{i=1\text{, }2\text{, }3\text{, }4}$ with%
\begin{equation}
\theta^{3}\left(  \tau;\alpha\right)  \equiv\theta^{1}\left(  \tau
;\alpha\right)  =\mu_{1}\left(  \tau;\alpha\right)  =\frac{A^{2}}{2\alpha
}\frac{1}{e^{-2\alpha\tau}+\frac{A^{2}}{8\alpha^{2}}}\text{,} \label{sette}%
\end{equation}%
\begin{equation}
\theta^{4}\left(  \tau;\alpha\right)  \equiv\theta^{2}\left(  \tau
;\alpha\right)  =A\frac{e^{-\alpha\tau}}{e^{-2\alpha\tau}+\frac{A^{2}}%
{8\alpha^{2}}}\text{.} \label{otto}%
\end{equation}
Note that because we will compare the two Jacobi fields $J_{\mathcal{M}%
_{s_{1}}}$ on $\mathcal{M}_{s_{1}}$ and $J_{\mathcal{M}_{s_{2}}}$ on
$\mathcal{M}_{s_{2}}$, we assume the same initial conditions as considered in
$\left(  4.1.3\right)  $. Recall that the non-vanishing Riemann tensor
components are $R_{1212}=-\frac{1}{\sigma_{1}^{4}}$ and $R_{3434}=-\frac
{1}{\sigma_{2}^{4}}$ given in (\ref{riemann (2)}). Therefore two of the four
differential equations describing the geodesic spread are,%
\begin{align}
&  \frac{d^{2}\left(  \delta\theta^{1}\right)  }{d\tau^{2}}+2\left[
\Gamma_{12}^{1}\frac{d\theta^{2}}{d\tau}\frac{d\left(  \delta\theta
^{1}\right)  }{d\tau}+\Gamma_{21}^{1}\frac{d\theta^{1}}{d\tau}\frac{d\left(
\delta\theta^{2}\right)  }{d\tau}\right]  +\partial_{2}\Gamma_{12}^{1}\left(
\frac{d\theta^{2}}{d\tau}\right)  ^{2}\delta\theta^{1}+\Gamma_{21}^{1}%
\Gamma_{22}^{2}\left(  \frac{d\theta^{2}}{d\tau}\right)  ^{2}\delta\theta
^{1}\nonumber\\
&  =\frac{1}{g_{11}}R_{1212}\frac{d\theta^{1}}{d\tau}\frac{d\theta^{2}}{d\tau
}\delta\theta^{2}-\frac{1}{g_{11}}R_{1212}\left(  \frac{d\theta^{2}}{d\tau
}\right)  ^{2}\delta\theta^{1}\text{,} \label{nove}%
\end{align}%
\begin{align}
&  \frac{d^{2}\left(  \delta\theta^{2}\right)  }{d\tau^{2}}+2\left[
\Gamma_{11}^{2}\frac{d\theta^{1}}{d\tau}\frac{d\left(  \delta\theta
^{1}\right)  }{d\tau}+\Gamma_{22}^{2}\frac{d\theta^{2}}{d\tau}\frac{d\left(
\delta\theta^{2}\right)  }{d\tau}\right]  +\partial_{2}\Gamma_{22}^{2}\left(
\frac{d\theta^{2}}{d\tau}\right)  ^{2}\delta\theta^{2}+\Gamma_{11}^{2}%
\Gamma_{12}^{1}\frac{d\theta^{2}}{d\tau}\frac{d\theta^{1}}{d\tau}\delta
\theta^{1}\nonumber\\
&  =\frac{1}{g_{22}}R_{1212}\frac{d\theta^{1}}{d\tau}\frac{d\theta^{2}}{d\tau
}\delta\theta^{1}-\frac{1}{g_{22}}R_{1212}\left(  \frac{d\theta^{1}}{d\tau
}\right)  ^{2}\delta\theta^{2}\text{.} \label{dieci}%
\end{align}
The other two equations can be obtained from (\ref{nove}) and (\ref{dieci})
substituting the index $1$ with $3$ and $2$ with $4$. Thus, we will limit our
considerations just to the above two equations. Using equations
(\ref{connection exp}), (\ref{riemann (2)}), (\ref{sette}) and (\ref{otto}) in
(\ref{nove}) and (\ref{dieci}) and considering the asymptotic limit
$\tau\rightarrow\infty$, the two equations of geodesic deviation become,%
\begin{equation}
\frac{d^{2}\left(  \delta\theta^{1}\right)  }{d\tau^{2}}+2\alpha\frac{d\left(
\delta\theta^{1}\right)  }{d\tau}+\frac{16\alpha^{2}}{A}e^{-\alpha\tau}%
\frac{d\left(  \delta\theta^{2}\right)  }{d\tau}+\left(  \alpha^{2}%
-\frac{8\alpha^{3}}{A}e^{-\alpha\tau}\right)  \delta\theta^{2}=0\text{,}
\label{undici}%
\end{equation}%
\begin{equation}
\frac{d^{2}\left(  \delta\theta^{2}\right)  }{d\tau^{2}}+2\alpha\frac{d\left(
\delta\theta^{2}\right)  }{d\tau}+\left(  \alpha^{2}-\frac{32\alpha^{4}}%
{A^{2}}e^{-2\alpha\tau}\right)  \delta\theta^{2}-\frac{8\alpha^{2}}%
{A}e^{-\alpha\tau}\frac{d\left(  \delta\theta^{1}\right)  }{d\tau}%
-\frac{8\alpha^{3}}{A}e^{-\alpha\tau}\delta\theta^{1}=0. \label{dodici}%
\end{equation}
Neglecting the exponentially decaying terms in $\delta\theta^{2}$ in
(\ref{undici}) and (\ref{dodici}) and assuming%
\begin{equation}
\underset{\tau\rightarrow\infty}{\lim}\left(  \frac{16\alpha^{2}}{A}%
e^{-\alpha\tau}\frac{d\left(  \delta\theta^{2}\right)  }{d\tau}\right)
=0\text{,}\underset{\tau\rightarrow\infty}{\lim}\left(  \frac{8\alpha^{2}}%
{A}e^{-\alpha\tau}\frac{d\left(  \delta\theta^{1}\right)  }{d\tau}\right)
=0\text{,}\underset{\tau\rightarrow\infty}{\lim}\left(  \frac{8\alpha^{3}}%
{A}e^{-\alpha\tau}\delta\theta^{1}\right)  =0 \label{limits (2)}%
\end{equation}
the geodesic deviation equations in (\ref{undici}) and (\ref{dodici}) become,%
\begin{equation}
\frac{d^{2}\left(  \delta\theta^{1}\right)  }{d\tau^{2}}+2\alpha\frac{d\left(
\delta\theta^{1}\right)  }{d\tau}+\alpha^{2}\delta\theta^{2}=0\text{, }%
\frac{d^{2}\left(  \delta\theta^{2}\right)  }{d\tau^{2}}+2\alpha\frac{d\left(
\delta\theta^{2}\right)  }{d\tau}+\alpha^{2}\delta\theta^{2}=0\text{.}
\label{sopra}%
\end{equation}
The consistency of the assumptions in (\ref{limits (2)}) will be checked
\textit{a posteriori} after integrating equations in (\ref{sopra}). It follows
that the geodesics spread on $\mathcal{M}_{s_{2}}$ is described by the
temporal evolution of the following deviation vector components,%
\begin{align}
\delta\mu_{1}\left(  \tau\right)   &  =\left(  a_{1}+a_{2}\tau\right)
e^{-\alpha\tau}-\frac{1}{2\alpha}a_{3}e^{-2\alpha\tau}+a_{4}\text{, }%
\delta\sigma_{1}\left(  \tau\right)  =\left(  a_{1}+a_{2}\tau\right)
e^{-\alpha\tau}\label{j(2)}\\
\text{ }\delta\mu_{2}\left(  \tau\right)   &  =\left(  a_{5}+a_{6}\tau\right)
e^{-\alpha\tau}-\frac{1}{2\alpha}a_{7}e^{-2\alpha\tau}+a_{8}\text{, }%
\delta\sigma_{2}\left(  \tau\right)  =\left(  a_{5}+a_{6}\tau\right)
e^{-\alpha\tau}\nonumber
\end{align}
where $a_{i}$, $i=1$,..., $8$ are integration constants. Note that $a_{4}$ and
$a_{8}$ in (\ref{j(2)}) equal $a_{6}$ in (\ref{simplemacro}). Furthermore,
note that these solutions above are compatible with the assumptions in
(\ref{limits (2)}). Finally, consider the Jacobi vector field intensity
$J_{\mathcal{M}_{s_{2}}}$ on $\mathcal{M}_{s_{2}}$,%
\begin{equation}
J_{\mathcal{M}_{s_{2}}}^{2}=\frac{1}{\sigma_{1}^{2}}\left(  \delta\mu
_{1}\right)  ^{2}+\frac{2}{\sigma_{1}^{2}}\left(  \delta\sigma_{1}\right)
^{2}+\frac{1}{\sigma_{2}^{2}}\left(  \delta\mu_{2}\right)  ^{2}+\frac
{2}{\sigma_{2}^{2}}\left(  \delta\sigma_{2}\right)  ^{2}\text{.}
\label{jsquare(2)}%
\end{equation}
Substituting (\ref{sette}), (\ref{otto}) and (\ref{j(2)}) in (\ref{jsquare(2)}%
) and keeping the leading term in the asymptotic expansion in $J_{\mathcal{M}%
_{s_{2}}}^{2}$, we obtain%
\begin{equation}
J_{\mathcal{M}_{s_{2}}}\overset{\tau\rightarrow\infty}{\approx}C_{\mathcal{M}%
_{s_{2}}}e^{\alpha\tau}\text{.} \label{asymptotic j2}%
\end{equation}
where the constant coefficient $C_{\mathcal{M}_{s_{2}}}=$ $\frac{Aa_{6}^{3}%
}{\sqrt{2}\alpha}\equiv2C_{\mathcal{M}_{s_{1}}}$ encodes information about
initial conditions and it depends on the model parameter $\alpha$. We conclude
that the geodesic spread on $\mathcal{M}_{s_{2}}$ is described by means of an
\textit{exponentially} divergent Jacobi vector field intensity $J_{\mathcal{M}%
_{s_{2}}}$.

\section{Statistical Curvature, Jacobi Field Intensity and Entropy-like
Quantities}

Many problems in mathematical statistics, information theory and in stochastic
processes can be tackled using differential geometric methods on curved
statistical manifolds. For instance, an important class of statistical
manifolds is that arising from the exponential family \cite{amari} and one
particular family is that of gamma probability distributions. These
distributions have been shown \cite{arwini} to have important uniqueness
properties for stochastic processes. In this Chapter, two statistical
manifolds of negative curvature $\mathcal{M}_{s_{1}}$ and $\mathcal{M}_{s_{2}%
}$ have been considered. They are representations of smooth families of
probability distributions (exponentials and Gaussians for $\mathcal{M}_{s_{1}%
}$, Gaussians for $\mathcal{M}_{s_{2}}$). They represent the "arena" where the
entropic dynamics takes place. The instability of the trajectories of the ED1
and ED2 on $\mathcal{M}_{s_{1}}$ and $\mathcal{M}_{s_{2}}$ respectively, has
been studied using the statistical weight $\left\langle V_{\mathcal{M}_{s}%
}\right\rangle _{\tau}$ defined on the curved manifold $\mathcal{M}_{s}$ and
the Jacobi vector field intensity $J_{\mathcal{M}_{s}}$. Does our analysis
lead to any possible further understanding of the role of statistical
curvature in physics and statistics? We argue that it does.

The role of curvature in physics is fairly well understood. It encodes
information about the field strengths for all the four fundamental
interactions in nature. The curvature plays a key role in the Riemannian
geometric approach to chaos \cite{casetti}. In this approach, the study of the
Hamiltonian dynamics is reduced to the investigation of geometrical properties
of the manifold on which geodesic flow is induced. For instance, the stability
or local instability of geodesic flows depends on the sectional curvature
properties of the suitable defined metric manifold. The sectional curvature
brings the same qualitative and quantitative information that is provided by
the Lyapunov exponents in the conventional approach. Furthermore, the
integrability of the system is connected with existence of Killing vectors on
the manifold. However, a rigorous relation among curvature, Lyapunov exponents
and Kolmogorov-Sinai entropy \cite{kawabe} is still under investigation. In
addition, there does not exist a well defined unifying characterization of
chaos in classical and quantum physics \cite{caves} due to fundamental
differences between the two theories. Even in other fields of research (for
instance, statistical inference) the role of curvature is not well understood.
The meaning of statistical curvature for a one-parameter model in inference
theory was introduced in \cite{efron}. Curvature served as an important tool
in the asymptotic theory of statistical estimation. The higher the scalar
curvature at a given point on the manifold, the more difficult it is to do
estimation at that point \cite{rodriguez}. Our analysis may be useful to shed
light in statistical inference theory as well.

Recall that the entropy-like quantity $\mathcal{S}$ is the asymptotic limit of
the natural logarithm of the average of the statistical volume $\left\langle
V_{\mathcal{M}_{s}}\right\rangle _{\tau}\ $associated to the evolution of the
geodesics on $\mathcal{M}_{s}$ . Considering equations
(\ref{asymptotic IG entropy}) and (\ref{IG (2)}), we obtain,%
\begin{equation}
\mathcal{S}_{\mathcal{M}_{s_{2}}}\overset{\tau\rightarrow\infty}{\approx
}2\mathcal{S}_{\mathcal{M}_{s_{1}}}\text{.} \label{imp1}%
\end{equation}
Furthermore, the relationship between the statistical curvatures on the curved
manifolds $\mathcal{M}_{s_{1}}$ and $\mathcal{M}_{s_{2}}$ is,
\begin{equation}
\mathcal{R}_{\mathcal{M}_{s_{2}}}=2\mathcal{R}_{\mathcal{M}_{s_{1}}}\text{.}
\label{imp2}%
\end{equation}
In view of (\ref{imp1}) and (\ref{imp2}), it follows that there is a direct
proportionality between the curvature $R_{\mathcal{M}_{s}}$ and the asymptotic
expression for the entropy-like quantity $S$ characterizing the ED on
manifolds $\mathcal{M}_{s_{i}}$ with $i=1$, $2$, namely%
\begin{equation}
\frac{\mathcal{R}_{\mathcal{M}_{s_{2}}}}{\mathcal{R}_{\mathcal{M}_{s_{1}}}%
}=\frac{\mathcal{S}_{\mathcal{M}_{s_{2}}}}{\mathcal{S}_{\mathcal{M}_{s_{1}}}%
}\text{.} \label{imp3}%
\end{equation}
Moreover, from (\ref{asymptotic j}) and (\ref{asymptotic j2}), we obtain the
following relation,%
\begin{equation}
J_{\mathcal{M}_{s_{2}}}\overset{\tau\rightarrow\infty}{\approx}2J_{\mathcal{M}%
_{s_{1}}}\text{.} \label{imp4}%
\end{equation}
The two manifolds $\mathcal{M}_{s_{1}}$ and $\mathcal{M}_{s_{2}}$ are
exponentially unstable and the intensity of Jacobi vector field
$J_{\mathcal{M}_{s_{2}}}$ of manifold $\mathcal{M}_{s_{2}}$ with curvature
$\mathcal{R}_{\mathcal{M}_{s_{2}}}=-2$ is asymptotically twice the intensity
of the Jacobi vector field $J_{\mathcal{M}_{s_{1}}}$of manifold $\mathcal{M}%
_{s_{1}}$with curvature $\mathcal{R}_{\mathcal{M}_{s_{1}}}=-1$. Considering
(\ref{imp2}) and (\ref{imp4}), we obtain
\begin{equation}
\frac{\mathcal{R}_{\mathcal{M}_{s_{2}}}}{\mathcal{R}_{\mathcal{M}_{s_{1}}}%
}=\frac{J_{\mathcal{M}_{s_{2}}}}{J_{\mathcal{M}_{s_{1}}}}\text{.} \label{imp5}%
\end{equation}
It seems there exists a direct proportionality between the curvature
$R_{\mathcal{M}_{s}}$ and the intensity of the Jacobi field $J_{\mathcal{M}%
_{s}}$ characterizing the degree of chaoticity of a statistical manifold of
negative curvature $\mathcal{M}_{s}$. Finally, comparison of (\ref{imp3}) and
(\ref{imp5}) leads to the formal link between curvature, entropy and
chaoticity:%
\begin{equation}
\mathcal{R}\sim\mathcal{S}\sim J\text{.}%
\end{equation}
Though several points need deeper understanding and analysis, we hope that our
work shows that this information-geometric approach may be useful in providing
a unifying framework to study chaos on statistical manifolds underlying
entropic dynamical models.

\section{Conclusions}

Two chaotic entropic dynamical models have been considered: a $3D$ and $4D$
statistical manifold $\mathcal{M}_{s_{1}}$ and $\mathcal{M}_{s_{2}}$
respectively. These manifolds serve as the stage on which the entropic
dynamics takes place. In the former case, macro-coordinates on the manifold
are represented by the expectation values of microvariables associated with
Gaussian and exponential probability distributions. In the latter case,
macro-coordinates are expectation values of microvariables associated with two
Gaussians distributions. The geometric structure of $\mathcal{M}_{s_{1}}$ and
$\mathcal{M}_{s_{2}}$ was studied in detail. It was shown that $\mathcal{M}%
_{s_{1}}$ is a curved manifold of constant negative curvature $-1$ while
$\mathcal{M}_{s_{2}}$ has constant negative curvature $-2$. The geodesics of
the ED models are hyperbolic curves on $\mathcal{M}_{s_{i}}$ $\left(
i=1\text{, }2\right)  $. A study of the stability of geodesics on
$\mathcal{M}_{s_{1}}$ and $\mathcal{M}_{s_{2}}$ was presented. The notion of
statistical volume elements was introduced to investigate the asymptotic
behavior of a one-parameter family of neighboring volumes $\mathcal{F}%
_{V_{\mathcal{M}_{s}}}\left(  \alpha\right)  \equiv\left\{  V_{\mathcal{M}%
_{s}}\left(  \tau\text{; }\alpha\right)  \right\}  _{\alpha\in%
\mathbb{R}
^{+}}$. \emph{An information-geometric analog of the Zurek-Paz chaos criterion
was presented}. It was shown that the behavior of geodesics is characterized
by exponential instability that leads to chaotic scenarios on the curved
statistical manifolds. These conclusions are supported by a study based on the
geodesic deviation equations and on the asymptotic behavior of the Jacobi
vector field intensity $J_{\mathcal{M}_{s}}$ on $\mathcal{M}_{s_{1}}$ and
$\mathcal{M}_{s_{2}}$. A Lyapunov exponent analog similar to that appearing in
the Riemannian geometric approach was suggested as an indicator of chaos. On
the basis of our analysis a relationship among an entropy-like quantity,
chaoticity and curvature in the two models ED1 and ED2 is proposed, suggesting
to interpret the statistical curvature as a measure of the entropic dynamical chaoticity.

The implications of this work are twofold. Firstly, it helps to understand the
possible future use of the statistical curvature in modelling real processes
by relating it to conventionally accepted quantities such as entropy (be it
the KS entropy, the Shannon entropy, the IGE entropy (that we have introduced
in this Chapter), the Kolmogorov complexity, the von Neumann entropy, the
quantum dynamical entropy, etc. etc.) and chaos. On the other hand, it serves
to cast what is already known in physics regarding curvature in a new light as
a consequence of its proposed link with inference. Finally we remark that
based on the results obtained from the chosen ED models, it is not
unreasonable to think that should the correct variables describing the true
degrees of freedom of a physical system be identified, perhaps deeper insights
into the foundations of models of physics and reasoning (and their
relationship to each other) may be uncovered.

\pagebreak

\begin{center}
{\LARGE Chapter 6: Information-constrained dynamics, Part II: Newtonian
entropic dynamics}
\end{center}

In collaboration with Prof. Ariel Caticha, I show that the ED formalism is not
purely a mathematical framework; it is indeed a general theoretical scheme
where conventional Newtonian dynamics is obtained as a special limiting case.
Newtonian dynamics is derived from prior information codified into an
appropriate statistical model. The basic assumption is that there is an
irreducible uncertainty in the location of particles so that the state of a
particle is defined by a probability distribution. The corresponding
configuration space is a statistical manifold the geometry of which is defined
by the information metric. The trajectory follows from a principle of
inference, the method of Maximum Entropy. No additional "physical" postulates
such as an equation of motion, or an action principle, nor the concepts of
momentum and of phase space, not even the notion of time, need to be
postulated. The resulting entropic dynamics reproduces the Newtonian dynamics
of any number of particles interacting among themselves and with external
fields. Both the mass of the particles and their interactions are explained in
terms of the underlying statistical manifold.

\section{Introduction}

In this Chapter, we use well established principles of inference to derive
Newtonian dynamics from relevant prior information codified into a statistical
model \cite{caticha-cafaro}. We do not assume equations of motion or
principles of least action. Moreover, neither the concept of momentum nor that
of the associated phase space is assumed. Indeed, not even the notion of an
absolute Newtonian time is postulated. Firstly, we construct a suitable
statistical model of the space of states of a system of particles. The
statistical configuration space is automatically endowed with a geometry and
this information geometry turns out to be unique \cite{amari, cencov}.
Secondly, we tackle the dynamics: given the initial and the final states, we
investigate what trajectory the system is expected to follow. In the
conventional approach one postulates an equation of motion or an action
principle that presumably reflects a "law of nature". However, in our
theoretical framework, the dynamics follows from a principle of inference, the
method of Maximum (relative) Entropy, ME \cite{catichaME}. We show that with a
suitable choice of the statistical manifold the resulting "entropic
dynamics"\ \cite{caticha02, caticha03} reproduces Newtonian dynamics, or more
properly, \emph{Newtonian entropic dynamics} (NED).

\section{Configuration space as a statistical manifold}

Let us consider a single particle moving in space $%
\mathbb{R}
^{3}$: the configuration space is a three dimensional manifold $\mathcal{M}%
_{s}$ with some unknown metric tensor $g_{ij}(\theta)$. Our main assumption is
that there is a certain fuzziness to space $%
\mathbb{R}
^{3}$; there is an irreducible uncertainty in the location of the particle.
Thus, the assertion \textquotedblleft the particle is at the point $\theta
$\textquotedblright\ means that its "true"\ position $x$ is somewhere in the
vicinity of $\theta$. This leads us to associate a probability distribution
$p(x|\theta)$ to each point $\theta$ and the configuration space is thus
transformed into a statistical manifold $\mathcal{M}_{s}$ : a point $\theta$
is no longer a structureless dot but a probability distribution. Remarkably
there is a \emph{unique} measure of the extent to which the distribution at
$\theta$ can be distinguished from the neighboring distribution at
$\theta+d\theta$. It is the information metric of Fisher and Rao \cite{amari}.
Thus, physical space, when viewed as a statistical manifold, inherits a metric
structure from the distributions $p(x|\theta)$. We will assume that the
originally unspecified metric $g_{ij}(\theta)$ is precisely the information
metric induced by the distributions $p(x|\theta)$.

\subsection{The Gaussian Model and the Covariance Problem}

Given a manifold $\mathcal{M}_{S}$ of probability distributions $\left\{
p\left(  x|\theta\right)  \right\}  $, the problem is to find the
corresponding information metric $g_{\mu\nu}\left(  \theta\right)  $. This is
commonly called the \emph{direct problem}. Its solution might be laborious but
it is quite mechanical. In this Chapter, we are tackling what is called the
\emph{inverse problem}: constructing a statistical manifold $\mathcal{M}_{S}$
with a given metric tensor $g_{\mu\nu}\left(  \theta\right)  $. In
\cite{caticha03}, it was proposed that a Gaussian model,
\begin{equation}
p(x|\theta)=\frac{\gamma^{1/2}(\theta)}{(2\pi)^{3/2}}\,\exp\left[
-\frac{\gamma_{ij}(\theta)(x^{i}-\theta^{i})(x^{j}-\theta^{j})}{2}\right]
\text{,} \label{Gaussian a}%
\end{equation}
where $\gamma=\det\gamma_{ij}$, encodes the physically relevant information,
which consists of an estimate of the particle position,
\begin{equation}
\langle x^{i}\rangle=%
{\textstyle\int}
dx\,p(x|\theta)\,x^{i}=\theta^{i}\text{\thinspace,} \label{exp position}%
\end{equation}
and of its uncertainty, given by the covariance matrix $\tilde{\gamma}%
^{ij}(\theta)$,
\begin{equation}
\left\langle (x^{i}-\theta^{i})(x^{j}-\theta^{j})\right\rangle =\tilde{\gamma
}^{ij}(\theta)~\text{,} \label{exp uncertainty}%
\end{equation}
where $\tilde{\gamma}^{ij}$ $=%
{\textstyle\int}
dx\,p(x|\theta)(x^{i}-\theta^{i})(x^{j}-\theta^{j})$ is the inverse of
$\gamma_{ij}$, $\tilde{\gamma}^{ik}\gamma_{kj}=\delta_{j}^{i}$. It is
worthwhile noticing that the expected values in eqs.(\ref{exp position}) and
(\ref{exp uncertainty}) are not covariant under coordinate transformations.
Indeed, the transformation $x^{\prime i}=f^{i}(x)$ does not lead to
$\theta^{\prime i}=f^{i}(\theta)$ because in general $\langle f(x)\rangle\neq
f(\langle x\rangle)$ except when uncertainties are small. Our Gaussian model
can at best be an approximation valid when $p(x|\theta)$ is sharply localized
in a very small region within which curvature effects are negligible.
Fortunately, this is all we need for our present purpose. The information
distance between $p(x|\theta)$ and $p(x|\theta+d\theta)$ is calculated from
(see e.g., \cite{amari})
\begin{equation}
d\ell^{2}=\mathcal{G}_{\mu\nu}\,d\theta^{\mu}d\theta^{\nu}\text{.}
\label{info metric(66)}%
\end{equation}
with $\mathcal{G}_{\mu\nu}=\int dx\,p(x|\theta)\partial_{\mu}\log
p(x|\theta)\partial_{\nu}\log p(x|\theta)$ and $\partial_{\mu}=\frac{\partial
}{\partial\theta^{\mu}}$. Consider the nine-dimensional space of Gaussians
\begin{equation}
p(x|\theta\text{, }\gamma)=\frac{\gamma^{1/2}}{(2\pi)^{3/2}}\exp\left[
-\frac{\gamma_{ij}(x^{i}-\theta^{i})(x^{j}-\theta^{j})}{2}\right]  ~\text{.}%
\end{equation}
Here the parameters $\theta^{\mu}$ include the three $x^{i}$ plus six
independent elements of the symmetric matrix $\gamma_{ij}$.
Eq.(\ref{info metric(66)}) defines the information distance between
$p(x|\theta$, $\gamma)$ and $p(x|\theta+d\theta$, $\gamma+d\gamma)$ as
\begin{equation}
d\ell^{2}=\mathcal{G}_{ij}d\theta^{i}d\theta^{j}+\mathcal{G}_{k}^{ij}%
d\gamma_{ij}d\theta^{k}+\mathcal{G}^{ij\,kl}d\gamma_{ij}d\gamma_{kl}~\text{,}%
\end{equation}
where $\mathcal{G}_{ij}=\gamma_{ij}\,$, $\mathcal{G}_{k}^{ij}=0\,$\ and,
$\mathcal{G}^{ij\,kl}=\frac{1}{4}(\tilde{\gamma}^{ik}\tilde{\gamma}%
^{jl}+\tilde{\gamma}^{il}\tilde{\gamma}^{jk})$ with $\tilde{\gamma}^{ik}%
\gamma_{kj}=\delta_{j}^{i}$. Therefore,
\begin{equation}
d\ell^{2}=\gamma_{ij}d\theta^{i}d\theta^{j}+\frac{1}{2}\tilde{\gamma}%
^{ik}\tilde{\gamma}^{jl}d\gamma_{ij}d\gamma_{kl}~\text{.}
\label{info metric 9d}%
\end{equation}
This is the metric of the full nine-dimensional manifold, but it is not what
we need. What we want is the metric of the embedded three-dimensional
submanifold where $\gamma_{ij}=\gamma_{ij}(\theta)$ is some function of
$\theta$. The tensorial behavior of $\gamma_{ij}$ in equation
(\ref{Gaussian a}) is our major concern. As we said, the notion of expected
value is not covariant. Our approach is an approximation that is valid when
the position uncertainty is much smaller than the local radius of curvature of
the manifold; or, alternatively, the effects of space curvature are negligible
within the region where $p(x|\theta)$ is appreciable. Moreover, $g_{ij}$ is a
tensor while $\tilde{\gamma}^{\mu\nu}$ does not behave as a tensor (in
general). However, it can be shown that special cases exist where the
tensorial behavior of $\gamma_{ij}$ can be restored. Namely, if we consider
exclusively linear change of coordinates\textit{\ }$\theta^{k}\rightarrow
\theta^{\prime r}=\theta^{\prime r}\left(  \theta^{k}\right)  $ such
that\textit{,}%
\begin{equation}
\frac{\partial^{2}\theta^{\prime k}}{\partial\theta^{i}\partial\theta^{j}}=0
\end{equation}
that is, if we assume that the new set of coordinates $\left\{  \theta^{\prime
r}\right\}  $ are not allowed to depend on the old set $\left\{  \theta
^{k}\right\}  $ in a nonlinear way, than we may conclude that $\gamma_{ij}$
transforms as a tensor. In these very restrictive conditions, we may conclude
that $\gamma_{ij}$\ has tensorial behavior provided that the new coordinates
are not allowed to depend on the old coordinates (source coordinates) in a
nonlinear way.

\subsection{The Gaussian-like Model and the Solution to the Covariance
Problem}

One of the major points in the indirect problem is choosing the correct
variance-covariance field tensor $\gamma_{ij}(\theta)$ that leads to a given
metric tensor $g_{ij}(\theta)$ on $\mathcal{M}_{S}$. We want to devise fully
covariant models. This can be achieved by constructing Gaussian-like
probability distributions that, in the limit of small uncertainties,
approximate the Gaussian distributions in (\ref{Gaussian a}). Such a
distribution may be defined as,%
\begin{equation}
p(x|\theta)=\frac{1}{\zeta}\gamma^{1/2}\left(  x\right)  \exp\left[
-\frac{\ell^{2}(x\text{, }\theta)}{2\sigma^{2}(\theta)}\right]  \text{.}
\label{zizi}%
\end{equation}
The quantity $\gamma\overset{\text{def}}{=}\det\left(  \gamma_{ij}\right)  $
in (\ref{zizi}) is the determinant of the positive definite metric tensor
field $\gamma_{ij}$ satisfying the relation,%
\begin{equation}
d\ell^{2}=\gamma_{ij}d\theta^{i}d\theta^{j}\text{.}%
\end{equation}
The quantity $\sigma(\theta)$ is a scalar field, $\ell(x$, $\theta)$ is the
$\gamma$-length along the $\gamma$-geodesic from the point $\theta$ to the
point $x$ and $\zeta$ is a normalization constant. The proposed probability
distribution in (\ref{zizi}) is a manifestly covariant quantity: the
normalization constant $\zeta$, the $\gamma$-length $\ell(x$, $\theta)$, the
scalar field $\sigma(\theta)$ and, $dx\,\gamma^{1/2}(x)$ are all invariant
quantities. We remark that the space of spherically symmetric Gaussians with
prescribed $\sigma(\theta)$,%
\begin{equation}
p(x|\theta)=\frac{1}{(2\pi\sigma^{2})^{3/2}}\exp\left[  -\frac{1}{2\sigma^{2}%
}\delta_{ij}(x^{i}-\theta^{i})(x^{j}-\theta^{j})\right]  \text{,}
\label{easycase}%
\end{equation}
is a special case of the distributions defined in (\ref{zizi}). In this case,
the variance-covariance matrix in (\ref{Gaussian a}) is diagonal and
proportional to the unit matrix,
\begin{equation}
\gamma_{ij}=\frac{1}{\sigma^{2}\left(  \theta\right)  }\delta_{ij}\text{,
}\tilde{\gamma}^{ij}=\sigma^{2}\left(  \theta\right)  \delta^{ij}%
\,\text{\ and, }\quad\gamma=\frac{1}{\sigma^{6}\left(  \theta\right)
}\text{.}%
\end{equation}
Differentiating $\gamma_{ij}$ and using the fact that $d\sigma=\partial
_{k}\sigma\left(  \theta\right)  d\theta^{k}$ with $\partial_{k}%
\overset{\text{def}}{=}\frac{\partial}{\partial\theta^{k}}$, we get%
\begin{equation}
d\gamma_{ij}=-\frac{2\delta_{ij}}{\sigma^{3}\left(  \theta\right)  }%
\partial_{k}\sigma\left(  \theta\right)  d\theta^{k}\text{.} \label{quest}%
\end{equation}
Substituting (\ref{quest}) in (\ref{info metric 9d}), we obtain
\begin{equation}
d\ell^{2}=\frac{1}{\sigma^{2}\left(  \theta\right)  }\delta_{ij}d\theta
^{i}d\theta^{j}+\frac{1}{2}\sigma^{4}\left(  \theta\right)  \delta^{ik}%
\delta^{jl}\frac{2\delta_{ij}}{\sigma^{3}\left(  \theta\right)  }\frac
{2\delta_{kl}}{\sigma^{3}\left(  \theta\right)  }\partial_{m}\sigma\left(
\theta\right)  \partial_{n}\sigma\left(  \theta\right)  d\theta^{m}d\theta^{n}%
\end{equation}
which, using $\delta^{ik}\delta^{jl}\delta_{ij}\delta_{kl}=\delta_{j}%
^{k}\delta_{k}^{j}=\delta_{k}^{k}=3$, simplifies to
\begin{equation}
d\ell^{2}=\frac{1}{\sigma^{2}(\theta)}\left[  \delta_{ij}+6\partial_{i}%
\sigma\left(  \theta\right)  \,\partial_{j}\sigma\left(  \theta\right)
\right]  d\theta^{i}d\theta^{j}\text{.}%
\end{equation}
In the case of probability distributions (\ref{easycase}), the metric tensor
$g_{ij}(\theta)$ on $\mathcal{M}_{S}$ becomes,%
\begin{equation}
g_{ij}\left(  \theta\right)  =\frac{1}{\sigma^{2}(\theta)}\left[  \delta
_{ij}+6\partial_{i}\sigma\left(  \theta\right)  \,\partial_{j}\sigma\left(
\theta\right)  \right]  \text{.}%
\end{equation}
If the $\sigma(\theta)$ field varies very slowly
\begin{equation}
\partial_{i}\sigma<<1\text{,} \label{limite}%
\end{equation}
then to first order the metric $d\ell^{2}$ is independent of the derivatives
$\partial_{i}\sigma\left(  \theta\right)  $,
\begin{equation}
d\ell^{2}\overset{\partial_{i}\sigma<<1}{\approx}\frac{1}{\sigma^{2}(\theta
)}\delta_{ij}d\theta^{i}d\theta^{j}\text{,}%
\end{equation}
and the metric tensor $g_{ij}(\theta)$ is just a conformal transformation of
flat metric $\delta_{ij}$,%
\begin{equation}
g_{ij}\left(  \theta\right)  \overset{\partial_{i}\sigma<<1}{\approx}\frac
{1}{\sigma^{2}(\theta)}\delta_{ij}\text{.}%
\end{equation}
We point out that condition (\ref{limite}) is compatible with the definition
given in (\ref{1-particle metric}), $\sigma^{2}(\theta)=$ $\frac{\sigma
_{0}^{2}}{\Phi(\theta)}$. Rewriting $\sigma\left(  \theta\right)  $ in terms
of $\Phi\left(  \theta\right)  $, equation (\ref{limite}) reads,%

\begin{equation}
\partial_{i}\sigma\equiv-\frac{1}{2}\sigma_{0}\Phi^{-\frac{3}{2}}\left(
\theta\right)  \partial_{i}\Phi\left(  \theta\right)  <<1\text{.}%
\end{equation}
Therefore, we are allowed to consider very small $\partial_{i}\sigma$ by
simply making the parameter $\sigma_{0}$ small and keeping $\partial_{i}%
\Phi\left(  \theta\right)  \sim F_{i}$ ($F$ would be the information geometric
analog of a conservative force appearing in Newton's second law) arbitrarily
big. For small $\sigma(\theta)$, the distribution in (\ref{zizi}) can be
written as,%
\begin{equation}
p(x|\theta)=\frac{1}{\zeta\left(  \theta\right)  }\gamma^{1/2}\left(
x\right)  \exp\left[  -\frac{\gamma_{ij}\left(  \theta\right)  }{2\sigma
^{2}(\theta)}(x^{i}-\theta^{i})(x^{j}-\theta^{j})\right]  \text{.}
\label{alis}%
\end{equation}
In local Cartesian coordinates (LCC), $\theta^{\left(  \text{old}\right)
}\overset{\text{LCC}}{\rightarrow}\theta^{\left(  \text{new}\right)  }%
\equiv\theta^{\prime}$, we obtain%
\begin{equation}
\gamma_{ij}^{\left(  \text{old}\right)  }\left(  \theta\right)  \overset
{\text{LCC}}{\rightarrow}\gamma_{ij}^{\left(  \text{new}\right)  }\left(
\theta^{\prime}\right)  =\delta_{ij}\text{ with }\partial_{i}^{\prime}%
\gamma_{ij}^{\left(  \text{new}\right)  }\left(  \theta^{\prime}\right)  =0
\end{equation}
and (\ref{alis}) becomes,%
\begin{equation}
p^{\left(  \text{old}\right)  }(x|\theta)\overset{\text{LCC}}{\rightarrow
}p^{\left(  \text{new}\right)  }(x^{\prime}|\theta^{\prime})\approx\frac
{1}{\zeta\left(  \theta^{\prime}\right)  }\gamma^{1/2}\left(  x^{\prime
}\right)  \exp\left[  -\frac{\delta_{ij}}{2\sigma^{2}(\theta^{\prime}%
)}(x^{\prime i}-\theta^{\prime i})(x^{\prime j}-\theta^{\prime j})\right]
\text{.} \label{bom}%
\end{equation}
In the case of probability distributions (\ref{bom}), the metric tensor
$g_{ij}(\theta)$ on $\mathcal{M}_{S}$ becomes,%
\begin{equation}
g_{ij}\left(  \theta\right)  \approx\frac{1}{\sigma^{2}(\theta)}\delta
_{ij}=\frac{1}{\sigma^{2}(\theta)}\gamma_{ij}\left(  \theta\right)  \text{.}
\label{papa}%
\end{equation}
Since (\ref{papa}) is a tensorial equation, its validity is preserved in
arbitrary coordinates system and,%
\begin{equation}
g_{ij}\left(  \theta\right)  \overset{\sigma-\text{small}}{\approx}\frac
{1}{\sigma^{2}(\theta)}\gamma_{ij}\left(  \theta\right)  \text{,}
\label{juppiter}%
\end{equation}
the tensorial relation between $g_{ij}\left(  \theta\right)  $ and
$\gamma_{ij}\left(  \theta\right)  $ is covariantly preserved. In conclusion,
in this subsection we provided a solution to the inverse problem suggesting a
fully covariant Gaussian-like model of probability distributions $\left\{
p(x|\theta)\text{ in (\ref{zizi})}\right\}  $ forming a curved statistical
manifold $\mathcal{M}_{S}$ with a given metric tensor given, in the limit of
small uncertainties, by $g_{ij}(\theta)$ in (\ref{juppiter}). The possibility
to extend this result in regions of high curvature, such as near
singularities, remains to be ascertained.

\section{Newtonian Entropic Dynamics of a single particle}

Our objective is to construct ED models on statistical manifolds of
Gaussian-like probability distributions presented in the previous section. We
follow the work presented in Chapter 3. Assume there exists a continuous path,
the key question is to find the trajectory the system is expected to follow,
given an initial and a final state. A large change is the result of a
succession of very many small changes and therefore we only need to determine
the properties of a short segment of the trajectory. The idea behind entropic
dynamics is that as the system moves from a point $\theta$ to a neighboring
point $\theta+\Delta\theta$ it must pass through a halfway point.
\cite{caticha02}. The basic dynamical question can now be rephrased as
follows: the system is initially described by the probability distribution
$p(x|\theta)$ and we are given the information that it has moved to one of the
neighboring states in the family $p(x|\theta^{\prime})$ where the
$\theta^{\prime}$ lie on the plane halfway between the initial $\theta$ and
the final $\theta+\Delta\theta$. Which $p(x|\theta^{\prime})$ do we select?
The answer is given by the method of maximum (relative) entropy, ME. The
selected distribution is that which maximizes the entropy of $p(x|\theta
^{\prime})$ relative to the prior $p(x|\theta)$ subject to the constraint that
$\theta^{\prime}$ is equidistant from $\theta$ and $\theta+\Delta\theta$. The
result is that the selected $\theta^{\prime}$ minimizes the distance to
$\theta$ and therefore the three points $\theta$, $\theta^{\prime}$ and
$\theta+\Delta\theta$ lie on a straight line. Since any three neighboring
points along the trajectory must line up, the trajectory predicted by entropic
dynamics is the geodesic that minimizes the length
\begin{equation}
\mathcal{J}=%
{\textstyle\int\limits_{\xi_{i}}^{\xi_{f}}}
d\xi\left[  g_{ij}\left(  \theta\right)  \dot{\theta}^{i}\text{ }\dot{\theta
}^{j}\right]  ^{1/2}\quad\text{with}\quad\dot{\theta}^{i}=\frac{d\theta^{i}%
}{d\xi}~\text{,} \label{Jacobi}%
\end{equation}
where $\xi$ is any parameter that labels points along the curve, $\theta
^{i}=\theta^{i}(\xi)$. In entropic dynamics, the minimal-length geodesics
represent the only family of curves that is singled out as special. The
construction of useful physics models does not require any additional
structure and therefore none will be introduced. The simplest statistical
model is a three-dimensional manifold of spherically symmetric Gaussians with
constant variance $\sigma_{0}^{2}$. From (\ref{papa}) follows that the
corresponding information metric is
\begin{equation}
g_{ij}^{(0)}(\theta)=\gamma_{ij}^{(0)}(\theta)=\frac{1}{\sigma_{0}^{2}}%
\delta_{ij}~\text{,} \label{euclidean  metric}%
\end{equation}
where $\delta_{ij}$ is the familiar metric of flat Euclidean space. We point
out that already in such a simple model entropic dynamics reproduces the
familiar straight line trajectories that are commonly associated with Galilean
inertial motion. However, non-trivial entropic dynamical models require some
curvature. For instance, consider the model of spherically symmetric Gaussians
where the variance is a non-uniform scalar field $\sigma^{2}(\theta)$. It is
convenient to write the corresponding information metric $g_{ij}(\theta)$ as
the Euclidean metric eq.(\ref{euclidean metric}) modulated by a positive
conformal factor $\Phi(\theta)$,
\begin{equation}
g_{ij}(\theta)=\gamma_{ij}(\theta)=\frac{\Phi(\theta)}{\sigma_{0}^{2}}%
\delta_{ij}~\text{, }\sigma^{2}(\theta)=\text{ }\frac{\sigma_{0}^{2}}%
{\Phi(\theta)}\text{.} \label{1-particle metric}%
\end{equation}
The conformal factor $\Phi(\theta)$ defines an angle-preserving transformation
(conformal transformation) and its effect is a local dilation. It is useful to
rewrite the length eq.(\ref{Jacobi}) with the metric (\ref{1-particle metric})
in the form%
\begin{equation}
\mathcal{J}=2^{1/2}%
{\textstyle\int\limits_{\xi_{i}}^{\xi_{f}}}
d\xi\,\mathcal{L}(\theta\text{, }\dot{\theta})~\text{,} \label{Jacobi a}%
\end{equation}
with a "Lagrangian"\ function $\mathcal{L}(\theta,\dot{\theta})$ given by,%
\begin{equation}
\mathcal{L}(\theta,\dot{\theta})=[\Phi(\theta)T_{\xi}(\dot{\theta}%
)]^{1/2}\text{, }T_{\xi}\overset{\text{def}}{=}\frac{1}{2\sigma_{0}^{2}}%
\delta_{ij}\frac{d\theta^{i}}{d\xi}\frac{d\theta^{j}}{d\xi}\text{.}
\label{one(6)}%
\end{equation}
The geodesics follow from the Lagrange equations,%
\begin{equation}
\frac{d}{d\xi}\frac{\partial\mathcal{L}(\theta\text{, }\dot{\theta})}%
{\partial\dot{\theta}^{i}}-\frac{\partial\mathcal{L}(\theta\text{, }%
\dot{\theta})}{\partial\theta^{i}}=0 \label{two(6)}%
\end{equation}
that is, substituting (\ref{one(6)}) in (\ref{two(6)}),%
\begin{equation}
\frac{1}{\sigma_{0}^{2}}\left(  \frac{\Phi}{T_{\xi}}\right)  ^{1/2}\frac
{d}{d\xi}\left[  \left(  \frac{\Phi}{T_{\xi}}\right)  ^{1/2}\frac{d\theta^{i}%
}{d\xi}\right]  -\frac{\partial\Phi}{\partial\theta^{i}}=0~\text{.}
\label{EQmotion}%
\end{equation}
These equations can be simplified considerably once we notice that the
parameter $\xi$ is arbitrary. Let us replace the original $\xi$ with a
new\ parameter $\tau$ defined as%
\begin{equation}
d\tau=\left(  \frac{T_{\xi}}{\Phi}\right)  ^{1/2}d\xi\quad\text{or}\quad
\frac{d}{d\tau}=\left(  \frac{\Phi}{T_{\xi}}\right)  ^{1/2}\frac{d}{d\xi
}~\text{.} \label{new t}%
\end{equation}
In terms of the new $\tau$ the equation of motion (\ref{EQmotion}) becomes,%
\begin{equation}
\frac{1}{\sigma_{0}^{2}}\frac{d^{2}\theta^{i}}{d\tau^{2}}-\frac{\partial\Phi
}{\partial\theta^{i}}=0~\text{.} \label{Newton a}%
\end{equation}
In order to allow us the possibility of considering the limit of small
uncertainties, $\sigma_{0}\ll1$, we redefine the relation between $\xi$ and
$\tau$ introducing a new parameter $\kappa$ ($\kappa$ has the dimensions of a
time) such that,%
\begin{equation}
d\tau=\kappa\left(  \frac{T_{\xi}}{\Phi}\right)  ^{1/2}d\xi\quad\text{or}%
\quad\kappa\frac{d}{d\tau}=\left(  \frac{\Phi}{T_{\xi}}\right)  ^{1/2}\frac
{d}{d\xi}~\text{.} \label{new t2}%
\end{equation}
From eq.(\ref{new t2}) the new $\tau$ is such that
\begin{equation}
\Phi=\kappa^{2}T_{\xi}\left(  \frac{d\xi}{d\tau}\right)  ^{2}=T_{\tau}\text{,
}T_{\tau}\overset{\text{def}}{=}\frac{1}{2}\frac{\kappa^{2}}{\sigma_{0}^{2}%
}\delta_{ij}\frac{d\theta^{i}}{d\tau}\frac{d\theta^{j}}{d\tau}~\text{.}
\label{E cons a}%
\end{equation}
Eqs.(\ref{Newton a}) and (\ref{E cons a}) are equivalent to Newtonian
dynamics. To make it explicit we introduce a "mass"\ $m$ and a
"potential"\ $\phi(\theta)$ through a mere change of notation,
\begin{equation}
m\overset{\text{def}}{=}\frac{\varepsilon\kappa^{2}}{\sigma_{0}^{2}}%
\quad\text{and}\quad\varepsilon\Phi(\theta)\overset{\text{def}}{=}%
E-\phi(\theta)\text{.}%
\end{equation}
The parameter $\varepsilon$ has been introduced because of dimensional
analysis convenience ($\varepsilon$ has the dimensions of an energy;
$\sigma_{0}$ has the dimensions of a length) and the constant $E$ reflects the
freedom to add a constant to the potential. The result is the Newtonian
information-dynamical equation,
\begin{equation}
m\frac{d^{2}\theta^{i}}{d\tau^{2}}+\frac{\partial\phi}{\partial\theta^{i}%
}=0~\text{,} \label{Newton b}%
\end{equation}
and energy conservation,%
\begin{equation}
\frac{1}{2}m\delta_{ij}\frac{d\theta^{i}}{d\tau}\frac{d\theta^{j}}{d\tau}%
+\phi(\theta)=E~\text{.} \label{E cons b}%
\end{equation}
Thus, the constant $E$ is interpreted as energy. We have just derived $F=ma$
purely from principles of inference applied to the relevant information
codified into a statistical model! From eq.(\ref{Jacobi}) onwards our
inference approach is formally identical to the Jacobi action principle of
classical mechanics \cite{lanczos} but we did not need to know this. Within
our theoretical construct, both the mass $m$ of the particles and their
interactions are explained in terms of an irreducible uncertainty of their
positions. Masses and interactions are features of the curved statistical
manifold underlying the information-dynamics. Even though our formalism
describes a non-relativistic model, there already appears a
"unification"\ between mass $m$ and potential energy $\phi(\theta)$: they are
different aspects of the same thing, the particle's "intrinsic" position
uncertainty $\sigma_{0}$ modulated throughout space by the field $\Phi
(\theta)$. The derivation presented in this section illustrates the main idea
but has two important limitations. First, it applies to a single particle with
a fixed constant energy $E$ and this means that we consider only isolated
systems. Second, even though we have identified a convenient parameter $\tau$,
we do not know that it actually represents "true"\ time. It could be that
$\tau$ is the universal Newtonian time. It could be that $\tau$ is just a
parameter that applies only to one particular isolated particle. The original
formulation in terms of the "Jacobi"\ action, eq.(\ref{Jacobi a}), is
completely timeless. Therefore the appearance of time is obscure. The solution
to both these problems emerges as we apply the formalism to the motion of the
only system known to be completely isolated: the whole universe. In this case,
the fact that the energy $E$ is a fixed constant does not represent a
restriction. Moreover, since the preferred time parameter would be associated
to the whole universe, it would not be at all inappropriate to call it the
\emph{universal} time.

\section{Newtonian Entropic Dynamics of the Whole Universe}

Our theoretical scheme may be generalized to arbitrary $N$-particles
interacting with external potentials and also with each other
\cite{caticha-cafaro}. However, to simplify our notation we will consider a
universe that consists of $N=2$ particles. For the $2$-particle system the
position $\theta=(\theta_{1}$, $\theta_{2})$ is denoted by $6$ coordinates
$\theta^{A}$ with $A=1$, $2$,.., $6$. Let $\theta^{A}=(\theta^{i_{1}}$,
$\theta^{i_{2}})$ with $i_{1}=1$, $2$, $3$ for particle 1 and $i_{2}=4$, $5$,
$6$ for particle 2. A point in the $N=2$ configuration space is a Gaussian
distribution,
\begin{equation}
p(x|\theta)=\frac{\gamma^{1/2}(\theta)}{(2\pi)^{3/2}}\,\exp\left[
-\frac{\gamma_{AB}(\theta)(x^{A}-\theta^{A})(x^{B}-\theta^{B})}{2}\right]
~\text{.} \label{Gaussian b}%
\end{equation}
The simplest model for two (possibly non-identical) particles assigns uniform
variances $\sigma_{1}^{2}$ and $\sigma_{2}^{2}$ to each particle. The
corresponding metric, analogous to eq.(\ref{euclidean metric}), is
\begin{equation}
g_{AB}^{(0)}=\gamma_{AB}^{(0)}=\tilde{m}_{AB}~\text{,}
\label{euclidean metric b}%
\end{equation}
where $m_{AB}$ is a constant $6\times6$ diagonal matrix,
\begin{equation}
\tilde{m}_{AB}\overset{\text{def}}{=}%
\begin{bmatrix}
\delta_{i_{1}j_{1}}/\sigma_{1}^{2} & 0\\
0 & \delta_{i_{2}j_{2}}/\sigma_{2}^{2}%
\end{bmatrix}
~\text{,}%
\end{equation}
where each entry represents a $3\times3$ matrix. The metric $m_{AB}$ describes
a flat space; the trajectories are familiar "straight"\ lines and the
particles move independently of each other; they do not interact. However,
non-trivial dynamics requires the introduction of curvature and the simplest
way to do this is through an overall conformal field $\Phi(\theta) $ with
$\theta=(\theta_{1}$, $\theta_{2})$. We propose
\begin{equation}
g_{AB}(\theta)=\gamma_{AB}(\theta)=\Phi(\theta)\tilde{m}_{AB}~\text{.}
\label{2-particle metric}%
\end{equation}
The equation of motion for the $N=2$ universe is the geodesic that minimizes
\begin{equation}
\mathcal{J}=2^{1/2}%
{\textstyle\int\limits_{\xi_{i}}^{\xi_{f}}}
d\xi\,\mathcal{L}(\theta_{1}\text{, }\theta_{2}\text{, }\dot{\theta}%
_{1}\text{, }\dot{\theta}_{2})~\text{,} \label{Jacobi b}%
\end{equation}
where $\mathcal{L}(\theta$, $\dot{\theta})=[\Phi(\theta)T_{\xi}(\dot{\theta
})]^{1/2}\quad$and$\quad T_{\xi}(\dot{\theta})=\frac{1}{2}\tilde{m}_{AB}%
\dot{\theta}^{A}\dot{\theta}^{B}$. The Lagrange equations yield,
\begin{equation}
\tilde{m}_{AB}\left(  \frac{\Phi}{T_{\xi}}\right)  ^{1/2}\frac{d}{d\xi}\left[
\left(  \frac{\Phi}{T_{\xi}}\right)  ^{1/2}\frac{d\theta^{B}}{d\xi}\right]
-\frac{\partial\Phi}{\partial\theta^{A}}=0~\text{,}%
\end{equation}
which suggests introducing a new parameter $\tau$ defined by
\begin{equation}
d\tau=\kappa\left(  \frac{T_{\xi}}{\Phi}\right)  ^{1/2}d\xi\quad\text{or}%
\quad\kappa\frac{d}{d\tau}=\left(  \frac{\Phi}{T_{\xi}}\right)  ^{1/2}\frac
{d}{d\xi}~\text{.} \label{time t}%
\end{equation}
In terms of the new parameter the equations of motion are
\begin{equation}
m_{AB}\frac{d^{2}\theta^{A}}{d\tau^{2}}-\varepsilon\frac{\partial\Phi
}{\partial\theta^{A}}=0~\text{.} \label{qq}%
\end{equation}
where $m_{AB}\overset{\text{def}}{=}\varepsilon\kappa^{2}\tilde{m}_{AB}$ is a
diagonal matrix. Equation (\ref{qq}) becomes
\begin{equation}
\frac{\varepsilon\kappa^{2}}{\sigma_{n}^{2}}\frac{d^{2}\theta^{i_{n}}}%
{d\tau^{2}}-\varepsilon\frac{\partial}{\partial\theta^{i_{n}}}\Phi(\theta
_{1}\text{, }\theta_{2})=0\text{~,}%
\end{equation}
for each of the particles, $n=1$, $2$. The motion of particle 1 depends on the
location of particle 2: \emph{these are interacting particles! }The new time
parameter $\tau$, eq.(\ref{time t}), is such that
\begin{equation}
\Phi=\kappa^{2}T_{\xi}\left(  \frac{d\xi}{d\tau}\right)  ^{2}=T_{\tau}\text{,
}T_{\tau}=\frac{1}{2}\kappa^{2}\tilde{m}_{AB}\frac{d\theta^{A}}{d\tau}%
\frac{d\theta^{B}}{d\tau}~\text{.}%
\end{equation}
As before, the equivalence to Newtonian dynamics is made explicit by a change
of notation,
\begin{equation}
\frac{\varepsilon\kappa^{2}}{\sigma_{n}^{2}}\overset{\text{def}}{=}m_{n}%
\quad\text{and}\quad\varepsilon\Phi(\theta)=E-\phi(\theta)\text{.}%
\end{equation}
The result is
\begin{equation}
m_{n}\frac{d^{2}\theta^{i_{n}}}{d\tau^{2}}+\frac{\partial}{\partial
\theta^{i_{n}}}\phi(\theta_{1}\text{, }\theta_{2})=0\text{,}\quad
\end{equation}
with%
\begin{equation}
\varepsilon\Phi\left(  \theta_{1}\text{, }\theta_{2}\right)  =\frac{1}%
{2}m_{AB}\frac{d\theta^{A}}{d\tau}\frac{d\theta^{B}}{d\tau}=E-\phi(\theta
_{1},\theta_{2})\text{.} \label{pizzaplace}%
\end{equation}
The constant $E$ in (\ref{pizzaplace}) is the total energy of the universe and
there are no restrictions on the energy of individual subsystems. The choice
for the conformal factor $\Phi(\theta_{1}$, $\theta_{2})$ is quite general.
For instance, we may consider $\Phi(\theta_{1}$, $\theta_{2})$,
\begin{equation}
\Phi(\theta_{1}\text{, }\theta_{2})=-V_{1}(\theta_{1})-V_{2}(\theta
_{2})-U(\theta_{1}\text{, }\theta_{2})+E~\text{,}%
\end{equation}
so the particles can interact with external potentials $V_{1}$ and $V_{2}$ and
also with each other through $U(\theta_{1}$, $\theta_{2})$. The definition of
the parameter $\tau$ requires taking into account all the particles in the
universe. We started with a completely timeless theory, eq.(\ref{Jacobi b}),
and in fact, no \emph{external }time has been introduced. What we have is a
convenient $\tau$ parameter associated to the change of the total system,
which in this case is the whole universe. The universe is its own clock and it
measures universal time. It is worthwhile noticing that the reparametrization
that allowed us to introduce a Newtonian time was possible only because the
same conformal factor $\Phi(\theta)$ applies equally to all particles.

\subsection{Remarks on Newtonian Entropic Dynamics}

Newtonian entropic dynamics offers a new perspective on the concept of mass
and interactions. In order to see this, notice that since $\gamma_{AB}$ in
(\ref{2-particle metric}) is diagonal the distribution (\ref{Gaussian b}) is a
product,
\begin{equation}
p(x|\theta)=p(x_{1}|\theta_{1}\text{, }\theta_{2})p(x_{2}|\theta_{1}\text{,
}\theta_{2})\text{.}%
\end{equation}
Although the model represents interacting particles, the distribution is a
product: the uncertain variables $x_{1}$ and $x_{2}$ are statistically
independent. The coupling arises through the conditioning on $\theta
=(\theta_{1}$, $\theta_{2})$. Focusing our attention on particle 1 (similar
remarks also apply to particle 2), we note the distribution $p(x_{1}%
|\theta_{1}$, $\theta_{2})$ is a spherically symmetric Gaussian,
\begin{equation}
p(x_{1}|\theta_{1}\text{, }\theta_{2})\propto\,\exp\left[  -\frac{\delta
_{ij}(x^{i}-\theta^{i})(x^{j}-\theta^{j})}{2\sigma_{1}^{2}(\theta_{1}\text{,
}\theta_{2})}\right]  \text{~.}%
\end{equation}
The uncertainty in the position of particle 1 is given by $\sigma_{1}%
(\theta_{1}$, $\theta_{2})=\left[  \Phi(\theta_{1}\text{, }\theta_{2}%
)m_{1}\right]  ^{-1/2}$. The mass $m_{1}$ is interpreted in terms of a uniform
background contribution to the uncertainty. Mass is a manifestation of an
uncertainty in location; higher mass reflects a lower uncertainty. On the
other hand, interactions arise from the non-uniformity of $\sigma_{1}%
(\theta_{1}$, $\theta_{2})$ that depends on the location of other particles
through the modulating field $\Phi(\theta_{1}$, $\theta_{2})$. It is amusing
to note that even though this is a non-relativistic model there already
appears a "unification" between mass and (potential) energy: they are
different aspects of the same thing, the position uncertainty.

\section{Conclusions}

In this Chapter, we have shown that the tools of inference- probability,
information geometry and entropy- are sufficiently rich that one can construct
entropic dynamics models that reproduce recognizable laws of physics. Indeed,
preliminary steps towards an entropic dynamics approach to general relativity
appeared in \cite{caticha03}. Moreover, an entropic dynamics approach to the
study of chaos has already lead to interesting results \cite{cafaro01,
cafaro02, cafaro03}. \ An extended version of these results will be presented
in Chapters 7 and 8. We emphasize that NED has limited applicability being a
nonrelativistic model. Our theoretical construct invokes two metrics: there is
the metric of flat three-dimensional Euclidean space, $\delta_{ij}$, that
appears in the kinetic energies and there is the information metric $g_{ij}$
that accounts for mass $m$ and interactions $\phi$ and applies to the curved
configuration space $\mathcal{M}_{s}$. This is a reflection of the fact that a
system of $N$ particles is described as a point in a $3N$-dimensional
configuration space $\mathcal{M}_{s}$ instead of $N$ points living within the
same evolving three-dimensional space $%
\mathbb{R}
^{3}$. Furthermore, we the choice of the modulating field $\Phi(\theta)$ does
not arise from first principles, it is merely an educated guess. This is no
different from Newton's dictum \emph{hypothesis non fingo}: a choice of
$\Phi(\theta)$ is justified by its explanatory success. Finally, at this very
premature point in the development of our Newtonian entropic dynamics, we do
not offer any physical insight about the underlying fuzziness of space. This
will be one of our major concerns in future works.

\pagebreak

\begin{center}
{\LARGE Chapter 7: Information geometrodynamical approach to chaos: An
Application}
\end{center}

I extend my study of chaotic systems (information geometrodynamical approach
to chaos, IGAC) to an ED Gaussian model describing an arbitrary system of $3N
$ degrees of freedom. It is shown that the hyperbolicity of a non-maximally
symmetric $6N$-dimensional statistical manifold $\mathcal{M}_{S}$ underlying
the ED Gaussian model leads to linear information-geometrodynamical entropy
growth and to exponential growth of the Jacobi vector field intensity, quantum
and classical features of chaos respectively. As a special physical
application, the information geometrodynamical scheme is applied to
investigate the chaotic properties of a set of $n$-uncoupled three-dimensional
anisotropic inverted harmonic oscillators (IHOs) characterized by an Ohmic
distributed frequency spectrum. I show that the asymptotic temporal behavior
of the information geometrodynamical entropy of such a system exhibits linear
growth and I suggest the system studied may be considered to be the classical
information-geometric analogue of the Zurek-Paz quantum chaos criterion in its
classical reversible limit.

\section{\textbf{Introduction}}

The lack of a unified characterization of chaos in classical and quantum
dynamics is well-known. In the Riemannian \cite{casetti} and Finslerian
\cite{cipriani} (a Finsler metric is obtained from a Riemannian metric by
relaxing the requirement that the metric be quadratic on each tangent space)
geometrodynamical approach to chaos in classical Hamiltonian systems, an
active field of research concerns the possibility of finding a rigorous
relation among the sectional curvature, the Lyapunov exponents, and the
Kolmogorov-Sinai dynamical entropy (i.e. the sum of positive Lyapunov
exponents) \cite{kawabe}. The largest Lyapunov exponent characterizes the
degree of chaoticity of a dynamical system and, if positive, it measures the
mean instability rate of nearby trajectories averaged along a sufficiently
long reference trajectory. Moreover, it is known that classical chaotic
systems are distinguished by their exponential sensitivity to initial
conditions and that the absence of this property in quantum systems has lead
to a number of different criteria being proposed for quantum chaos.
Exponential decay of fidelity, hypersensitivity to perturbation, and the
Zurek-Paz quantum chaos criterion of linear von Neumann's entropy growth
\cite{zurek} are some examples \cite{caves}. These criteria accurately predict
chaos in the classical limit, but it is not clear that they behave the same
far from the classical realm. The present work makes use of Entropic Dynamics
(ED) \cite{caticha1}. ED is a theoretical framework that arises from the
combination of inductive inference (Maximum relative Entropy Methods,
\cite{caticha2}) and Information Geometry (Riemannian geometry applied to
probability theory) (IG) \cite{amari}. As such, ED is constructed on
statistical manifolds. It is developed to investigate the possibility that
laws of physics - either classical or quantum - might reflect laws of
inference rather than laws of nature. This Chapter contains works that follow
up a series of my works \cite{cafaro1, cafaro2, cafaro3}. In this Chapter, the
ED theoretical framework is used to explore the possibility of constructing a
unified characterization of classical and quantum chaos. We investigate a
system whose microstates $\left\{  \vec{X}\right\}  $ are characterized by
$3N$ degrees of freedom $\left\{  x_{a}^{\left(  \alpha\right)  }\right\}
_{a=1\text{, }2\text{, }3}^{\alpha=1\text{,..,}n}$. Each degree of freedom is
Gaussian-distributed and it is described by two pieces of relevant
information, its mean expected value and its variance. This leads to consider
an ED model on a non-maximally symmetric $6N$-dimensional statistical manifold
$\mathcal{M}_{s}$. It is shown that $\mathcal{M}_{s}$ possesses a constant
negative Ricci curvature that is proportional to the number of degrees of
freedom of the system, $\mathcal{R}_{\mathcal{M}_{s}}=-3N$. It is shown that
the system explores statistical volume elements on $\mathcal{M}_{s}$ at an
exponential rate. We define a dynamical information-geometric entropy
$\mathcal{S}_{\mathcal{M}_{s}}$\ of the system and we show it increases
linearly in time (statistical evolution parameter) and is moreover,
proportional to the number of degrees of freedom of the system. The geodesics
on $\mathcal{M}_{s}$ are hyperbolic trajectories. Using the Jacobi-Levi-Civita
(JLC) equation for geodesic spread, it is shown that the Jacobi vector field
intensity $J_{\mathcal{M}_{s}}$ diverges exponentially and is proportional to
the number of degrees of freedom of the system. Thus, $\mathcal{R}%
_{\mathcal{M}_{s}}$, $\mathcal{S}_{\mathcal{M}_{s}} $ and $J_{\mathcal{M}_{s}%
}$ are proportional to the number of Gaussian-distributed microstates of the
system. This proportionality leads to conclude there is a substantial link
among these information-geometric indicators of chaoticity. Finally, as a
special physical application, the information geometrodynamical scheme is
applied to investigate the chaotic properties of a set of $n$-uncoupled
three-dimensional anisotropic inverted harmonic oscillators (IHOs)
characterized by an Ohmic distributed frequency spectrum. We study the
three-dimensional IHOs for the sake of generality. However we also illustrate
the main idea in a simpler example studying the IGAC (Information
Geometrodynamical Approach to Chaos)\ of two uncoupled inverted
one-dimensional harmonic oscillators. I show that the asymptotic temporal
behavior of the information geometrodynamical entropy of such a system
presents linear growth and I suggest the system studied may be considered the
classical information-geometric analogue of the Zurek-Paz quantum chaos
criterion in its classical reversible limit.

\section{Specification of the Gaussian ED-model}

Maximum relative Entropy (ME) methods are used to construct an ED model that
follows from an assumption about what information is relevant to predict the
evolution of the system. Given a known initial macrostate (probability
distribution) and that the system evolves to a final known macrostate, the
possible trajectories of the system are examined. A notion of
\textit{distance} between two probability distributions is provided by IG. As
shown in \cite{fisher, rao} this distance is quantified by the Fisher-Rao
information metric tensor.

We consider an ED model whose microstates span a $3N$-dimensional space
labelled by the variables $\left\{  \vec{X}\right\}  =\left\{  \vec
{x}^{\left(  1\right)  }\text{, }\vec{x}^{\left(  2\right)  }\text{,....,
}\vec{x}^{\left(  N\right)  }\right\}  $ with $\vec{x}^{\left(  \alpha\right)
}\equiv\left(  x_{1}^{\left(  \alpha\right)  }\text{, }x_{2}^{\left(
\alpha\right)  }\text{, }x_{3}^{\left(  \alpha\right)  }\right)  $, $\alpha
=1$,...., $N$ and $x_{a}^{\left(  \alpha\right)  }\in%
\mathbb{R}
$ with $a=1$, $2$, $3$. We assume the only testable information pertaining to
the quantities $x_{a}^{\left(  \alpha\right)  }$ consists of the expectation
values $\left\langle x_{a}^{\left(  \alpha\right)  }\right\rangle $ and
variance $\Delta x_{a}^{\left(  \alpha\right)  }=\sqrt{\left\langle \left(
x_{a}^{\left(  \alpha\right)  }-\left\langle x_{a}^{\left(  \alpha\right)
}\right\rangle \right)  ^{2}\right\rangle }$. The set of these expectation
values define the $6N$-dimensional space of macrostates of the system. A
measure of distinguishability among the states of the ED model is obtained by
assigning a probability distribution $P\left(  \vec{X}\left\vert \vec{\Theta
}\right.  \right)  $ to each macrostate $\vec{\Theta}$ where $\left\{
\vec{\Theta}\right\}  =\left\{  ^{\left(  1\right)  }\theta_{a}^{\left(
\alpha\right)  }\text{, }^{\left(  2\right)  }\theta_{a}^{\left(
\alpha\right)  }\right\}  $ with $\alpha=1$, $2$,$....$, $N$ and $a=1$, $2$,
$3$. The process of assigning a probability distribution to each state endows
$\mathcal{M}_{S}$ with a metric structure. Specifically, the Fisher-Rao
information metric defined in (\ref{fisher-rao}) is a measure of
distinguishability among macrostates. It assigns an IG to the space of states.

\subsection{The Gaussian statistical manifold $\mathcal{M}_{S}$}

We consider an arbitrary system evolving over a $3N$-dimensional space.\ The
variables $\left\{  \vec{X}\right\}  =\left\{  \vec{x}^{\left(  1\right)
}\text{, }\vec{x}^{\left(  2\right)  }\text{,...., }\vec{x}^{\left(  N\right)
}\right\}  $ label the $3N$-dimensional space of microstates of the system.
All information relevant to the dynamical evolution of the system is assumed
to be contained in the probability distributions. For this reason, no other
information is required. Each macrostate may be viewed as a point of a
$6N$-dimensional statistical manifold with coordinates given by the numerical
values of the expectations $^{\left(  1\right)  }\theta_{a}^{\left(
\alpha\right)  }=\left\langle x_{a}^{\left(  \alpha\right)  }\right\rangle $
and $^{\left(  2\right)  }\theta_{a}^{\left(  \alpha\right)  }=\Delta
x_{a}^{\left(  \alpha\right)  }\equiv\sqrt{\left\langle \left(  x_{a}^{\left(
\alpha\right)  }-\left\langle x_{a}^{\left(  \alpha\right)  }\right\rangle
\right)  ^{2}\right\rangle }$. The available information is contained in the
following $6N$ information constraint equations,%
\begin{equation}%
\begin{array}
[c]{c}%
\left\langle x_{a}^{\left(  \alpha\right)  }\right\rangle =%
{\displaystyle\int\limits_{-\infty}^{+\infty}}
dx_{a}^{\left(  \alpha\right)  }x_{a}^{\left(  \alpha\right)  }P_{a}^{\left(
\alpha\right)  }\left(  x_{a}^{\left(  \alpha\right)  }\left\vert ^{\left(
1\right)  }\theta_{a}^{\left(  \alpha\right)  }\text{,}^{\left(  2\right)
}\theta_{a}^{\left(  \alpha\right)  }\right.  \right)  \text{,}\\
\\
\Delta x_{a}^{\left(  \alpha\right)  }=\left[
{\displaystyle\int\limits_{-\infty}^{+\infty}}
dx_{a}^{\left(  \alpha\right)  }\left(  x_{a}^{\left(  \alpha\right)
}-\left\langle x_{a}^{\left(  \alpha\right)  }\right\rangle \right)  ^{2}%
P_{a}^{\left(  \alpha\right)  }\left(  x_{a}^{\left(  \alpha\right)
}\left\vert ^{\left(  1\right)  }\theta_{a}^{\left(  \alpha\right)  }%
\text{,}^{\left(  2\right)  }\theta_{a}^{\left(  \alpha\right)  }\right.
\right)  \right]  ^{\frac{1}{2}}\text{,}%
\end{array}
\label{constraint1}%
\end{equation}
where $^{\left(  1\right)  }\theta_{a}^{\left(  \alpha\right)  }=\left\langle
x_{a}^{\left(  \alpha\right)  }\right\rangle $ and $^{\left(  2\right)
}\theta_{a}^{\left(  \alpha\right)  }=\Delta x_{a}^{\left(  \alpha\right)  }$
with $\alpha=1$, $2$,$....$, $N$ and $a=1$, $2$, $3$. The probability
distributions $P_{a}^{\left(  \alpha\right)  }$ are constrained by the
conditions of normalization,%
\begin{equation}%
{\displaystyle\int\limits_{-\infty}^{+\infty}}
dx_{a}^{\left(  \alpha\right)  }P_{a}^{\left(  \alpha\right)  }\left(
x_{a}^{\left(  \alpha\right)  }\left\vert ^{\left(  1\right)  }\theta
_{a}^{\left(  \alpha\right)  }\text{,}^{\left(  2\right)  }\theta_{a}^{\left(
\alpha\right)  }\right.  \right)  =1\text{.} \label{constraint2}%
\end{equation}
The Gaussian distribution is identified by information theory as the maximum
entropy distribution if only the expectation value and the variance are known.
ME methods allows to associate a probability distribution $P\left(  \vec
{X}\left\vert \vec{\Theta}\right.  \right)  $ to each point in the space of
states $\vec{\Theta}$. The distribution that best reflects the information
contained in the prior distribution $m\left(  \vec{X}\right)  $ updated by the
information $\left(  \left\langle x_{a}^{\left(  \alpha\right)  }\right\rangle
\text{, }\Delta x_{a}^{\left(  \alpha\right)  }\right)  $ is obtained by
maximizing the relative entropy
\begin{equation}
S\left(  \vec{\Theta}\right)  =-\int\limits_{\left\{  \vec{X}\right\}  }%
d^{3N}\vec{X}P\left(  \vec{X}\left\vert \vec{\Theta}\right.  \right)
\log\left(  \frac{P\left(  \vec{X}\left\vert \vec{\Theta}\right.  \right)
}{m\left(  \vec{X}\right)  }\right)  \text{.} \label{entropy}%
\end{equation}
As a working hypothesis, the prior $m\left(  \vec{X}\right)  $ is set to be
uniform since we assume the lack of prior available information about the
system (postulate of equal \textit{a priori} probabilities). Upon maximizing
(\ref{entropy}), given the constraints (\ref{constraint1}) and
(\ref{constraint2}), we obtain%
\begin{equation}
P\left(  \vec{X}\left\vert \vec{\Theta}\right.  \right)  =%
{\displaystyle\prod\limits_{\alpha=1}^{N}}
{\displaystyle\prod\limits_{a=1}^{3}}
P_{a}^{\left(  \alpha\right)  }\left(  x_{a}^{\left(  \alpha\right)
}\left\vert \mu_{a}^{\left(  \alpha\right)  }\text{, }\sigma_{a}^{\left(
\alpha\right)  }\right.  \right)  \label{prob}%
\end{equation}
where%
\begin{equation}
P_{a}^{\left(  \alpha\right)  }\left(  x_{a}^{\left(  \alpha\right)
}\left\vert \mu_{a}^{\left(  \alpha\right)  }\text{, }\sigma_{a}^{\left(
\alpha\right)  }\right.  \right)  =\left(  2\pi\left[  \sigma_{a}^{\left(
\alpha\right)  }\right]  ^{2}\right)  ^{-\frac{1}{2}}\exp\left[
-\frac{\left(  x_{a}^{\left(  \alpha\right)  }-\mu_{a}^{\left(  \alpha\right)
}\right)  ^{2}}{2\left(  \sigma_{a}^{\left(  \alpha\right)  }\right)  ^{2}%
}\right]  \label{gaussian}%
\end{equation}
and $^{\left(  1\right)  }\theta_{a}^{\left(  \alpha\right)  }=\mu
_{a}^{\left(  \alpha\right)  }$, $^{\left(  2\right)  }\theta_{a}^{\left(
\alpha\right)  }=\sigma_{a}^{\left(  \alpha\right)  }$. For the rest of the
Chapter, unless stated otherwise, the statistical manifold $\mathcal{M}_{S}$
will be defined by the following expression,%
\begin{equation}
\mathcal{M}_{S}=\left\{  P\left(  \vec{X}\left\vert \vec{\Theta}\right.
\right)  \text{ in (\ref{prob})}:\vec{X}\in%
\mathbb{R}
^{3N}\text{, }\vec{\Theta}\in\mathcal{D}_{\Theta}=\left[  \left(
-\infty\text{, }+\infty\right)  _{\mu}\times\left(  0\text{, }+\infty\right)
_{\sigma}\right]  ^{3N}\right\}  \text{.} \label{manifold}%
\end{equation}
The probability distribution (\ref{prob}) encodes the available information
concerning the system. Note we assumed uncoupled constraints among
microvariables $x_{a}^{\left(  \alpha\right)  }$. In other words, we assumed
that information about correlations between the microvariables need not to be
tracked. This assumption leads to the simplified product rule (\ref{prob}).
However, coupled constraints would lead to a generalized product rule in
(\ref{prob}) and to a metric tensor (\ref{fisher-rao}) with non-trivial
off-diagonal elements (covariance terms). For instance, the total probability
distribution $P\left(  x\text{, }y|\mu_{x}\text{, }\sigma_{x}\text{, }\mu
_{y}\text{, }\sigma_{y}\right)  $ of two dependent Gaussian distributed
microvariables $x$ and $y$ reads%
\begin{align}
&  P\left(  x\text{, }y|\mu_{x}\text{, }\sigma_{x}\text{, }\mu_{y}\text{,
}\sigma_{y}\right)  =\frac{1}{2\pi\sigma_{x}\sigma_{y}\sqrt{1-r^{2}}}%
\times\label{corr-prob}\\
&  \times\exp\left\{  -\frac{1}{2\left(  1-r^{2}\right)  }\left[
\frac{\left(  x-\mu_{x}\right)  ^{2}}{\sigma_{x}^{2}}-2r\frac{\left(
x-\mu_{x}\right)  \left(  y-\mu_{y}\right)  }{\sigma_{x}\sigma_{y}}%
+\frac{\left(  y-\mu_{y}\right)  ^{2}}{\sigma_{y}^{2}}\right]  \right\}
\text{,}\nonumber
\end{align}
where $r\in\left(  -1\text{, }+1\right)  $ is the correlation coefficient
given by%
\begin{equation}
r=\frac{\left\langle \left(  x-\left\langle x\right\rangle \right)  \left(
y-\left\langle y\right\rangle \right)  \right\rangle }{\sqrt{\left\langle
x-\left\langle x\right\rangle \right\rangle }\sqrt{\left\langle y-\left\langle
y\right\rangle \right\rangle }}=\frac{\left\langle xy\right\rangle
-\left\langle x\right\rangle \left\langle y\right\rangle }{\sigma_{x}%
\sigma_{y}}\text{.}%
\end{equation}
The metric induced by (\ref{corr-prob}) is obtained by use of
(\ref{fisher-rao}), the result being%
\begin{equation}
g_{ij}=\left[
\begin{array}
[c]{cccc}%
-\frac{1}{\sigma_{x}^{2}\left(  r^{2}-1\right)  } & 0 & \frac{r}{\sigma
_{x}\sigma_{y}\left(  r^{2}-1\right)  } & 0\\
0 & -\frac{2-r^{2}}{\sigma_{x}^{2}\left(  r^{2}-1\right)  } & 0 & \frac{r^{2}%
}{\sigma_{x}\sigma_{y}\left(  r^{2}-1\right)  }\\
\frac{r}{\sigma_{x}\sigma_{y}\left(  r^{2}-1\right)  } & 0 & -\frac{1}%
{\sigma_{y}^{2}\left(  r^{2}-1\right)  } & 0\\
0 & \frac{r^{2}}{\sigma_{x}\sigma_{y}\left(  r^{2}-1\right)  } & 0 &
-\frac{2-r^{2}}{\sigma_{y}^{2}\left(  r^{2}-1\right)  }%
\end{array}
\right]  \text{,} \label{corr-metric}%
\end{equation}
where $i$, $j=1$, $2$, $3$, $4$. The Ricci curvature scalar associated with
manifold characterized by (\ref{corr-metric}) is given by%
\begin{equation}
\mathcal{R}=g^{ij}R_{ij}=-\frac{8\left(  r^{2}-2\right)  +2r^{2}\left(
3r^{2}-2\right)  }{8\left(  r^{2}-1\right)  }\text{.}%
\end{equation}
It is clear that in the limit $r\rightarrow0$, the off-diagonal elements of
$g_{ij}$ vanish and the scalar $\mathcal{R}$ reduces to the result obtained in
\cite{cafaro2}, namely $\mathcal{R}=-2<0$. Correlation terms may be
fictitious. They may arise for instance from coordinate transformations. On
the other hand, correlations may arise from external fields in which the
system is immersed. In such situations, correlations among $x_{a}^{\left(
\alpha\right)  }$ effectively describe interaction between the microvariables
and the external fields. Such generalizations would require more delicate
analysis. Before proceeding, a comment is in order. Most probability
distributions arise from the maximum entropy formalism as a result of simple
statements concerning averages (Gaussians, exponential, binomial, etc.). Not
all distribution are generated in this manner however. Some distributions are
generated by combining the results of simple cases (multinomial from a
binomial) while others are found as a result of a change of variables (Cauchy
distribution). For instance, the Weibull and Wigner-Dyson distributions can be
obtained from an exponential distribution as a result of a power law
transformation \cite{wigner-dyson}.

\subsubsection{Metric structure of $\mathcal{M}_{S}$}

We cannot determine the evolution of microstates of the system since the
available information is insufficient. Not only is the information available
insufficient but we also do not know the equation of motion. In fact there is
no standard "equation of motion".\ Instead we can ask: how close are the two
total distributions with parameters $(\mu_{a}^{\left(  \alpha\right)  }$,
$\sigma_{a}^{\left(  \alpha\right)  })$ and $(\mu_{a}^{\left(  \alpha\right)
}+d\mu_{a}^{\left(  \alpha\right)  }$, $\sigma_{a}^{\left(  \alpha\right)
}+d\sigma_{a}^{\left(  \alpha\right)  })$? Once the states of the system have
been defined, the next step concerns the problem of quantifying the notion of
change from the state $\vec{\Theta}$ to the state $\vec{\Theta}+d\vec{\Theta}%
$. A convenient measure of change is distance. The measure we seek is given by
the dimensionless \textit{distance} $ds$ between $P\left(  \vec{X}\left\vert
\vec{\Theta}\right.  \right)  $ and $P\left(  \vec{X}\left\vert \vec{\Theta
}+d\vec{\Theta}\right.  \right)  $,%
\begin{equation}
ds^{2}=g_{\mu\nu}d\Theta^{\mu}d\Theta^{\nu}\text{ with }\mu\text{, }%
\nu=1\text{, }2\text{,.., }6N\text{,} \label{line-element}%
\end{equation}
where%
\begin{equation}
g_{\mu\nu}=\int d\vec{X}P\left(  \vec{X}\left\vert \vec{\Theta}\right.
\right)  \frac{\partial\log P\left(  \vec{X}\left\vert \vec{\Theta}\right.
\right)  }{\partial\Theta^{\mu}}\frac{\partial\log P\left(  \vec{X}\left\vert
\vec{\Theta}\right.  \right)  }{\partial\Theta^{\nu}} \label{fisher-rao}%
\end{equation}
is the Fisher-Rao information metric. Substituting (\ref{prob}) into
(\ref{fisher-rao}), the metric $g_{\mu\nu}$ on $\mathcal{M}_{s}$ becomes a
$6N\times6N$ matrix $M$ made up of $3N$ blocks $M_{2\times2}$ with dimension
$2\times2$ given by,%
\begin{equation}
M_{2\times2}=\left(
\begin{array}
[c]{cc}%
\left(  \sigma_{a}^{\left(  \alpha\right)  }\right)  ^{-2} & 0\\
0 & 2\times\left(  \sigma_{a}^{\left(  \alpha\right)  }\right)  ^{-2}%
\end{array}
\right)  \label{m-matrix}%
\end{equation}
with $\alpha=1$, $2$,$....$, $N$ and $a=1,2,3$. From (\ref{fisher-rao}), the
"length" element (\ref{line-element}) reads,%
\begin{equation}
ds^{2}=%
{\displaystyle\sum\limits_{\alpha=1}^{N}}
{\displaystyle\sum\limits_{a=1}^{3}}
\left[  \frac{1}{\left(  \sigma_{a}^{\left(  \alpha\right)  }\right)  ^{2}%
}d\mu_{a}^{\left(  \alpha\right)  2}+\frac{2}{\left(  \sigma_{a}^{\left(
\alpha\right)  }\right)  ^{2}}d\sigma_{a}^{\left(  \alpha\right)  2}\right]
\text{.}%
\end{equation}
We bring attention to the fact that the metric structure of $\mathcal{M}_{s}$
is an emergent (not fundamental) structure. It arises only after assigning a
probability distribution $P\left(  \vec{X}\left\vert \vec{\Theta}\right.
\right)  $ to each state $\vec{\Theta}$.

\subsubsection{Curvature of\ $\mathcal{M}_{s}$}

Given the Fisher-Rao information metric, we use standard differential geometry
methods applied to the space of probability distributions to characterize the
geometric properties of $\mathcal{M}_{s}$. Recall that the Ricci scalar
curvature $R$ is given by,%
\begin{equation}
R=g^{\mu\nu}R_{\mu\nu}\text{,} \label{ricci-scalar}%
\end{equation}
where $g^{\mu\nu}g_{\nu\rho}=\delta_{\rho}^{\mu}$ so that $g^{\mu\nu}=\left(
g_{\mu\nu}\right)  ^{-1}$. The Ricci tensor $R_{\mu\nu}$ is given by,%
\begin{equation}
R_{\mu\nu}=\partial_{\gamma}\Gamma_{\mu\nu}^{\gamma}-\partial_{\nu}\Gamma
_{\mu\lambda}^{\lambda}+\Gamma_{\mu\nu}^{\gamma}\Gamma_{\gamma\eta}^{\eta
}-\Gamma_{\mu\gamma}^{\eta}\Gamma_{\nu\eta}^{\gamma}\text{.}%
\end{equation}
The Christoffel symbols $\Gamma_{\mu\nu}^{\rho}$ appearing in the Ricci tensor
are defined in the standard manner as,
\begin{equation}
\Gamma_{\mu\nu}^{\rho}=\frac{1}{2}g^{\rho\sigma}\left(  \partial_{\mu
}g_{\sigma\nu}+\partial_{\nu}g_{\mu\sigma}-\partial_{\sigma}g_{\mu\nu}\right)
. \label{christoffel}%
\end{equation}
Using (\ref{m-matrix}) and the definitions given above, we can show that the
Ricci scalar curvature becomes%
\begin{equation}
R_{\mathcal{M}_{s}}=R_{\text{ }\alpha}^{\alpha}=\sum_{\rho\neq\sigma}K\left(
e_{\rho}\text{, }e_{\sigma}\right)  =-3N<0\text{.} \label{Ricci}%
\end{equation}
The scalar curvature is the sum of all sectional curvatures of planes spanned
by pairs of orthonormal basis elements $\left\{  e_{\rho}=\partial
_{\Theta_{\rho}(p)}\right\}  $ of the tangent space $T_{p}\mathcal{M}_{s}$
with $p\in\mathcal{M}_{s}$,
\begin{equation}
K\left(  a\text{, }b\right)  =\frac{R_{\mu\nu\rho\sigma}a^{\mu}b^{\nu}a^{\rho
}b^{\sigma}}{\left(  g_{\mu\sigma}g_{\nu\rho}-g_{\mu\rho}g_{\nu\sigma}\right)
a^{\mu}b^{\nu}a^{\rho}b^{\sigma}}\text{, }a=\sum_{\rho}\left\langle a\text{,
}h^{\rho}\right\rangle e_{\rho}\text{,} \label{sectionK}%
\end{equation}
where $\left\langle e_{\rho}\text{, }h^{\sigma}\right\rangle =\delta_{\rho
}^{\sigma}$. Notice that the sectional curvatures completely determine the
curvature tensor. From (\ref{Ricci}) we conclude that $\mathcal{M}_{s}$ is a
$6N$-dimensional statistical manifold of constant negative Ricci scalar
curvature. A detailed analysis on the calculation of Christoffel connection
coefficients using the ED formalism for a four-dimensional manifold of
Gaussians can be found in \cite{cafaro2}.

\subsubsection{Anisotropy and Compactness}

It can be shown that $\mathcal{M}_{s}$ is not a maximally symmetric
multidimensional manifold. The first way this can be understood is from the
fact that the Weyl Projective curvature tensor \cite{goldberg} (or the
anisotropy tensor) $W_{\mu\nu\rho\sigma}$ defined by%
\begin{equation}
W_{\mu\nu\rho\sigma}=R_{\mu\nu\rho\sigma}-\frac{\mathcal{R}_{\mathcal{M}_{s}}%
}{n\left(  n-1\right)  }\left(  g_{\nu\sigma}g_{\mu\rho}-g_{\nu\rho}%
g_{\mu\sigma}\right)  \text{,} \label{Weyl}%
\end{equation}
with $n=6N$ in the present case, is non-vanishing. In (\ref{Weyl}), the
quantity $R_{\mu\nu\rho\sigma}$ is the Riemann curvature tensor defined in the
usual manner by%
\begin{equation}
R^{\alpha}\,_{\beta\rho\sigma}=\partial_{\sigma}\Gamma_{\text{ \ }\beta\rho
}^{\alpha}-\partial_{\rho}\Gamma_{\text{ \ }\beta\sigma}^{\alpha}%
+\Gamma^{\alpha}\,_{\lambda\sigma}\Gamma^{\lambda}\,_{\beta\rho}%
-\Gamma^{\alpha}\,_{\lambda\rho}\Gamma^{\lambda}\,_{\beta\sigma}\text{.}%
\end{equation}
Considerations regarding the negativity of the Ricci curvature as a
\textit{strong criterion} of dynamical instability and the necessity of
\textit{compactness} of $\mathcal{M}_{s}$\textit{\ }in "true" chaotic
dynamical systems is under investigation \cite{cafaro4}.

The issue of symmetry of $\mathcal{M}_{s}$ can alternatively be understood
from consideration of the sectional curvature. In view of (\ref{sectionK}),
the negativity of the Ricci scalar implies the existence of expanding
directions in the configuration space manifold $\mathcal{M}_{s}$. Indeed, from
(\ref{Ricci}) one may conclude that negative principal curvatures (extrema of
sectional curvatures) dominate over positive ones. Thus, the negativity of the
Ricci scalar is only a \textit{sufficient} (not necessary) condition for local
instability of geodesic flow. For this reason, the negativity of the scalar
provides a \textit{strong }criterion of local instability. Scenarios may arise
where negative sectional curvatures are present, but the positive ones could
prevail in the sum so that the Ricci scalar is non-negative despite the
instability in the flow in those directions. Consequently, the signs of the
sectional curvatures are of primary significance for the proper
characterization of chaos.

Yet another useful way to understand the anisotropy of the $\mathcal{M}_{s}$
is the following. It is known that in $n$ dimensions, there are at most
$\frac{n\left(  n+1\right)  }{2}$ independent Killing vectors (directions of
symmetry of the manifold). Since $\mathcal{M}_{s}$ is not a pseudosphere, the
information metric tensor does not admit the maximum number of Killing vectors
$K_{\nu}$ defined as%
\begin{equation}
\mathcal{L}_{K}g_{\mu\nu}=D_{\mu}K_{\nu}+D_{\nu}K_{\mu}=0\text{,}%
\end{equation}
where $D_{\mu}$, given by%
\begin{equation}
D_{\mu}K_{\nu}=\partial_{\mu}K_{\nu}-\Gamma_{\nu\mu}^{\rho}K_{\rho}\text{,}%
\end{equation}
is the covariant derivative operator with respect to the connection $\Gamma$
defined in (\ref{christoffel}). The Lie derivative $\mathcal{L}_{K}g_{\mu\nu}$
of the tensor field $g_{\mu\nu}$ along a given direction $K$ measures the
intrinsic variation of the field along that direction (that is, the metric
tensor is Lie transported along the Killing vector) \cite{clarke}. Locally, a
maximally symmetric space of Euclidean signature is either a plane, a sphere,
or a hyperboloid, depending on the sign of $\mathcal{R}$. In our case, none of
these scenarios occur. As will be seen in what follows, this fact has a
significant impact on the integration of the geodesic deviation equation on
$\mathcal{M}_{s}$. At this juncture, we emphasize it is known that the
anisotropy of the manifold underlying system dynamics plays a crucial role in
the mechanism of instability. In particular, fluctuating sectional curvatures
require also that the manifold be anisotropic. However, the connection between
curvature variations along geodesics and anisotropy is far from clear and is
currently under investigation.

Krylov was the first to emphasize \cite{krylov} the use of $\mathcal{R}<0$ as
an instability criterion in the context of an $N$-body system (a gas)
interacting via Van der Waals forces, with the ultimate hope to understand the
relaxation process in a gas. However, Krylov neglected the problem of
compactness of the configuration space manifold which is important for making
inferences about exponential mixing of geodesic flows \cite{pellicott}. Mixing
provides statistical independence of different parts of a trajectory. This is
the condition for application of probability theory that allows to calculate
statistical properties such as diffusion, relaxation etc. Why is compactness
so significant in the characterization of chaos? \textit{True} chaos should be
identified by the occurrence of two crucial features: 1) strong dependence on
initial conditions and exponential divergence of the Jacobi vector field
intensity, i.e., \textit{stretching} of dynamical trajectories; 2) compactness
of the configuration space manifold, i.e., \textit{folding} of dynamical
trajectories. Compactness \cite{cipriani, jost} is required in order to
discard trivial exponential growths due to the unboundedness of the "volume"
available to the dynamical system. In other words, the folding is necessary to
have a dynamics actually able to mix the trajectories, making practically
impossible, after a finite interval of time, to discriminate between
trajectories which were very nearby each other at the initial time. When the
space is not compact, even in presence of strong dependence on initial
conditions, it could be possible in some instances (though not always), to
distinguish among different trajectories originating within a small distance
and then evolved subject to exponential instability.

The statistical manifold $\mathcal{M}_{s}$ defined in (\ref{manifold}) is
compact provided that the parameter space $\mathcal{D}_{\Theta}$ is compact.
This can be seen as follows. It is known from IG that there is a one-to-one
relation between elements of the statistical manifold and the parameter space.
More precisely, the statistical manifold $\mathcal{M}_{s}$ is
\textit{homeomorphic} to the parameter space $\mathcal{D}_{\Theta}$. This
implies the existence of a continuous, bijective map $h_{\mathcal{M}%
_{s}\text{, }\mathcal{D}_{\Theta}}$,%
\begin{equation}
h_{\mathcal{M}_{s}\text{, }\mathcal{D}_{\Theta}\text{ }}:\mathcal{M}_{S}\ni
P\left(  \vec{X}\left\vert \vec{\Theta}\right.  \right)  \rightarrow
\vec{\Theta}\in\mathcal{D}_{\Theta}%
\end{equation}
where $h_{\mathcal{M}_{s}\text{, }\mathcal{D}_{\Theta}\text{ }}^{-1}\left(
\vec{\Theta}\right)  =P\left(  \vec{X}\left\vert \vec{\Theta}\right.  \right)
$. The inverse image $h_{\mathcal{M}_{s}\text{, }\mathcal{D}_{\Theta}\text{ }%
}^{-1}$ is the so-called homeomorphism map. In addition, since homeomorphisms
preserve compactness, it is sufficient to restrict ourselves to a compact
subspace of the parameter space $\mathcal{D}_{\Theta}$ in order to ensure that
$\mathcal{M}_{S}$ is itself compact. In our specific case, the accessible
region of the statistical manifold $\mathcal{M}_{s}$ is given by%
\begin{equation}
\mathcal{M}_{S}=\left\{  P\left(  \vec{X}\left\vert \vec{\Theta}\right.
\right)  \text{ in (\ref{prob}): }\vec{X}\in%
\mathbb{R}
^{3N},\vec{\Theta}\in\mathcal{D}_{\Theta}\right\}  \text{,}%
\end{equation}
where the $6N$ dimensional $\mathcal{D}_{\Theta}$ is defined as
\begin{equation}
\mathcal{D}_{\Theta}\overset{\text{def}}{=}I_{\mu}\times I_{\sigma}=\left[
\mu_{\text{min}}\text{, }\mu_{\text{max}}\right]  ^{3N}\times\left[
\sigma_{\text{min}}\text{, }\sigma_{\text{max}}\right]  ^{3N}\text{.}%
\end{equation}
Assuming $\tau\in%
\mathbb{R}
^{+}$, from equations (\ref{doubleG}), we obtain%
\begin{equation}
\mu_{\text{min}}=\frac{4\beta B^{2}}{8\beta^{2}+B^{2}}+C\text{, }%
\mu_{\text{max}}=4\beta+C\text{, }\sigma_{\text{min}}=D\text{, }%
\sigma_{\text{max}}=\sqrt{2}\beta+D\text{. }%
\end{equation}
In conclusion, $\mathcal{D}_{\Theta}$ is compact provided we restrict
ourselves to consider only the accessible macrostates on $\mathcal{M}_{s}$.

\section{Canonical formalism for the Gaussian ED-model}

Jacobi was the first one to develop a technique of classical mechanics where a
Hamiltonian system is geometrized by transforming it into a geodesic flow on a
suitable manifold with a convenient Riemannian metric \cite{jacobi}. The two
key steps in obtaining the geometrization of a Hamiltonian system are the
introduction of a conformal transformation of the metric and the rescaling of
the time parameter \cite{biesiada1} (and, for more details, Chapter 6). The
reformulation of dynamics in terms of a geodesic problem allows the
application of a wide range of well-known geometrical techniques in the
investigation of the solution space and properties of equations of motions.
The power of the Jacobi reformulation is that all of the dynamical information
is collected into a single geometric object - the manifold on which geodesic
flow is induced - in which all the available manifest symmetries are retained.
For instance, integrability of the system is connected with the existence of
Killing vectors and tensors on this manifold \cite{biesiada2, uggla}.

In this section we study the trajectories of the system on $\mathcal{M}_{s}$.
We emphasize ED can be derived from a standard principle of least action (of
Maupertuis-Euler-Lagrange-Jacobi type) \cite{caticha1, arnold}. The main
differences are that the dynamics being considered here, namely ED, is defined
on a space of probability distributions $\mathcal{M}_{s}$, not on an ordinary
vectorial space $V$ and the standard coordinates $q_{\mu}$ of the system are
replaced by statistical macrovariables $\Theta^{\mu}$. The geodesic equations
for the macrovariables of the Gaussian ED model are given by,%
\begin{equation}
\frac{d^{2}\Theta^{\mu}}{d\tau^{2}}+\Gamma_{\nu\rho}^{\mu}\frac{d\Theta^{\nu}%
}{d\tau}\frac{d\Theta^{\rho}}{d\tau}=0 \label{geodesic}%
\end{equation}
with $\mu=1$, $2$,..., $6N$. The geodesic equations are\textit{\ nonlinear}
second order coupled ordinary differential equations. They describe a
\textit{reversible} dynamics whose solution is the trajectory between an
initial and a final macrostate. The trajectory can be equally well traversed
in both directions.

\subsection{Geodesics on $\mathcal{M}_{s}$}

We determine the explicit form of (\ref{geodesic}) for the pairs of
statistical coordinates $(\mu_{a}^{\left(  \alpha\right)  }$, $\sigma
_{a}^{\left(  \alpha\right)  })$. Substituting the expression of the
Christoffel connection coefficients into (\ref{geodesic}), the geodesic
equations for the macrovariables $\mu_{a}^{\left(  \alpha\right)  }$ and
$\sigma_{a}^{\left(  \alpha\right)  }$ associated to the microstate
$x_{a}^{\left(  \alpha\right)  }$ become,%
\begin{equation}
\frac{d^{2}\mu_{a}^{\left(  \alpha\right)  }}{d\tau^{2}}-\frac{2}{\sigma
_{a}^{\left(  \alpha\right)  }}\frac{d\mu_{a}^{\left(  \alpha\right)  }}%
{d\tau}\frac{d\sigma_{a}^{\left(  \alpha\right)  }}{d\tau}=0\text{, }%
\frac{d^{2}\sigma_{a}^{\left(  \alpha\right)  }}{d\tau^{2}}-\frac{1}%
{\sigma_{a}^{\left(  \alpha\right)  }}\left(  \frac{d\sigma_{a}^{\left(
\alpha\right)  }}{d\tau}\right)  ^{2}+\frac{1}{2\sigma_{a}^{\left(
\alpha\right)  }}\left(  \frac{d\mu_{a}^{\left(  \alpha\right)  }}{d\tau
}\right)  ^{2}=0\text{,} \label{totti}%
\end{equation}
with $\alpha=1$, $2$,$....$, $N$ and $a=1$, $2$, $3$. The $3N$ Gaussians are
not coupled to each other, however each Gaussian is characterized by coupled
macrovariables $\mu_{a}^{\left(  \alpha\right)  }$ and $\sigma_{a}^{\left(
\alpha\right)  }$. Equation (\ref{totti}) describes a set of coupled ordinary
differential equations, whose solutions are%
\[
\mu_{a}^{\left(  \alpha\right)  }\left(  \tau\right)  =\frac{\left(
B_{a}^{\left(  \alpha\right)  }\right)  ^{2}}{2\beta_{a}^{\left(
\alpha\right)  }}\frac{1}{\cosh\left(  2\beta_{a}^{\left(  \alpha\right)
}\tau\right)  -\sinh\left(  2\beta_{a}^{\left(  \alpha\right)  }\tau\right)
+\frac{\left(  B_{a}^{\left(  \alpha\right)  }\right)  ^{2}}{8\left(
\beta_{a}^{\left(  \alpha\right)  }\right)  ^{2}}}+C_{a}^{\left(
\alpha\right)  }\text{,}%
\]%
\begin{equation}
\sigma_{a}^{\left(  \alpha\right)  }\left(  \tau\right)  =B_{a}^{\left(
\alpha\right)  }\frac{\cosh\left(  \beta_{a}^{\left(  \alpha\right)  }%
\tau\right)  -\sinh\left(  \beta_{a}^{\left(  \alpha\right)  }\tau\right)
}{\cosh\left(  2\beta_{a}^{\left(  \alpha\right)  }\tau\right)  -\sinh\left(
2\beta_{a}^{\left(  \alpha\right)  }\tau\right)  +\frac{\left(  B_{a}^{\left(
\alpha\right)  }\right)  ^{2}}{8\left(  \beta_{a}^{\left(  \alpha\right)
}\right)  ^{2}}}+D_{a}^{\left(  \alpha\right)  }\text{.} \label{doubleG}%
\end{equation}
The quantities $B_{a}^{\left(  \alpha\right)  }$, $C_{a}^{\left(
\alpha\right)  }$, $D_{a}^{\left(  \alpha\right)  }$, $\beta_{a}^{\left(
\alpha\right)  }$ are \textit{real} integration constants that can be
evaluated upon specification of boundary conditions. We are interested in the
stability of the trajectories on $\mathcal{M}_{s}$. It is known \cite{arnold}
that the Riemannian curvature of a manifold is intimately related to the
behavior of geodesics on it. If the Riemannian curvature of a manifold is
negative, geodesics (initially parallel) rapidly diverge from one another. We
observe that since every maximal geodesic (one that cannot be extended to any
larger interval) is well-defined for all temporal parameters $\tau$,
$\mathcal{M}_{s}$ constitute a geodesically complete manifold \cite{lee}. It
is therefore a natural setting within which one may consider \textit{global}
questions and search for a \textit{weak criterion} of chaos \cite{cipriani}.

\section{Exponential divergence of the Jacobi vector field intensity}

The actual interest of the Riemannian formulation of the dynamics stems from
the possibility of studying the instability of natural motions through the
instability of geodesics of a suitable manifold, a circumstance that has
several advantages. First of all a powerful mathematical tool exists to
investigate the stability or instability of a geodesic flow: the
Jacobi-Levi-Civita equation for geodesic spread \cite{carmo}. The JLC-equation
describes covariantly how nearby geodesics locally scatter. It is a familiar
object both in Riemannian geometry and theoretical physics (it is of
fundamental interest in experimental General Relativity). Moreover the
JLC-equation relates the stability or instability of a geodesic flow with
curvature properties of the ambient manifold, thus opening a wide and largely
unexplored field of investigation of the connections among geometry, topology
and geodesic instability, hence chaos.

Consider the behavior of the one-parameter (at fixed $a$ and $\alpha$,
$\lambda_{a}^{\left(  \alpha\right)  }\equiv\lambda$) family of neighboring
geodesics $\mathcal{F}_{G_{\mathcal{M}_{s}}}\left(  \lambda\right)
\equiv\left\{  \Theta_{\mathcal{M}_{s}}^{\mu}\left(  \tau\text{; }%
\lambda\right)  \right\}  _{\lambda\in%
\mathbb{R}
^{+}}^{\mu=1\text{,.., }6N}$ where%
\begin{align}
\mu_{a}^{\left(  \alpha\right)  }\left(  \tau\text{; }\lambda\right)   &
=\frac{\Lambda_{a}^{\left(  \alpha\right)  2}}{2\lambda_{a}^{\left(
\alpha\right)  }}\frac{1}{\cosh\left(  2\lambda_{a}^{\left(  \alpha\right)
}\tau\right)  -\sinh\left(  2\lambda_{a}^{\left(  \alpha\right)  }\tau\right)
+\frac{\Lambda_{a}^{\left(  \alpha\right)  2}}{8\lambda_{a}^{\left(
\alpha\right)  2}}}+C_{a}^{\left(  \alpha\right)  }\text{,}\nonumber\\
& \label{solns}\\
\sigma_{a}^{\left(  \alpha\right)  }\left(  \tau\text{; }\lambda\right)   &
=\Lambda_{a}^{\left(  \alpha\right)  }\frac{\cosh\left(  \lambda_{a}^{\left(
\alpha\right)  }\tau\right)  -\sinh\left(  \lambda_{a}^{\left(  \alpha\right)
}\tau\right)  }{\cosh\left(  2\lambda_{a}^{\left(  \alpha\right)  }%
\tau\right)  -\sinh\left(  2\lambda_{a}^{\left(  \alpha\right)  }\tau\right)
+\frac{\Lambda_{a}^{\left(  \alpha\right)  2}}{8\lambda_{a}^{\left(
\alpha\right)  2}}}+D_{a}^{\left(  \alpha\right)  }\text{.}\nonumber
\end{align}
with $\alpha=1$, $2$,$....$, $N$ and $a=1$, $2$, $3$. The relative geodesic
spread on a (non-maximally symmetric) curved manifold as $\mathcal{M}_{s}$ is
characterized by the Jacobi-Levi-Civita equation, the natural tool to tackle
dynamical chaos \cite{clarke, carmo},%
\begin{equation}
\frac{D^{2}\delta\Theta^{\mu}}{D\tau^{2}}+R_{\nu\rho\sigma}^{\mu}%
\frac{\partial\Theta^{\nu}}{\partial\tau}\delta\Theta^{\rho}\frac
{\partial\Theta^{\sigma}}{\partial\tau}=0 \label{gen-geoDev}%
\end{equation}
where the Jacobi vector field $J^{\mu}$ is defined as,%
\begin{equation}
J^{\mu}\equiv\delta\Theta^{\mu}\overset{\text{def}}{=}\delta_{\lambda
_{a}^{\left(  \alpha\right)  }}\Theta^{\mu}=\left.  \left(  \frac
{\partial\Theta^{\mu}\left(  \tau\text{; }\lambda\right)  }{\partial
\lambda_{a}^{\left(  \alpha\right)  }}\right)  \right\vert _{\tau
=\text{const}}\delta\lambda_{a}^{\left(  \alpha\right)  }\text{.}
\label{jacobi}%
\end{equation}
Notice that the JLC-equation appears intractable already at rather small $N$.
For isotropic manifolds, the JLC-equation can be reduced to the simple form,%
\begin{equation}
\frac{D^{2}J^{\mu}}{D\tau^{2}}+KJ^{\mu}=0\text{, }\mu=1\text{,...., }6N
\label{geo-deviation}%
\end{equation}
where $K$ is the constant value assumed throughout the manifold by the
sectional curvature. The sectional curvature of manifold $\mathcal{M}_{s}$ is
the $6N$-dimensional generalization of the Gaussian curvature of
two-dimensional surfaces of $%
\mathbb{R}
^{3}$. If $K<0$, unstable solutions of equation (\ref{geo-deviation}) assumes
the form%
\begin{equation}
J\left(  \tau\right)  =\frac{1}{\sqrt{-K}}\omega\left(  0\right)  \sinh\left(
\sqrt{-K}\tau\right)
\end{equation}
once the initial conditions are assigned as $J\left(  0\right)  =0$,
$\frac{dJ\left(  0\right)  }{d\tau}=\omega\left(  0\right)  $ and $K<0$.
Equation (\ref{gen-geoDev}) forms a system of $6N$ coupled ordinary
differential equations \textit{linear} in the components of the deviation
vector field (\ref{jacobi}) but\textit{\ nonlinear} in derivatives of the
metric (\ref{fisher-rao}). It describes the linearized geodesic flow: the
linearization ignores the relative velocity of the geodesics. When the
geodesics are neighboring but their relative velocity is arbitrary, the
corresponding geodesic deviation equation is the so-called generalized Jacobi
equation \cite{chicone, hodgkinson}. The nonlinearity is due to the existence
of velocity-dependent terms in the system. Neighboring geodesics accelerate
relative to each other with a rate directly measured by the curvature tensor
$R_{\alpha\beta\gamma\delta}$. Substituting (\ref{solns}) in (\ref{gen-geoDev}%
) and neglecting the exponentially decaying terms in $\delta\Theta^{\mu}$ and
its derivatives, integration of (\ref{gen-geoDev}) leads to the following
asymptotic expression of the Jacobi vector field intensity,%
\begin{equation}
J_{\mathcal{M}_{S}}=\left\Vert J\right\Vert =\left(  g_{\mu\nu}J^{\mu}J^{\nu
}\right)  ^{\frac{1}{2}}\overset{\tau\rightarrow\infty}{\approx}%
{\displaystyle\sum\limits_{\alpha=1}^{N}}
{\displaystyle\sum\limits_{a=1}^{3}}
\exp\left(  \lambda_{a}^{\left(  \alpha\right)  }\tau\right)  \text{.}
\label{enzo}%
\end{equation}
As a side remark, we point out that if we consider the special case where
different Gaussians are characterized by the same initial conditions leading
to the same $\lambda\equiv\lambda_{a}^{\left(  \alpha\right)  }=$
$\lambda_{a^{\prime}}^{\left(  \alpha^{\prime}\right)  }$ $\forall a$,
$a^{\prime}=1$, $2$, $3$ and $\forall\alpha$, $\alpha^{\prime}=1$,.., $N$,
equation (\ref{enzo}) becomes,%
\begin{equation}
J_{\mathcal{M}_{S}}\overset{\tau\rightarrow\infty}{\approx}3N\exp\left(
\lambda\tau\right)  \text{.}%
\end{equation}
We conclude that the geodesic spread on $\mathcal{M}_{s}$ is described by
means of an \textit{exponentially} \textit{divergent} Jacobi vector field
intensity $J_{\mathcal{M}_{s}}$, a \textit{classical} feature of chaos. In our
approach the quantity $\lambda_{J}$,%
\begin{equation}
\lambda_{J}\overset{\text{def}}{=}\underset{\tau\rightarrow\infty}{\lim}%
\frac{1}{\tau}\ln\left[  \frac{\left\Vert J_{_{\mathcal{M}_{S}}}\left(
\tau\right)  \right\Vert }{\left\Vert J_{_{\mathcal{M}_{S}}}\left(  0\right)
\right\Vert }\right]
\end{equation}
would play the role of the conventional Lyapunov exponents.

\section{Linearity of the information geometrodynamical entropy}

The statistical manifold $\mathcal{M}_{s}$ is the space of probability
distributions $\left\{  P\left(  \vec{X}\left\vert \vec{\Theta}\right.
\right)  \right\}  $ labeled by $6N$ statistical parameters $\vec{\Theta}$.
These parameters are the coordinates for the point $P$, and in these
coordinates a volume element $dV_{\mathcal{M}_{s}}$ reads,
\begin{equation}
dV_{\mathcal{M}_{S}}=\sqrt{g}d^{6N}\vec{\Theta}=%
{\displaystyle\prod\limits_{\alpha=1}^{N}}
{\displaystyle\prod\limits_{a=1}^{3}}
\frac{\sqrt{2}}{\left(  \sigma_{a}^{\left(  \alpha\right)  }\right)  ^{2}}%
d\mu_{a}^{\left(  \alpha\right)  }d\sigma_{a}^{\left(  \alpha\right)
}\text{.}%
\end{equation}
The volume of an extended region $\Delta V_{\mathcal{M}_{s}}\left(
\tau\text{; }\lambda\right)  $ of $\mathcal{M}_{s}$ is defined by,%
\begin{equation}
\Delta V_{\mathcal{M}_{s}}\left(  \tau\text{; }\lambda\right)  \overset
{\text{def}}{=}%
{\displaystyle\prod\limits_{\alpha=1}^{N}}
{\displaystyle\prod\limits_{a=1}^{3}}
\int\nolimits_{\mu_{a}^{\left(  \alpha\right)  }\left(  0\right)  }^{\mu
_{a}^{\left(  \alpha\right)  }\left(  \tau\right)  }\int\nolimits_{\sigma
_{a}^{\left(  \alpha\right)  }\left(  0\right)  }^{\sigma_{a}^{\left(
\alpha\right)  }\left(  \tau\right)  }\frac{\sqrt{2}}{\left(  \sigma
_{a}^{\left(  \alpha\right)  }\right)  ^{2}}d\mu_{a}^{\left(  \alpha\right)
}d\sigma_{a}^{\left(  \alpha\right)  }%
\end{equation}
where $\mu_{a}^{\left(  \alpha\right)  }\left(  \tau\right)  $ and $\sigma
_{a}^{\left(  \alpha\right)  }\left(  \tau\right)  $ are given in
(\ref{solns}). The quantity that encodes relevant information about the
stability of neighboring volume elements is the average volume $\mathcal{V}%
_{\mathcal{M}_{s}}\left(  \tau\text{; }\lambda\right)  \equiv\left\langle
\Delta V_{\mathcal{M}_{s}}\left(  \tau\text{; }\lambda\right)  \right\rangle
_{\tau}$,
\begin{equation}
\mathcal{V}_{\mathcal{M}_{s}}\left(  \tau\text{; }\lambda\right)
\overset{\text{def}}{=}\frac{1}{\tau}%
{\displaystyle\int\limits_{0}^{\tau}}
\Delta V_{\mathcal{M}_{s}}\left(  \tau^{\prime}\text{; }\lambda\right)
d\tau^{\prime}\overset{\tau\rightarrow\infty}{\approx}\exp\left[
{\displaystyle\sum\limits_{\alpha=1}^{N}}
{\displaystyle\sum\limits_{a=1}^{3}}
\lambda_{a}^{\left(  \alpha\right)  }\tau\right]  \text{.} \label{avg-vol}%
\end{equation}
Again, as a side remark, we point out that if we consider the special case
where different Gaussians are characterized by the same initial conditions,
equation (\ref{enzo}) becomes,%
\begin{equation}
\mathcal{V}_{\mathcal{M}_{s}}\left(  \tau\text{; }\lambda\right)
\overset{\tau\rightarrow\infty}{\approx}\exp\left(  3N\lambda\tau\right)
\end{equation}
This asymptotic regime of evolution in (\ref{avg-vol}) describes the
exponential increase of average volume elements on $\mathcal{M}_{s}$. The
exponential instability characteristic of chaos forces the system to rapidly
explore large areas (volumes) of the statistical manifold. It is interesting
to note that this asymptotic behavior appears also in the conventional
description of quantum chaos where the entropy increases linearly at a rate
determined by the Lyapunov exponents \cite{ruelle}. The linear increase of
entropy as a quantum chaos criterion was introduced by Zurek and Paz
\cite{zurek}. In our information-geometric approach a relevant quantity that
can be useful to study the degree of instability characterizing the ED model
is the information-geometric entropy defined as,%
\begin{equation}
\mathcal{S}_{\mathcal{M}_{s}}\overset{\text{def}}{=}\underset{\tau
\rightarrow\infty}{\lim}\log\mathcal{V}_{\mathcal{M}_{s}}\left(  \tau\text{;
}\lambda\right)  \text{.} \label{asym-ent}%
\end{equation}
Substituting (\ref{avg-vol}) in (\ref{asym-ent}), we obtain%
\begin{equation}
\mathcal{S}_{\mathcal{M}_{s}}=\underset{\tau\rightarrow\infty}{\lim}%
\log\left\{  \frac{1}{\tau}%
{\displaystyle\int\limits_{0}^{\tau}}
\left[
{\displaystyle\prod\limits_{\alpha=1}^{N}}
{\displaystyle\prod\limits_{a=1}^{3}}
\int\nolimits_{\mu_{a}^{\left(  \alpha\right)  }\left(  0\right)  }^{\mu
_{a}^{\left(  \alpha\right)  }\left(  \tau^{\prime}\right)  }\int
\nolimits_{\sigma_{a}^{\left(  \alpha\right)  }\left(  0\right)  }^{\sigma
_{a}^{\left(  \alpha\right)  }\left(  \tau^{\prime}\right)  }\frac{\sqrt{2}%
}{\left(  \sigma_{a}^{\left(  \alpha\right)  }\right)  ^{2}}d\mu_{a}^{\left(
\alpha\right)  }d\sigma_{a}^{\left(  \alpha\right)  }\right]  d\tau^{\prime
}\right\}  \overset{\tau\rightarrow\infty}{\approx}%
{\displaystyle\sum\limits_{\alpha=1}^{N}}
{\displaystyle\sum\limits_{a=1}^{3}}
\lambda_{a}^{\left(  \alpha\right)  }\tau\text{.} \label{ent-Ms}%
\end{equation}
In the special case of same initial conditions for different Gaussians,
(\ref{ent-Ms}) becomes,%
\begin{equation}
\mathcal{S}_{\mathcal{M}_{s}}\left(  \tau\text{; }\lambda\right)
\overset{\tau\rightarrow\infty}{\approx}3N\lambda\tau\text{.}%
\end{equation}
The entropy $S_{\mathcal{M}_{s}}$ in (\ref{ent-Ms}) is the asymptotic limit of
the natural logarithm of the statistical weight $\left\langle \Delta
V_{\mathcal{M}_{s}}\right\rangle _{\tau}$ defined on $\mathcal{M}_{s}$. Its
linear growth in time is reminiscent of the aforementioned quantum chaos
criterion. Indeed, equation (\ref{ent-Ms}) may be considered the
information-geometric analog of the Zurek-Paz chaos criterion.

In conclusion, we have shown that%
\begin{equation}
\mathcal{R}_{\mathcal{M}_{s}}=-3N\text{, }J_{\mathcal{M}_{S}}\left(
\tau\text{; }\lambda\right)  \overset{\tau\rightarrow\infty}{\approx}%
{\displaystyle\sum\limits_{\alpha=1}^{N}}
{\displaystyle\sum\limits_{a=1}^{3}}
\exp\left(  \lambda_{a}^{\left(  \alpha\right)  }\tau\right)  \text{,
}\mathcal{S}_{\mathcal{M}_{s}}\left(  \tau\text{; }\lambda\right)
\overset{\tau\rightarrow\infty}{\approx}%
{\displaystyle\sum\limits_{\alpha=1}^{N}}
{\displaystyle\sum\limits_{a=1}^{3}}
\lambda_{a}^{\left(  \alpha\right)  }\tau\text{.}%
\end{equation}
Each indicator of chaos behaves as expected: $\mathcal{R}_{\mathcal{M}_{s}}$
is negative (this is a sufficient but not necessary condition for chaos),
$J_{\mathcal{M}_{S}}$ grows \emph{exponentially} in $\tau$ and, $\mathcal{S}%
_{\mathcal{M}_{s}}$ grows \emph{linearly} in $\tau$ and is proportional to the
sum of positive Lyapunov exponents of the system. Furthermore, it is
reasonable to state that the temporal complexity (chaoticity) of a system
ought to grow linearly as a function of the number of its variables
\cite{feldy} and it seems reasonable to assume that this complexity should not
depend on the special choice of the initial conditions of the system but only
on its dynamical evolution. The selection of a special set of initial
conditions should not affect the degree of chaoticity of a dynamical system.
Because of these considerations, we are allowed to choose any special set of
convenient initial conditions and evaluate the behavior of the indicators of
chaos in such special case. As we have showed, assuming the same initial
conditions for different Gaussians, we obtain
\begin{equation}
\mathcal{R}_{\mathcal{M}_{s}}=-3N\text{, }\mathcal{S}_{\mathcal{M}_{s}%
}\overset{\tau\rightarrow\infty}{\approx}3N\lambda\tau\text{, }J_{\mathcal{M}%
_{S}}\overset{\tau\rightarrow\infty}{\approx}3Ne^{\lambda\tau}\text{.}%
\end{equation}
The Ricci scalar curvature $\mathcal{R}_{\mathcal{M}_{s}}$ grows as a function
of the number of the microvariables of the system, the information-geometric
entropy $S_{\mathcal{M}_{s}}$ grows linearly as a function of the
microvariables of the system and the Jacobi vector field intensity
$J_{\mathcal{M}_{S}}$ grows exponentially as a function of the microvariables
of the system: $\mathcal{R}_{\mathcal{M}_{s}}$, $S_{\mathcal{M}_{s}}$ and
$J_{\mathcal{M}_{S}}$ behave as proper indicators of chaoticity and are
proportional to the number of Gaussian-distributed microstates of the system.
This proportionality leads to the conclusion that there exists a formal link
among these information-geometric measures of chaoticity. Formally,%
\begin{equation}
\mathcal{R}_{\mathcal{M}_{s}}\sim\mathcal{S}_{\mathcal{M}_{s}}\sim
J_{\mathcal{M}_{S}}\text{.} \label{cool-relation}%
\end{equation}
Equation (\ref{cool-relation}), together with the information-geometric analog
of the Zurek-Paz quantum chaos criterion, equation (\ref{ent-Ms}), represent
the fundamental results of this work. We are aware that equation
(\ref{cool-relation}) is reliable in the restrictive assumption of Gaussianity
and for very special initial conditions. However, we believe that with some
additional technical machinery, more general conclusions can be achieved and
this connection among indicators of chaoticity may be strengthened.
Furthermore, we believe our theoretical modelling scheme may be used to
describe actual systems where transitions from quantum to classical chaos
scenario occur, but this requires additional analysis. In the following
section, we briefly consider some similarities among the von Neumann,
Kolmogorov-Sinai and Information-Geometrodynamical entropies.

\section{On the von Neumann, Kolmogorov-Sinai and information
geometrodynamical Entropies}

In conventional approaches to chaos, the notion of entropy is introduced, in
both classical and quantum physics, as the missing information about the
systems fine-grained state \cite{jaynes, caves}. Following the first work in
reference \cite{caves}, we consider a classical system and suppose that the
phase space is partitioned into very fine-grained cells of uniform volume
$\Delta v$, labelled by an index $j$. If one does not know which cell the
system occupies, one assigns probabilities $p_{j}$ to the various cells;
equivalently, in the limit of infinitesimal cells, one can use a phase-space
density $\rho\left(  X_{j}\right)  =\frac{p_{j}}{\Delta v}$. Then, in a
classical chaotic evolution, the asymptotic expression of the information
needed to characterize a particular coarse-grained trajectory out to time
$\tau$ is given by the Shannon information entropy (measured in bits)
\cite{caves},%
\begin{equation}
\mathcal{S}_{\text{classical}}^{\left(  \text{chaotic}\right)  }=-\int
dX\rho\left(  X\right)  \log_{2}\left(  \rho\left(  X\right)  \Delta v\right)
=-\sum_{j}p_{j}\log_{2}p_{j}\sim\mathcal{K}\tau\text{.} \label{chao-classEnt}%
\end{equation}
where $\rho\left(  X\right)  $ is the phase-space density and $p_{j}%
=\frac{v_{j}}{\Delta v}$ is the probability for the corresponding
coarse-grained trajectory. $\mathcal{S}_{\text{classical}}^{\left(
\text{chaotic}\right)  }$ is the missing information about which fine-grained
cell the system occupies. Equation (\ref{chao-classEnt}) can be explained with
the following approximate reasoning \cite{caves}: the number of pieces in the
partition of the evolved pattern grows as $2^{\mathcal{K}\tau}$ (i.e.,
$\#\left\{  j\right\}  \approx$ $2^{\mathcal{K}\tau}$), each piece having
approximately the same phase-space volume and, therefore, the same probability
$p_{j}=\frac{v_{j}}{\Delta v}\approx2^{-\mathcal{K}\tau}$. That said, equation
(\ref{chao-classEnt}) follows in a straightforward way. However, we think this
picture is not the most clear one. We believe that within our IGAC, a better
and clearer understanding of what actually is happening may be achieved. The
quantity $\mathcal{K}$ represents the linear rate of information increase and
it is called the Kolmogorov-Sinai entropy (or metric entropy). $\mathcal{K}$
quantifies the degree of classical chaos (for a more detailed discussion about
$\mathcal{K}$, see Chapter 4). It is worthwhile emphasizing that the quantity
that grows asymptotically as $\mathcal{K}\tau$ is really the average of the
information on the left side of equation (\ref{chao-classEnt}). This
distinction can be ignored however, if we assume that the chaotic system has
roughly constant Lyapunov exponents over the accessible region of phase space.

In quantum mechanics the fine-grained alternatives are normalized state
vectors in Hilbert space. From a set of probabilities for various state
vectors, one can construct a density operator
\begin{equation}
\widehat{\rho}=\sum_{j}\lambda_{j}\left\vert \psi_{j}\right\rangle
\left\langle \psi_{j}\right\vert \text{, }\widehat{\rho}\left\vert \psi
_{j}\right\rangle =\lambda_{j}\left\vert \psi_{j}\right\rangle \text{.}%
\end{equation}
The normalization of the density operator, $tr\left(  \widehat{\rho}\right)
=1$, implies that the eigenvalues make up a normalized probability
distribution. The von Neumann entropy (natural generalization of both
Boltzmann's and Shannon's entropy) of the density operator $\widehat{\rho}$
(measured in bits) \cite{stenholm, caves2},%
\begin{equation}
S_{\text{quantum}}^{\left(  \text{chaotic}\right)  }=-tr\left(  \widehat{\rho
}\log_{2}\widehat{\rho}\right)  =-\sum_{j}\lambda_{j}\log_{2}\lambda_{j}%
\sim\mathcal{K}_{q}\tau\label{squantum}%
\end{equation}
can be thought of as the missing information about which eigenvector the
system is in. Entropy quantifies the degree of unpredictability about the
system's fine-grained state. In quantum mechanics, the von Neumann entropy
plays a role analogous to that played by the Shannon entropy in classical
probability theory. They are both monotone functionals of the state. The von
Neumann entropy reduces to the Shannon entropy for diagonal density matrices.
However, in general the von Neumann entropy is a subtler object than its
classical counterpart. The quantity $\mathcal{K}_{q}$\ in (\ref{squantum}) can
be interpreted as the non-commutative (quantum theory is a non-commutative
probability theory) quantum analog of the Kolmogorov-Sinai dynamical entropy,
the so-called quantum dynamical entropy \cite{benny}. Examples of quantum
dynamical entropies applied to quantum chaos and quantum information theory
are the Alicki-Fannes (AF) \cite{alicki} entropy and the
Connes-Narnhofer-Thirring (CNT) \cite{connes} entropy. Both the AF and CNT
entropy coincide with the KS entropy on classical dynamical systems. They also
coincide on finite-dimensional quantum systems. However, they differ when
moving from finite to infinite quantum systems. Furthermore, recall that
decoherence is the loss of phase coherence between the set of preferred
quantum states in the Hilbert space of the system due to the interaction with
the environment. Moreover, decoherence induces transitions from quantum to
classical systems. Therefore, classicality is an emergent property of an open
quantum system. Motivated by such considerations, Zurek and Paz investigated
implications of the process of decoherence for quantum chaos.

They considered a chaotic system, a single unstable harmonic oscillator
characterized by a potential $V\left(  x\right)  =-\frac{\omega^{2}x^{2}}{2}$
($\lambda$ is the Lyapunov exponent), coupled to an external environment. In
the \textit{reversible classical limit }\cite{zurek2}, the von Neumann entropy
of such a system increases linearly at a rate determined by the Lyapunov
exponent,%
\begin{equation}
\mathcal{S}_{\text{quantum}}^{\left(  \text{chaotic}\right)  }\left(
\text{Zurek-Paz}\right)  \overset{\tau\rightarrow\infty}{\sim}\omega
\tau\text{.}%
\end{equation}
Notice that the consideration of $3N$ uncoupled identical unstable harmonic
oscillators characterized by potentials $V_{i}\left(  x\right)  =-\frac
{\omega_{i}^{2}x^{2}}{2}$ $\left(  \omega_{i}=\omega_{j}\equiv\omega\text{;
}i\text{, }j=1\text{, }2\text{,..., }3N\right)  $ would simply lead to%
\begin{equation}
\mathcal{S}_{\text{quantum}}^{\left(  \text{chaotic}\right)  }\left(
\text{Zurek-Paz}\right)  \overset{\tau\rightarrow\infty}{\sim}3N\omega
\tau\text{.} \label{other-ent}%
\end{equation}
The resemblance of equations (\ref{ent-Ms}) and (\ref{other-ent}) is
remarkable. In what follows, we apply our information geometrical method to a
set of two ($n=2$) uncoupled inverted anisotropic harmonic oscillators and
show we obtain asymptotic linear IGE\ growth. The case for an arbitrary
$n$-set in three dimensions is presented in the Appendix.

\section{The information geometry of a set of $2$-uncoupled inverted harmonic
oscillators (IHO)}

In this section, our objective is to characterize chaotic properties of a set
of two one-dimensional inverted harmonic oscillators, each with different
frequency $\omega_{1}\neq\omega_{2}$. We will study the asymptotic behavior of
the geometrodynamical entropy and the functional dependence of the Ricci
scalar curvature of the $2$-dimensional manifold $\mathcal{M}_{IHO}^{\left(
2\right)  }$ underlying the ED model of the IHOs on the frequencies
$\omega_{i\text{ }}$, $i=1$, $2$. In Chapter 6, we explored the possibility of
using well established principles of inference to derive Newtonian dynamics
from relevant prior information codified into an appropriate statistical
manifold \cite{cafaro5}. In what follows, we introduce the basics of the
general formalism for a set of $n$ IHOs. This approach is\ similar
(mathematically but not conceptually) to the geometrization of Newtonian
dynamics used in the Riemannian geometrodynamical to chaos \cite{casetti,
biesiada3}.

\subsection{Informational geometrization of Newtonian dynamics}

In what follows, we apply the general formalism developed in Chapter 6 to our
specific problem under consideration. The system under investigation has $n$
degrees of freedom and a point on the $n$ dimensional configuration space
manifold $\mathcal{M}_{IHO}^{\left(  n\right)  }$ is parametrized by the $n$
Lagrangian coordinates $\left(  \theta_{1}\text{,...., }\theta_{n}\right)  $.
Moreover, the system under investigation is described by the Lagrangian
$\mathcal{L}$,%
\begin{equation}
\mathcal{L}=T\left(  \dot{\theta}_{1}\text{,..,}\dot{\theta}_{n}\right)
-\Phi\left(  \theta_{1}\text{,.., }\theta_{n}\right)  =\frac{1}{2}\delta
_{ij}\dot{\theta}_{i}\dot{\theta}_{j}+\frac{1}{2}\overset{n}{\underset
{j=1}{\sum}}\omega_{j}^{2}\theta_{j}^{2}%
\end{equation}
so that the Hamiltonian function $\mathcal{H=}T+\Phi\equiv E$ is a constant of
motion. For the sake of simplicity, let us set $E=1$. According to the
principle of stationary action - in the form of Maupertuis - among all the
possible isoenergetic paths $\gamma\left(  \tau\right)  $ with fixed end
points, the paths that make vanish the first variation of the action
functional%
\begin{equation}
\mathcal{I}=\int_{\gamma\left(  \tau\right)  }\frac{\partial\mathcal{L}%
}{\partial\dot{\theta}_{i}}\dot{\theta}_{i}d\tau
\end{equation}
are natural motions. As the kinetic energy $T$ is a homogeneous function of
degree two, we have $2T=\dot{\theta}_{i}\frac{\partial\mathcal{L}}%
{\partial\dot{\theta}_{i}}$, and Maupertuis' principle reads%
\begin{equation}
\delta\mathcal{I}=\delta\int_{\gamma\left(  \tau\right)  }2Td\tau=0\text{.}%
\end{equation}
The manifold $\mathcal{M}_{IHO}^{\left(  n\right)  }$ is naturally given a
proper Riemannian structure. In fact, let us consider the matrix%
\begin{equation}
g_{ij}\left(  \theta_{1}\text{,.., }\theta_{n}\right)  =\left[  1-\Phi\left(
\theta_{1}\text{,.., }\theta_{n}\right)  \right]  \delta_{ij}
\label{iho-metric2}%
\end{equation}
so that Maupertuis' principle becomes%
\begin{align}
\delta\int_{\gamma\left(  \tau\right)  }Td\tau &  =\delta\int_{\gamma\left(
\tau\right)  }\left(  T^{2}\right)  ^{\frac{1}{2}}d\tau=\delta\int
_{\gamma\left(  \tau\right)  }\left\{  \left[  1-\Phi\left(  \theta
_{1}\text{,..,}\theta_{n}\right)  \right]  \delta_{ij}\dot{\theta}_{i}%
\dot{\theta}_{j}\right\}  ^{\frac{1}{2}}\nonumber\\
&  =\delta\int_{\gamma\left(  \tau\right)  }\left(  g_{ij}\dot{\theta}_{i}%
\dot{\theta}_{j}\right)  ^{\frac{1}{2}}d\tau=\delta\int_{\gamma\left(
s\right)  }ds=0\text{, }ds^{2}=g_{ij}d\theta^{i}d\theta^{j}%
\end{align}
thus motions are geodesics of $\mathcal{M}_{IHO}^{\left(  n\right)  }$,
provided we define $ds$ as its arclength. The metric tensor $g\left(
\cdot\text{, }\cdot\right)  $ of $\mathcal{M}_{IHO}^{\left(  n\right)  }$ is
then defined by%
\begin{equation}
g=g_{ij}d\theta^{i}\otimes d\theta^{j} \label{iho-metric}%
\end{equation}
where $\left(  d\theta^{1}\text{,....., }d\theta^{n}\right)  $ is a natural
base of $T_{\theta}^{\ast}\mathcal{M}_{IHO}^{\left(  n\right)  }$ - the
cotangent space at the point $\theta$ - in the local chart $\left(  \theta
^{1}\text{,...., }\theta^{n}\right)  $. This is known as the Jacobi metric (or
kinetic energy metric). Denoting by $\nabla$ the canonical Levi-Civita
connection, the geodesic equation is defined as \cite{stewart},%
\begin{equation}
\nabla_{\dot{\gamma}}\dot{\gamma}=0\text{,} \label{superior}%
\end{equation}
where $\dot{\gamma}$ is the tangent vector of the allowed paths $\gamma$ at
constant energy $E$. In the local chart $\left(  \theta^{1}\text{,....,
}\theta^{n}\right)  $, equation (\ref{superior}) becomes,%
\begin{equation}
\frac{d^{2}\theta^{i}}{ds^{2}}+\Gamma_{jk}^{i}\frac{d\theta^{j}}{ds}%
\frac{d\theta^{k}}{ds}=0
\end{equation}
where the Christoffel coefficients are the components of $\nabla$ defined by%
\begin{equation}
\Gamma_{jk}^{i}=\left\langle d\theta^{i}\text{, }\nabla_{j}e_{k}\right\rangle
=\frac{1}{2}g^{im}\left(  \partial_{j}g_{km}+\partial_{k}g_{mj}-\partial
_{m}g_{jk}\right)  \text{,} \label{iho-connection}%
\end{equation}
with $\partial_{i}=\frac{\partial}{\partial\theta^{i}}$. Since $g_{ij}\left(
\theta_{1}\text{,.., }\theta_{n}\right)  =\left[  1-\Phi\left(  \theta
_{1}\text{,.., }\theta_{n}\right)  \right]  \delta_{ij}$, from the geodesic
equation we obtain%
\begin{equation}
\frac{d^{2}\theta^{i}}{ds^{2}}+\frac{1}{2\left(  1-\Phi\right)  }\left[
2\frac{\partial\left(  1-\Phi\right)  }{\partial\theta_{j}}\frac{d\theta^{j}%
}{ds}\frac{d\theta^{i}}{ds}-g^{ij}\frac{\partial\left(  1-\Phi\right)
}{\partial\theta_{j}}g_{km}\frac{d\theta^{k}}{ds}\frac{d\theta^{m}}%
{ds}\right]  =0\text{,} \label{inter}%
\end{equation}
whereupon using $ds^{2}=\left(  1-\Phi\right)  ^{2}d\tau^{2}$, we verify that
(\ref{inter}) reduces to%
\begin{equation}
\frac{d^{2}\theta^{i}}{d\tau^{2}}+\frac{\partial\Phi\left(  \theta
_{1}\text{,.., }\theta_{n}\right)  }{\partial\theta_{i}}=0\text{,
}i=1\text{,..., }n\text{.} \label{Newton}%
\end{equation}
Equation (\ref{Newton}) are Newton's equations. It is worthwhile emphasizing
that the transformation to geodesic motion on a curved statistical manifold is
obtained in two key steps: the \textit{conformal transformation of the
metric}, $\delta_{ij}\rightarrow g_{ij}=\left(  1-\Phi\right)  $ $\delta_{ij}$
and, the \textit{rescaling of the temporal evolution parameter}, $d\tau
^{2}\rightarrow ds^{2}=2\left(  1-\Phi\right)  ^{2}d\tau^{2}$.

\subsection{Two uncoupled inverted one-dimensional harmonic oscillators}

As a simple physical example, we examine the IG associated with a set of two
one dimensional IHOs. In this case, the metric tensor $g_{ij}$ appearing in
(\ref{iho-metric2}) takes the form%
\begin{equation}
g_{ij}\left(  \theta_{1}\text{, }\theta_{2}\right)  =\left[  1-\Phi\left(
\theta_{1}\text{, }\theta_{2}\right)  \right]  \delta_{ij}\text{ with
}i\text{, }j=1\text{, }2\text{.}%
\end{equation}
where the function $\Phi\left(  \theta_{1}\text{, }\theta_{2}\right)  $ is
given by,%
\begin{equation}
\Phi\left(  \theta_{1}\text{, }\theta_{2}\right)  =\overset{2}{\underset
{j=1}{\sum}}\Phi_{j}\left(  \theta_{j}\right)  \text{, }\Phi_{j}\left(
\theta_{j}\right)  =-\frac{1}{2}\omega_{j}^{2}\theta_{j}^{2}\text{.}%
\end{equation}
Hence the metric tensor $g_{ij}$ on $\mathcal{M}_{IHO}^{\left(  2\right)  }$
becomes,%
\begin{equation}
g_{ij}=\left(
\begin{array}
[c]{cc}%
1+\frac{1}{2}\left(  \omega_{1}^{2}\theta_{1}^{2}+\omega_{2}^{2}\theta_{2}%
^{2}\right)  & 0\\
0 & 1+\frac{1}{2}\left(  \omega_{1}^{2}\theta_{1}^{2}+\omega_{2}^{2}\theta
_{2}^{2}\right)
\end{array}
\right)  \text{.}%
\end{equation}
Using the standard definition of the Ricci scalar (\ref{ricci-scalar}), we
obtain%
\begin{equation}
\mathcal{R}_{\mathcal{M}_{IHO}^{\left(  2\right)  }}\left(  \omega_{1}\text{,
}\omega_{2}\right)  =\frac{4\left(  \theta_{1}^{2}\omega_{1}^{4}+\theta
_{2}^{2}\omega_{2}^{4}\right)  -4\left(  \theta_{1}^{2}+\theta_{2}^{2}\right)
\omega_{1}^{2}\omega_{2}^{2}-8\left(  \omega_{1}^{2}+\omega_{2}^{2}\right)
}{\left(  \theta_{1}^{2}\omega_{1}^{2}+\theta_{2}^{2}\omega_{2}^{2}+2\right)
^{3}}\text{.} \label{ricciS-iho}%
\end{equation}
For $\omega_{1}=\omega_{2}=\omega$, the scalar curvature (\ref{ricciS-iho}) is
always negative,
\begin{equation}
\mathcal{R}_{\mathcal{M}_{IHO}^{\left(  2\right)  }}\left(  \omega\right)
=\frac{-16\omega^{2}}{\left[  2+\left(  \theta_{1}^{2}+\theta_{2}^{2}\right)
\omega^{2}\right]  ^{3}}<0\text{, }\forall\omega\geq0\text{.}%
\end{equation}
However, in presence of distinct frequency values, $\omega_{1}\neq\omega_{2}$,
it is possible to properly choose the $\omega$'s so that $\mathcal{R}%
_{\mathcal{M}_{IHO}^{\left(  2\right)  }}\left(  \omega_{1}\text{, }\omega
_{2}\right)  $ becomes either negative or positive. In addition, we notice
that the manifold underlying the IHO model is anisotropic since its associated
Weyl projective curvature tensor components are non-vanishing. For the special
case, $\omega_{1}=\omega_{2}$, we obtain%
\begin{equation}
W_{1212}\left(  \omega\right)  =\frac{8\omega^{4}\left(  \theta_{1}^{2}%
+\theta_{2}^{2}\right)  +2\omega^{6}\left(  \theta_{1}^{4}+\theta_{2}%
^{4}\right)  +4\omega^{6}\theta_{1}^{2}\theta_{2}^{2}}{\left(  \theta_{1}%
^{2}\omega^{2}+\theta_{2}^{2}\omega^{2}+2\right)  ^{3}}\text{.}%
\end{equation}
Clearly, the frequency parameter $\omega$ drives the degree of anisotropy of
the statistical manifold $\mathcal{M}_{IHO}^{\left(  2\right)  }$ and, as
expected, in the limit of vanishing $\omega$, we recover the flat
($\mathcal{R}=0$), isotropic ($W=0$) Euclidean manifold characterized by
metric $\delta_{ij}$. This result is a concrete example of the fact that
conformal transformations change the degree of anisotropy of the ambient
statistical manifold underlying the Newtonian dynamics. Our only remaining
task is to compute the information geometrodynamical entropy $\mathcal{S}%
_{\mathcal{M}_{IHO}^{\left(  2\right)  }}\left(  \tau\text{; }\omega
_{1}\text{, }\omega_{2}\right)  $, defined as%
\begin{equation}
\mathcal{S}_{\mathcal{M}_{IHO}^{\left(  2\right)  }}\left(  \tau\text{;
}\omega_{1}\text{, }\omega_{2}\right)  \overset{\text{def}}{=}\underset
{\tau\rightarrow\infty}{\lim}\log\left[  \left\langle \Delta V_{\mathcal{M}%
_{IHO}^{\left(  2\right)  }}\left(  \tau\text{; }\omega_{1}\text{, }\omega
_{2}\right)  \right\rangle _{\tau}\right]  \text{.} \label{iho-entropy}%
\end{equation}
The quantity $\left\langle \Delta V_{\mathcal{M}_{IHO}^{\left(  2\right)  }%
}\left(  \tau\text{; }\omega_{1}\text{, }\omega_{2}\right)  \right\rangle
_{\tau}$ appearing in (\ref{iho-entropy}) is the average volume element,
defined by%
\begin{equation}
\left\langle \Delta V_{\mathcal{M}_{IHO}^{\left(  2\right)  }}\left(
\tau\text{; }\omega_{1}\text{, }\omega_{2}\right)  \right\rangle _{\tau}%
=\frac{1}{\tau}%
{\displaystyle\int\limits_{0}^{\tau}}
\Delta V_{\mathcal{M}_{IHO}^{\left(  2\right)  }}\left(  \tau\text{; }%
\omega_{1}\text{, }\omega_{2}\right)  d\tau^{\prime}\text{,}%
\end{equation}
with the statistical volume element $\Delta V_{\mathcal{M}_{IHO}^{\left(
2\right)  }}$ given by%
\begin{align}
\Delta V_{\mathcal{M}_{IHO}^{\left(  2\right)  }}\left(  \tau\text{; }%
\omega_{1}\text{, }\omega_{2}\right)   &  =\underset{\left\{  \vec{\theta
}^{\prime}\right\}  }{\int}\left[  1+\frac{1}{2}\left(  \omega_{1}^{2}%
\theta_{1}^{\prime2}+\omega_{2}^{2}\theta_{2}^{\prime2}\right)  \right]
d\theta_{1}^{\prime}d\theta_{2}^{\prime}\label{delV}\\
& \nonumber\\
&  \overset{\tau\rightarrow\infty}{\approx}\frac{1}{6}\theta_{1}^{\prime
}\theta_{2}^{\prime}\left(  \omega_{1}^{2}\theta_{1}^{\prime2}+\omega_{2}%
^{2}\theta_{2}^{\prime2}\right)  \text{.}\nonumber
\end{align}
Recall that the two Newtonian equations of motion for each inverted harmonic
oscillator are given by,%
\begin{equation}
\frac{d^{2}\theta_{j}}{d\tau^{2}}-\omega_{j}^{2}\theta_{j}=0\text{, }\forall
j=1\text{, }2\text{.}%
\end{equation}
Hence, the asymptotic behavior of such macrovariables on manifold
$\mathcal{M}_{IHO}^{\left(  2\right)  }$ is given by,%
\begin{equation}
\theta_{j}\left(  \tau\right)  \overset{\tau\rightarrow\infty}{\approx}\Xi
_{j}e^{\omega_{j}\tau}\text{, }\Xi_{j}\in%
\mathbb{R}
\text{, }\forall j=1\text{, }2\text{.}%
\end{equation}
Substituting $\theta_{1}\left(  \tau^{\prime}\right)  =\Xi_{1}e^{\omega
_{1}\tau^{\prime}}$ and $\theta_{2}\left(  \tau^{\prime}\right)  =\Xi
_{2}e^{\omega_{2}\tau^{\prime}}$ into (\ref{delV}), we obtain%
\begin{equation}
\Delta V_{\mathcal{M}_{IHO}^{\left(  2\right)  }}\left(  \tau\text{; }%
\omega_{1}\text{, }\omega_{2}\right)  \overset{\tau\rightarrow\infty\text{ }%
}{\approx}\frac{\Xi_{1}\Xi_{2}}{6}e^{\left(  \omega_{1}+\omega_{2}\right)
\tau}\left(  \Xi_{1}^{2}e^{2\omega_{1}\tau}\omega_{1}^{2}+\Xi_{2}%
^{2}e^{2\omega_{2}\tau}\omega_{2}^{2}\right)  \text{.} \label{above}%
\end{equation}
By direct computation, we find the average of (\ref{above}) is given by,%
\begin{equation}
\left\langle \Delta V_{\mathcal{M}_{IHO}^{\left(  2\right)  }}\left(
\tau\text{; }\omega_{1}\text{, }\omega_{2}\right)  \right\rangle _{\tau
}\overset{\tau\rightarrow\infty\text{ }}{\approx}\frac{1}{\tau}%
{\displaystyle\int\limits_{0}^{\tau}}
\left[  \frac{\Xi_{1}\Xi_{2}}{6}e^{\left(  \omega_{1}+\omega_{2}\right)
\tau^{\prime}}\left(  \Xi_{1}^{2}e^{2\omega_{1}\tau^{\prime}}\omega_{1}%
^{2}+\Xi_{2}^{2}e^{2\omega_{2}\tau^{\prime}}\omega_{2}^{2}\right)  \right]
d\tau^{\prime}\text{.}%
\end{equation}
Assuming as a working hypothesis that $\Xi_{1}=\Xi_{2}=\Xi$, we obtain%
\begin{equation}
\frac{1}{\tau}%
{\displaystyle\int\limits_{0}^{\tau}}
\left[  \frac{\Xi_{1}\Xi_{2}}{6}e^{\left(  \omega_{1}+\omega_{2}\right)
\tau^{\prime}}\left(  \Xi_{1}^{2}e^{2\omega_{1}\tau^{\prime}}\omega_{1}%
^{2}+\Xi_{2}^{2}e^{2\omega_{2}\tau^{\prime}}\omega_{2}^{2}\right)  \right]
d\tau^{\prime}=\left\{
\begin{array}
[c]{c}%
\frac{1}{12}\Xi^{6}\omega\frac{\exp\left(  4\omega\tau\right)  }{\tau}\text{,
if }\omega_{1}=\omega_{2}\text{,}\\
\frac{1}{18}\Xi^{6}\omega_{1}\frac{\exp\left(  3\omega_{1}\tau\right)  }{\tau
}\text{, if }\omega_{1}\gg\omega_{2}\text{,}\\
\frac{1}{18}\Xi^{6}\omega_{2}\frac{\exp\left(  3\omega_{2}\tau\right)  }{\tau
}\text{, if }\omega_{2}\gg\omega_{1}\text{.}%
\end{array}
\right.  \text{.} \label{this}%
\end{equation}
Finally, substituting (\ref{this}) in (\ref{iho-entropy}), we obtain%
\begin{equation}
\mathcal{S}_{\mathcal{M}_{IHO}^{\left(  2\right)  }}\left(  \tau\text{;
}\omega_{1}\text{, }\omega_{2}\right)  \overset{\tau\rightarrow\infty}%
{\propto}\left\{
\begin{array}
[c]{c}%
2\omega\tau\text{, if }\omega_{1}=\omega_{2}\text{,}\\
\omega_{1}\tau\text{, if }\omega_{1}\gg\omega_{2}\text{,}\\
\omega_{2}\tau\text{, if }\omega_{2}\gg\omega_{1}\text{.}%
\end{array}
\right.  \text{.} \label{iho-entropy1}%
\end{equation}
It is clear that the information-geometrodynamical entropy $S_{\mathcal{M}%
_{IHO}^{\left(  2\right)  }}\left(  \tau\text{; }\omega_{1}\text{, }\omega
_{2}\right)  $ exhibits classical linear behavior in the asymptotic limit,
with proportionality coefficient $\Omega=$ $\omega_{1}+\omega_{2}$,%
\begin{equation}
\mathcal{S}_{\mathcal{M}_{IHO}^{\left(  2\right)  }}\left(  \tau\text{;
}\omega_{1}\text{, }\omega_{2}\right)  \overset{\tau\rightarrow\infty}%
{\propto}\Omega\tau\text{.} \label{sopra(7)}%
\end{equation}
Equation (\ref{sopra(7)}) expresses the asymptotic linear growth of our
information geometrodynamical entropy for the IHO system considered. This
result (for $n=2$) extends the result of Zurek-Paz (\ref{other-ent}) in a
classical information-geometric setting. This result, together with my
previous works \cite{cafaro2, cafaro3} lend substantial support for the IGAC
approach advocated in the present Chapter.

\section{Conclusions}

A Gaussian ED statistical model has been constructed on a $6N$-dimensional
statistical manifold $\mathcal{M}_{s}$. The macro-coordinates on the manifold
are represented by the expectation values of microvariables associated with
Gaussian distributions. The geometric structure of $\mathcal{M}_{s}$ was
studied in detail. It was shown that $\mathcal{M}_{s}$ is a curved manifold of
constant negative Ricci curvature $-3N$ . The geodesics of the ED model are
hyperbolic curves on $\mathcal{M}_{s}$. A study of the stability of geodesics
on $\mathcal{M}_{s}$ was presented. The notion of statistical volume elements
was introduced to investigate the asymptotic behavior of a one-parameter
family of neighboring volumes $\mathcal{F}_{V_{\mathcal{M}_{s}}}\left(
\lambda\right)  \equiv\left\{  V_{\mathcal{M}_{s}}\left(  \tau\text{; }%
\lambda\right)  \right\}  _{\lambda\in%
\mathbb{R}
^{+}}$. An information-geometric analog of the Zurek-Paz chaos criterion was
suggested. It was shown that the behavior of geodesics is characterized by
exponential instability that leads to chaotic scenarios on the curved
statistical manifold. These conclusions are supported by a study based on the
geodesic deviation equations and on the asymptotic behavior of the Jacobi
vector field intensity $J_{\mathcal{M}_{s}}$ on $\mathcal{M}_{s}$. A Lyapunov
exponent analog similar to that appearing in the Riemannian geometric approach
to chaos was suggested as an indicator of chaoticity. On the basis of our
analysis a relationship among an entropy-like quantity, chaoticity and
curvature is proposed, suggesting to interpret the statistical curvature as a
measure of the entropic dynamical chaoticity.

The results obtained in this work are significant, in our opinion, since a
rigorous relation among curvature, Lyapunov exponents and Kolmogorov-Sinai
entropy is still under investigation \cite{kawabe}. In addition, there does
not exist a well defined unifying characterization of chaos in classical and
quantum physics \cite{caves} due to fundamental differences between the two
theories. In addition, the role of curvature in statistical inference is even
less understood. The meaning of statistical curvature for a one-parameter
model in inference theory was introduced in \cite{efron}. Curvature served as
an important tool in the asymptotic theory of statistical estimation.
Therefore the implications of this work is twofold. Firstly, it helps
understanding possible future use of the statistical curvature in modelling
real processes by relating it to conventionally accepted quantities such as
entropy and chaos. On the other hand, it serves to cast what is already known
in physics regarding curvature in a new light as a consequence of its proposed
link with inference.

As a simple physical example, we considered the information-geometry
$\mathcal{M}_{IHO}^{\left(  2\right)  }$ associated with a set of two inverted
harmonic oscillators. It was determined that in the limit of a flat frequency
spectrum ($\omega_{1}=\omega_{2}=\omega$), the scalar curvature $\mathcal{R}%
_{\mathcal{M}_{IHO}^{\left(  2\right)  }}\left(  \omega_{1}\text{, }\omega
_{2}\right)  $ is constantly negative. In the case of distinct frequencies,
i.e., $\omega_{1}\neq\omega_{2}$, it is possible - for appropriate choices of
$\omega_{1}$ and $\omega_{2}$ - to obtain either negative or positive values
of $\mathcal{R}_{\mathcal{M}_{IHO}^{\left(  2\right)  }}\left(  \omega
_{1}\text{, }\omega_{2}\right)  $. Moreover, it was shown that $\mathcal{M}%
_{IHO}^{\left(  2\right)  }$ is an anisotropic manifold since the Weyl
projective curvature tensor has a non-vanishing component $W_{1212}$. It was
found that the information geometrodynamical entropy of the IHO system
exhibits asymptotic linear growth. This IHO example is generalized to
arbitrary values of $n$ in the Appendix.

The descriptions of a classical chaotic system of arbitrary interacting
degrees of freedom, deviations from Gaussianity and chaoticity arising from
fluctuations of positively curved statistical manifolds are being investigated
\cite{cafaro4}.

\section{Appendix}

\subsection{The set of $n$ uncoupled inverted anisotropic three-dimensional
harmonic oscillators}

\subsubsection{Ohmic frequency spectrum}

We now generalize the results obtained in this Chapter for a set of $n$ IHOs.
The information metric on the $3n$-dimensional statistical manifold
$\mathcal{M}_{IHO}^{\left(  3n\right)  }$ is given by%
\begin{equation}
g_{ij}\left(  \theta_{1}\text{,...., }\theta_{3n}\right)  =\left[
1-\Phi\left(  \theta_{1}\text{,...., }\theta_{3n}\right)  \right]  \delta
_{ij}\text{,}%
\end{equation}
where%
\begin{equation}
\Phi\left(  \theta_{1}\text{,...., }\theta_{3n}\right)  =\overset
{3n}{\underset{j=1}{\sum}}\Phi_{j}\left(  \theta_{j}\right)  \text{, }\Phi
_{j}\left(  \theta_{j}\right)  =-\frac{1}{2}\omega_{j}^{2}\theta_{j}%
^{2}\text{.}%
\end{equation}
The information geometrodynamical entropy $S_{\mathcal{M}_{IHO}^{\left(
3n\right)  }}\left(  \tau\text{; }\omega_{1}\text{,.., }\omega_{3n}\right)  $
is defined as%
\begin{equation}
\mathcal{S}_{\mathcal{M}_{IHO}^{\left(  3n\right)  }}\left(  \tau\text{;
}\omega_{1}\text{,.., }\omega_{3n}\right)  \overset{\text{def}}{=}%
\underset{\tau\rightarrow\infty}{\lim}\log\left[  \left\langle \Delta
V_{\mathcal{M}_{IHO}^{\left(  3n\right)  }}\left(  \tau\text{; }\omega
_{1}\text{,.., }\omega_{3n}\right)  \right\rangle _{\tau}\right]  \text{,}
\label{gen-ent}%
\end{equation}
where the average volume element $\Delta V_{\mathcal{M}_{IHO}^{\left(
3n\right)  }}$ is given by%
\begin{equation}
\left\langle \Delta V_{\mathcal{M}_{IHO}^{\left(  3n\right)  }}\left(
\tau\text{; }\omega_{1}\text{,.., }\omega_{3n}\right)  \right\rangle _{\tau
}=\frac{1}{\tau}%
{\displaystyle\int\limits_{0}^{\tau}}
\Delta V_{\mathcal{M}_{IHO}^{\left(  3n\right)  }}\left(  \tau^{\prime}\text{;
}\omega_{1}\text{,.., }\omega_{3n}\right)  d\tau^{\prime}\text{,}
\label{inter2}%
\end{equation}
and the statistical volume element $\Delta V_{\mathcal{M}_{IHO}^{\left(
3n\right)  }}$ is defined as%
\begin{equation}
\Delta V_{\mathcal{M}_{IHO}^{\left(  3n\right)  }}\left(  \tau^{\prime}\text{;
}\omega_{1}\text{,.., }\omega_{n}\right)  =\underset{\left\{  \vec{\theta
}^{\prime}\right\}  }{\int}d^{3n}\vec{\theta}^{\prime}\left(  1+\frac{1}%
{2}\underset{j=1}{\overset{3n}{\sum}}\omega_{j}^{2}\theta_{j}^{\prime
2}\right)  ^{\frac{3n}{2}}\text{.} \label{inter3}%
\end{equation}
Substituting (\ref{inter2}) and (\ref{inter3}) in (\ref{gen-ent}) we obtain
the general expression for $\mathcal{S}_{\mathcal{M}_{IHO}^{\left(  3n\right)
}}\left(  \tau\text{; }\omega_{1}\text{,.., }\omega_{3n}\right)  $,
\begin{equation}
\mathcal{S}_{\mathcal{M}_{IHO}^{\left(  3n\right)  }}\left(  \tau\text{;
}\omega_{1}\text{,.., }\omega_{3n}\right)  \overset{\text{def}}{=}%
\underset{\tau\rightarrow\infty}{\lim}\log\left\{  \frac{1}{\tau}\int
_{0}^{\tau}\left[  \underset{\left\{  \vec{\theta}^{\prime}\right\}  }{\int
}d^{3n}\vec{\theta}^{\prime}\left(  1+\frac{1}{2}\underset{j=1}{\overset
{3n}{\sum}}\omega_{j}^{2}\theta_{j}^{\prime2}\right)  ^{\frac{3n}{2}}\right]
d\tau^{\prime}\right\}  \text{.} \label{inter4}%
\end{equation}
To evaluate (\ref{inter4}) we observe $\Delta V_{\mathcal{M}_{IHO}^{\left(
3n\right)  }}$ can be written as%
\begin{align}
\Delta V_{\mathcal{M}_{IHO}^{\left(  3n\right)  }}\left(  \tau^{\prime}\text{;
}\omega_{1}\text{,.., }\omega_{3n}\right)   &  =\underset{\left\{  \vec
{\theta}^{\prime}\right\}  }{\int}d^{3n}\vec{\theta}^{\prime}\left(
1+\frac{1}{2}\underset{j=1}{\overset{3n}{\sum}}\omega_{j}^{2}\theta
_{j}^{\prime2}\right)  ^{\frac{3n}{2}}\text{,}\nonumber\\
& \nonumber\\
&  =\int d\theta_{1}^{\prime}\int d\theta_{2}^{\prime}\text{...}\int
d\theta_{3n-1}^{\prime}\left[  \int\left(  1+\frac{1}{2}\underset
{j=1}{\overset{3n}{\sum}}\omega_{j}^{2}\theta_{j}^{\prime2}\right)
^{\frac{3n}{2}}d\theta_{3n}^{\prime}\right]  \text{,}\nonumber\\
& \nonumber\\
&  \overset{\text{ }}{\approx}\frac{1}{3n}\frac{1}{2^{\frac{3n}{2}}}\left(
\overset{3n}{\underset{i=1}{\Pi}}\theta_{i}^{\prime}\right)  \left[
\underset{j=1}{\overset{3n}{\sum}}\omega_{j}^{2}\theta_{j}^{\prime2}\right]
^{\frac{3n}{2}}\text{.}%
\end{align}
Since the $n$-Newtonian equations of motions for each IHO are given by%
\begin{equation}
\frac{d^{2}\theta_{j}}{d\tau^{2}}-\omega_{j}^{2}\theta_{j}=0\text{, }\forall
j=1\text{,..., }3n\text{,}%
\end{equation}
the asymptotic behavior of such macrovariables on manifold $\mathcal{M}%
_{IHO}^{\left(  3n\right)  }$ is given by%
\begin{equation}
\theta_{j}\left(  \tau\right)  \overset{\tau\rightarrow\infty}{\approx}\Xi
_{j}e^{\omega_{j}\tau}\text{, }\Xi_{j}\in%
\mathbb{R}
\text{, }\forall j=1\text{,..., }3n\text{.}%
\end{equation}
We therefore obtain%
\begin{equation}
\Delta V_{\mathcal{M}_{IHO}^{\left(  3n\right)  }}\left(  \tau\text{; }%
\omega_{1}\text{,...., }\omega_{3n}\right)  \overset{\tau\rightarrow
\infty\text{ }}{\approx}\frac{1}{3n}\frac{1}{2^{\frac{3n}{2}}}\left(
\underset{i=1}{\overset{3n}{\Pi}}\Xi_{i}\right)  \cdot\exp\left(  \overset
{3n}{\underset{i=1}{\sum}}\omega_{i}\tau\right)  \left[  \underset
{j=1}{\overset{3n}{\sum}}\Xi_{j}^{2}e^{2\omega_{j}\tau}\omega_{j}^{2}\right]
^{\frac{3n}{2}}\text{.} \label{inter5}%
\end{equation}
Upon averaging (\ref{inter5}) we find%
\begin{equation}
\left\langle \Delta V_{\mathcal{M}_{IHO}^{\left(  3n\right)  }}\left(
\tau\text{; }\omega_{1}\text{,...., }\omega_{3n}\right)  \right\rangle _{\tau
}\overset{\tau\rightarrow\infty\text{ }}{\approx}\frac{1}{\tau}%
{\displaystyle\int\limits_{0}^{\tau}}
\left\{  \frac{1}{3n}\frac{1}{2^{\frac{3n}{2}}}\left(  \underset{i=1}%
{\overset{3n}{\Pi}}\Xi_{i}\right)  \cdot\exp\left(  \Omega\tau^{\prime
}\right)  \left[  \underset{j=1}{\overset{3n}{\sum}}\Xi_{j}^{2}e^{2\omega
_{j}\tau^{\prime}}\omega_{j}^{2}\right]  ^{\frac{3n}{2}}\right\}
d\tau^{\prime}\text{.}%
\end{equation}
where $\Omega=\overset{3n}{\underset{i=1}{\sum}}\omega_{i}$. As a working
hypothesis, we assume $\Xi_{i}=\Xi_{j}\equiv\Xi$ $\forall i$, $j=1$,.., $3n$.
Furthermore, assume that $n\rightarrow\infty$ so that the spectrum of
frequencies becomes continuum and, as an additional working hypothesis, assume
this spectrum is linearly distributed (Ohmic frequency spectrum),%
\begin{equation}
\rho_{\text{Ohmic}}\left(  \omega\right)  =\frac{2}{\Omega_{\text{cut-off}%
}^{2}}\omega\text{ with}\underset{0}{\overset{\Omega_{\text{cut-off}}}{\int}%
}\rho_{\text{Ohmic}}\left(  \omega\right)  d\omega=1\text{, }\Omega
_{\text{cut-off}}=\xi\Omega\text{, }\xi\in%
\mathbb{R}
\text{. }%
\end{equation}
Therefore, we obtain
\begin{equation}
\left\langle \Delta V_{\mathcal{M}_{IHO}^{\left(  3n\right)  }}\left(
\tau\text{; }\omega_{1}\text{,...., }\omega_{3n}\right)  \right\rangle _{\tau
}\overset{\tau\rightarrow\infty\text{ }}{\approx}\frac{1}{3n}\frac{1}%
{2^{\frac{3n}{2}}}\Xi^{6n}\left(  \frac{\xi^{2}\Omega^{2}}{2}\right)
^{\frac{3n}{2}}\frac{\exp\left(  \frac{3}{2}n\xi\Omega\tau\right)  }{\tau
}\text{.} \label{inter6}%
\end{equation}
Finally, substituting (\ref{inter6}) into (\ref{gen-ent}), we obtain the
remarkable result%
\begin{equation}
\mathcal{S}_{\mathcal{M}_{IHO}^{\left(  3n\right)  }}\left(  \tau\text{;
}\omega_{1}\text{,.., }\omega_{3n}\right)  \overset{\tau\rightarrow\infty
}{\propto}\Omega\tau\text{, }\Omega=\overset{3n}{\underset{i=1}{\sum}}%
\omega_{i}\text{.} \label{Fin}%
\end{equation}
Equation (\ref{Fin}) displays the asymptotic, linear information
geometrodynamical entropy growth of the generalized $n$-set of inverted
harmonic oscillators and extends the result of Zurek-Paz to an arbitrary set
of anisotropic inverted harmonic oscillators \cite{zurek} in a classical
information-geometric setting.

The Ohmic frequency spectrum case leads to asymptotic IGE growth. However, in
this case, we are not able to compactify the parameter space of our
statistical model. The compactification of the parameter space (and therefore
of the statistical manifold) is required for true chaos where the folding
mechanism must be present (indeed, the lack of such feature has been one of
the most important criticisms to the Zurek-Paz model). The folding mechanism
required for true chaos is not even restored in a statistical sense (averaging
over $\omega$ and $\tau$),%
\begin{equation}
\bar{\theta}_{\text{Ohmic}}\left(  \tau\text{; }\omega\right)  \overset
{\text{def}}{=}\underset{\tau\text{, }\omega\rightarrow\infty}{\lim}\left\{
\frac{1}{\tau}\overset{t}{\underset{0}{\int}}d\tau^{\prime}\left[
\overset{\omega}{\underset{0}{\int}}d\omega^{\prime}\rho_{\text{Ohmic}}\left(
\omega^{\prime}\right)  \theta\left(  \tau^{\prime}\text{; }\omega^{\prime
}\right)  \right]  \right\}  \approx\underset{\tau\text{, }\omega
\rightarrow\infty}{\lim}\frac{\exp\left(  \omega\tau\right)  }{\tau^{2}%
}\rightarrow\infty\text{.}%
\end{equation}
This missing feature may lead us to consider in the near future the
possibility of considering other frequency spectra.

\pagebreak

\begin{center}
{\LARGE Chapter 8: Concluding Remarks and Future Research Directions}
\end{center}

I present concluding remarks emphasizing strengths and weakness of my approach
and I address possible further research directions.

\section{Concluding Remarks}

In this doctoral dissertation, I considered two important questions:

First, are laws of physics practical rules to process information about the
world using geometrical methods? Are laws of physics rules of inference?

Second, since a unifying framework to describe chaotic dynamics in classical
and quantum domains is missing, is it possible to construct a new
information-geometric model, to develop new tools so that a unifying framework
is provided or, at least, new insights and new understandings are given?

After setting the scene of my thesis and after stating the problem and its
motivations, I reviewed the basic elements of the maximum relative entropy
formalism (ME method) and recall the basics of Riemannian geometry with
special focus to its application to probability theory (this is known as
Information Geometry, IG). IG and ME are the fundamental tools that Prof.
Ariel Caticha has used to build a form of information-constrained dynamics on
statistical manifolds to investigate the possibility that Einstein's general
theory of gravity (or any classical or quantum theory of physics) may emerge
as a macroscopic manifestation of an underlying microscopic statistical
structure. This dynamics is known in the literature as Entropic
Dynamics\ (ED). Therefore, since ED was an important element of this thesis, I
reviewed the key-points of such dynamics, emphasizing the most relevant points
that I used in my own information geometrodynamical approach to chaos (IGAC).
Of course, before introducing my IGAC, I briefly reviewed the basics of the
conventional Riemannian geometrodynamics approach to chaos and discussed the
notion of chaos in physics in general. After this long background information
that was needed because of the originality and novelty of these topics, I
started with my original contributions. Two entropic dynamical models are
considered. The geometric structure of the statistical manifolds underlying
these models is studied. It is found that in both cases, the resulting metric
manifolds are negatively curved. Moreover, the geodesics on each manifold are
described by hyperbolic trajectories. A detailed analysis based on the
Jacobi-Levi-Civita equation for geodesic spread (JLC equation) is used to show
that the hyperbolicity of the manifolds leads to chaotic exponential
instability. A comparison between the two models leads to a relation among
scalar curvature of the manifold ($\mathcal{R}$), Jacobi field intensity ($J$)
and information geometrodynamical entropy (IGE, $\mathcal{S}_{\mathcal{M}}$).
The IGE entropy is proposed as a brand new measure of chaoticity.

\bigskip

\begin{description}
\item[First Contribution] \cite{cafaro1, cafaro2, cafaro3}: \textit{I suggest
that these three quantities, }$\mathcal{R}$\textit{, }$J$\textit{, and
}$\mathcal{S}_{\mathcal{M}}$\textit{\ are useful indicators of chaoticity for
chaotic dynamical systems on curved statistical manifolds. Furthermore, I
suggest a classical \ information-geometric criterion of linear information
geometrodynamical entropy growth in analogy with the Zurek-Paz quantum chaos
criterion.}
\end{description}

\bigskip

In collaboration with Prof. Ariel Caticha, I show that the ED formalism is not
purely a mathematical framework; it is indeed a general theoretical scheme
where conventional Newtonian dynamics can be obtained as a special limiting
case. Newtonian dynamics is derived from prior information codified into an
appropriate statistical model. The basic assumption is that there is an
irreducible uncertainty in the location of particles so that the state of a
particle is defined by a probability distribution. The corresponding
configuration space is a statistical manifold the geometry of which is defined
by the information metric. The trajectory follows from a principle of
inference, the method of Maximum Entropy. No additional "physical" postulates
such as an equation of motion, or an action principle, nor the concepts of
momentum and of phase space, not even the notion of time, need to be
postulated. The resulting entropic dynamics reproduces the Newtonian dynamics
of any number of particles interacting among themselves and with external
fields. Both the mass of the particles and their interactions are explained as
a consequence of the underlying statistical manifold.

\bigskip

\begin{description}
\item[Second Contribution] \cite{caticha-cafaro}: \textit{The derivation of
the Newtonian dynamics from first principles of probable inference and
information geometric methods is another original contribution of my work in
collaboration with Prof. Ariel Caticha}.
\end{description}

\bigskip

Third, I extend my study of chaotic systems (information geometrodynamical
approach to chaos, IGAC) to an ED Gaussian model describing an arbitrary
system of $3N$ degrees of freedom. It is shown that the hyperbolicity of a
non-maximally symmetric $6N$-dimensional statistical manifold $\mathcal{M}%
_{\mathcal{S}}$ underlying the ED Gaussian model leads to linear
information-geometrodynamical entropy growth and to exponential divergence of
the Jacobi vector field intensity. As a special physical application, the
information geometrodynamical scheme is applied to investigate the chaotic
properties of a set of $n$-uncoupled three-dimensional anisotropic inverted
harmonic oscillators coupled to an internal environment and I show that the
asymptotic behavior of the information-geometrodynamical entropy is
characterized by linear growth. Finally, considerations concerning the
anisotropy of the statistical manifold underlying such physical system and its
relationship with the spectrum of frequencies of the oscillators are carried out.

\bigskip

\begin{description}
\item[Third Contribution] \cite{cafaro4}: \textit{I compute the asymptotic
temporal behavior of the information geometrodynamical entropy of a set of
}$n$\textit{-uncoupled three-dimensional anisotropic inverted harmonic
oscillators (IHOs) characterized by an Ohmic distributed frequency spectrum
and I suggest the classical information-geometric analogue of the Zurek-Paz
quantum chaos criterion in its classical reversible limit}.
\end{description}

\bigskip

I am aware that several points in my IGAC need deeper understanding and
analysis, however I hope that my work convincingly shows that this
information-geometric approach may be useful in providing a unifying criterion
of chaos, thus deserving further research and developments.

\section{Future Research Directions}

It is evident that my work requires additional improvements and deeper
understanding of some of its results in view of its comparison to more recent
results obtained in more orthodox approaches.

I am working on the possibility of extending the IGAC to chaotic systems with
an arbitrary number of interacting microscopic degrees of freedom. This would
lead to the problem of considering an information-geometric line element with
non trivial off-diagonal matrix elements. This is relevant because it allow us
to study not only macroscopic chaos but also microscopic dynamics
characterizing several known chaotic Newtonian dynamical systems. In addition,
this would increase the understanding of the relationship between the
microscopic and macroscopic behaviors of a physical system.

I am considering the IGAC arising from arbitrary (non-uniform) prior
probability distributions together with possible deviations from Gaussianity
and chaoticity arising from fluctuations of positively curved statistical
manifolds. I would like to take into consideration the study of more realistic
physical, biological, complex systems in general (brain dynamics, finance,
etc.). I am searching for a better understanding of the possible relevance of
my formalism to key concepts in modern quantum information theory: chaos,
decoherence and entanglement. I am thinking about the possibility of employing
my IGAC to describe situations of transitions from quantum to classical
chaotic physical systems.

These last two points require a generalization of the IGAC in an appropriate
manner to facilitate the study of quantum mechanical systems. In collaboration
with Dr. Saleem Ali, I already started some additional works in this
direction. We are trying to construct a quantum Hilbert space from a classical
curved statistical manifold. This extension is implemented via the
introduction of complex and symplectic structure tensors that are compatible
with the underlying Riemannian geometry induced by the Fisher-Rao information
metric. Vectors on the resulting manifold $\mathcal{H}$ are interpreted as
state vectors associated with points (i.e., probability density functions) on
$\mathcal{M}$.

As a final remark, I would like to point out that my primary concern in the
near future is refining some of my latest \emph{controversial} work on regular
and chaotic quantum energy level statistics and, possibly, extending the
application of the IGAC to soft chaos regimes \cite{cafaro5}.

\pagebreak

\begin{acknowledgments}
There are many people I wish to thank. Friends, relatives, colleagues,
professors and financial sponsors: I am grateful to you all. I am deeply
indebted to my supervisor, Prof. Ariel Caticha, for allowing me the freedom to
pursue my own interests. I am grateful to all the dissertation Committee
members: Prof. Ariel Caticha, Prof. Akira Inomata, Prof. John Kimball, Prof.
Kevin Knuth and Prof. Carlos Rodriguez. I thank Prof. Ali Mohammad-Djafari
(Paris, France), Prof. Kevin Knuth (Saratoga, USA), Prof. Stefano Mancini and
Prof. Fabio Marchesoni (Erice, Italy) for allowing me the opportunity to
present my work on chaos at their Conferences. I thank Prof. Carlo Bradaschia
and Dr. Giancarlo Cella (my former advisor and co-advisor at the University of
Pisa, Italy) for their interest in my current research and for having me back
there to present my works. I am grateful to Prof. Frederic Barbaresco (Thales
Air Defence, Surface Radar Business Line, France), Prof. Rafael Gutierrez
(Director of the Centre for Complex Systems, Columbia-South America), Prof.
Michael Frey (Department of Mathematics, Bucknell University, Pennsylvania-
USA) and Prof. Yasunori Nishimori (National Institute of Advanced Industrial
Science and Technology, Tsukaba-Japan) for their deep interest in my work
presented in this thesis. I thank the co-authors of some of my works, Dr.
Saleem Ali, Prof. Salvatore Capozziello, Dr. Christian Corda and Mr. Adom
Giffin. Finally, I wish to express my immense gratitude to the people who were
with me from the beginning of this long journey: my unconditioned sponsors, my
parents Peppino (from Bellona, Italy) and Carmelina (from Vitulazio, Italy)
and, my brother Joe.
\end{acknowledgments}

\pagebreak


\begin{thebibliography}{999}
\bibitem {note 3}There is a second solution to (\ref{minimi}), $\eta=(1-\xi)$.
It corresponds to minimizing rather than maximizing $S$.
\end{thebibliography}
\end{document}